\documentclass[%
    reprint,
    amsmath,
    amssymb,
    aps,
    pra,
    floatfix,
    natbib,
]{revtex4-2}

\usepackage[T1]{fontenc}

\usepackage{appendix}
\usepackage{multirow}
\usepackage{amsmath}
\usepackage{amssymb}
\usepackage{mathtools}

\usepackage{subcaption}
\usepackage{ragged2e} 

\usepackage{paralist}
\usepackage{booktabs}
\usepackage{adjustbox}

\usepackage[colorlinks]{hyperref}
\hypersetup{
    colorlinks=true,
    linkcolor=blue,
    citecolor=blue,
    urlcolor=blue,
}
\usepackage[capitalise]{cleveref} 
\usepackage{comment}

\makeatletter
\def\namelabel#1#2{\@bsphack
  \begingroup
  \UseHookWithArguments{label}{1}{#1}%
  \protected@write\@auxout{}%
         {\string\newlabel{#1}{{\@currentlabel}{\thepage}%
          {#2}{\@currentHref}{\@kernel@reserved@label@data}}}%
  \endgroup
  \@esphack}
\makeatother

\makeatletter
\let\frontmatter@abstractwidth\linewidth
\makeatother

\usepackage{color}
\usepackage{xcolor}
\usepackage{graphicx} 
\usepackage{qcircuit}
\usepackage{tikz}
\usetikzlibrary{external}
\usetikzlibrary{decorations.pathmorphing}
\tikzset{snake it/.style={decorate, decoration=snake}}
\usetikzlibrary{math}

\usepackage{placeins}

\definecolor{ibmOrange}{HTML}{FFB000}
\definecolor{ibmRed}{HTML}{FE6100}
\definecolor{ibmMagenta}{HTML}{DC267F}
\definecolor{ibmViolet}{HTML}{785EF0}
\definecolor{ibmBlue}{HTML}{648FFF}
\definecolor{armygreen}{HTML}{009F6B}

\newcommand{\mk}[1]{{\color{magenta}MK: #1}}

\newcommand{\evolutiontime}{t}

\newcommand{\nshots}{N_{s}}

\newcommand{\ntwirling}{N_{t}}

\newcommand{\neel}{N\'eel\ }
\newcommand{\mpsim}{Majorana propagation simulation }
\newcommand{\mpsims}{Majorana propagation simulations }

\usepackage{bbm}

\newcommand{\bra}[1]{\mathinner{\langle #1\rvert}}
\newcommand{\ket}[1]{{\lvert #1\rangle}}

\newcommand{\ketbra}[3][]{%
  \def\@tempa{#1}%
  \ifx\@tempa\@empty\relax%
    \mathinner{\lvert#2\rangle\langle #3\rvert}%
  \else%
    \mathinner{\lvert#2\rangle\langle #3\rvert}_{#1}%
  \fi}
\newcommand{\Ketbra}[3][]{%
  \def\@tempa{#1}%
  \ifx\@tempa\@empty\relax%
    \left|#2\middle>\middle<#3\right|%
  \else%
    \left|#2\middle>\middle<#3\right|_{#1}%
  \fi}

\newcommand{\braket}[2][]{%
  \Braket@syntax#2|\@nil[#1]{}{}{}{\langle}{\rangle}}
\newcommand\Braket[2][]{%
  \Braket@syntax#2|\@nil[#1]{\left}{\middle}{\right}{\langle}{\rangle}}
\newcommand\bigBraket[2][]{%
  \Braket@syntax#2|\@nil[#1]{\bigl}{\bigm}{\bigr}{\langle}{\rangle}}
\newcommand\BigBraket[2][]{%
  \Braket@syntax#2|\@nil[#1]{\Bigl}{\Bigm}{\Bigr}{\langle}{\rangle}}
\newcommand\biggBraket[2][]{%
  \Braket@syntax#2|\@nil[#1]{\biggl}{\biggm}{\biggr}{\langle}{\rangle}}
\newcommand\BiggBraket[2][]{%
  \Braket@syntax#2|\@nil[#1]{\Biggl}{\Biggm}{\Biggr}{\langle}{\rangle}}

\def\Braket@syntax#1|#2\@nil{%
  \def\@tempa{#2}%
  \ifx\@tempa\@empty%
    \def\@tempb{\Braket@args{#1}}%
  \else%
    \def\@tempb{\@Braket#1|#2\@nil}%
  \fi%
  \@tempb}

\def\Braket@args#1[#2]#3#4#5#6#7#8{%
  \Braket@two{#1}{#8}[#2]{#3}{#4}{#5}{#6}{#7}}

\def\@Braket#1|#2|#3\@nil{%
  \def\@tempa{#3}%
  \ifx\@tempa\@empty%
    \def\@tempb{\Braket@two{#1}{#2}}%
  \else%
    \def\@tempb{\@Braket@three{#1}{#2}#3\@nil}%
  \fi\@tempb}
\def\@Braket@three#1#2#3|\@nil{%
  \Braket@three{#1}{#2}{#3}}%

\def\Braket@two#1#2[#3]#4#5#6#7#8{
  \def\@tempa{#3}%
  \ifx\@tempa\@empty%
    #4#7#1#5|#2#6#8%
  \else%
    #4#7#1#5|#2#6#8_{\hspace{-0.1em}#3}%
  \fi}

\def\Braket@three#1#2#3[#4]#5#6#7#8#9{%
  \def\@tempa{#4}%
  \ifx\@tempa\@empty%
    #5#8#1\vphantom{#2#3}#7|#2%
    #5\vert#3\vphantom{#1#2}#7#9%
  \else%
    #5#8#1\vphantom{#2#3}#7|#2%
    #5\vert#3\vphantom{#1#2}#7#9_{\hspace{-0.1em}#4}%
  \fi}

\newcommand\braXket[4][]{\mathinner{\langle#2\vert#3\vert#4\rangle}_{#1}}
\newcommand\BraXket[4][]{%
  \Braket@three{#2}{#3}{#4}[#1]{\left}{\middle}{\right}{\langle}{\rangle}}
\newcommand\bigBraXket[4][]{%
  \Braket@three{#2}{#3}{#4}[#1]{\bigl}{\bigm}{\bigr}{\langle}{\rangle}}
\newcommand\BigBraXket[4][]{%
  \Braket@three{#2}{#3}{#4}[#1]{\Bigl}{\Bigm}{\Bigr}{\langle}{\rangle}}
\newcommand\biggBraXket[4][]{%
  \Braket@three{#2}{#3}{#4}[#1]{\biggl}{\biggm}{\biggr}{\langle}{\rangle}}
\newcommand\BiggBraXket[4][]{%
  \Braket@three{#2}{#3}{#4}[#1]{\Biggl}{\Biggm}{\Biggr}{\langle}{\rangle}}

\newcommand\braKet{%
  \message{\string\braKet\space is obsolete
    - use \string\braket\space or \string\braXket\space instead}
  \braXket}
\newcommand\BraKet{%
  \message{\string\BraKet\space is obsolete
    - use \string\Braket\space or \string\BraXket\space instead}
  \BraXket}

\crefname{section}{Appendix}{Appendices}

\makeatletter
\def\l@subsubsection#1#2{}
\makeatother

\let\doaddcontentsline\addcontentsline
\newcommand{\noaddcontentsline}[3]{}

\newcommand{\tocoff}{\let\addcontentsline\noaddcontentsline}
\newcommand{\tocon}{\let\addcontentsline\doaddcontentsline}

\newcommand{\notocline}[2]{%
  \let\addcontentsline\noaddcontentsline%
  #1{#2}%
  \let\addcontentsline\doaddcontentsline%
}

\tocoff

\begin{document}

\title{\texorpdfstring{Programmable digital quantum simulation of 2D Fermi-Hubbard dynamics\\ using 72 superconducting qubits}{Programmable Digital Quantum Simulation of 2D Fermi-Hubbard Dynamics using 72 Superconducting Qubits}}

\renewcommand{\andname}{with}

\author{Phasecraft}\email{info@phasecraft.io}
\author{Google collaborators{\hyperlink{author_dagger}{$^\dagger$}}}

\noaffiliation

\date{\today}

\begin{abstract}
{\bf
    Simulating the time-dynamics of quantum many-body systems was the original use of quantum computers proposed by Feynman~\cite{Feynman}, motivated by the critical role of quantum interactions between electrons in the properties of materials and molecules.
    Accurately simulating such systems remains one of the most promising applications of general-purpose digital quantum computers, in which all the parameters of the model can be programmed and any desired physical quantity output.
    However, performing such simulations on today's quantum computers at a scale beyond the reach of classical methods requires advances in the efficiency of simulation algorithms and error mitigation techniques.
    Here we demonstrate programmable digital quantum simulation of the dynamics of the 2D Fermi-Hubbard model -- one of the best-known simplified models of electrons in crystalline solids -- at a scale beyond exact classical state-vector simulation.
    We successfully implement simulations of this model on lattice sizes up to $\mathbf{6\times 6}$ using 72 qubits on Google's Willow quantum processor, across a range of physical parameters, including different on-site electron-electron interaction strengths and magnetic flux values,
    and study phenomena including formation of magnetic polarons~\cite{Alexandrov2009-mz} (charge carriers surrounded by local magnetic polarisation), dynamical symmetry-breaking in stripe-ordered states~\cite{Zaanen2001},
    attraction of charge carriers on an entangled background state known as a valence bond solid~\cite{ANDERSON1973}, and
    the approach to equilibrium through thermalisation~\cite{Rigol2008-pe}.
    We validate our results against exact calculations in parameter regimes where these are feasible, and compare them to approximate classical simulations performed using tensor network and operator propagation methods.
    Our results demonstrate that meaningful programmable digital quantum simulation of many-body interacting electron models is now feasible on state-of-the-art quantum hardware.
}
\end{abstract}
\maketitle

Accurate digital simulations of the properties of interacting fermions, modelling electrons in crystalline solids or in molecules, have the potential to accelerate the discovery of novel materials and chemicals on which modern technology such as batteries~\cite{NASAJPOURESFAHANI2024114783,D5YA00065C,booth_perspectives_2021, urban_computational_2016} and photovoltaics~\cite{park2015perovskite} depends. Critically, as in other areas where computer simulation has revolutionised the way we model and design physical artefacts, accurate digital quantum simulation promises the ability to rapidly model and calculate properties of a wide range of candidate materials or molecules, significantly faster and more flexibly than synthesising samples in the laboratory.

Simulating the time-dynamics of quantum systems evolving under a many-body Hamiltonian -- time-dynamics simulation for short -- is one of the most extensively studied families of quantum algorithms~\cite{Lloyd}, with a history dating back to the advent of the field~\cite{Feynman}, and many important algorithmic developments since~\cite{Gily_n_2019,Childs_2021}. It has been proven that time-dynamics simulation can be performed efficiently on a quantum computer, and that classical computers cannot perform this task efficiently~\cite{jordan2018bqp} -- barring the unlikely scenario that they are somehow able to simulate all quantum computers efficiently.

This is also borne out in practice. Whilst classical algorithms such as density functional theory, quantum Monte-Carlo and tensor network methods are often very effective at computing equilibrium properties of quantum many-body systems, simulating the late-time dynamics of such systems is more computationally challenging on classical computers.
Exact state-vector simulations -- wherein the full quantum state is represented and operated upon in memory -- are limited to small sizes. Meanwhile, approximate simulation methods such as tensor networks and operator-propagation methods can achieve larger scales, but their accuracy diminishes as entanglement and correlations build up in the model.
Quantum computers hold the promise of significantly outperforming classical methods, while still offering the versatility of programmable digital simulation.

\begin{figure*}[!htbp]
    \centering
    \includegraphics[width=\linewidth]{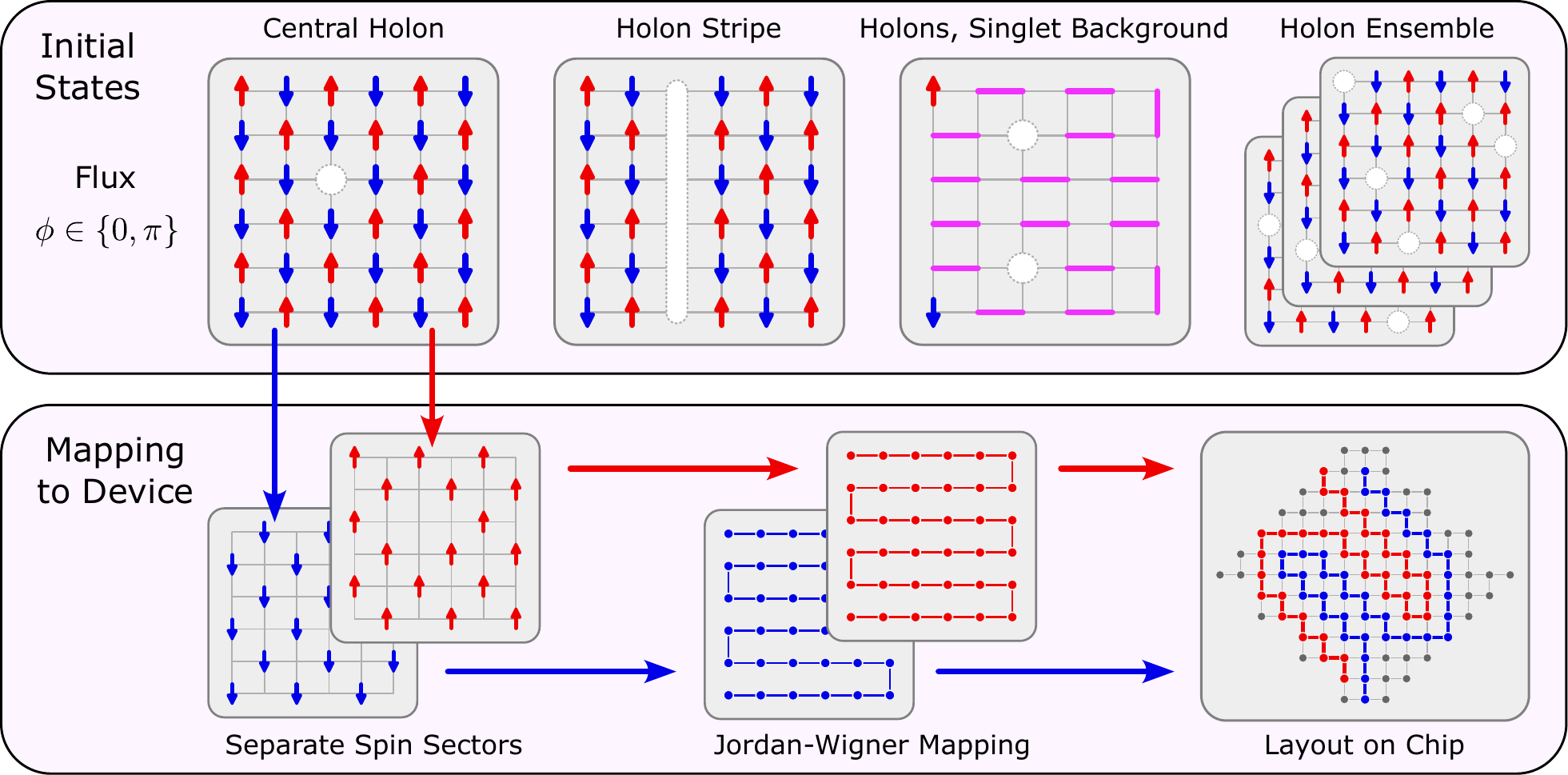}
    \caption{\justifying
    \textbf{Selection of initial states considered and their representation on hardware.} We use various initial states to probe different physical phenomena. A N\'eel-ordered background with one or two holons is used to explore magnetic polaron formation and holon attraction, respectively. The holon stripe is used to probe percolation. Two holons on a singlet background is used to investigate spin-charge separation. Finally, an ensemble of states with varying numbers of holons at random positions is used to explore thermalisation as a function of doping.
    Each spin sector of the model is separately mapped to a Jordan-Wigner representation in which qubits are ordered along a line. Time dynamics is simulated via fermionic swap networks along this line. The two representations are then arranged on the hardware in an interlocking pattern. We evolve up to time 1.2 in units of inverse hopping strength, and measure the time-evolved quantum state in the computational basis. We then compute expectation values of observables, and analyse signals based directly on the measured bit-strings. We perform approximate classical simulations to benchmark the results.
    }
    \label{fig:main_figure}
\end{figure*}

Modelling the electronic structure of periodic crystalline materials is one of the more promising near-term applications of digital quantum simulation~\cite{cao2019quantum,mcardle2020quantum,Magritte}. Due to symmetries and local structure, these systems can be compressed into comparatively small circuits~\cite{Magritte}. 
But time-dynamics simulation of real crystalline materials systems is still beyond the reach of current quantum computers.

However, the simulation of simplified theoretical models of electronic lattice systems, at scales beyond what is possible for exact state-vector simulation, 
is now within reach of state-of-the-art quantum hardware. Amongst the simplest and best-studied of these is the Fermi-Hubbard model, a single-band approximation of transition metals with $3d$ orbitals exhibiting local repulsion. Despite its apparent simplicity, this model has a complex phase diagram, non-trivial dynamics and rich physics phenomenology~\cite{FH_book}.

The Fermi-Hubbard Hamiltonian
\begin{align}\label{eq:FH_Ham_intro}
H =  -J \hspace{-.1cm}\sum_{\langle i,j\rangle,\sigma}(e^{i\phi_{i,j}}c_{i,\sigma}^\dagger c_{j,\sigma}+h.c.)+U\sum_in_{i,\uparrow}n_{i,\downarrow}
\end{align}
represents electrons (fermions) with spin $\sigma =\ \uparrow$ or $\downarrow$, hopping between neighbouring sites $i$ and $j$ on a 2D square lattice with hopping strength $-J$, and interacting locally with strength $U$ when two electrons meet at the same site. Here, $c_{i,\sigma}^\dagger$ and $c_{i,\sigma}$ are fermionic creation and annihilation operators for fermions with spin $\sigma$ at site $i$, and the local density operator $n_{i,\sigma} = c_{i,\sigma}^\dagger c_{i,\sigma}$ counts the number of such fermions. The Peierls phases $\phi_{i,j}$ model an external magnetic flux through the system, and we choose them in a translationally invariant pattern (see \cref{sec:model_details,sec:circuits}). As only the ratio $U/J$ matters, we use the standard convention of setting $J=1$. We use natural units where Planck's constant $\hbar=1$ throughout.

In this work we realise a fully programmable digital quantum simulation of the time dynamics of the 2D Fermi-Hubbard model on a square lattice, using Google's Willow superconducting quantum processor~\cite{google2025quantum}, at a scale that is beyond direct classical state-vector simulation. 
We use this digital simulator to explore several aspects of Fermi-Hubbard dynamics on a 2D square lattice. Specifically, we consider Fermi-Hubbard lattice sizes up to $6\times 6$ sites, requiring a total of up to 72 qubits.
Using different initial states (Fig.~\ref{fig:main_figure}), we study the formation of magnetic polarons emerging from a single holon  (a quasiparticle indicating absence of electrons on a site)  on a chequerboard pattern of up and down spins, known as a N\'{e}el state; the dynamical breaking of bipartite sublattice symmetry on a hole-doped state;
the confinement of charge carriers in a valence bond solid state, a state where neighbouring spins combine into spin-zero objects (singlets);
and the thermalisation of the system as a function of interaction strength and of hole doping.
We explore these dynamical phenomena over a range of interaction strengths, and for systems with zero and $\pi$ magnetic flux per plaquette. (See below and Fig.~\ref{fig:phys_headline_1} for a discussion of the physics simulation results.)

We use a second-order Trotter algorithm~\cite{Childs_2021} with up to 3 Trotter steps (with a similar structure to 6 first-order steps, but with better accuracy than directly applying first-order bounds would suggest) to produce a discretised approximation to the continuous-time dynamics under the Fermi-Hubbard Hamiltonian. (The approximation error from this discretisation is commonly referred to as the Trotter error; a discussion of this can be found in \cref{sec:trotter_error}.) We are able to resolve a signal from such quantum circuits with 4372 two-qubit gates in total and circuit depths of up to 294 for the $6\times6$ lattice, allowing time-dynamics simulation out to time $1.2$ (in natural units of inverse hopping strength).
We illustrate the layout of the experiment on the device in Fig.~\ref{fig:main_figure}.

Such deep circuits are made possible by two complementary technical developments.
First, continued improvements in the performance and scale of superconducting quantum chips~\cite{google2025quantum}, combined with their high clock speed.
Second, efficient design of simulation algorithms, circuit implementations, and error mitigation techniques tailored to the target application and hardware.
Specifically, we achieve significant constant-factor efficiency savings over standard implementations of fermionic quantum simulations by tailoring the fermion representation to the qubit layout on the device, and using gate implementations of the Fermi-Hubbard Hamiltonian terms that are also optimised for this.
(See \nameref{sec:methods} and~\cref{sec:circuits} for details.)

We also apply training by fermionic linear optics~(TFLO)~\cite{montanaro2021}, a training-based error mitigation method which exploits the fact that interacting fermion models such as Fermi-Hubbard become classically exactly solvable when $U=0$, whilst the quantum circuit structure -- and hence the hardware noise -- changes very modestly as the parameter $U$ is varied, only requiring modifying the rotation angles of certain single-qubit gates.
This error mitigation partially corrects systematic components of the effects of noise on few-body observables. For quantities that cannot be reduced to few-body observables, we also implement a novel TFLO-based maximum entropy shot-reweighting method to correct for noise directly on the bitstrings (``shots'') sampled from the hardware. We combine these with a suite of other standard methods, including noise averaging and suppression, post-selection on fermion-number conservation, symmetry averaging, Gaussian process regression (GPR) on time series data, and more. (See \nameref{sec:methods} for an overview, and \cref{sec:error_mitigation} for extensive discussion and performance comparison of different noise mitigation methods.)

To the best of our knowledge, all previous digital quantum simulations of the dynamics of the spinful Fermi-Hubbard model on quantum computers are either 1D instances~\cite{bespalova2025simulating,arute20,Vilchez_Estevez_2025,chowdhury2025} (which are tractable by tensor network methods)
or relatively small 2D instances~\cite{evered25, H_mery_2024} which are amenable to exact state-vector simulation.

A primary motivation for simulating the Fermi-Hubbard model on quantum hardware is that, despite the importance of the model, classical methods still struggle with this task. Although significant progress has been made in recent decades in approximate simulation of equilibrium properties, such as ground states and phase diagrams -- with modern supercomputers enabling good numerical results on lattices up to $60\times 60$~\cite{jiang2024ground,sun2025boosting} -- late-time non-equilibrium and dynamical properties of the Fermi-Hubbard model remain challenging to simulate on classical computers.

Exact state-vector simulations are limited by available memory, with the largest to date able to simulating gate operations on at most $45$ qubits~\cite{H_ner_2017, amazon44}.
On the Fermi-Hubbard model, this corresponds to systems slightly larger than $4\times5$.
Approximate classical computations can be performed at larger lattice sizes. For example, tensor-network-based simulations of Fermi-Hubbard dynamics, which approximate the quantum state of the system via classically efficient representations, have been demonstrated to approximate the dynamics of $6\times 5$ systems for single-site observables out to time $1$~\cite{thompson2025}.
Meanwhile, Majorana propagation methods can compute expectation values of time-evolved few-body observables under complex quantum circuits~\cite{miller2025}. However, as these methods involve approximations, just as for quantum computation their accuracy must be validated separately. Comparisons between the leading approximate classical methods and the quantum computation are described below. A more extensive discussion can be found in the appendices.

Analogue quantum simulation is an alternative quantum simulation method, where the many-body quantum Hamiltonian of interest is engineered directly into the interactions between e.g.\ cold atoms trapped in an optical lattice~\cite{browaeys2020many}, or between the ions in an ion trap~\cite{monroe2021programmable}. This has been performed successfully at scale for multiple decades now.
These approaches can implement the Fermi-Hubbard model on thousands of sites~\cite{shao2024antiferromagnetic, hartke2023direct}, well beyond the number of qubits currently available on state-of-the-art digital quantum computers,
and recent implementations have been able to simulate dynamical properties as well as equilibrium quantities~\cite{Xu2025,Guardado2020Subdiffusion, Brown2019Bad,Nichols2019Spin, Bakr_2025_microscopy, Bohrdt_2021_expolaration}.
However, pure analogue simulation approaches have drawbacks compared to digital simulation. In particular, the fact that (by definition) arbitrary quantum circuits cannot be implemented in purely analogue quantum simulators limits somewhat which states the simulation can be initialised in, and also restricts which observables can be accessed at the end of the simulation.
Hybrid analogue-digital simulations afford more flexibility and allow a similar range of initial states to be prepared as digital quantum circuits, as demonstrated e.g.\ in work by Google~\cite{andersen2024}.
However, unlike some analogue simulations which have physical fermions in the simulator itself, hybrid analogue-digital simulations on Google's superconducting hardware are currently restricted to using the underlying bosonic degrees of freedom.

Digital quantum simulation allows circuits that prepare different initial states, and circuits that extract different output observables, to be flexibly programmed in and run at will. This allows a rich variety of regimes and physics to be simulated, as the range of results discussed in the following section attests to.

\section*{Physics simulation results}

\begin{figure*}[ht!]
    \centering
    \includegraphics{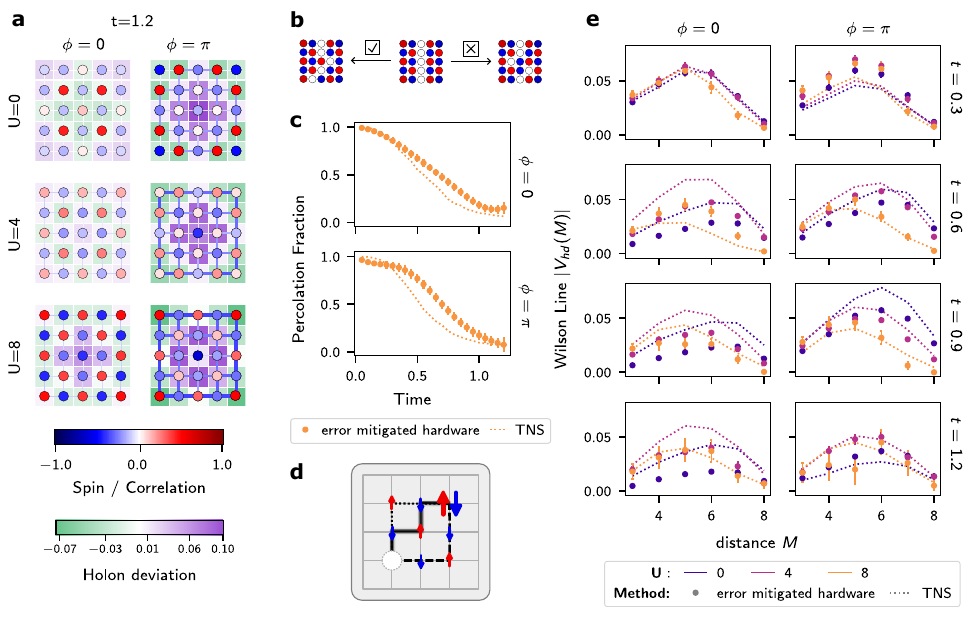}
    \caption{\justifying
    \textbf{Fermi-Hubbard dynamics phenomenology explored at the $5\times 5$ scale.}
    \textbf{a)}~Deviation from mean holon density (squares in the background), local spin (circles in the foreground) and nearest-neighbour spin correlations (links).
    We observe that a localisation of the excess holon density (purple squares) correlates with the strongest development of nearest-neighbour correlations as the interaction and flux increase. Note that for non-zero flux and on-site interaction strength $U$, a local spin moment is trapped in the middle, where a lack of neighbour correlations is observed.
    \textbf{b)}~Illustration of paths used to compute the percolation fraction.
    \textbf{c)}~Probability as a function of time for a stripe of holons to connect two opposite sides of the lattice, computed by: post-selecting on shots with less than four doublons and with negative spin-spin correlation across sites, and applying error-mitigation.
     While mean-field approaches~\cite{Zaanen2001} assume that the main dynamics of a stripe is to oscillate around its initial position, we observe that the stripe quickly breaks apart, even for the largest interaction strengths considered ($U=8$).
     The largest tensor network simulation that we considered ($\chi=2048$) agrees with this result.
     \textbf{d)}~Estimating a Wilson line between a holon (open circle) and a doublon (site with both up and down arrows): this is computed by finding shots with a holon and a doublon at Manhattan distance $M$, and computing $\sum_{m,n\in\gamma} S_n^z S^z_m$ for consecutive sites along a path~$\gamma$. We sum over all possible paths of the same length. Here we show three paths for $M=4$ as dotted, dashed and bold lines.
     \textbf{e)}~Time-snapshots of the Wilson line between a holon and a doublon as a function of their separation, generated by the spin background.
     We observe that the potential decreases as a function of time and increases as a function of distance up to a separation~$\sim 6$, with the experimental results showing the strongest suppression at zero interaction and flux, as expected for the non-interacting system.
     The classical simulations seem to converge to a different value.
     The decrease of the Wilson line for larger distances appears to be a finite-size effect, as going to even larger system sizes we see a movement of the peak to larger separations (see \cref{sec:spin_charge_u1}).}
    \label{fig:phys_headline_1}
\end{figure*}

The programmability of digital quantum simulations on general-purpose quantum computing hardware, combined with the high gate speed of superconducting qubits, enables iterating over a wide range of physically interesting scenarios with minimal modifications to the circuits.

The hole-doped Fermi-Hubbard model at intermediate coupling strengths hosts a variety of complex quantum phases and phenomena that may be observed in real materials. Dynamical effects can shed light on the mechanisms underlying these phenomena~\cite{fava2024magnetic}.
We are therefore interested in the non-equilibrium dynamics of the Fermi-Hubbard model from physically relevant initial states that can be prepared with low-depth quantum circuits.
In the limit of very large interaction strength and close to half filling, the presence of doubly-occupied sites (doublons) is energetically penalised. At exactly half filling, and $U\gg |J|$,
the Hamiltonian (\cref{eq:FH_Ham_intro}) is well approximated by the antiferromagnetic Heisenberg Hamiltonian~\cite{FH_book}
\begin{align}
H_{\rm Heis}=\frac{4J^2}{U}\sum_{\langle i,j\rangle}\left(\mathbf{S}_i\cdot \mathbf{S}_j-\frac{1}{4}\right),
\end{align}
where $\mathbf{S}_j$ is the spin operator at site $j$, so low-energy states favour anti-correlation of neighbouring spins. For this reason, we choose the N\'{e}el state as one of the initial states, and the valence-bond solid as another: an entangled initial state consisting of a singlet covering of the lattice. With these motivations in mind, our experiments are performed at small hole-doping on the $L_x\times L_y$ lattice, where we perturb the initial state by removing particles. We consider on-site interaction strengths $U=0,4,8$, and several initial states with different hole-dopings ranging from 1~to $\min(L_x,L_y)$.
Each initial state has a given number of fermions $N_e$ and total spin $S^z=0,1$, depending on the parity of $L_x\times L_y -N_e$. The initial state configuration corresponds to either a N\'{e}el background with holons placed at various locations, or a singlet covering with two holons. All these states have an energy density corresponding to a temperature $T\sim J$~\cite{LeBlanc2013} (see \cref{sec:model_details}).
Finally, for some of these configurations, we study the evolution of the Fermi-Hubbard model with both zero and $\pi$ flux per plaquette. While the simulation of the Fermi-Hubbard model with and without magnetic flux differs at the circuit level just by some single-qubit gate parameters, the physics of these two models differs dramatically.
At half filling and without interactions, the model with zero flux is a typical metal at low energies, while the $\pi$ flux system is a Dirac semimetal~\cite{Young_2015}.

Away from half filling, the interplay between spin and charge degrees of freedom in the dynamics of the system is not well understood as the interaction strength increases. We study the formation and dependence on flux of a magnetic polaron~\cite{Alexandrov2009-mz}, a quasiparticle  -- previously observed in analogue simulations~\cite{Koepsell2019, Koepsell_2021, ji2021coupling} -- believed to be intimately linked to the mechanism of high-temperature superconductivity in cuprates~\cite{MIHAILOVIC2023}, by evolving a system with initially one holon in the middle of an otherwise N\'{e}el-ordered state.
As shown in Fig.~\ref{fig:phys_headline_1}a, correlations between nearest neighbours build up over time in the $\pi$ flux system, compared with the zero flux case. Around the original position of the holon, the magnetic order is suppressed, and the local spin moment is trapped.
This effect is further explored in  in \cref{sec:single_holon}.

As the electron density is reduced from half filling, several phases are expected to appear as a function of (hole) doping at low energy in the Fermi-Hubbard model~\cite{bennemann2014novel}. In particular, for hole-doping $n_h\sim 1/8$, stripe order is expected to occur~\cite{Zheng2017,Liu_2025}. We introduce a stripe of holons across the N\'{e}el ordered background and investigate the stability of this holon stripe over time. In the high interaction strength regime, the mean-field description of the stripe evolution hinges on the existence of bipartite lattice symmetry \cite{Zaanen2001}. We study how this symmetry breaks dynamically due to the proliferation of doublon-holon pairs.
Starting from a stripe of holons, we study the sampled configurations as a function of time, and count the number of shots with a path connecting opposite ends (see \cref{fig:phys_headline_1} b).  Holon-doublon pair proliferation has two mechanisms that affect the survival of bipartite lattice symmetry. Firstly, the emergence of more holons increases the probability of seeing a holon stripe, but is likely to break the bipartite symmetry. Secondly, a more subtle effect is due to the motion of these charge carriers. They disturb the spin background leading to a melting of anti-ferromagnetic order. Therefore, to counteract these two effects and increase the likelihood of conserving the bipartite symmetry we post-select on shots which have three or fewer doublons and negative spin-spin correlation. We observe that, even within this subset, the bipartite symmetry is broken. This is a consequence of virtual holon-doublon pairs that appear, evolve and disappear in the system. This phenomenon seems insensitive to the flux (zero or $\pi$) considered. Several other quantities related to this effect are discussed in \cref{sec:stripe}.

We also consider an initial entangled state, corresponding to an almost columnar valence-bond solid where the electrons between some edges in the lattice form singlets. This is motivated by the expectation that quantum fluctuations can prevent the development of magnetic long-range order, so $\langle S^z_i \rangle=0$~\cite{Senthil_2004,Senthil_2004b,Wang2015-ky}. These states on the square lattice spontaneously break translational invariance. In our setting, we choose the arrangement of singlets in a way that minimises the circuit cost whilst retaining a columnar order in the bulk. The configuration of singlets is shown in Fig.~\ref{fig:main_figure}, bottom left, and its circuit construction is discussed in \cref{sec:singlets}.
Trivially, in these states, the spin and charge propagation differ as the charge evolves to the homogeneous state, but the local spin expectation vanishes.  We study the evolution of the potential between holons and doublon pairs at a (Manhattan) distance $M$ mediated by the spinons (fermion quasiparticles carrying spin but no charge).
This potential appears naturally in a description of the Fermi-Hubbard model in terms of holons, doublons and spinons (see \cref{sec:spin_charge_u1}) in terms of extended objects known as Wilson lines \cite{Wilson1974} (see \cref{fig:phys_headline_1} d).
As we observe for a $5\times 5$ system in Fig.~\ref{fig:phys_headline_1}e), the expectation of the potential is roughly independent of the interactions at early times, for both the zero- and $\pi$ flux systems, separating for longer times into increasing potentials at longer distances, with a smaller potential appearing in the non-interacting regime in the zero flux case. For $\pi$ flux, we do not see a significant decrease of the potential at longer times, even for zero interactions. The decrease in the magnitude of the potential for the biggest distances can be attributed to finite-size effects, as -- going to larger system sizes -- the peak of the curve moves to larger distances. Note that these predictions differ from the ones obtained through classical simulation. This signal (e.g. $t=1.2$ in panel \cref{fig:phys_headline_1}e)) indicates that the quantum device is predicting that separating a doublon from a holon requires little energy when the system is non-interacting ($U=0$) and without flux, while requiring more energy in the interacting case (yellow and purple markers). Tensor network simulation of this predicts the same trends, but quantitatively different signals.
A full discussion of this appears in \cref{sec:spin_charge_u1}.

\begin{figure*}[!htbp]
    \centering
    \includegraphics[width=\linewidth]{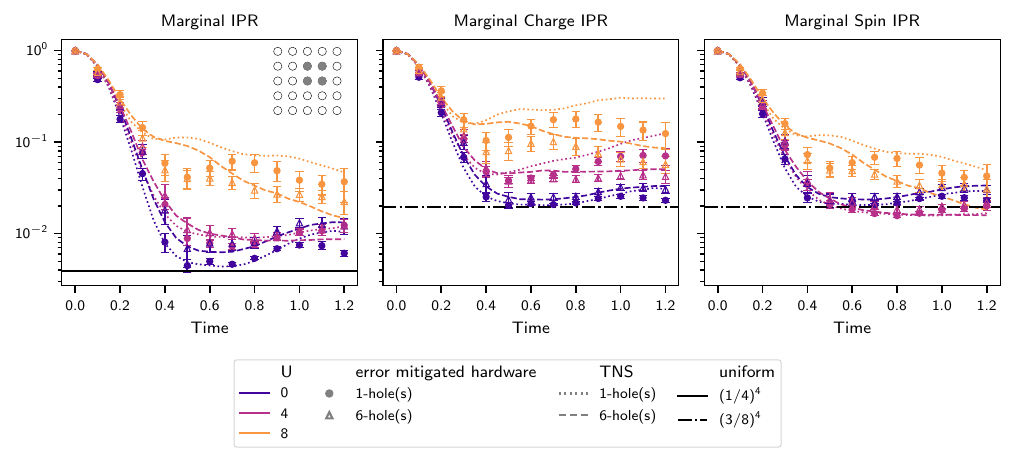}
    \caption{\justifying
    \textbf{IPR results.} Full, charge, and spin versions of the marginal inverse participation ratio (IPR) for an $m=4$-site region of the $5\times5$ system, constructed from different local basis states.
     The ``full'' marginal IPR uses the bitstrings directly.
    The charge IPR is obtained by grouping states with definite charge, while the spin IPR uses a basis grouped by definite spin labels (see \cref{sec:ipr} for a full discussion).
    Results are shown for initial states with one and with six randomly placed holes on top of a N\'eel background, each averaged over five configurations.
    The marginal IPR quantifies the delocalisation of the wavefunction in the corresponding local basis: it is equal to~$1$ for a state localised on a single basis configuration, and it becomes exponentially small in the marginal size (here $m=4$) for a delocalised state.
    For an on-site uniform distribution, the corresponding baseline values are $(1/4)^m$ for the full IPR and $(3/8)^m$ for the charge and spin IPRs. We use the convergence to the estimate on the uniform distribution as a measure of thermalisation. In all cases the marginal IPR decays in time, and this decay becomes slower as $U$ increases.
    The spin IPR approaches its baseline value more rapidly than the charge IPR, indicating faster local equilibration of the spin sector compared to the charge sector.}
    \label{fig:main_ipr}
\end{figure*}

As is clear from the experimental results, the system is far from equilibrated. We study the approach to equilibration by considering the wavefunction's spread in the computational basis, using the inverse participation ratio (IPR) on a four-qubit subsystem, defined as ${\rm IPR}=\sum_ip_i^2$ where $p_i$ is the probability of observing the bitstring $i$. In \cref{fig:main_ipr} we show the results for the evolution of the IPR considering the bit-string, charge and spin basis, for different interaction strengths and hole-dopings. For a fully delocalised state, corresponding to an equal superposition over all possible basis states, the IPR becomes $1/D$ where $D$ is the Hilbert space dimension. On the other hand, for a state with support on a single basis state the IPR takes the value~1. In \cref{fig:main_ipr} we also show the IPR estimate corresponding to the uniform distributions in the bit-string, charge and spin bases. We observe that while the IPR is not qualitatively sensitive to the doping, it is strongly dependent on the interaction. For larger interaction strengths, the thermalisation -- as measured by the proximity to the value on the uniform distribution -- slows down. Among the three different bases that we considered, the marginal spin IPR appears to thermalise faster, something that also is captured by the classical methods.
Further results on the different subsystems' IPRs that we consider, in terms of charge, spin and occupation and for different sizes, appear in \cref{subsec:app:thermalisation}.

Finally, we explore the evolution of the mean distance between two holons placed in a N\'{e}el ordered state.
As with the other input states, the dynamics induce holon-doublon pair proliferation. We find that, while the average distance of doublons rapidly converges to the value expected by homogeneous random placement, the distance between holons shows a non-trivial dependence on the interactions. This is discussed in detail in \cref{sec:holon_pair}.

\section*{Validation of the simulation results}

\begin{figure*}[ht!]
    \centering
    \includegraphics{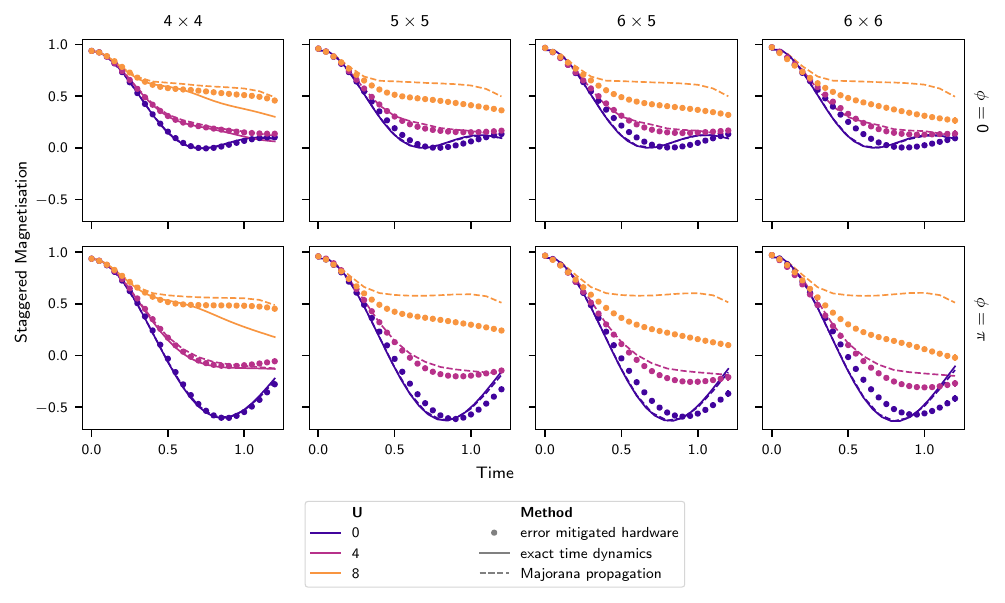}
    \caption{\justifying
    \textbf{Dynamics of the staggered magnetisation at increasing scale.}
    The staggered magnetisation is $M_s = \frac{1}{L_xL_y}\sum_i (-1)^i S^z_i$ where $S^z_j$ is the spin in the $z$ direction at site $j$, and $L_x$, $L_y$ are the system dimensions, ranging from $4\times 4$ to $6 \times 6$, with and without a magnetic flux $\phi$ passing through the plane of the model.
    Different interaction strengths $U$ are indicated by colour.
    Error-mitigated hardware data is given by points, with solid curves corresponding to smoothed fits.
    For the $4\times 4$ case, exact state vector simulations are possible at all values of $U$, indicated by the solid lines in that panel.
    For larger lattice sizes, exact classical simulations computed by FLO are shown with dashed lines for the non-interacting case ($U=0$), and Majorana propagation simulations are shown with dashed lines. At $U=0$ the Majorana propagation simulations are expected to be accurate simulations of the ideal quantum circuit output, so differences to the exact time dynamics are indicative of Trotter error. Similar plots illustrating performance with scaling can be found in \cref{sec:further_hw_plots}.
}
    \label{fig:staggered_magnetisation_scales}
\end{figure*}

There is an inherent challenge in validating the output of quantum simulations in regimes that cannot be computed exactly classically.
A striking benefit of simulating fermionic systems, such as the Fermi-Hubbard model, is that they naturally afford access to a parameter regime that can be verified classically; but at the same time, changing only single-qubit gate parameters in the quantum circuit gives access to computations in a regime that is classically hard to access. Specifically, for $U=0$ (but not for $U\neq 0$), the exact dynamics on many initial states can be computed efficiently on a classical computer using Fermionic Linear Optics (FLO)~\cite{terhal2002} for all system sizes. Indeed, for $U=0$ this method allows us to efficiently compute both the true, continuous-time evolution under the Fermi-Hubbard Hamiltonian, and to exactly simulate the quantum circuit (which is a discretised approximation to the continuous-time dynamics). Meanwhile, the difference between the time-dynamics simulation circuit with $U=0$ and the circuit with $U\neq 0$ is just a change to the rotation angles of a set of single-qubit gates. This minor change is not expected to have a large impact on the performance of these gates. It is therefore reasonable to expect that the variation of the interaction strength $U$ in the Fermi-Hubbard simulation does not impact the performance of the underlying digital simulation circuit in a substantial way. Thus, we can build evidence of the correct performance of the quantum computation in the $U=0$ regime, whilst being able to access the harder-to-verify $U\neq 0$ regime via a minor change to the circuit parameters.
(It should be noted, however, that the TFLO error mitigation used to attenuate the systematic components of device noise has been trained on $U=0$ data, so care is required in interpreting the experimental fit to the $U=0$ curves as experimental validation. See \nameref{sec:methods}.)

We find that, up to lattices of size $6\times 6$ (72 qubits), the evolution of a number of observables can be accurately retrieved by the quantum computer. We have validated our results against exact dynamics, where applicable, as well as both approximate simulations of our circuits with Majorana propagation and approximate simulations of the continuous-time Hamiltonian dynamics with Matrix Product States evolved with the time-dependent variational principle algorithm (TDVP)\cite{haegeman2011time,Haegeman_2016_unifying}.

For larger lattice sizes where exact state-vector simulation is prohibitive, we can only compare with exact calculations for $U=0$. Changing $U$ does induce a substantial change in the dynamics, which qualitatively matches what is expected on physical grounds (see physics results discussion, above). For $U>0$ we can still compare to Majorana propagation calculations, which are expected to be accurate for quantities such as the staggered magnetisation which depend only on 1-body observables. But as an approximate method its correctness cannot be guaranteed in general.

In Fig.~\ref{fig:staggered_magnetisation_scales} we show an example of the evolution of staggered magnetisation, a quantity which can be compared against exact state-vector simulation of the continuous-time dynamics for $4\times4$ and against Majorana propagation of the target circuit, as well as FLO simulations for $U=0$. For $U=0$ and~$4$, where the Trotter error is expected to be smallest, we see close numerical agreement between Majorana propagation and hardware data accross all system sizes, and with exact dynamics where available. For $U=8$, we see a deviation between hardware and Majorana propagation at late times for all system sizes, and also deviation from the exact continuous-time dynamics at late times for~$4 \times 4$ -- indicating that both Trotter error and hardware noise are playing a role in this case.

The quantum circuit depth is kept identical across all times $t>0.2$ and for values of $U$ at a given system size, so the underlying hardware noise is not expected to differ significantly at each fixed size across all the majority of the data shown. However, it is likely that the TFLO noise-mitigation method becomes less effective at values of $U$ far from~0, 
which may account for the larger numerical deviations for the staggered magnetisation  at $U=8$. See \cref{sec:physics_results} for an extensive comparison across a range of other physical quantities.

In addition, we use cross-entropy benchmarking for statistical validation of the correctness of the quantum hardware. Our results (\cref{sec:cross-entropy}) indicate that error-mitigated hardware samples achieve a linear XEB fidelity of at least 1\% with the ideal state at late evolution times, for a range of initial states and system sizes up to the largest circuits with 72~qubits.

\section*{Benchmarking against approximate classical simulation methods}

Quantum circuits on 72~qubits are currently well beyond the largest exact state-vector simulations. Nevertheless, approximate classical methods exist that are able to approximately simulate the quantum dynamics of significantly larger systems.
In particular, state-of-the-art tensor network methods can accurately approximate the time-evolved state of many-body quantum systems on hundreds of qubits~\cite{berezutskii2025tensor, patra2024efficient}, as long as quantum correlations and entanglement do not build up too rapidly as the state evolves.
Amongst tensor network methods, those based on matrix product states evolved in time with the TDVP algorithm have been found previously to perform particularly well for approximately simulating the exact time evolution of many-body systems~\cite{hemery2019matrix, leviatan2017}, and in parallel work targetting the Fermi-Hubbard model specifically~\cite{alam2025fermionic} we also found that TDVP performed the most reliably. We therefore take TDVP~\cite{tenpy2024} here as the main tensor-network comparator. (See \cref{sec:tensor_networks} and~\cite{alam2025fermionic} for more extensive discussion and comparison with other tensor network methods.)

Complementing these tensor network approaches, operator propagation methods such as Pauli propagation~\cite{rall2019simulation} and Majorana propagation~\cite{miller2025} are capable of approximating expectation values of time-evolved observables even when the state is highly entangled, as long as the time-evolved observable can be expressed as a sum of a small number of terms. Majorana propagation performs particularly well in simulating layers of fermionic Trotter dynamics~\cite{miller2025}.
Therefore, we have also implemented Majorana propagation computations explicitly targeting our Trotterised quantum circuits, which approximate the continuous-time dynamics of the system.
(See \cref{sec:majorana_propagation} for more detailed discussion of Majorana propagation computations.)

To benchmark our digital quantum simulations against leading classical methods, we carried out extensive classical simulations using TDVP and (where applicable) Majorana propagation to mirror every configuration of the Fermi-Hubbard model and all data taken on the quantum simulator.
These classical computations took over $100$ CPU years on the Google Cloud Platform and generated over 25TB of data.

Quantities that are derived from expectation values of small numbers of few-body observables are expected to be reproduced accurately by both Majorana propagation and by tensor network methods -- including many of the quantities discussed in the physics results above. This is borne out in the results, where we see (e.g.\ in Figs.~\ref{fig:phys_headline_1} and~\ref{fig:staggered_magnetisation_scales}) that the classical simulations and error-mitigated quantum simulations are in qualitative -- and in many cases quantitative -- agreement with one another and with the exact dynamics at $U=0$.
Indeed, because the quantum simulation is subject to hardware noise and errors, whereas the classical approximate computations are not, we see that the classical methods often achieve better agreement with the exact $U=0$ calculations than the quantum simulation for such quantities. On the other hand, for more complex many-body quantities (see e.g.\ Fig.~\ref{fig:phys_headline_1}e and \cref{sec:spin_charge_u1}), the classical approximate methods do not always achieve as good quantitative agreement with the exact $U=0$ dynamics as the hardware; it is therefore unclear which method is more accurate for $U\neq0$ where no comparison to exact results is possible.



\section*{Conclusions}



In this work, we have presented the first successful digital quantum simulation of the seminal spinful 2D Fermi-Hubbard model on lattice sizes up to $6\times 6$ (72 qubits). This is enabled by efficient algorithm implementations that minimize circuit depth and gate count for the target hardware and application; and by novel error mitigation methods that train on non-interacting fermion models where exact classical simulations are possible.


The wide range of Hubbard dynamics phenomenology we investigate is also accessible via modern, heuristic classical computational approaches, including tensor network and operator propagation methods.
We have benchmarked the quantum simulation against these classical methods by running medium- and large-scale numerical computations of the quantities studied in this work.
In some cases, the classical methods do not match the results obtained on the quantum hardware.
We do, however, observe that the quality of the quantum simulation gradually degrades with system size and quantum gate count, as expected on pre-fault-tolerant quantum hardware.

These results demonstrate that meaningful, programmable digital quantum simulations of many-body interacting electron models can now be performed on state-of-the-art quantum hardware using state-of-the-art algorithmic techniques, heralding the era of general-purpose digital quantum physics simulation.
Further algorithmic and implementation improvements are feasible. In particular, exploiting continuously tunable gate parameters would give a significant additional reduction in circuit depth, and further improvements can be made in particular to the fermionic training-based noise mitigation methods that may enhance their performance.
As quantum hardware and algorithmic implementations continue to improve, we expect more complex electronic structure models related to periodic crystalline materials to come within reach.

\section*{Data availability}
Raw data were taken from Google's Willow quantum processors. 
Those data are available at \cite{zenodo_dataset}, along with expectation values of $Z$ and $ZZ$ observables, computed from shots as well as using other numerical techniques.

\section*{Acknowledgements}
The hardware data presented in this work was taken remotely on 105-qubit and 72-qubit processors, with access provided via Google’s Quantum Engine. Calibration and support were provided by the Quantum Hardware Residency Program. We thank the Google Quantum AI team for providing the quantum systems and support that enabled these results.
The views expressed in this work are solely those of the authors and do not reflect the policy of Google or the Google Quantum AI team.
We thank Zlatko Minev for advice on PEA.  We also thank Andrew Childs for discussions on time-dynamic simulations.
This work received funding from the European Research Council (ERC) under the European Union's Horizon 2020 research and innovation programme (grant agreement No.\ 817581).

\clearpage
\onecolumngrid
{
\setlength{\parindent}{0pt}

\hypertarget{author_dagger}{
\noindent {$^\dagger$} \textbf{Authors}\\
}

\noindent \textbf{Phasecraft} \\
Faisal Alam$^1$, 
Jan Lukas Bosse$^1$, 
Ieva \v{C}epait\.{e}$^1$, 
Adrian Chapman$^2$, 
Laura Clinton$^1$, 
Marcos Crichigno$^2$, 
Elizabeth Crosson$^2$, 
Toby Cubitt$^{1,3}$, 
Charles Derby$^1$, 
Oliver Dowinton$^1$, 
Norhan Eassa$^{2,9}$, 
Paul K. Faehrmann$^{1,4}$, 
Steve Flammia$^{2,5}$, 
Brian Flynn$^1$, 
Filippo Maria Gambetta$^1$, 
Ra\'ul Garc\'ia-Patr\'on$^{1,6}$, 
Max Hunter Gordon$^1$, 
Glenn Jones$^1$, 
Abhishek Khedkar$^1$, 
Joel Klassen$^1$, 
Michael Kreshchuk$^2$, 
Edward Harry McMullan$^1$, 
Lana Mineh$^1$, 
Ashley Montanaro$^{1,7}$, 
Caterina Mora$^1$, 
John J. L. Morton$^{1,8}$, 
Alberto Nocera$^1$,
Dhrumil Patel$^{2,5}$, 
Pete J. Rolph$^1$, 
Raul A. Santos$^1$, 
James R. Seddon$^1$, 
Evan Sheridan$^1$, 
Wilfrid Somogyi$^1$, 
Marika Svensson$^1$, 
Niam Vaishnav$^1$, 
Sabrina Yue Wang$^1$, 
Gethin Wright$^1$\\
\\
\textbf{Google collaborators} \\
Dmitry Abanin$^9$, 
Mohammed Alghadeer$^9$, 
Jaehong Choi$^9$, 
Tyler Cochran$^9$, 
Gaurav Gyawali$^9$, 
Shashwat Kumar$^9$, 
Ricky Oliver$^9$, 
Eliott Rosenberg$^9$, 
Pedram Roushan$^9$\\

\noindent \textbf{Author contributions} \newline
JK, RAS, TC conceived the project and provided overall leadership and oversight. JK, MHG, CD, JLB, EHM, RAS, NV designed the quantum circuits. MHG, EHM, JLB implemented and executed the circuits on hardware. MHG, EHM, JLB, LC, JRS, JK, AM, TC, SYW, EC developed and applied the error mitigation techniques. BF, SYW, JK, GW, RAS, FMG, AK, PJR, LM performed the tensor network simulations. SYW, JLB, BF, AM performed the FLO simulations. JLB implemented the Majorana propagation. LC, JK performed the Trotter Error analysis. EC, TC, SYW, MHG, MS, BF, NV performed the XEB analysis. RAS, SYW, FMG, MK, MHG, JK, TC, EC performed the physics analysis. MHG, BF, EHM, JLB, PJR, GW developed the underpinning methods and infrastructure used for data analysis. 
ER and PR. liaised between the Google and Phasecraft teams and provided project consultation. DA consulted on Fermi-Hubbard physics. MA, JC, TC, NE, GG, SK, and RO performed hardware calibrations.
All authors contributed to the writing and revision of the manuscript and the Supplementary Information.

\vfill
\begin{center}
\rule{0.25\textwidth}{0.4pt}  
\end{center}

$^1$ Phasecraft Ltd, London, UK\\
$^2$ Phasecraft Inc, Washington DC, USA\\
$^3$ Department of Computer Science, University College London, UK\\
$^4$ Dahlem Center for Complex Quantum Systems, Freie Universit\"{a}t Berlin, 14195 Berlin, Germany\\
$^5$ Department of Computer Science, Virginia Tech, USA \\
$^6$ School of Informatics, QSL, University of Edinburgh, UK \\
$^7$ University of Bristol, UK\\
$^8$ Department of Electrical and Electronic Engineering, UCL, London, UK\\
$^9$ Google Quantum AI, Santa Barbara, California, USA
\\

}

\clearpage
\twocolumngrid
\section*{Methods}
\namelabel{sec:methods}{Methods}

We give here a high-level overview of the implementation of the experiment. This includes:
the algorithm and circuits employed for the dynamical simulation of the model on hardware; the error-mitigation techniques employed; the classical simulation techniques used to compare against; and the statistical benchmarking techniques used to indicate high-quality performance of the device. A brief discussion of the hardware used is given in \cref{sec:hardware}.

\subsection*{Algorithm and Circuits}


The time evolution of the system is approximated by three layers of second-order Trotterised time dynamics for nearly all times -- excluding the the first four time points which used $1$ and $2$ layers to improve signal quality in all experiments except thermalisation with doping. The maximum simulation time attained is $t=1.2$, with a maximum time step per Trotter layer of $0.4$, when using $3$ trotter layers.

The comparison of simulated quantum dynamics against ground truth suggests that the Trotter error remains small for $U=0$ at all times simulated here. This is corroborated by full state-vector simulations of the dynamics of sub $4 \times 4$ systems, where we see that the Trotter error averaged over up to all weight~4 observables remains low for $U=0$ and $U=4$, but becomes significantly more pronounced for $U=8$ at intermediate- to late-times. Additionally, simulations of $4 \times 4$ dynamics for a number of target physics observables suggest that for $U=0$, Trotter error is negligible at all times, for $U=4$, Trotter error begins to play an important role at approximately $t=0.9$, and for $U=8$, Trotter error begins to play an important role at about $t=0.5$. We discuss this in more detail in \cref{sec:trotter_error}.

Each spin sector of the fermionic system is represented by a separate Jordan-Wigner (JW)~\cite{jordan1928} transform arranged in a zig-zag pattern following the square connectivity of the device.
The arrangement of the JW transforms ensures that the spin-up and spin-down modes associated with the same site are as close as possible, given the hardware constraints. The smallest simulation requires 34~qubits, 1220~two-qubit gates and a circuit depth of~198, to simulate the $4\times4$ system. While the largest simulation requires 74 qubits, 4372 two-qubit gates and a circuit depth of 294 to simulate the $6\times6$ system -- the maximum we could reasonably fit on the hardware devices given layout constraints. We employ a fermionic swap (FSWAP) network~\cite{cade2020strategies,stanisic2022observing}, which dynamically rearranges the mode ordering of the JW transform in order to bring vertically adjacent modes next to one another. 
This amortizes the cost of executing vertical hopping terms across these modes, which would otherwise be much more costly to execute.

Since (2-qubit) gates are a major source of error, and the device has a limited coherence time, improving the circuit's gate count and depth are both critical to reducing noise in the computation. Here we achieve this by adopting a range of circuit optimisation techniques, some specifically tailored to fermionic simulation, including:
\begin{itemize}
 \item employing the term ordering symmetry of the second-order Trotter layer to minimize fermionic swaps;
\item merging terms within and between second-order Trotter layers;
\item identifying opportunities to replace qubit swaps with fermionic swaps, which require fewer 2-qubit gates to implement on the hardware;
\item merging fermionic hopping terms with fermionic swaps;
\item making a judicious choice of JW layout to ensure minimal cost of spin-spin interactions.
\end{itemize}
Whilst the impact of each of the above is incremental, the combined effect of the entire set of optimisations results in a significant reduction in gate count. A full account of the circuit constructions and costs is given in \cref{sec:circuits}.

\subsection*{Error mitigation}

Hardware noise and errors remain a major obstacle to the extraction of scientifically meaningful results from any current digital quantum simulation. In order to achieve  meaningful results, it is necessary to combine multiple algorithmic design techniques (described above, and in more detail in \cref{sec:circuits}) with suitable error-mitigation methods.

We employ several error-mitigation techniques to significantly improve our signal. The most impactful of these are (i)~Post-selection by occupation sector, (ii)~Training with Fermionic Linear Optics (TFLO), (iii)~Maximum Entropy Shot-Reweighting (MESR) and (iv)~Gaussian process regression (GPR).

Post-selection by occupation sectors exploits the fact that the dynamics of the system conserves particle numbers in both the spin-up and spin-down sectors. Whenever density measurements are performed in all modes, we can therefore verify whether the spin-up and spin-down counts have been preserved. Whenever that is not the case, we can know with certainty that there has been a significant error in the computation and discard that run.

TFLO is a training-based noise mitigation method \cite{Czarnik2021, Lowe2021, Bultrini2023} that exploits the fact that the Fermi-Hubbard model reduces to a free-fermion model for $U=0$, hence can be classically efficiently simulated using FLO. The quantum circuit structure at the gate-level remains very similar for $U\neq 0$, and so we can expect the underlying hardware noise profile to be similar.
The deviation of the noisy signal from the ground truth at $U=0$ is used to generate a linear fit on a set of observables corresponding to those being targeted for correction. The linear fit is then employed to apply linear noise correction for observables measured at non-zero values of~$U$. We discuss this method in more detail in \cref{sec:tflo}. We note here that a critical motivating factor in deploying a fixed number of Trotter layers for all time steps, despite fewer layers in principle being required at low times, is to achieve consistency in training data for TFLO. An added benefit of this is that the contribution of device noise to the signal is largely decorrelated from the time parameter of the dynamics, as the 2-qubit circuit depth and gate count remain fixed across all data points.

For a number of signals, we are interested in retrieving statistics over patterns found in shots. We introduce and employ in this work a Maximum-Entropy Shot-Reweighting (MESR) technique, which mirrors the power of TFLO in a shot-based setting. This technique tilts the distribution of shots with importance weights that maximize entropy subject to the constraint that the reweighted shots produce the corrected or improved local expectation values from FLO or TFLO.  Maximizing entropy corresponds to minimally adjusting the sample weights according to the information gained by the corrected local observables, and also leads to a practically efficient solution in terms of Lagrange multipliers~\cite{jaynes1957information}, as discussed in more detail in \cref{sec:shot_reweighting}.

Unless otherwise stated, ``error-mitigated'' in this paper means mitigated by TFLO or MESR.

Notably, the deployment of all the above methods is bespoke to interacting fermion models, such as the Fermi-Hubbard model, targeting key specific properties of the system. This makes them particularly effective in mitigating errors for the simulations of interest.

These experiments target dynamical properties of the system, and aim to produce observables that vary smoothly with time. We can leverage this assumption of continuity for error mitigation via Gaussian process regression (GPR). GPR is a standard method for generating smooth curves from data points under an assumption of continuity and a maximum rate of change. We discuss this method in more detail in \cref{sec:GPR}.

Additional error mitigation methods we employ include Pauli- and readout-twirling,
dynamical decoupling, and averaging over the symmetries of the Hamiltonian,
circuit and initial states. All are discussed in more detail in \cref{sec:error_mitigation}.

The set of techniques we have described here, and the details of their implementation, were settled upon after an exhaustive iterative process of comparisons against alternative implementations and techniques. We do not include the full account of these here. However, we note that we did compare our methods to another prominent error mitigation technique called probabilistic error amplification (PEA) with zero noise extrapolation (ZNE), and found TFLO to perform more favourably in our setting. We include a discussion of this in \cref{sec:PEA}.

One concern regarding employing TFLO and MESR as error mitigation techniques is that we risk confounding our ability to verify the performance of our simulations when comparing against $U=0$ dynamics. This is because the signal is trained on the behaviour of $U=0$, and although the aim is to use this to learn the generic impact of hardware noise on our signals, we may instead simply learn the specific impact that the combination of hardware noise and algorithmic errors has on shifting the $U=0$ signal. Although we can not guarantee that this effect plays no role in the performance of our $U=0$ signal, there are a number of factors that allow us, with care, to still validate the simulation performance on $U=0$ results.

First, it is important to dispel the impression that TFLO is simply taking the curves and directly shifting them to align with $U=0$. The linear fit training data is taken over all times, and the varying signals we observe require significantly more parameters to specify than are present in the linear fit. Second, in the case of a $4\times4$ system (see Fig.~\ref{fig:central_hole_4x4_global_observables_alt} in \cref{sec:further_hw_plots}), where the ground truth is available via exact state-vector simulations, we see that TFLO performs well for all values of $U$, even when the behaviour of the signal is dramatically different from that at $U=0$. This suggests that the dominant effect of TFLO is not to retrieve a $U=0$ signal, but rather to undo the effect of hardware noise. Finally, TFLO is trained on few-body observables. Where we consider many-body observables with support on a large number of qubits, we use MESR with FLO or TFLO training data. This performs well, despite these many-body quantities being in principle independent of the 1- and 2-body observables being trained on.


\subsection*{Classical simulation}

Some of the most performant of the tensor network methods targeting dynamical simulation include time-evolving matrix product states via the time-evolving block decimation~\cite{vidal2004efficient} or TDVP~\cite{haegeman2011time,Haegeman_2016_unifying} algorithms, or time-dependent density matrix renormalisation group (tDMRG) for matrix product states~\cite{provazza_fast_2024}, and recent circuit-model DMRG simulators~\cite{Ayral_2023_dmrg, thompson2025} and projected entangled pair states (PEPS) algorithms~\cite{Tindall_2024_efficient, Begusic_2024_fast, patra2024efficient}.

In parallel work~\cite{alam2025fermionic}, we have tested multiple tensor network techniques: direct tensor contraction; time-evolving matrix product states via TDVP method; and tDMRG for matrix product states. These tests indicate that TDVP performs most reliably, and so for our main comparisons, we include TDVP simulations of the dynamics of the system. Importantly, TDVP simulates continuous time dynamics, not circuit dynamics directly. It does this by iteratively evolving the Schr\"{o}dinger equation for short time steps, and at each step projecting the state onto the subset of MPS states of a given bond dimension. We perform simulations up to a bond dimension of $2048$ with time steps of $\Delta t = 0.01$, and also perform bond-dimension extrapolation analysis in some cases.
See \cref{sec:tensor_networks} for a more detailed description of these simulations.

Majorana propagation (MP) is an approximate simulation method wherein the target observables are evolved under the circuit dynamics in the Heisenberg picture, and tracked as an operator expansion in Majorana monomials (products of individual Majorana fermion operators). Under this evolution, the operator expansion can balloon into an exponential number of monomials, which must be truncated in order to remain efficient. The truncation criteria employed here are: number of Majoranas in the monomial, fixed at $10$; and coefficient magnitude, fixed at $3 \times 10^{-5}$. We have found Majorana propagation to be a particularly performant method for simulating multiple layers of fermionic Trotter dynamics. We employ this method throughout to approximate the exact circuit dynamics.
See \cref{sec:majorana_propagation} for a more detailed description of our simulations.

\subsection*{Validation and benchmarking against exact and approximate classical methods}

In \cref{sec:cross-entropy} we adapt the method of linear XEB benchmarking~\cite{arute2019quantum,morvan2024phase,zlokapa2023boundaries,cheng2025generalized,decross2025computational} to the setting of FLO circuits, and compare both unmitigated and MESR-mitigated hardware samples against a noiseless, untrotterised FLO simulator, to provide efficient statistical evidence of the global overlap of the ideal state with the output of the $U=0$ digital simulation of the Fermi-Hubbard dynamics. In contrast with applications of linear XEB to random quantum circuits, our verification protocol utilizes samples generated in physically meaningful FLO circuits, allowing for efficient usage of hardware samples and end-to-end benchmarking of the digital simulation against the ideal target physical state.


\clearpage
\appendix
\newpage
\onecolumngrid
\setcounter{page}{1}  

\tocon
\tableofcontents
\clearpage

\section{Fermi-Hubbard physics at zero and \texorpdfstring{$\pi$}{pi} flux}\label{sec:model_details}

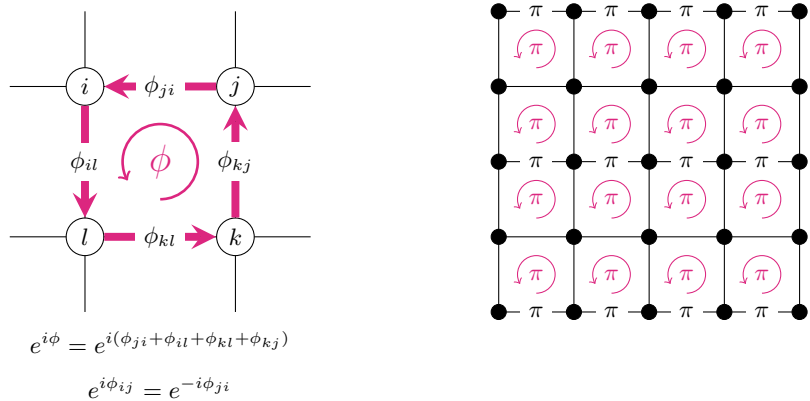
\begin{figure}
    \centering
\begin{tikzpicture}[]
    \foreach \x in {0,1} {
        \draw (-1,-2*\x)--(3,-2*\x);
        \draw (2*\x,1)--(2*\x,-3);
    }

        \node (i) at (0,0) [circle,draw,fill=white,minimum size=5mm] {};
        \node at (0,0) {$i$};
        \node (j) at (2,0) [circle,draw,fill=white,minimum size=5mm] {};
        \node at (2,0) {$j$};
        \node (k) at (2,-2) [circle,draw,fill=white,minimum size=5mm] {};
        \node at (2,-2) {$k$};
        \node (l) at (0,-2) [circle,draw,fill=white,minimum size=5mm] {};
        \node at (0,-2) {$l$};

        \draw[line width=3,->,>=stealth,ibmMagenta] (j)--(i);
        \draw[line width=3,->,>=stealth,ibmMagenta] (i)--(l);
        \draw[line width=3,->,>=stealth,ibmMagenta] (l)--(k);
        \draw[line width=3,->,>=stealth,ibmMagenta] (k)--(j);

        \node at (1,0) [fill=white] {$\phi_{ji}$};
        \node at (2,-1) [fill=white] {$\phi_{kj}$};
        \node at (1,-2) [fill=white] {$\phi_{kl}$};
        \node at (0,-1) [fill=white] {$\phi_{il}$};

        \node (flux) at (1,-1) [ibmMagenta,font=\Large] {$\phi$};
        \draw[->,ibmMagenta,line width=1] (1,-1.5) arc (-90:210:.5);

        \node at (1,-3.4) [] {$e^{i\phi}=e^{i(\phi_{ji}+\phi_{il}+\phi_{kl}+\phi_{kj})}$};
        \node at (1,-4) [] {$e^{i\phi_{ij}}=e^{-i\phi_{ji}}$};

        \begin{scope}[shift={(5.5,-3)}]
            \foreach \x in {0,1,2,3,4}{
            \draw (0,\x)--(4,\x);
            \draw (\x,0)--(\x,4);
                \foreach \y in {0,1,2,3,4}{
                \node at (\x,\y) [circle,draw,fill,minimum size=2mm, inner sep=0pt] {};
                }
            }

            \foreach \y in {0,2,4}{
                \foreach \x in {.5,1.5,2.5,3.5}{
                \node at (\x,\y) [fill=white] {$\pi$};
                }
            }

            \foreach \x in {.5,1.5,2.5,3.5}{
                \foreach \y  in {.5,1.5,2.5,3.5}{
                    \node at (\x,\y) [ibmMagenta] {$\pi$};
                    \draw[->,ibmMagenta] (\x,\y-0.25) arc (-90:210:.25);
                }
            }
        \end{scope}
\end{tikzpicture}
    \caption{\justifying Flux is represented on the model by attaching Peierls phases $\phi_{ij}$ to lattice edges $(i,j)$ and dressing the Fermi-Hubbard hopping terms as in \cref{eq:FH_Ham}, the phases are directional with $\phi_{ij}=-\phi_{ji}$. When a charged particle moves in a closed loop the state picks up a phase equal to the sum of phases for each edge crossed, as illustrated in the left hand figure. The total phase $\phi$ is proportional to the magnetic flux through the loop, analogous to the Aharonov-Bohm effect with the Peierls phases taking the role of the vector potential.
    Accordingly, a choice of phases $\phi_{ij}$ specifies a flux profile up to some gauge freedom, on the right we show the phases we use  
    to induce a $\pi$ flux through every plaquette (blank edges indicate a trivial phase). The directionality of the $\pi$ phases for each edge is arbitrary as the effect is the same either way. %
    }
    \label{fig:peierls phases}
\end{figure}

In this work, we study the time evolution of several states under the Fermi-Hubbard Hamiltonian
\begin{align}\label{eq:FH_Ham}
H = -J\sum_{\langle i,j\rangle\in \mathcal{L},\sigma}(e^{i\phi_{ij}}c_{i,\sigma}^\dagger c_{j,\sigma}+{\rm h.c.})+U\sum_{i\in \mathcal{L}}n_{i,\uparrow}n_{i,\downarrow}.
\end{align}
Here, the first term represents the hopping (with amplitude $-J$) of electrons on a lattice $\mathcal{L}$ between nearest neighbour sites, while the second represents the (repulsive) interaction between electrons occupying the same position. The fermionic operators $c^\dagger_{i,\sigma}$ $(c_{i,\sigma})$ create (destroy) an electron at lattice site $i=(i_x,i_y)$ of spin $\sigma= \{\uparrow, \downarrow\} $ and satisfy the usual anticommutation relations $\{c_{i,\sigma},c_{j,\sigma'}^\dagger\}=\delta_{ij}\delta_{\sigma\sigma'}$. The operator $n_{j,\sigma}=c_{i,\sigma}^\dagger c_{i,\sigma}$ is the local electron density at site $i$ of spin $\sigma$. The Peierls phases $\phi_{ij}$ encode a magnetic flux $\phi$ in a direction perpendicular to the sample by enforcing that the sum of Peierls phases around each individual square face (plaquette) of the lattice add up to $\phi$, see \cref{fig:peierls phases} for more detail. In our experimental setup, we consider rectangular lattices of $L_y\times L_x$ sites with open boundary conditions. 

This system has several conserved quantities, which we employ with varying degrees to improve the experimental signal.
The most useful conserved quantity is total number and spin conservation: The transformation $c_{j,\sigma}\rightarrow e^{i\alpha_\sigma}c_{j,\sigma}$ leaves the Hamiltonian invariant. The conserved quantities associated with this $U(1)\times U(1)$ symmetry are total number $N= N_\uparrow +N_\downarrow$ and total spin $S^z=N_\uparrow-N_\downarrow$, where $N_\sigma = \sum_i n_{i,\sigma}$.

\begin{figure*}[!htbp]
    \centering
    \includegraphics[width=\linewidth]{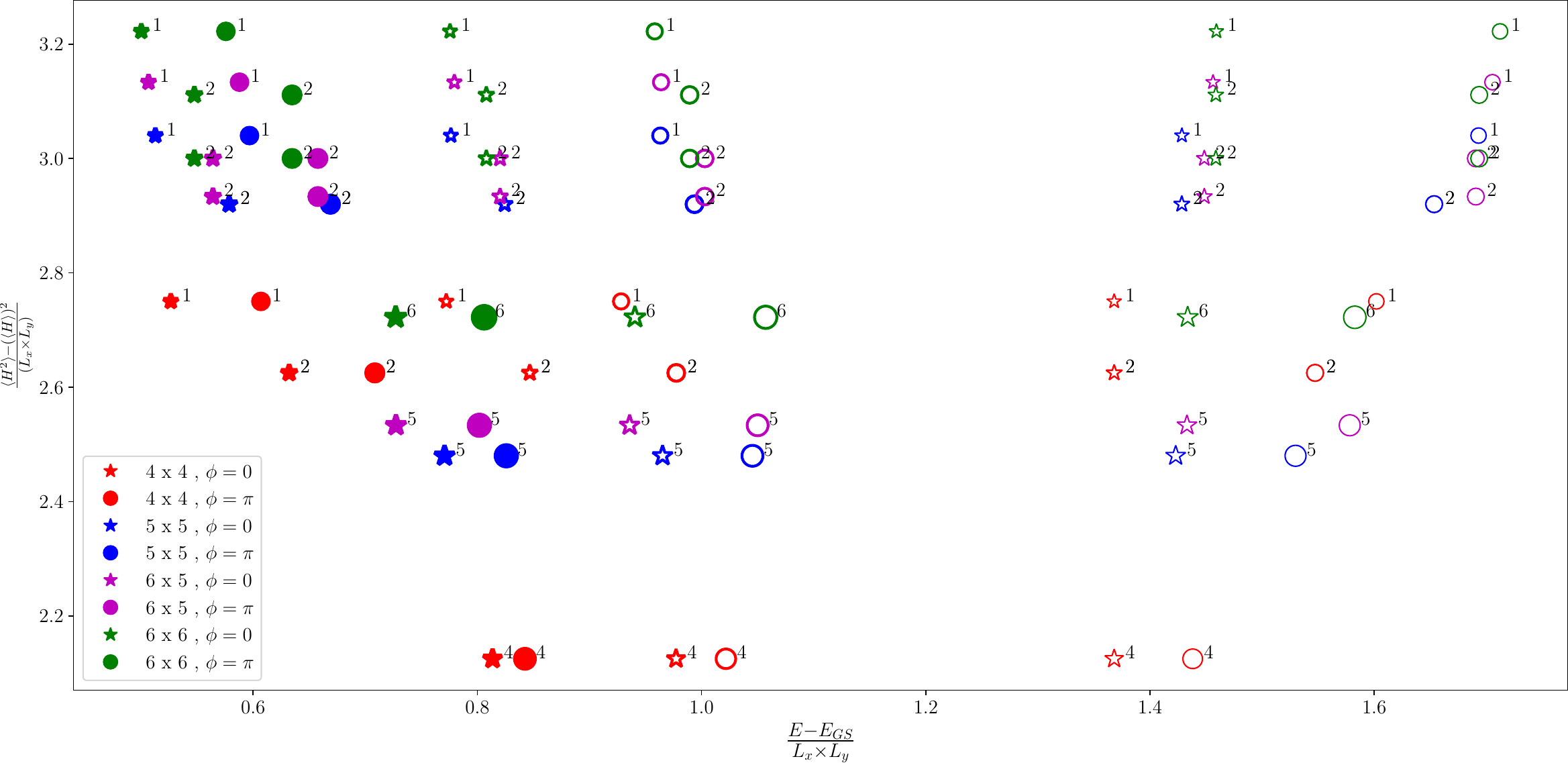}
    \caption{\justifying Landscape of the systems studied in this work. Here we plot the energy of an initial state measured with respect to the ground state energy (obtained with DMRG) in the same number of particle, flux sector and its energy variance, both normalised by the system's size. Different colours represent different sizes. The numbers close to each marker indicate the number of holes in the system. $U=0$ ($U=4$) systems are shown with thin (thick) lines. Solid markers represent $U=8$. Circles correspond to systems without magnetic flux, while stars represent systems with flux.}
    \label{fig:landscape_of_initial_states}
\end{figure*}

We are interested in understanding the quenched dynamics of physically relevant low-energy states that can be prepared with minimal circuit cost. In the limit of very large interactions and near half filling, the presence of doubly occupied sites (doublons) is energetically penalised. At exactly half filling, and $U\gg |J|$, the low-energy manifold of the Hamiltonian (\cref{eq:FH_Ham}) is well approximated (in the sense that corrections are of order $(|J|/U)^2$) by the antiferromagnetic Heisenberg Hamiltonian
\begin{align}
H_{\text{Heis}}=\frac{4J^2}{U}\sum_{\langle i,j\rangle}\left(\mathbf{S}_i\cdot \mathbf{S}_j-\frac{1}{4}\right)
\end{align}
where $\mathbf{S}_j$ is the spin operator at site $j$, so low-energy states favour anti-correlation of neighboring spins. For this reason, we choose the N\'{e}el state as one of the initial states.  With these motivations in mind, our experiments are performed at different levels of small hole-doping, where we perturb the initial state by removing particles. We consider interactions $U=0,4,8$ in units of the hopping strength $J$, which we set to one.
We consider several initial states with different hole-dopings, ranging from one to $\min(L_x,L_y)$. Each initial state has a given number of electrons $N_e$ and total spin $S^z=0,1$, depending on the parity of $L_x\times L_y -N_e$. The initial state configuration corresponds to either a N\'{e}el background with the some holons placed at various locations, or a singlet covering with two holons. A singlet across two sites $i$ and $j$ is defined as
\begin{equation}
    \centering
    \ket{\text{singlet}} = \frac{c^{\dagger}_{i,\uparrow} c^{\dagger}_{j,\downarrow} - c^{\dagger}_{i,\downarrow} c^{\dagger}_{j,\uparrow}}{\sqrt{2}} \ket{\Omega}
\end{equation}
where $\ket{\Omega}$ is the vacuum state, and a singlet covering is simply the tensor product over all singlet pairs, $\otimes_{s=1}^{N_{\text{singlets}}} \ket{\text{singlet}_{s}}$. In Fig.~\ref{fig:landscape_of_initial_states}, we show the variance and energy  (with respect to the ground state of the corresponding  $N_e$, $S^z$ sector) per site of each of the states considered. Using the energy density of these states, we can assign a corresponding temperature, understood as the temperature of a Gibbs state with the same energy, using the equation of state of the model \cite{LeBlanc2013}. All these state have an energy density of the order of the hopping $J$. We observe here a manifestation of Lieb's $\pi$ flux phase theorem \cite{Lieb1994}, namely that, for all configurations of fixed system size, interaction, and filling, the one that attains the lowest energy is the $\pi$ flux system. For the states with two holons, the singlet states have either strictly smaller variance (for sizes $6\times 5$ and $6\times 6$) or equal energy variance (sizes $4\times 4$ and $5\times 5$) compared with the corresponding N\'{e}el ordered state.

\begin{figure*}[!htbp]
    \centering
    \includegraphics[width=0.48\linewidth]{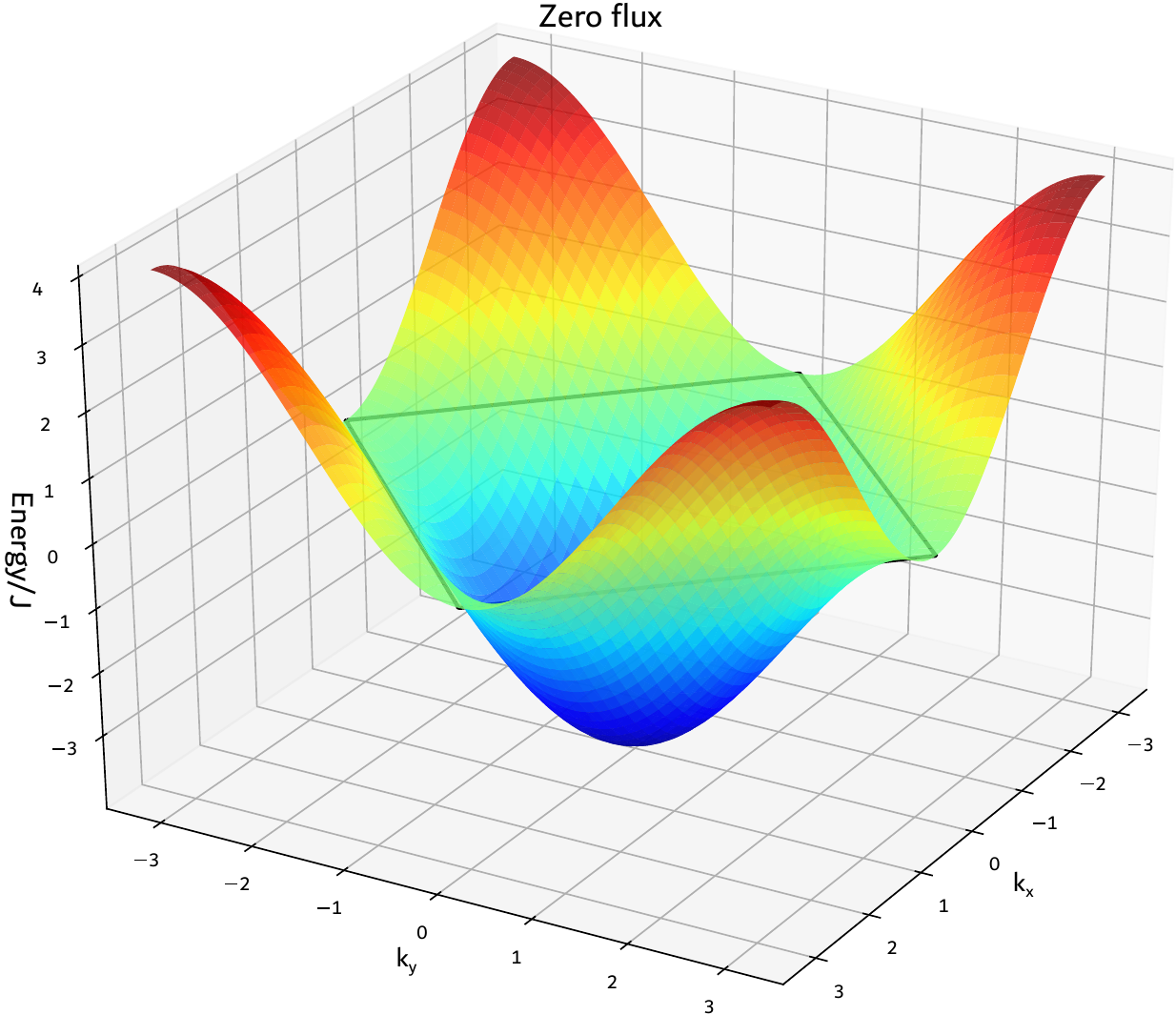}
    \includegraphics[width=0.48\linewidth]{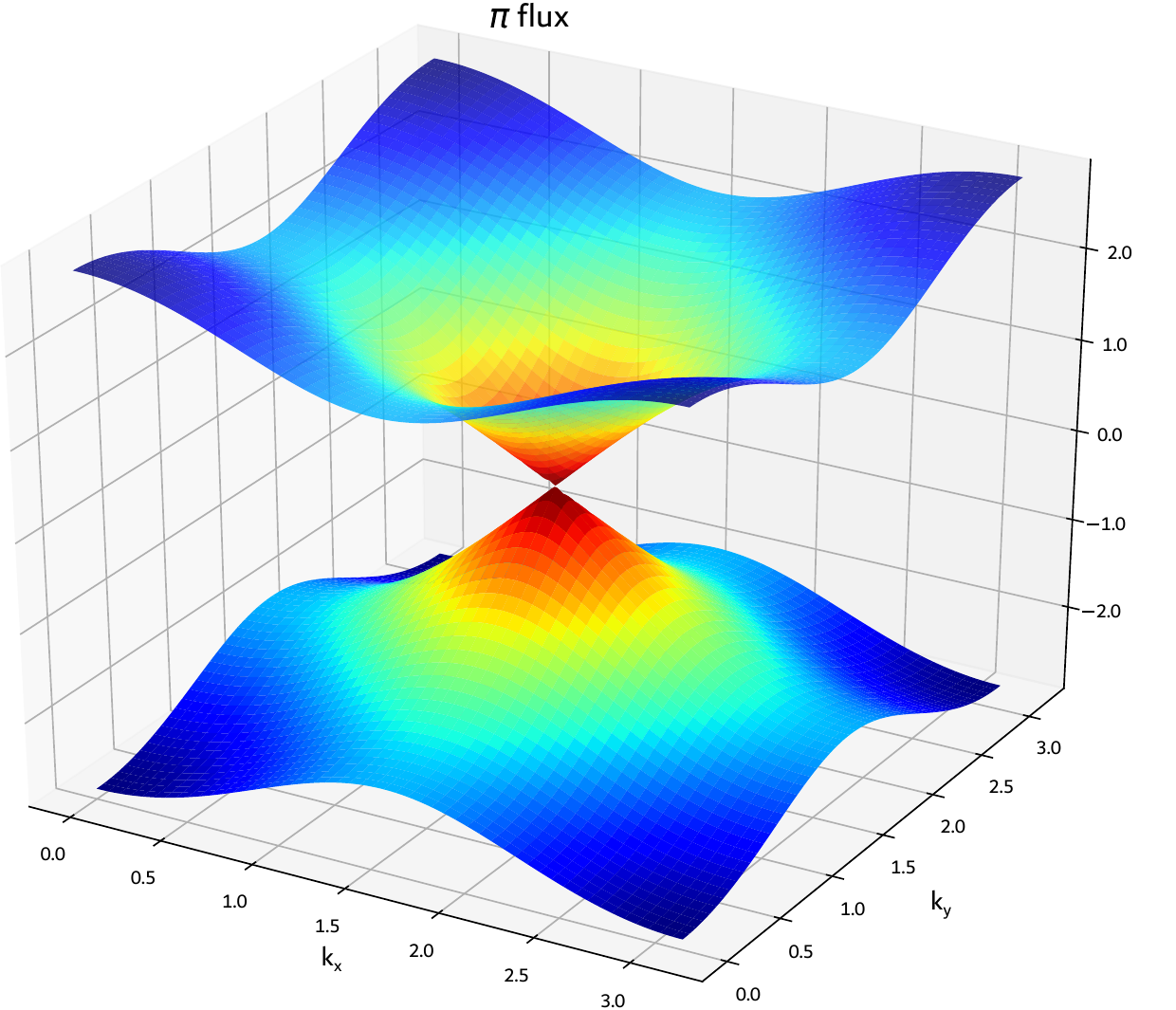}
    \caption{\justifying Single particle band in the zero flux (left) and $\pi$ flux regime (right, only one Dirac point shown) at $U=0$ for the Fermi-Hubbard model. The dispersion relation for the zero flux case is given by $E(k)_{\phi=0}=-2\cos(k_x)-2\cos(k_y)$, while for the $\pi$ flux case it is $E(k)_{\phi=\pi}=\pm 2\sqrt{\cos^2(k_x)+\cos^2(k_y)}$, for wavenumbers $k_x$ and $k_y$. At half filling, the system with no flux has a square Fermi surface, while the $\pi$ flux system has only four points. This state is also known as a Dirac semimetal.}
    \label{fig:bands}
\end{figure*}

While the simulation of the Fermi-Hubbard model with and without magnetic flux differs at the circuit level just in some single-qubit gates, the physics of these two models is completely different at low energies. As an illustration, we show the band dispersion of the Fermi-Hubbard model at zero and $\pi$ flux in Fig.~\ref{fig:bands} (left and right).

At zero interaction, the ground state of the Fermi-Hubbard model with and without flux corresponds to two copies of spinless electrons filling the single-particle bands. For zero flux, the system has a cosine band with a minimum around zero momentum. At half filling, the Fermi surface is a square and the system (square contour in Fig.~\ref{fig:bands} left) is very unstable to interactions as Umklapp processes are allowed by momentum conservation. At $\pi$ flux the unit cell is enlarged and the system becomes a Dirac semimetal, violating Luttinger's theorem \cite{Luttinger_1960}. At half filling, the Fermi surface becomes 4 points, where the dispersion relation around each has a conical singularity (see Fig.~\ref{fig:bands} right) and at low energy is described by two-dimensional Dirac fermions.

\section{Experimental Results} \label{sec:physics_results}

In this section, we collate the experimental results obtained. Unless otherwise specified, all results were obtained using error mitigation as described in \cref{sec:error_mitigation}.

\subsection{Dynamics of a single holon}\label{sec:single_holon}

At exactly half filling, the ground state has antiferromagnetic correlations. A holon moving around flips the spins in its path, creating aligned neighbours that break the N\'{e}el order pattern. The energy cost of this makes the movement of the holon different from a freely propagating one. In this section, we study the movement of charge and spin on an initial state where a single holon is placed in the center on an otherwise N\'{e}el background.

\begin{figure*}[!htbp]
    \centering
    \includegraphics[width=\linewidth]{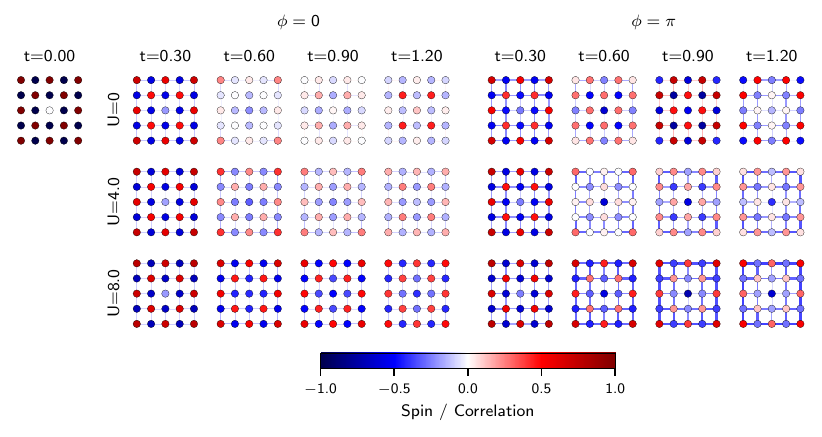}
    \caption{\justifying
    Spin dynamics for $\phi=0$ (left insert) and $\phi=\pi$ (right insert) for the $5\times5$ lattice with a central hole.
    Showing the evolution of spins \ref{eq:Local_spin} at each site (circles) and nearest neighbour spin correlators \ref{eq:NN_spin} (links) in a $5\times 5$ system. Obtained from mitigated experimental results.
    }
    \label{fig:central_hole_5x5_spins}
\end{figure*}

\begin{figure*}[!htbp]
    \centering
    \includegraphics[width=\linewidth]{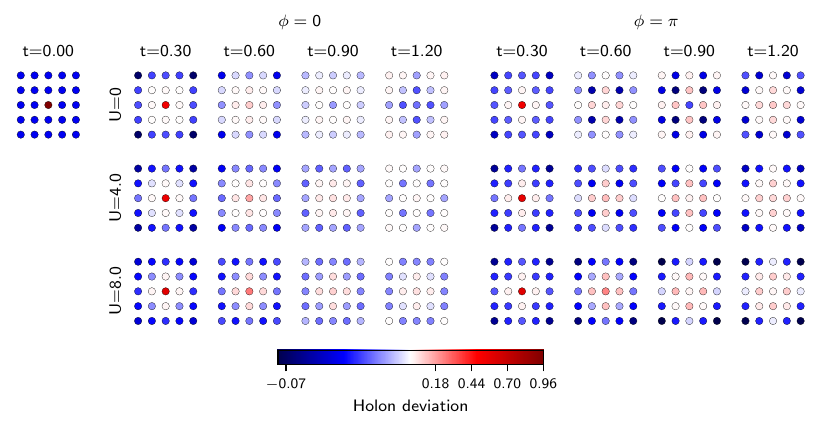}
    \caption{\justifying Deviation from the average holon density in  \cref{eq:holon_deviation}  as a function of time for $\phi=0$ (left insert) and $\phi=\pi$ (right insert) for the $5\times5$ lattice with a central hole. Obtained from TFLO-mitigated experimental results (for more details see \cref{sec:error_mitigation}). Note the nonlinear colour scale, chosen for better contrast.}
    \label{fig:central_hole_5x5_holon_deviation}
\end{figure*}

\begin{figure*}[!htbp]
    \centering
    \includegraphics[width=\linewidth]{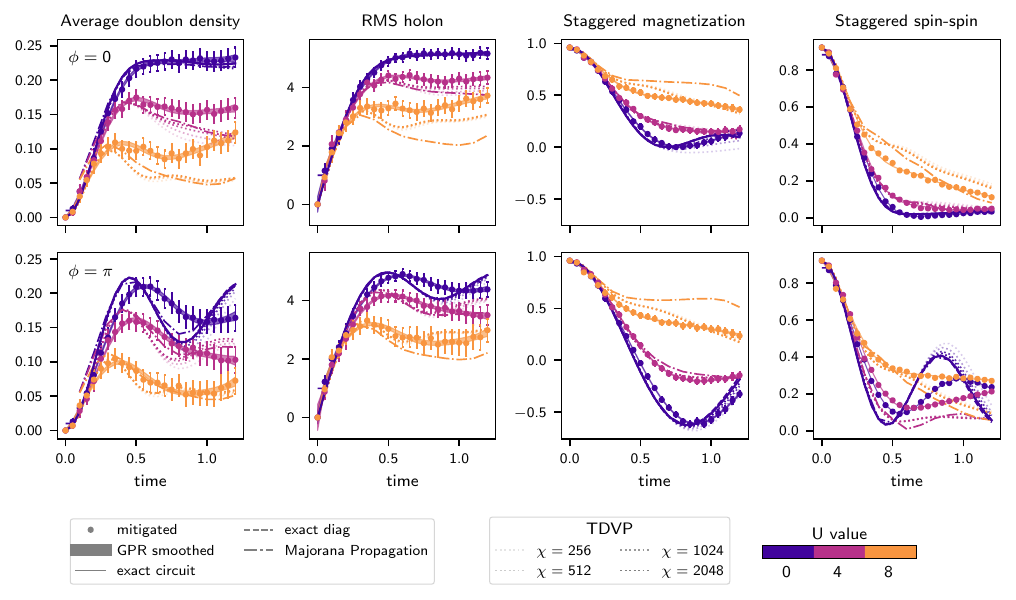}
    \caption{\justifying Average doublon density \cref{eq:doublon_density}, RMS holon \cref{eq:holon_RMS}, staggered magnetisation \ref{eq:stag_mag} and staggered spin-spin \cref{eq:stag_spin_spin} as a function of time for $\phi=0$ (top row) and $\phi=\pi$ (bottom row) for the $5\times5$ lattice with a central hole. Using $62$ qubits and $3172$ two-qubit gates. Here we also plot the results obtained using tensor network simulations at different bond dimensions, and exact circuit and TDS evolution when possible $(U=0)$.}
    \label{fig:central_hole_5x5_global_observables}
\end{figure*}

First, we show the dynamics of the error-mitigated local spin
\begin{align}\label{eq:Local_spin}
    \langle S^z_i(t)\rangle&=\langle n_{i\uparrow}(t)-n_{i\downarrow}(t)\rangle,
\end{align}
and nearest neighbour spin correlation
\begin{align}\label{eq:NN_spin}
     C_{i,e}(t)&:=\langle S^z_i(t)S^z_{i+e}(t)\rangle,
\end{align}
where $e$ is a primitive lattice vector. We observe that for zero flux, the spin dynamics slows down as the interaction increases, and the original N\'{e}el order gradually melts towards the uniform local spin zero state at a rate that decreases as the interaction increases.
For non-zero flux, the reduction of the magnetic correlations slows down with increasing interactions, as in the case of zero flux, but some striking differences appear. At zero interaction, the dynamics is fast and we see a revival of the spin density with time, which is expected from the exact $U=0$ evolution. As the interaction increases, we see two phenomena occurring simultaneously, the melting of the N\'{e}el order and the increase of antiferromagnetic correlations between neighbours. For $U=0$, note that the N\'{e}el order flips around $t=0.9$, with the roles of spin up and spin down interchanged with respect to the original configuration. Antiferromagnetic order between sites that have the same sign of the spin density appears around the edge. At the position where the holon was initially located, we see a suppression of the magnetic correlations between neighbours. We can investigate how this signal correlates with the presence of a holon at the same place and time where the magnetic distortion is observed. A straightforward measure of the local holon density $\langle h_i(t)\rangle =\langle (1-n_{i\uparrow})(1-n_{i\downarrow})\rangle$ is not sufficient, as the number of holon-doublon pairs increases in time (see e.g.\ Fig.~\ref{fig:central_hole_5x5_global_observables} left most panel). We measure instead the deviation from the average holon density
\begin{align}\label{eq:holon_deviation}
    \langle h_i(t)\rangle_{\rm dev}:= \langle (1-n_{i\downarrow}(t))(1-n_{i\uparrow}(t)\rangle - \frac{1}{L_xL_y}\sum_j \langle (1-n_{j\downarrow}(t))(1-n_{j\uparrow}(t)\rangle,
\end{align}
The resulting signal is shown in Fig.~\ref{fig:central_hole_5x5_holon_deviation}. At early times (up to $t\sim 0.3$), the evolution is essentially ballistic and independent of interactions and flux, which is a result that also appears in Fig.~\ref{fig:central_hole_5x5_global_observables} (left and center left). For both zero and nonzero flux, the effect of interactions in the holon density deviation  $\langle h_i(t)\rangle_{\text{dev}}$ is not substantial. The effect of flux, on the other hand, is more profound. For $\pi$ flux, we observe that the approach to the average uniform value is faster than for the zero flux case. Precisely in the region where there is an excess of holon density, we see a perturbation of the nearest neighbour correlation. This signals the appearance of a magnetic polaron, which has been observed in the dynamics of the Fermi-Hubbard model in analogue experiments \cite{Koepsell2019}.

\begin{figure*}[!htbp]
    \centering
    \includegraphics[width=0.32\linewidth]{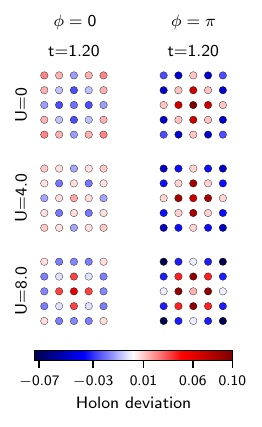}
    \caption{\justifying Deviation from the average holon density in \cref{eq:holon_deviation} for $\phi=0$ (left column) and $\phi=\pi$ (right column) for the $5\times5$ lattice with a central hole at the last time point, isolated for better contrast. These results are obtained from TFLO-mitigated experimental results (for more details, see \cref{sec:error_mitigation}). }
    \label{fig:central_hole_5x5_holon_deviation_last_time_point_circle_last_time_point_circles}
\end{figure*}

Finally, we look at some global observables for the evolution of this initial state. An observable that we will use to assess global properties of all the initial states presented is the average doublon density, $n_d$, defined as
\begin{align}\label{eq:doublon_density}
    n_d = \frac{1}{L_xL_y}\left\langle \sum_j n_{j\uparrow}n_{j\downarrow}\right\rangle,
\end{align}
which is shown in Fig.~\ref{fig:central_hole_5x5_global_observables} (left).  Note that the number of holons minus the number of doublons is conserved, as they are created in pairs.
Increasing the (repulsive) interaction, the total doublon number decreases as generating doublons is energetically more costly, which translates into a faster time oscillation of the components in the evolution of the initial state with more doublons, decreasing their contribution.
To quantify the delocalisation of the original holon, we compute the root-mean-squared holon observable, defined as
\begin{align}\label{eq:holon_RMS}
    {\rm holon}_{\rm RMS}:=\sqrt{\sum_{i}|i|^2\langle (1-n_{i\uparrow})(1-n_{i\downarrow})\rangle},
\end{align}
where the (Euclidean) distances are measured with respect to the initial position of the holon at $t=0$. This signal measures the spread of the holon's size in time. As mentioned before, we see (Fig.~\ref{fig:central_hole_5x5_global_observables} center left) that the dynamics of the holon is universal (i.e $U$ independent) and ballistic for small times, as expected from the Heisenberg equation of motion $\dot{n}_d=i\langle [H,n_d] \rangle =i\langle [H_0,n_d]\rangle $ where the derivative at $t=0$ only depends on the quadratic part of the Hamiltonian $H_0$ as witnessed by the doublon number and RMS holon observables.

The staggered magnetisation, shown in Fig.~\ref{fig:central_hole_5x5_global_observables} (center right) is defined as
\begin{align}\label{eq:stag_mag}
    M_s = \frac{1}{L_xL_y}\sum_i (-1)^i S^z_i,
\end{align}
and together with the staggered spin-spin correlation Fig.~\ref{fig:central_hole_5x5_global_observables} (right)
\begin{align}\label{eq:stag_spin_spin}
    (S^zS^z)_{\rm stag} :=&\frac{1}{(L_xL_y)^2}
        \sum_{i,j}
        (-1)^{|i-j|} \langle S^z_i S^z_j \rangle,
\end{align}
quantifies the decay of the initial N\'{e}el order. Looking at these global observables, we see that the effect of the flux is to promote a faster decay in general and a revival of the magnetic order in the non-interacting case.

We assess the quality of the quantum simulation using several methods. In Fig.~\ref{fig:central_hole_5x5_global_observables} we include exact dynamics and exact circuit simulation for $U=0$ to gain some understanding of the Trotter error of the simulation (a thorough analysis of Trotter errors appears in \cref{sec:trotter_error}). For all the parameters considered (flux and interaction) we also perform tensor network simulations based on matrix product states, evolving the system using the time dependent variational principle (TDVP). We scale the bond dimension $\chi$ in these and all results shown below from 256 to 2048 (for more details see \cref{sec:tensor_networks}).
Another classical simulation technique that we explore to benchmark our results is Majorana propagation \cite{miller2025} (more details in \cref{sec:majorana_propagation}). We observe that the mitigated experimental results in general agree qualitatively with the classical benchmarks.

\newpage

\subsection{Evolution of holon stripe }\label{sec:stripe}

As the electron density is reduced from half filling, several orders are expected to appear as a function of (hole) doping at low energy in the Fermi-Hubbard model \cite{bennemann2014novel}. In particular, for hole-doping $n_h\sim 1/8$, \cite{Zheng2017, Liu_2025} stripe order is expected to appear.
At high interaction, disturbing the N\'{e}el background is energetically costly and the main mode that a stripe can evolve is by fluctuating around its position. The mean-field approach to this dynamics has been studied in \cite{Zaanen2001}, where the concept of bipartite symmetry of the spin background plays a central role. In this approach, the magnetic order is studied once the holes have been removed, using the concept of squeezed space \cite{ogata1990bethe}. At finite interactions, different mechanisms can disrupt the stripe dynamics. One is the proliferation of doublon-holon pairs, that change the number of available holons and make the percolation of a stripe more probable. Another effect is the fluctuation of the magnetic order, that obscures the bipartite lattice symmetry order considered in \cite{Zaanen2001}. In this section, we study the fluctuations away from the mean-field dynamics of the stripe.

\begin{figure}[!htbp]
\centering
\includegraphics[width=\textwidth]{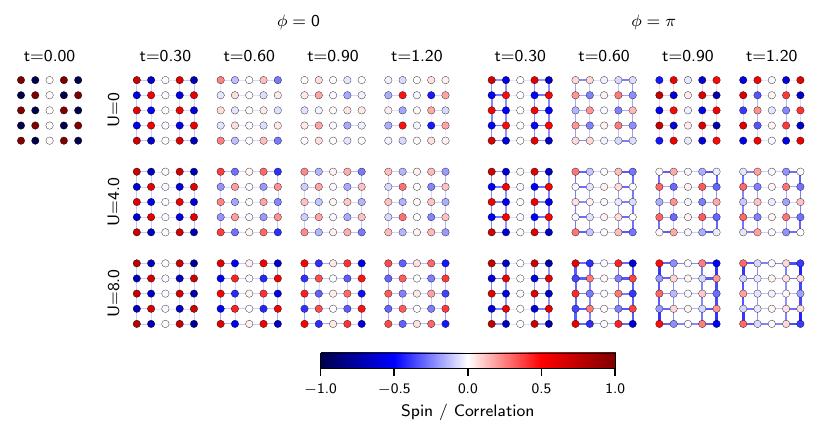}
\caption{\justifying
Spin dynamics for $\phi=0$ (left insert) and $\phi=\pi$ (right insert) for the $5\times5$ lattice with a central holon stripe. This displays the evolution of spins defined in \cref{eq:Local_spin} at each site (circular nodes) and nearest neighbour spin correlators defined in \cref{eq:NN_spin} (whereby the magnitude of the negative correlation is denoted in the links' thickness) in a $5\times 5$ system. Obtained from TFLO mitigated experimental results.
Observe the enhancement of spin correlations in the Dirac semimetal regime ($\phi=\pi$).}
\label{fig:5x5_densities_allshots}
\end{figure}

We analyse the dynamics of a stripe by evolving an initial state consisting of a N\'{e}el ordered background with a stripe of holons in the middle running from top to bottom as shown schematically in Fig.~\ref{fig:5x5_densities_allshots} (leftmost panel) for a $5\times 5$ system. We do this in two different regimes, evolving with the Fermi Hubbard Hamiltonian with zero and $\pi$ flux per plaquette.
Note that the N\'{e}el pattern that we choose in the initial state is such that in the squeezed space, where we remove the holons, the system is truly bipartite.

\begin{figure}[ht] 
\centering
\includegraphics[width=\textwidth]{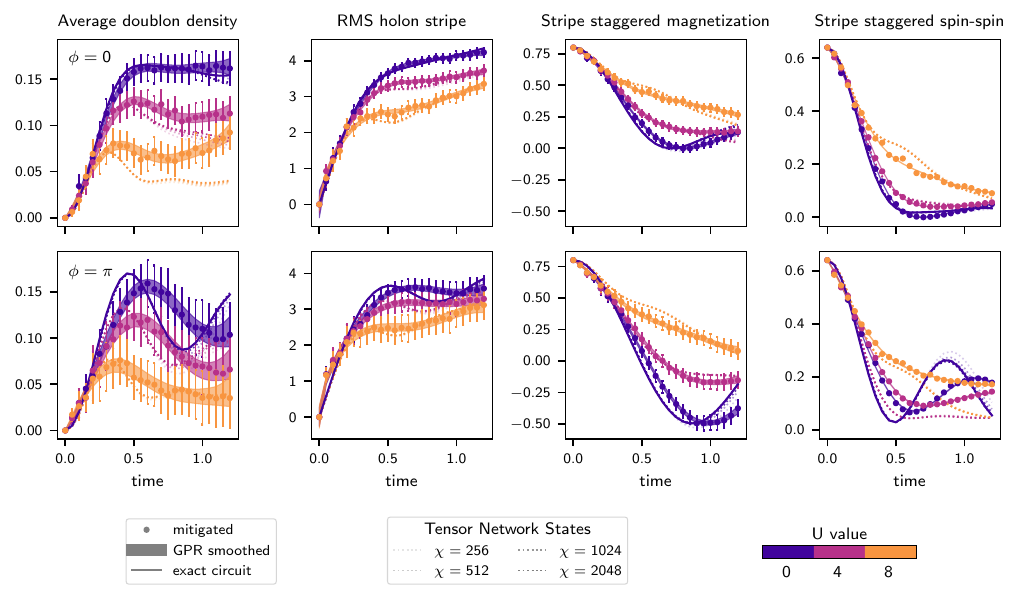}
\caption{\justifying Average doublon density \ref{eq:doublon_density}, RMS holon stripe \ref{eq:rms_stripe}, stripe staggered magnetisation \ref{eq:stripe_stag_mag} and stripe staggered spin-spin as a function of time for $\phi=0$ (top row) and $\phi=\pi$ (bottom row) for the $5\times5$ lattice with a holon stripe down the middle. Using $51$ qubits and $2626$ two-qubit gates. Here we also plot the results obtained using tensor network simulations at different bond dimensions, and exact circuit and TDS evolution when possible $(U=0)$ .
}
\label{fig:global_holon}
\end{figure}

In Fig.~\ref{fig:5x5_densities_allshots} we show the (experimental, error mitigated) average spin density \cref{eq:Local_spin} and the nearest-neighbour spin correlation \cref{eq:NN_spin}
The effect of interactions is radically different for zero and $\pi$ flux. At zero flux, the initial configuration melts more slowly as the interaction increases, with the spin density tending to the uniform zero magnetisation value. For the Dirac semimetal regime $\phi=\pi$, the situation is somewhat different. For zero interaction, we see a fast melting and revival of the antiferromagnetic order, with the holon stripe reappearing. For increasing interaction, a slowdown of the dynamics is apparent, accompanied by an increase in the neighbour spin anti-correlation as time evolves. To complement this qualitative description, we study global properties that quantify these effects. Together with the error-mitigated experimental signals, we show the corresponding results obtained using classical tensor networks simulations with the TVDP algorithm (see \cref{sec:tensor_networks} for more details) and exact simulation for zero interaction. In Fig.~\ref{fig:global_holon} (left), we plot the evolution of the average density of doublons in the system for $\phi=0,\pi$. The proliferation of doublons (and holons) depends strongly on the interaction, as expected. The asymptotic value of the doublon density approaches its maximum value \begin{align}
n_d(U=0,t\rightarrow\infty)=\frac{1}{4}\left(\frac{L_x-1}{L_x}\right)^2,
\end{align}
at $U=0$ and zero flux, where $L_x$ is the side length of the system.

 Note that this trend also appears for non-zero flux, but the experimental data becomes less reliable in this scenario as the error bars are amplified.

\begin{figure}[!htbp]
\centering
\includegraphics[width=\textwidth]{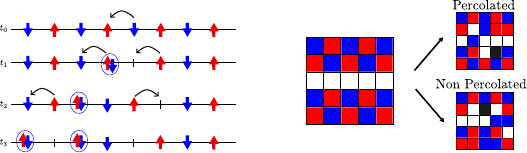}
\caption{\justifying Left: Illustration of the melting of antiferromagnetic order in the presence of mobile charges. The hopping of electrons in the N\'{e}el ordered background produces doublon-holon pairs of net zero spin. Their movement reduces the value of the staggered magnetisation and spin-spin correlations. Right: The evolution of the initial holon-stripe generates states that, when sampled in the computational basis, can be qualified as percolated or non-percolated. We use the percolated subset of shots to analyse the breakdown of the mean-field description of the stripe dynamics. Black sites denote the presence of a doublon.}
\label{fig:sketch_dilution_doublons}
\end{figure}

In Fig.~\ref{fig:global_holon} (center left), we show the dynamics of the root-mean-square (RMS) holon density. This observable is defined as
\begin{align}\label{eq:rms_stripe}
     {\rm stripe}_{\rm RMS} = \sqrt{\sum_{i}(i_x-i_{x_0})^2\langle n^h_{i}\rangle},
\end{align}
where we sum over the distance perpendicular to the initial stripe located at $x_0$. It measures the spread of the initial holon stripe in time. Here we see that interactions reduce the spread, in line with the general expectation that interactions reduce the effective velocity of propagation of mobile quasiparticles. Including $\pi$ flux, the trend remains. The role of magnetisation is captured by the stripe staggered magnetisation $M^{\rm stripe}_{\rm stag}$ (Fig.~\ref{fig:global_holon}, center right) and the stripe staggered spin-spin $(S^zS^z)_{\rm stag}$ (Fig.~\ref{fig:global_holon}, right) observables, defined respectively as
\begin{align}\label{eq:stripe_stag_mag}
M^{\rm stripe}_{\rm stag} &: = \frac{1}{L_x L_y}\sum_{i} (-1)^{i_x+i_y}{\rm sign}(i_x-i_{x_0})\langle S_z^i\rangle \\
(S^z S^z)_{\text{stag}}&:= \frac{1}{(L_x L_y)^2}
        \sum_{i,j}
        (-1)^{|i-j|} \langle S^z_i S^z_j \rangle
\end{align}
$M^{\rm stripe}_{\rm stag}$
measures the magnetisation, taking into account the initial state configuration, where the type of bipartition in the N\'{e}el order flips crossing the holon stripe. $(S^zS^z)_{\rm stag}$, on the other hand, captures the correlation between spins, with the initial staggered order considered between all pairs. We see that these signals dilute in time as expected due to the presence of mobile particles. A sketch of this process is shown in Fig.~\ref{fig:sketch_dilution_doublons} (left).

\begin{figure}[ht] 
\centering
\includegraphics[width=\textwidth]{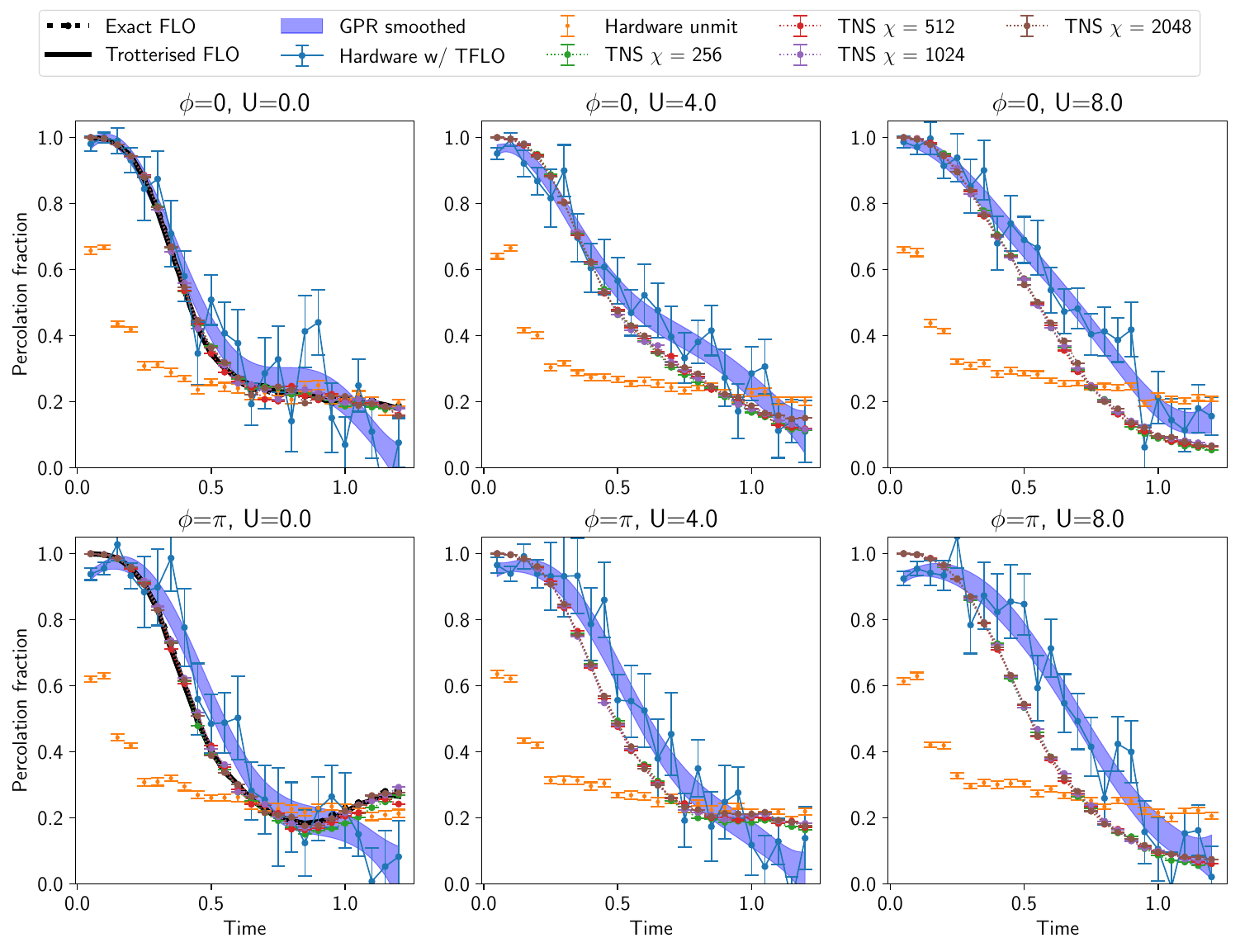}
\caption{\justifying Percolation fraction for a $5\times 5$ system. Note that unmitigated hardware data is postselected by antiferromagnetic order measured by AFM $\in [-1,0]$ (see \cref{eq:afm}, number of doublons $\mathcal{D}=[0,1,2,3]$ present and particle number conservation. TFLO is then applied to this set of shots to produce the Hardware w/o TFLO line, and finally GPR is applied to give the 68$\%$ one sigma region (for more details, see \cref{sec:post_selection}, \cref{sec:GPR} and \cref{sec:tflo}).}
\label{fig:5x5_percolation_postselection}
\end{figure}

We want to investigate the departure from the mean field description of the stripe dynamics, which assumes a stationary N\'{e}el background and only movement of the stripe such that it fluctuates around its initial position. The rationale behind this is based on the strong interaction picture, where the creation of doublons is forbidden and breaking the N\'{e}el pattern is very costly energetically.
Under strict enforcement of these conditions, the only soft mode of the stripe is oscillation around its ``centre of mass'' -- preserving the bipartition of the \neel pattern.
We have already observed that both conditions are broken in practice for the interactions that we discussed. Motivated by this result, we study how likely it is for a configuration of holons to percolate from one edge of the lattice to the other with non-zero background magnetisation. Assuming the mean-field description, all sampled configurations should satisfy these criteria.
In particular, we define the percolation fraction as the number of samples where a continuous line of holons crosses the lattice from one side to the opposite, divided by the total number of samples at a given time. A path is continuous if any holon in the path has a neighbour on its side or along the diagonal. An illustration of a percolated and a non-percolated path is shown in Fig.~\ref{fig:sketch_dilution_doublons} (right).
In Fig.~\ref{fig:5x5_percolation_postselection}, we show the percolation fraction postselected by number of doublons and by antiferromagnetic order. The first filters the obfuscation introduced by doublon-holon pairs appearing in the dynamics, while the second selects the states with a stronger N\'{e}el pattern. We defined the antiferromagnetic order in this case as
\begin{align}\label{eq:afm}
{\rm AFM} = \frac{1}{N_{\rm shots}(\mathcal D)}\sum_{\substack{s\in{\rm shots}\\ N_d\in \mathcal{D}}}\sum_{\langle i,j\rangle}\frac{S_i^z(s) S_j^z(s)}{n_{\rm pairs}(s)}
\end{align}
where ${\rm shots}(\mathcal D)$ is the number of experimental shots with doublons in the range $\mathcal{D}$, and $n_{\rm pairs}(s)$ is the number of nearest neighbour pairs in the shot $s$ with nonzero correlation $S_i^z(s) S_j^z(s)$.
This observable essentially measures the antiferromagnetic correlation of the background of a postselected shot, according to the number of doublons present, and is normalised such that AFM $\in [-1,1]$. In Fig.~\ref{fig:5x5_percolation_postselection}, we show the evolution of the percolation fraction in time for zero and $\pi$ flux. In this figure, we also include the result of sampling from the MPS produced by TDVP at different bond dimensions.
Since the initial state is Gaussian, we can efficiently sample from the $U=0$ experiment to compute the true expected percolation fraction and then apply TFLO corrections to our hardware data. We observe some interesting behaviour. Apart from the non-interacting case, the effect of flux is not important. Moreover, the late time (postselected) percolation fraction {\it decreases} with interaction, in a completely opposite trend to the one expected by the mean-field description of the stripe dynamics. A likely explanation is provided by the disordering effect of the moving doublon-holon pairs which, although they are being postselected to a low number (in Fig.~\ref{fig:5x5_percolation_postselection} $\mathcal{D}=[0,1,2,3]$ doublons), are still present in the dynamics. Consider a virtual transition of a state with zero doublons, where a doublon-holon pair is created, they move and then they annihilate. In this process, the background could be perturbed away from the exact bipartite symmetry, reducing the antiferromagnetic order.

\begin{figure}[!htbp]
\centering
\includegraphics[width=\textwidth]{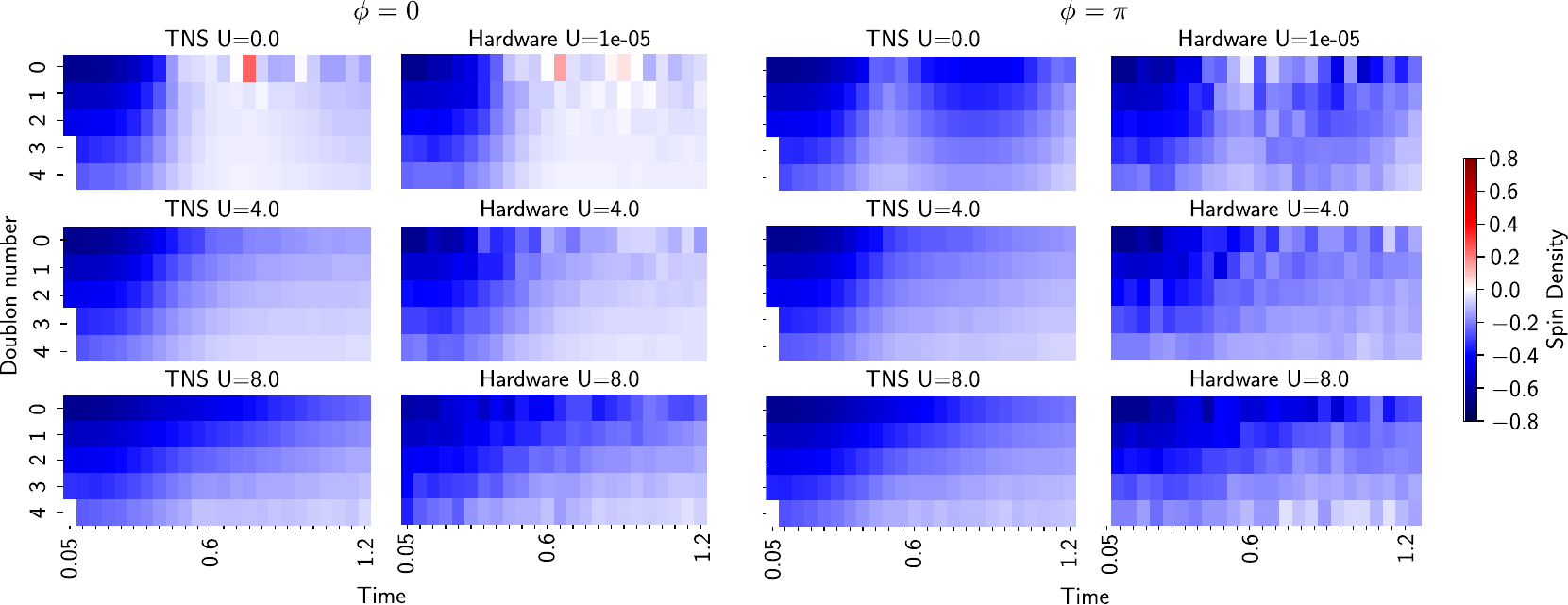}
\caption{\justifying Evolution of antiferromagnetic order, defined as $\sum_{\langle ij\rangle}S^z_jS^z_j$ for different fluxes (left two columns $\phi=0$, right two columns $\phi=\pi$, for AFM range $\in [-1,0]$, doublons included $\in [0,1,2,3]$. TNS represent classical results using tensor network simulation (see \cref{sec:tensor_networks}), while hardware results employ the MESR error mitigation discussed in appendix \ref{sec:shot_reweighting}).}
\label{fig:5x5_heatmap_postselection}
\end{figure}

To gain an understanding of this phenomenon, we analyse the evolution of the magnetisation $\sum_{\langle ij\rangle}S^z_iS_j^z$ in time, over postselected shots with a doublon number in Fig.~\ref{fig:5x5_heatmap_postselection}, where we employ MESR as the error mitigation technique. For more information on the FLO sampler, see \cref{sec:flo_sampling}. For the TNS, we can also exactly and efficiently sample from the MPS state in the TeNPy TDVP simulations.

We observe that indeed a lower number of doublons leads to a slower melting of the antiferromagnetic order as the interaction increases. We plot the comparison between this signal obtained from sampling the classical tensor network state at the largest bond dimension considered $\chi=2048$, against the results obtained from the device, where the experimental shots have been mitigated according to the shot mitigation procedure outlined in \cref{sec:shot_reweighting}. We see qualitative agreement between the TN and the experimental pictures.

For similar plots for other system sizes, see \cref{app:additional_stripe_info}.

\subsection{Spin-charge separation as a function of doping and magnetic field}\label{sec:spin_charge_u1}

Spin charge separation is a phenomenon wherein the electron's charge and spin fractionalise into independent degrees of freedom. In one dimension, this separation was predicted from the exact solution \cite{ogata1990bethe} and it has since been observed in many scenarios \cite{arute20,Vijayan2020timeresolved}. In higher dimensions, it has been conjectured that above the ground state and around half filling, the electron fractionalises into spinons and chargons (holons and doublons) of fermionic and bosonic character respectively, where the bosons are the carriers of charge \cite{Anderson1995charge, Lee2006doping}. We study this fractionalisation by looking at the equivalent description of the Fermi-Hubbard model as a dual gauge theory in terms of the chargons and spinons. We do this using the  (modified) Kotliar-Ruckenstein representation \cite{Kotliar1986new} of the electron operator
\begin{align}\label{eq:mKR}
c_{i\sigma}^{\dagger}=f_{i\sigma}^{\dagger}h_{i}+\sigma f_{i\bar{\sigma}}d_{i}^{\dagger},
\end{align}
where $\sigma = \{\uparrow,\downarrow\} = \{1,-1\}$ is the spin index. In this representation, the doublon $d^\dagger_i:= \frac{1}{2}(X_i^d+iY_i^d)$ (holon $h^\dagger_i:= \frac{1}{2}(X_i^h+iY_i^h)$  satisfies the usual algebra inherited from the Pauli matrices $(X,Y,Z)$. These quasiparticles commute at different sites. The spinon operator, on the other hand, satisfies fermionic anticommutation relations $\{f_{i\sigma},f_{j\sigma'}\}=0$ and $\{f_{i\sigma},f^\dagger_{j\sigma'}\}=\delta_{ij}\delta_{\sigma\sigma'}$. As the representation \cref{eq:mKR} enlarges the local Hilbert space, a local constraint is needed to recover the physical subspace. The constraint of exactly one quasiparticle per site
\begin{align}
    C_i:=d_i^\dagger d_i + f^\dagger_{i\uparrow} f_{i\uparrow} + f^\dagger_{i\downarrow} f_{i\downarrow} + h_i^\dagger h_i =1
\end{align}
defines the physical states and fixes the correct commutation relations of the physical fermion operators, i.e for all states satisfying $C_i|{\rm phys}\rangle = |{\rm phys}\rangle $ $\forall i$, the operator $c_{i\sigma}$ satisfies
\begin{align}
    \{c_{i\sigma},c^\dagger_{j\sigma'}\}|{\rm phys}\rangle = |{\rm phys}\rangle
\end{align}
thus defining a fermionic operator. The projection $\mathcal{P}=
\prod_j\frac{C_{j}(2-C_{j})(3-C_{j})(4-C_{j})}{6}$ into the physical space commutes with $c_{i\sigma}$ as $[C_i,c_{j\sigma}]=0$. This implies that the action of $c_{j\sigma}$ in a physical state produces another physical state.

The Hamiltonian acting on the physical space becomes
 $H=H_{n}+H_{s}+H_{{\rm int}}$ with
\begin{align}\label{eq:H_normal}
H_{n} & =-J\sum_{\sigma \langle ij\rangle }e^{i\phi_{ij}}(f_{i\sigma}^{\dagger}(h_{i}h_{j}^{\dagger})f_{j\sigma}+f_{i{\sigma}}(d_{i}^{\dagger}d_{j})f_{j{\sigma}}^{\dagger})+{\rm h.c.},\\
H_{s} & =-J\sum_{\langle ij\rangle}e^{i\phi_{ij}}[(f_{i\uparrow}^{\dagger}f_{j\downarrow}^{\dagger}-f_{i\downarrow}^{\dagger}f_{j\uparrow}^{\dagger})h_{i}d_{j}+(f_{i\downarrow}f_{j\uparrow}-f_{i\uparrow}f_{j\downarrow})d_{i}^{\dagger}h_{j}^{\dagger}]+{\rm h.c.},\\
H_{{\rm int}} & =U\sum_{i}d_{i}^{\dagger}d_{i}.
\end{align}
where $e^{i\phi_{ij}}$ represents flux through the system, as considered in some of our experiments.  This extended Hamiltonian has a local $U(1)$ gauge symmetry generated by $e^{i\theta C_j}$ as $e^{i\theta C_j}f_{i,\sigma}^\dagger e^{-i\theta C_j} = e^{i\theta}f_{i\sigma}^\dagger$, and similarly for the doublon and holon quasiparticles operators. This local gauge symmetry motivates the definition of two extended operators, the Wilson loop
\begin{align}
  \mathcal{W}_{d}(\mathcal{C}):=\prod_{j\in\mathcal{C}}(1-n_{j,\uparrow}n_{j,\downarrow}),
\end{align}
that measures the deconfinement transition of the gauge field mediating the interaction between spinons, and the Wilson lines between chargons (holons and doublons)
\begin{equation}
    \label{eq:app:wilson_lines}
    V_{ab}(M):=\sum_{\substack{\textnormal{sites }  i, j:\\ |i-j| = M}}  \frac{n^a_{i} n^b_{j}}{N_{\rm pairs}}
    \sum_{\substack{\textnormal{paths } \gamma:\\ \textnormal{ from } i \textnormal{ to } j }}\frac{\sum_{(m, n) \in \gamma} S_{m}^{z}S_{n}^{z}}{N_{\rm paths}(i,j)}
\end{equation}
where in the inner sum $m,n$ are consecutive positions along the path $\gamma$. The labels $a$ and $b$ can be either a holon or a doublon label, with $n_j^d=n_{j\uparrow}n_{j\downarrow}$ and $n_j^h=(1-n_{j\uparrow})(1-n_{j\downarrow})$.

\begin{figure}[ht] 
    \centering
    \includegraphics[width=0.8\linewidth]{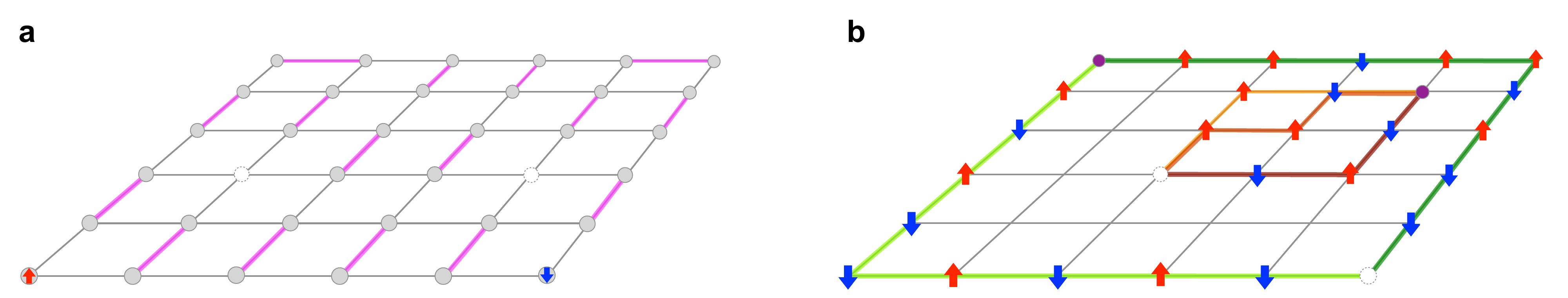}
    \caption{\justifying \textbf{a} Sketch of the initial state with two holes (empty circles) and singlet covering (magenta links) for a $6\times6$ lattice. \textbf{b} Various open Wilson lines between pairs with one holon (empty circle) and one doublon (purple circle). Here, the green paths highlight the two only possible paths with Manhattan distance $M=10$, while the red paths show three of the six possible paths with Manhattan distance $M = 6$. To compute the value of $V_{hd}(M)$ on a single sample, one has to average the spin correlations between neighbours along all the possible paths between all the possible doublon-holon pairs with Manhattan distance $M$.}
    \label{fig:wilson_lines_sketch}
\end{figure}

In Fig.~\ref{fig:wilson_lines_sketch}, we show the initial configuration we consider, namely a half-filled lattice with 2 holes and a singlet covering. Some examples of Wilson lines with fixed Manhattan distance between holon-doublon pairs entering the computation of $V_{hd}(M)$ are shown in Fig.~\ref{fig:wilson_lines_sketch}b.

\begin{figure}[ht] 
    \centering
    \includegraphics[width=\linewidth]{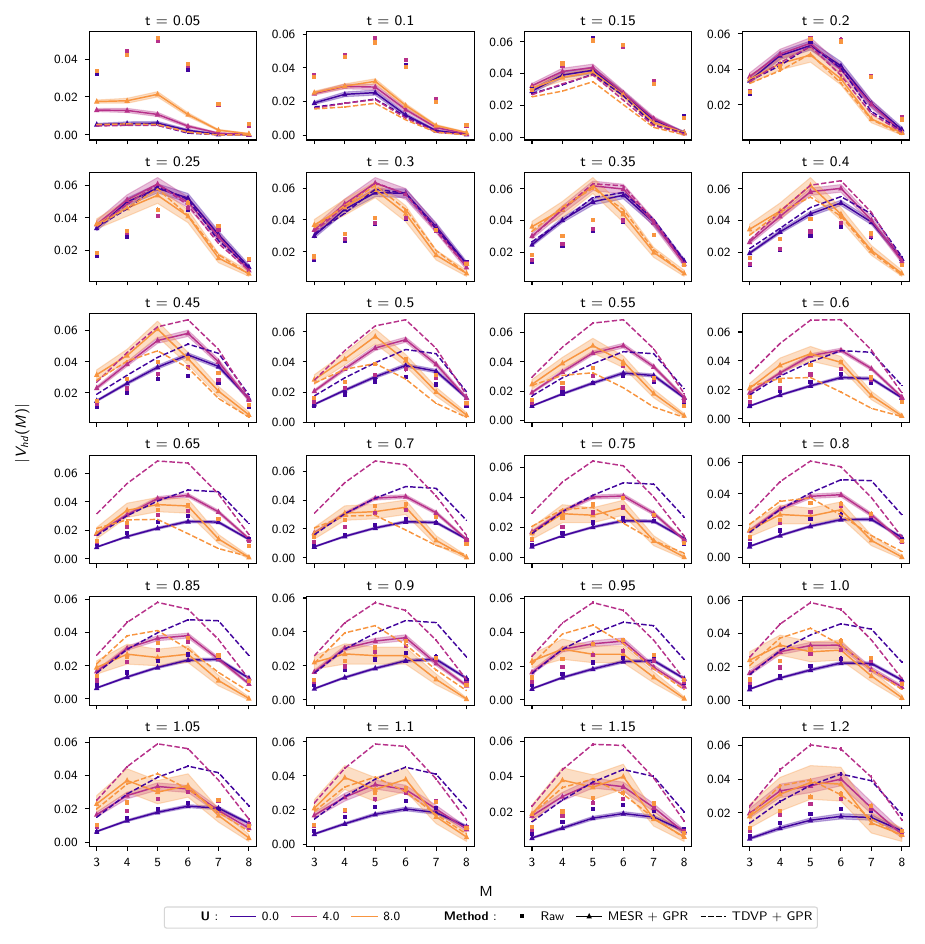}
    \caption{\justifying Expectation value of open Wilson lines $V_{hd}(M)$ for a $5\times5$ instance initialised in the state with singlets covering and two holes with zero flux $(\phi=0)$ as a function of the Manhattan distance $M$ between holon-doublon pairs for different values of time. Dashed lines have been obtained using 10000 samples from a TDVP simulation with $\chi = 2048$. Solid lines have been obtained using the maximum entropy shot-reweighting (MESR) procedure described in  \cref{sec:shot_reweighting}. Both experimental and TDVP data have been smoothed using GPR on the time-dependent signal.}
\label{fig:wilson_lines_5x5_no_flux}
\end{figure}

\begin{figure}[ht] 
    \centering
    \includegraphics[width=\linewidth]{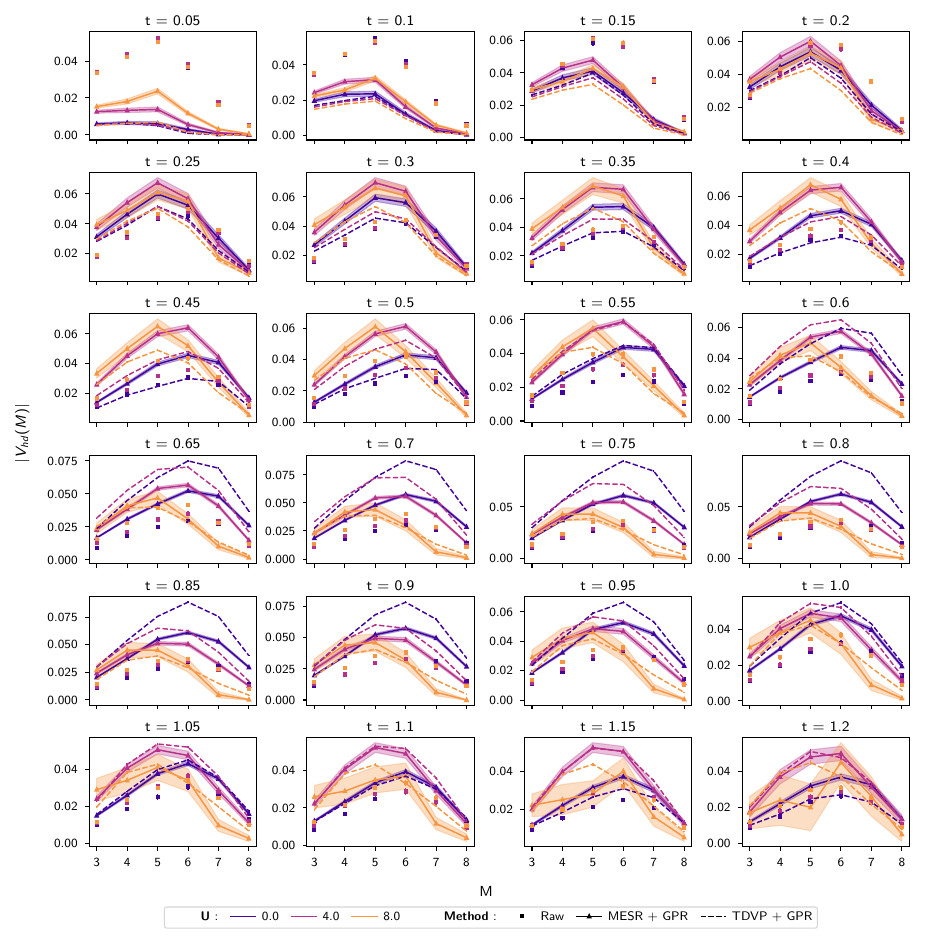}
    \caption{\justifying Same as Fig.~\ref{fig:wilson_lines_5x5_no_flux} for $\phi = \pi$.}
\label{fig:wilson_lines_5x5_w_flux}
\end{figure}

\begin{figure}[ht] 
    \centering
    \includegraphics[width=\linewidth]{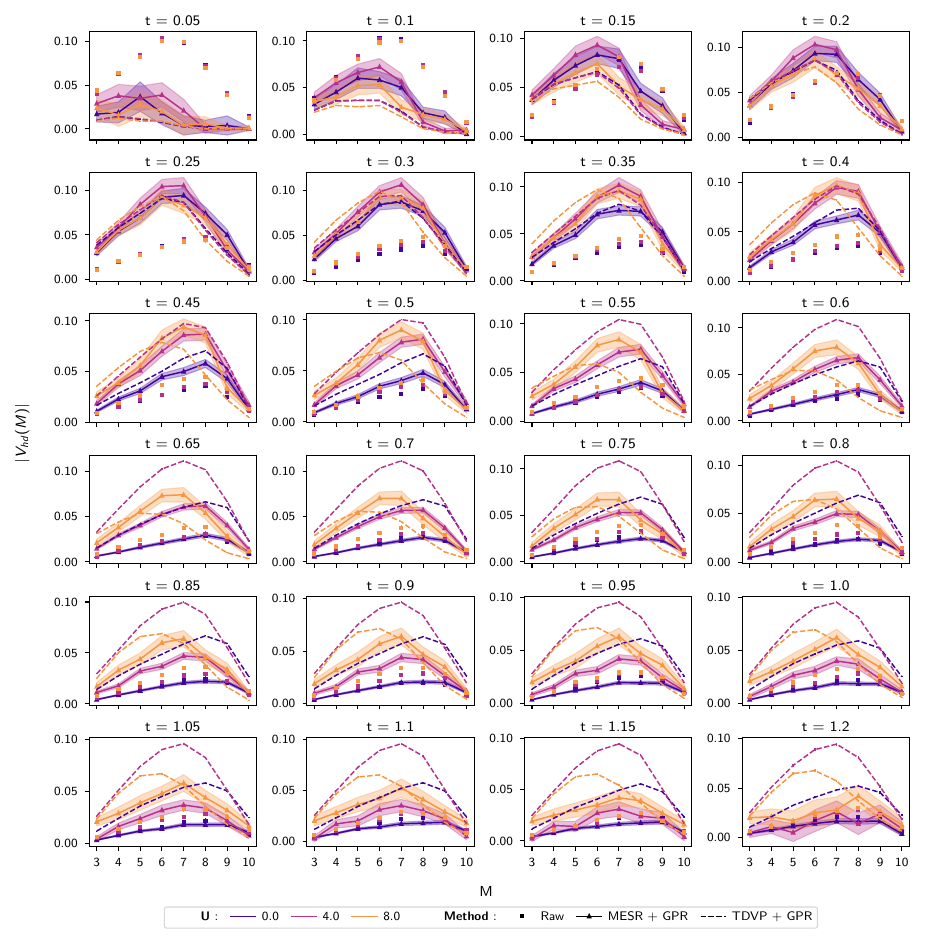}
    \caption{\justifying Expectation value of open Wilson lines $V_{hd}(M)$ for a $6\times6$ instance initialised in the state sketched in Fig.~\ref{fig:wilson_lines_sketch}a with zero flux $(\phi=0)$ as a function of the Manhattan distance $M$ between holon-doublon pairs for different values of time. Dashed lines have been obtained using 10000 samples from a TDVP simulation with $\chi = 2048$. Solid lines have been obtained using the maximum entropy shot-reweighting (MESR) procedure described in  \cref{sec:shot_reweighting}. Both experimental and TDVP data have been smoothed using GPR on the time-dependent signal.}
\label{fig:wilson_lines_no_flux}
\end{figure}

\begin{figure}[ht] 
    \centering
    \includegraphics[width=\linewidth]{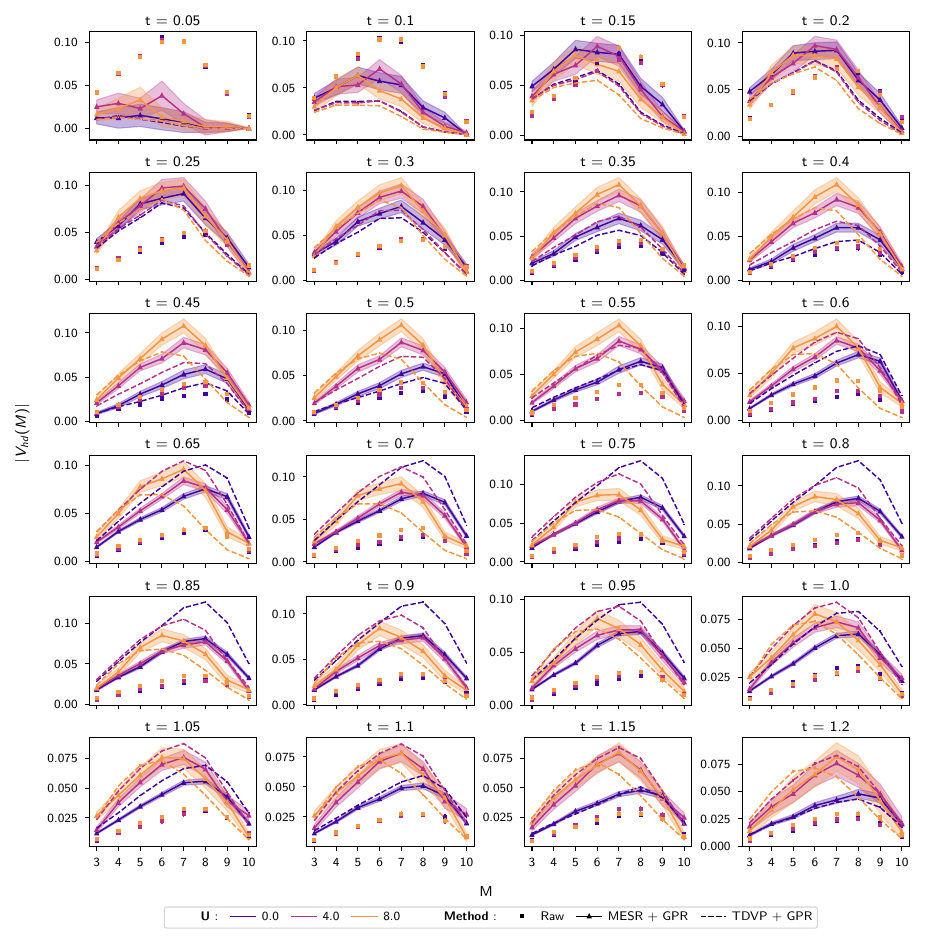}
    \caption{\justifying Same as Fig.~\ref{fig:wilson_lines_no_flux} for $\phi = \pi$.}
\label{fig:wilson_lines_w_flux}
\end{figure}

Figs.~\ref{fig:wilson_lines_5x5_no_flux} and~\ref{fig:wilson_lines_5x5_w_flux} show results for $V_{hd}(M)$ for a $5\times 5$ lattice initialised in a state with two holes and singlet covering with and without magnetic flux, respectively. Figs.~\ref{fig:wilson_lines_no_flux} and~\ref{fig:wilson_lines_w_flux} show the same for a $6\times 6$ lattice. In the presence of confinement between chargons, we expect $V_{hd}(M)$ to grow with the (Manhattan) distance between the holon and the doublon of a given pair. In the absence of magnetic flux, both Figs.~\ref{fig:wilson_lines_5x5_no_flux} and \ref{fig:wilson_lines_no_flux} show that after an initial transient, the value of $V_{hd}(M)$ for $U = 0$ decreases faster than in the presence of interactions. This suggests the absence of any confinement between chargons in the non-interacting regime with $\phi =0$.  We ascribe any residual value to a lack of thermalisation (see also \cref{subsec:app:thermalisation}). Interestingly, despite the MESR and TDVP signals agreeing well at short times, the experimental data predict a faster decay of $V_{hd}(M)$ at longer times.

On the other hand, as shown in Figs.~\ref{fig:wilson_lines_5x5_w_flux} and \ref{fig:wilson_lines_w_flux}, a magnetic flux makes $V_{hd}(M)$ non-vanishing even for $U=0$, at least up to the largest time we can access in our experiment. This suggests that a magnetic flux induces a stronger confinement between chargons.
In this case as well, MESR and TDVP signals agree well at short times and start to deviate from each other at intermediate times. However, in contrast with the case with $\phi = 0$, at longer times the agreement between the two signals improves again. From the global observables, like the staggered magnetisation shown in Fig.~\ref{fig:6b6global} below, we observe a similar revival in the non-interacting case, signalling that in this case the system is far away from equilibration and the signal is showing transient behaviour.

In all the figures discussed above, we observe that $V_{hd}(M)$ saturates at a value of $M$ which depends on the strength of the interactions. While a decay at larger $M$ can be ascribed to combinatorics effects (e.g., as shown in Fig.~\ref{fig:wilson_lines_sketch}b, there are only a few holon-doublon configurations with Manhattan distance $M = 10$), we defer the analysis of this phenomenon to future work.

 \begin{figure}[ht] 
    \centering
    \includegraphics[width=\linewidth]{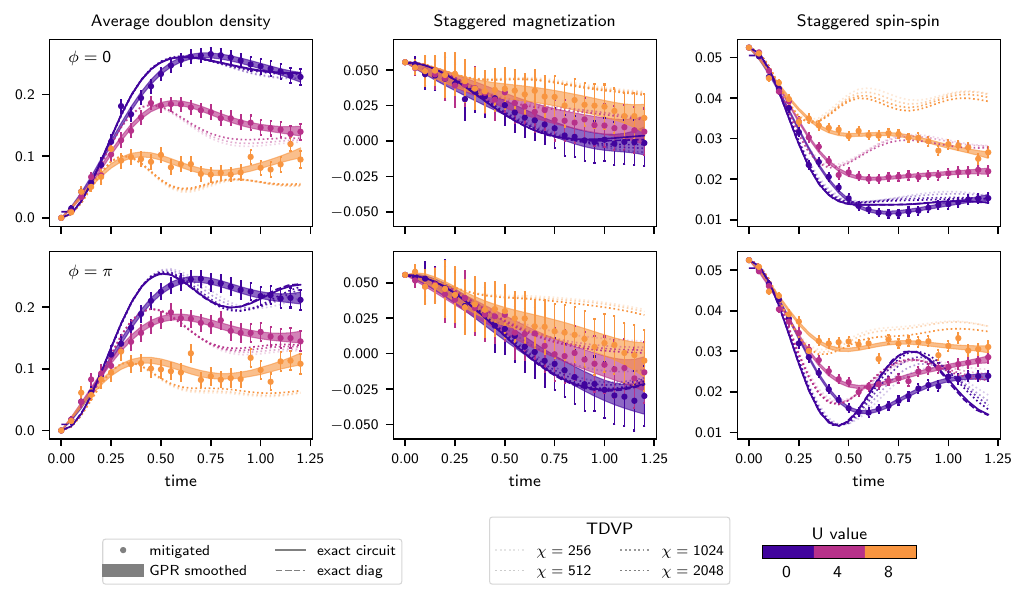}
    \caption{\justifying Global observables evolved from the $6\times 6$ initial state with two holons. Note that the tiny values of the staggered magnetisation are expected due to the initial singlet configuration. The revival seen in the staggered spin-spin observable indicates that the system is far away from a steady state (equilibrated).}
\label{fig:6b6global}
\end{figure}

\clearpage

\subsection{Thermalisation}
\label{subsec:app:thermalisation}

As we have seen so far from the global observables on different initial states, the behaviour of signals between them is qualitatively similar, showing, e.g. an increase in the number of doublons as time progresses, seemingly approaching an asymptotic value depending on the interaction strength and the initial occupation.
Under generic conditions, it is expected \cite{Deutsch,Srednicki_1994, Gogolin2016equilibration} that the nonequilibrium dynamics relaxes to states where macroscopic quantities are stationary, universal with respect to initial conditions sharing the same global quantum numbers, and predictable through (quantum) statistical mechanics.
The eigenstate thermalisation hypothesis \cite{Rigol2008-pe, DAlessio2016chaos} provides a mechanism for this relaxation to happen. It implies that expectation values of all local observables reach their asymptotic values given by the \emph{diagonal ensemble} (in the eigenstate basis) for that initial state and this value is given by the expectation in the microcanonical ensemble on a narrow energy range around the energy of the initial state.

The timescale that we consider in our evolution is not enough to see full thermalisation, as several observables have nontrivial time dependence even at the largest time explored.
Nevertheless, the trend of approaching the thermal asymptotic value can be distinguished in terms of different initial quantum numbers. In this section, we explore the approach to thermalisation in the Fermi-Hubbard model as we change the number of electrons. We do this by looking at observables obtained by averaging five different initial states with the same number of holons, ranging from one to six, placed on random positions on an otherwise N\'{e}el state with zero flux.

\subsubsection{Doublon Number}

\begin{figure}[!htbp]
  \centering

    \begin{subfigure}[t]{0.43\linewidth}
      \centering
      \includegraphics[width=\linewidth]{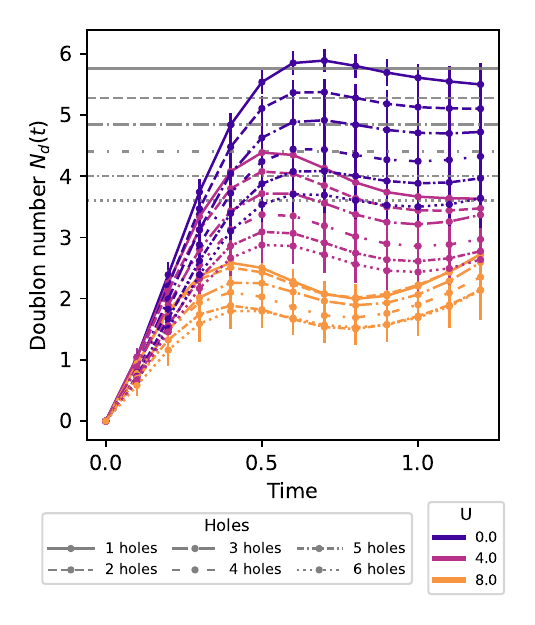}
      \caption{\justifying Doublon number for the $5\times5$ system. Grey lines show the long-time asymptotic values of doublon number for the free fermion system, $N_\uparrow N_\downarrow / L$.}
      \label{fig:doublon_tds}
    \end{subfigure}
    \hfill
    \begin{subfigure}[t]{0.53\textwidth}
        \centering
        \includegraphics[width=\linewidth]{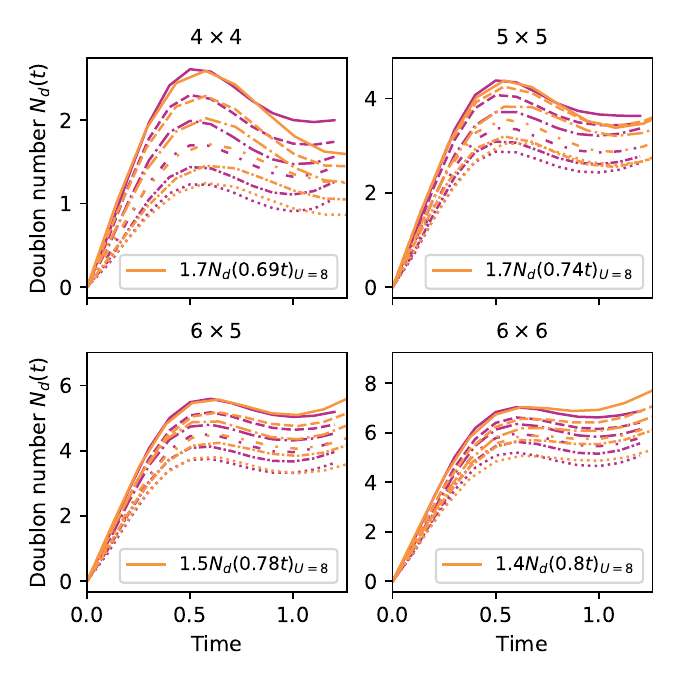}
        \caption{\justifying Doublon number collapse for $4\times 4$, $5\times5$, $6\times 5$, and $6\times6$ system. Shown are the original $U=4$ line and the rescaled $U=8$ line. }
        \label{fig:doublon_rescaling}
    \end{subfigure}
  \caption{\justifying
  Doublon number. All lines are obtained by averaging over five initial configurations with a given number of randomly placed holons on top of the N\'eel background.}
  \label{fig:doublon_number}
\end{figure}

In Fig.~\ref{fig:doublon_tds}, we see the evolution of the doublon number (given by the density \cref{eq:doublon_density} times the system's size, $N_d = \bigl\langle \sum_j n_{j\uparrow}n_{j\downarrow}\bigr\rangle$) in time, for different hole-doped states in a $5\times 5$ system. It shows the expected dependence of $N_d(t)$ on $U$ (slower dynamics for larger repulsion) and doping (more holes in the initial state result in fewer doublons).
The asymptotic long-time values of $N_d(t)$ for the non-interacting case are given by $N_\uparrow N_\downarrow / L$, and are shown on the plot with grey lines.
The shape of the curves for $U=4$ and $U=8$ suggests the possibility of collapse upon rescaling, which has been investigated in Fig.~\ref{fig:doublon_rescaling}, where the $U=4$ curve is shown in its original form while the $U=8$ curve is rescaled along both axes, with coefficients dependent on the problem size.

We can understand this scaling as follows.
Consider the Fermi-Hubbard Hamiltonian $H=H_0+UH_I$, where $H_0$ is the non-interacting Hamiltonian and $H_I$ parametrises the interactions. Here we make explicit the dependence on the interaction parameter $U$. An operator $\mathcal{O}$ made of densities commutes with the interactions, so going to the interaction picture with respect to $H_I$ gives
\begin{align}
\langle \mathcal{O}(t)\rangle  =\left\langle e^{it(H_{0}+UH_{I})}\mathcal{O}e^{-it(H_{0}+UH_{I})}\right\rangle =\left\langle \mathcal{T}e^{i\int_{0}^{t}H_{0}(\tau)}\mathcal{O}\mathcal{T}e^{-i\int_{0}^{t}H_{0}(\tau)}\right\rangle
\end{align}
where $H_{0}(\tau)=e^{iU\tau H_{I}}H_{0}e^{-iU\tau H_{I}}$. Now using
the fermion commutation relations we have
\begin{align}
e^{iU\tau H_{I}}c_{j\sigma}e^{-iU\tau H_{I}} =e^{iUtn_{j\sigma}n_{j\bar{\sigma}}}c_{j\sigma}e^{-iUtn_{j\sigma}n_{j\bar{\sigma}}}
 =c_{j\sigma}e^{-iUtn_{j\bar{\sigma}}}
\end{align}
so the time dependent Hamiltonian is $H_{0}(\tau)=\sum_{\langle jk\rangle}c_{k\sigma}^{\dagger}c_{j\sigma}e^{-iUt(n_{j\bar{\sigma}}-n_{k\bar{\sigma}})}$.
The expectation value takes the form
\begin{align}
\langle \mathcal{O}(t)\rangle_{U}=\left\langle \mathcal{T}e^{i\int_{0}^{t}\sum_{\langle jk\rangle}c_{k\sigma}^{\dagger}c_{j\sigma}e^{-iU\tau(n_{j\bar{\sigma}}-n_{k\bar{\sigma}})}d\tau}\mathcal{O}\mathcal{T}e^{-i\int_{0}^{t}\sum_{\langle jk\rangle}c_{k\sigma}^{\dagger}c_{j\sigma}e^{-iU\tau(n_{j\bar{\sigma}}-n_{k\bar{\sigma}})}d\tau}\right\rangle
\end{align}
where we use the subscript $U$ to emphasise that this is an expectation
of a time evolved operator where the interaction has strength
$U$. The integral in the time-ordered product can be written as
\begin{align}
\int_{0}^{t}\sum_{\langle jk\rangle}c_{k\sigma}^{\dagger}c_{j\sigma}e^{-iU\tau(n_{j\bar{\sigma}}-n_{k\bar{\sigma}})}d\tau=\frac{1}{U}\int_{0}^{Ut}\sum_{\langle jk\rangle}c_{k\sigma}^{\dagger}c_{j\sigma}e^{-is(n_{j\bar{\sigma}}-n_{k\bar{\sigma}})}ds
\end{align}
so expanding in powers of $1/U$, we have
\begin{align}
\langle \mathcal{O}(t)\rangle_{U} & =\langle\mathcal{O}(0)\rangle + i\frac{1}{U}\sum_{\langle jk\rangle,\sigma}\int_{0}^{Ut}e^{-is(n_{j\bar{\sigma}}-n_{k\bar{\sigma}})} d s\left\langle \left[c_{k\sigma}^{\dagger}c_{j\sigma},\mathcal{O}\right]\right\rangle \\
&+ \frac{1}{U^2}\sum_{\langle jk\rangle,\langle lm \rangle, \sigma}\int_{0}^{Ut} e^{-is(n_{j\bar{\sigma}}-n_{k\bar{\sigma}})} e^{-is'(n_{l\bar{\sigma}}-n_{m\bar{\sigma}})} ds ds' \left( \left\langle c^\dagger_{k\sigma}c_{j\sigma} \mathcal{O} c^\dagger_{m\sigma}c_{l\sigma}\right\rangle - \frac{1}{2} \left\langle \left\{ c^\dagger_{k\sigma}c_{j\sigma} c^\dagger_{m\sigma}c_{l\sigma} , \mathcal{O} \right\} \right\rangle \right) + O(t^3) \nonumber
\end{align}
For every operator $\mathcal{O}$ made of densities, $\mathcal{O}\ket{\psi} = n_{\mathcal{O}}\ket{\psi}$ where $n_{\mathcal{O}}$ is an integer which implies that the $O(t)$ term is always zero as $\langle[c_{k\sigma}^{\dagger}c_{j\sigma},\mathcal{O}]\rangle = \langle c_{k\sigma}^{\dagger}c_{j\sigma}\mathcal{O} \rangle - \langle \mathcal{O}c_{k\sigma}^{\dagger}c_{j\sigma} \rangle = n_{\mathcal{O}}(\langle c_{k\sigma}^{\dagger}c_{j\sigma} \rangle - \langle c_{k\sigma}^{\dagger}c_{j\sigma} \rangle) = 0$. This equation can be further simplified for our case, where the initial state is a product state with the number of doublons $\mathcal{O} = N_d = 0$. The expectation of the anti-commutator also becomes zero since $n_{\mathcal{O}} = 0$ and following the same argument as the commutator. Additionally, since the initial state is a product state, the only non-zero terms occur when $m=j$ and $l=k$. So, for early times, the doublon number satisfies the scaling relation
\begin{align*}
\langle N_{d}(t)\rangle_{U_{1}} & = \frac{1}{U_1^2}\sum_{\langle jk\rangle,\sigma}\int_{0}^{U_1t} e^{-i(s-s')(n_{j\bar{\sigma}}-n_{k\bar{\sigma}})} ds ds' \left\langle c^\dagger_{k\sigma}c_{j\sigma} N_d c^\dagger_{j\sigma}c_{k\sigma}\right\rangle  + O(t^3) \\
&= \left(\frac{U_2}{U_1}\right)^2 \frac{1}{U_2^2} \sum_{\langle jk\rangle,\sigma}\int_{0}^{U_2(U_1/U_2)t} e^{-i(s-s')(n_{j\bar{\sigma}}-n_{k\bar{\sigma}})} ds ds' \left\langle c^\dagger_{k\sigma}c_{j\sigma} N_d c^\dagger_{j\sigma}c_{k\sigma}\right\rangle + O(t^3) \\
&= \left(\frac{U_2}{U_1}\right)^2 \left\langle N_d\left(\frac{U_1}{U_2} t\right) \right\rangle_{U_2} + O(t^3).
\end{align*}

We observe this same scaling relation with renormalised parameters, shown in the insets of Fig.~\ref{fig:doublon_rescaling}. Note that the collapse deteriorates for larger times $t>1$.

\begin{figure}[!htbp]
    \centering
    \begin{subfigure}{0.44\textwidth}
        \centering

        \includegraphics[width=\linewidth]{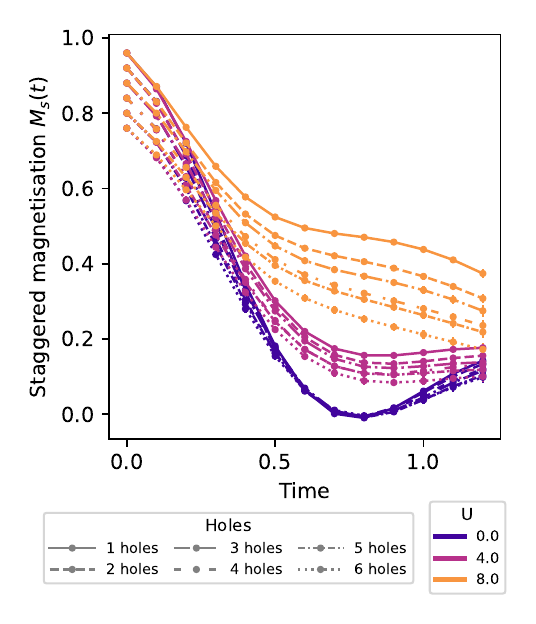}
        \caption{\justifying Staggered magnetisation for the $5\times 5$ system, shown for different interactions and dopings.}
        \label{fig:stag_mag5x5}
    \end{subfigure}
    \
    \begin{subfigure}{0.473\textwidth}
        \centering
        \includegraphics[width=\linewidth]{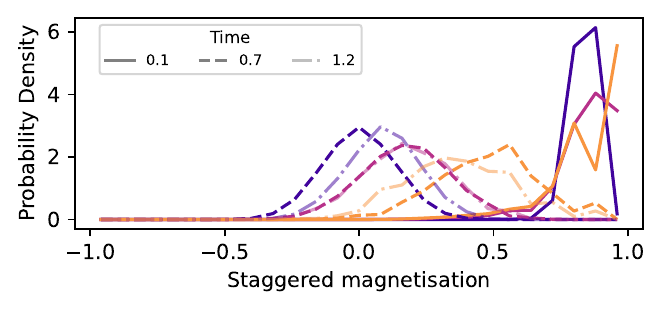}
        \caption{\justifying Evolution of the staggered magnetisation distribution, averaged over all one hole states.}
        \label{fig:stag_mag_dist_time}
        \includegraphics[width=\linewidth]{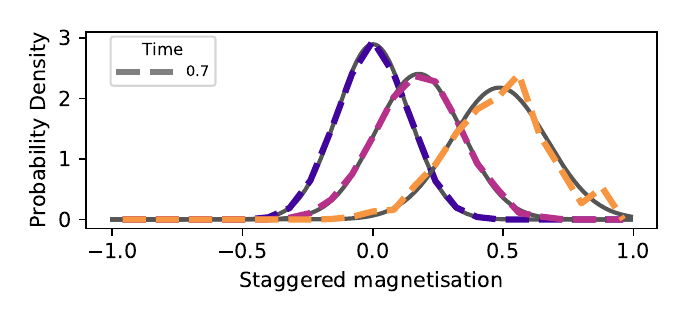}
        \caption{\justifying Staggered magnetisation distribution quickly acquires \mbox{nearly Gaussian shape (grey line), same initial state as in (b).}}
        \label{fig:stag_mag_dist_gaussian}
    \end{subfigure}
    \
    \begin{subfigure}[t]{0.49\textwidth}
        \centering
        \includegraphics[width=\linewidth]{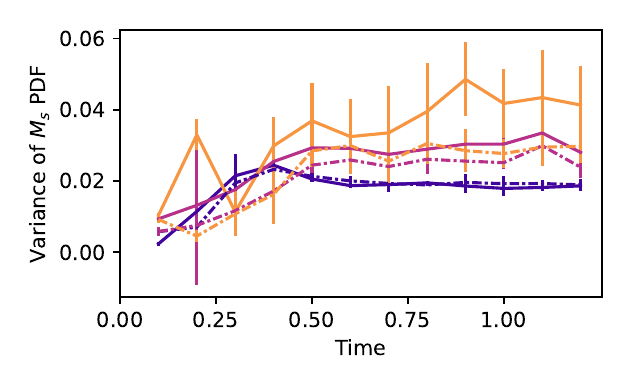}
        \caption{\justifying Evolution of the variance of the staggered magnetisation distribution, shown for initial states with one and five holons.}
        \label{fig:stag_mag_dist_sigma}
    \end{subfigure}
    \hfill
    \begin{subfigure}[t]{0.49\textwidth}
        \centering
        \includegraphics[width=\linewidth]{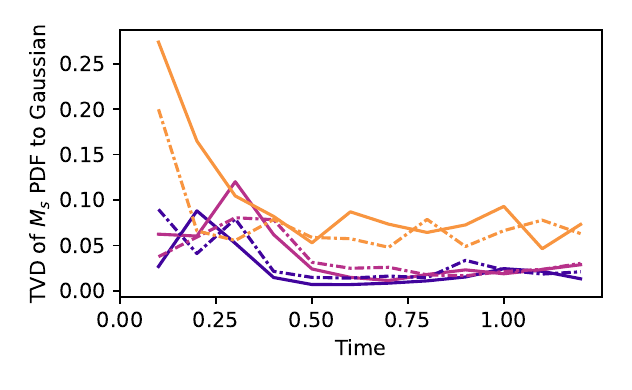}
        \caption{\justifying Total Variational Distance between the staggered magnetisation distribution and a normal distribution with the same mean and variance.}
        \label{fig:stag_mag_dist_tvd}
    \end{subfigure}
    \caption{\justifying
    Staggered magnetisation
    for $5\times 5$ system, for various numbers of holes on top of the N\'eel background, averaged over five random states for each number of holes.}
    \label{fig:stag_mag_doublon}
\end{figure}


\subsubsection{Staggered Magnetisation}

In the spin sector, we study the staggered magnetisation \cref{eq:stag_mag} which is shown in Fig.~\ref{fig:stag_mag5x5}  as a function of doping for different values of interactions $U=0,4,8$ in a $5\times 5$ system, from error-mitigated experimental results (see \cref{sec:error_mitigation}). As we have observed before,  increasing the interaction strength $U$ slows down the dynamics of the system. The magnetic order generally decreases with time as the system evolves, while a lower number of electrons leads monotonically to a lower magnetisation. The degree to which the reduction in magnetisation happens at a fixed time as we vary the doping is controlled by the interaction.

All lines show non-trivial behaviour far from infinite-time thermal values ($\langle M_s\rangle_\infty=0$ for the even number of sites $L$ and $\langle M_s\rangle_\infty=(N_\uparrow-N_\downarrow)/L$ for odd system size $L=L_x\times L_y$).

A more refined observable allowing one to track the system's departure from the antiferromagnetic order is the probability distribution of $M_s$ defined as~\cite{collura2020order}
\begin{equation}
    P_{M_s}(m,t) = \langle\delta(M_s-m)\rangle\, ,
\end{equation}
where $m$ are the eigenvalues of $M_s$, whose details are presented in~Fig.~\ref{fig:stag_mag_dist_time}.
In particular, Fig.~\ref{fig:stag_mag_dist_time} shows the profile of $P_{M_s}(m,t)$ at different moments of time and for various values of $U$.
These plots illustrate in more detail how larger values of $U$ slow down the spin exchange by suppressing the doublon formation.

We approximate the evolution of $P_{M_s}(m,t)$ by the Gaussian distribution
\begin{equation}
\label{eq:pms_gaussian}
P_{M_s}(m, t) =
\frac{1}{\sqrt{2\pi \sigma(t)}}
\exp \left\{ -\frac{[m - M_s(t)]^{2}}{2\sigma(t)} \right\}.
\end{equation}
where the variance $\sigma(t)$ is a (time-dependent) fitting parameter.

The distributions $P_{M_s}(m,t)$ rapidly approach the Gaussian shape \eqref{eq:pms_gaussian}, as is evident from Fig.~\ref{fig:stag_mag_dist_gaussian}.
The variances of distributions from Fig.~\ref{fig:stag_mag_dist_time} are shown in Fig.~\ref{fig:stag_mag_dist_sigma} from which the dependence on interaction and doping can be inferred.
Stronger interactions maintain configurations with large staggered magnetization, leading to a broader distribution.
Increasing the number of holes introduces mobile charge carriers that disrupt long-range spin correlations, accelerating relaxation.
The convergence of experimental probability distributions $P_{M_s}(m,t)$ to the Gaussian shape can be characterised by their total variation distance, which is shown in Fig.~\ref{fig:stag_mag_dist_tvd}.
Larger values of $U$ lead to a slower convergence toward the Gaussian distribution, with $U=8$ not showing signs of convergence, a feature that remains consistent across different system sizes.
We observe that while the variance is not strongly dependent on the number of holes present, it does have a monotone dependence on the interaction strength.

\subsubsection{Inverse Participation Ratio\label{sec:ipr}}

So far, we have discussed the approach to thermalisation through the behaviour of local few-body observables. In this section, we discuss a more distributed measure of thermalisation, the inverse participation ratio (IPR) \cite{Kramer1993localization, DAlessio2016chaos}.
This measure, first used in the study of Anderson localisation of non-interacting fermions \cite{Thouless1974electrons, Kramer1993localization, Evers2008anderson}, quantifies the degree to which the wavefunction is spread in some particular basis.
For a state $\ket{\psi}=\sum_i c_i\ket{i}$ expanded in the basis $|i\rangle$, the IPR is defined as
\begin{equation}
    \label{eq:ipr}
    \mathrm{IPR} = \sum_i |c_i|^4 = \sum_i p_i^2.
\end{equation}
where $p_i$ is the probability of observing the outcome $i$ measuring the state $|\psi\rangle$ in the basis $|i\rangle$.
For a fully delocalised state, corresponding to an equal superposition over all possible basis states, the IPR becomes $1/D$ where $D$ is the Hilbert space dimension. On the other hand, for a state with support on a single basis state, the IPR takes the value one.
Since in a given basis $\mathrm{IPR}=\sum_i(\rho^2)_{ii}$, it can be viewed as a basis-dependent ``classical'' version of purity $\mathrm{Tr}(\rho^2)$, which always provides a lower bound on it.

Calculating the full IPR defined in~\cref{eq:ipr} requires a full characterisation of the probabilities $p_i$, which is unfeasible at the system sizes that we consider.   Instead, we study a \emph{marginal} version of the IPR, which for a state  $\ket{\psi}=\sum_{ij} c_{ij}\ket{i}_A\ket{j}_B$ defined on a bipartition $A$ and $B$
corresponds to the usual IPR over the marginalised probabilities,~i.e.
\begin{equation}
    \mathrm{IPR}_A := \sum_i p_i^2, \qquad p_i := \sum_j |c_{ij}|^2.
\end{equation}
For a marginal on $m$ sites, this quantity can be estimated experimentally with sample overhead that scales at most with the subsystem dimension $4^m$, providing a measure of thermalisation in that particular basis as well as a lower bound on the subsystem's purity.
As suggested by classical state vector simulations on smaller systems, the marginal IPR exhibits behaviour qualitatively similar to that of the full-system IPR: decay in time as the state becomes delocalised in the Fock basis, with decay rates that can be related to other observables like time-dependent holon number as described below.

For the Fermi--Hubbard model, we construct several versions of the marginal inverse participation ratio (IPR) using different single-site bases, each corresponding to a distinct assignment of spin and charge quantum numbers to the original local Fock states $ \{\ket{0}, \ket{\uparrow}, \ket{\downarrow}, \ket{\uparrow\downarrow}\}$.
The full single-site IPR employs the complete alphabet $\{0,\uparrow,\downarrow,\uparrow\downarrow\}$ to characterise the basis at each site.
The charge version reduces this alphabet to $\{H,S,D\}$ by merging the spin-resolved singly occupied states into a single ``singly-occupied'' class $S$, while keeping the empty (hole, $H$) and doubly occupied (doublon, $D$) states distinct.
Conversely, the spin version isolates the spin degrees of freedom by merging the empty and doublon states into a single ``spinless'' category $\varnothing$, while preserving the distinction between up ($U$) and down ($D$) singly occupied states.
These mappings define coarse-graining functions $C(\cdot)$ and $S(\cdot)$ on the single-site states:
\begin{subequations}
\begin{alignat}{9}
    &C(0)=H,\quad C(\uparrow)=C(\downarrow)=S,\quad C(\uparrow\downarrow)=D,\\
& S(0)=S(\uparrow\downarrow)=\varnothing,\quad S(\uparrow)=U,\quad S(\downarrow)=D.
\end{alignat}
\end{subequations}

For a single-site marginal, the three IPR variants are:
\begin{subequations}
\label{eq:IPR_one}
\begin{alignat}{20}
    \mathrm{IPR}_1^{\mathrm{full}}
        &= p_0^2 &&+ p_{\uparrow}^2 &&+ p_{\downarrow}^2 &&+ p_{\uparrow\downarrow}^2\,, \\
     \mathrm{IPR}_1^{\mathrm{charge}}
        &= p_H^2 &&+ p_S^2 &&+ p_D^2
        &&= p_0^2 + (p_{\uparrow} + p_{\downarrow})^2 + p_{\uparrow\downarrow}^2\,, \\
    \mathrm{IPR}_1^{\mathrm{spin}}
        &= p_\varnothing^2 &&+ p_U^2 &&+ p_D^2
        &&= (p_0 + p_{\uparrow\downarrow})^2 + p_{\uparrow}^2 + p_{\downarrow}^2\,.
\end{alignat}
\end{subequations}

These definitions naturally extend to $m$-site marginals. Let $p_{i_1\ldots i_m}$ denote the joint probability of observing the configuration $(i_1,\ldots,i_m)$ in the original four-state Fock basis across $m$ sites. The $m$-site IPRs are defined as the sum of squared probabilities of the relevant marginal distributions. For the full IPR, this uses the original four-state Fock basis; for the charge and spin variants, it uses coarse-grained probabilities obtained by summing over all fine-grained configurations consistent with the charge ($C$) or spin ($S$) mappings:
\begin{subequations}
\label{eq:IPR_m}
\begin{align}
\mathrm{IPR}_m^{\mathrm{full}} &= \sum_{i_1,\ldots,i_m \in \{0,\uparrow,\downarrow,\uparrow\downarrow\}} \left(p_{i_1\ldots i_m}\right)^2, \\
\mathrm{IPR}_m^{\mathrm{charge}} &= \sum_{c_1,\ldots,c_m \in \{H,S,D\}} \left( \sum_{\substack{i_1,\ldots,i_m \\ C(i_k)=c_k \,\forall k}} p_{i_1\ldots i_m} \right)^{\!2}, \\
\mathrm{IPR}_m^{\mathrm{spin}} &= \sum_{s_1,\ldots,s_m \in \{\varnothing,U,D\}} \left( \sum_{\substack{i_1,\ldots,i_m \\ S(i_k)=s_k \,\forall k}} p_{i_1\ldots i_m} \right)^{\!2}.
\end{align}
\end{subequations}

\begin{figure}[!htbp]
  \centering
    \includegraphics[width=0.98\linewidth]{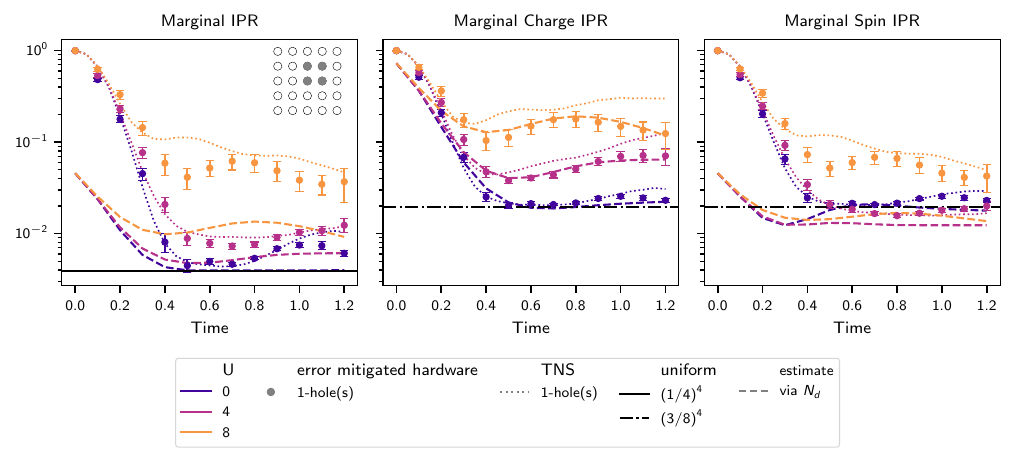}
    \caption{\justifying
  Marginal inverse participation ratios (IPRs) for an $m=4$-site region of the $5\times5$ system, as defined in \cref{eq:IPR_m}, averaged over all single-hole initial states.
We show the marginal IPR in the full local Fock basis $\{0,\uparrow,\downarrow,\uparrow\downarrow\}$, in the charge basis $\{H,S,D\}$ (empty, singly occupied, doublon), and in the spin basis $\{\varnothing,U,D\}$ (spinless, up, down).
Markers denote experimental data; curves indicate theoretical predictions (TN simulations and the independent-site IID approximation based on the doublon density).
Horizontal lines mark the baseline values for an on-site distribution with equal weights $p_i=1/4$ (which translates to $(1/4)^m$ for the full IPR and $(3/8)^m$ for the charge and spin IPRs).
The IID approximation reproduces the marginal charge IPR very accurately across times and interaction strengths, but provides only a loose description of the full and spin IPRs, especially at larger $U$, indicating that local charge statistics are closer to an independent-site picture than the corresponding spin and full Fock configurations.
  }
  \label{fig:mipr}
\end{figure}

\begin{figure*}[!htbp]
    \centering
    \includegraphics[width=\linewidth]{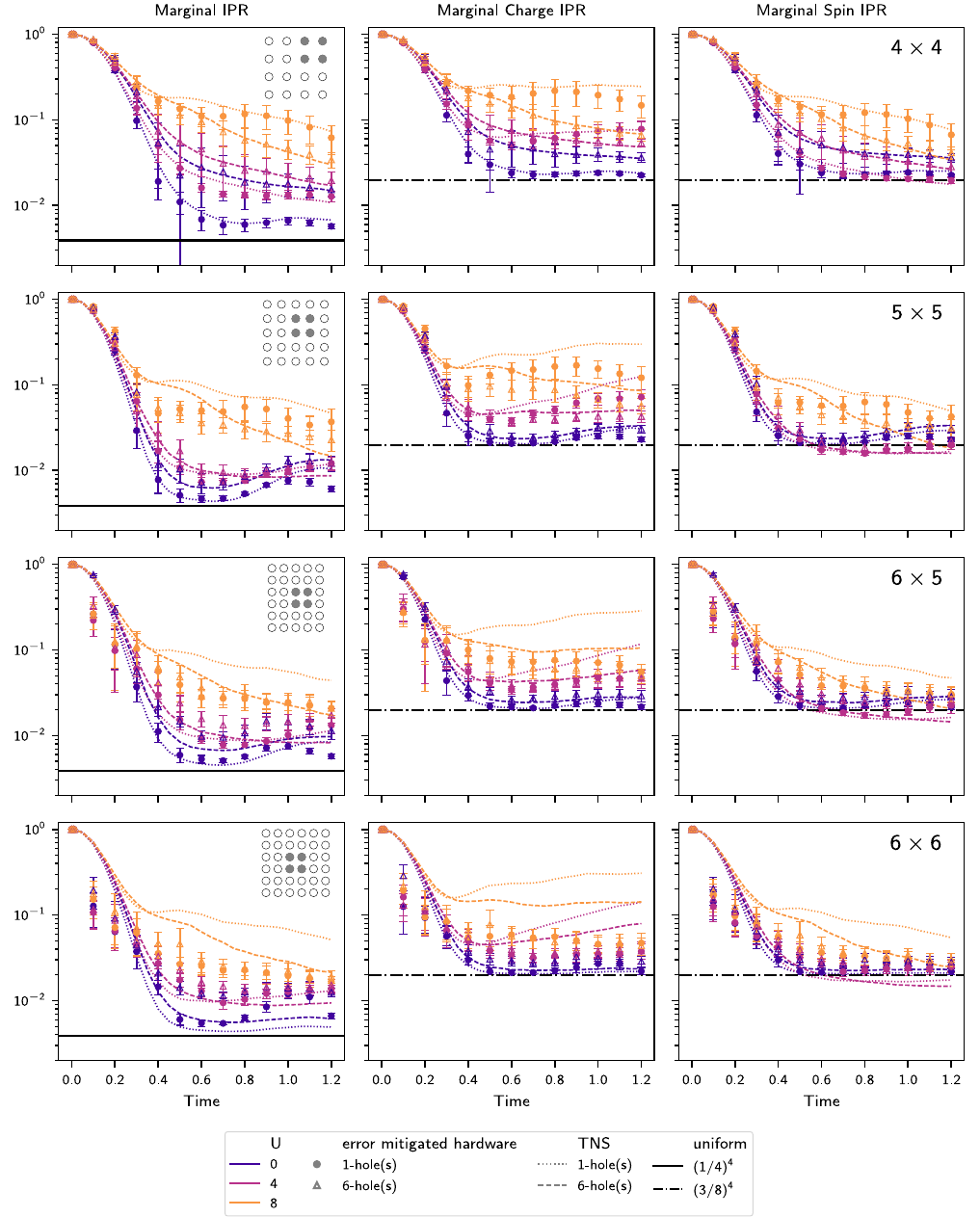}
    \caption{\justifying
    Full, charge, and spin versions of the marginal inverse participation ratio (IPR) for an $m=4$-site region and varying system's size. While the general trend of slower thermalisation with increasing interaction remains, the quality of the signal degrades with increasing system's size. }
    \label{fig:ipr_system_size}
\end{figure*}

All three IPR variants are directly computable from (mitigated) experimental measurement shots: one first constructs the $m$-site marginal by counting the relative frequencies of observed configurations, and then -- depending on the desired IPR -- either uses the raw Fock outcomes (for the full IPR) or applies the charge ($C$) or spin ($S$) mapping to each site before aggregating probabilities and evaluating the sum of squared frequencies. The resulting IPRs for 4-site marginals are shown in Figs.~\ref{fig:mipr} and \ref{fig:ipr_system_size}.

The absolute lower bound for the full mIPR is obtained when each mode on each site is occupied with probability $1/4$, so for an $m$-site marginal IPR this floor is given by $(1/4)^m$.
The value of the charge and spin IPRs in this case would asymptote to $(3/8)^m$ (while they are lower bounded by $(1/3)^m$).

Under the independent and identically distributed (IID)-site approximation, the instantaneous doublon count $N_d(t)$ can be used to estimate all $m$-site probabilities required for the marginal IPRs.
The number of sites occupied with spin up and down fermions (but not doublons), as well as holons at a given moment $t$ can be found as
\begin{align}
    N_0(t) = L - N_{\uparrow} - N_{\downarrow} + N_d(t)\,,\quad
    N_{\uparrow}(t) = N_{\uparrow} - N_d(t)\,, \qquad
    N_{\downarrow}(t) = N_{\downarrow} - N_d(t)\,.
\end{align}
where $L$ is the system's size. Dividing by $L$ gives the corresponding probabilities $p_{\alpha}(t)=N_{\alpha}(t)/L$ for $\alpha\in\{0,\uparrow,\downarrow,\uparrow\downarrow\}$ gives the full $m$-site marginal IPR estimate:
\begin{equation}\label{eq:IPR_fock}
    \mathrm{IPR}_{m, \rm est}^{\mathrm{full}}(t) :=
        \left[
        \left(\tfrac{N_H(t)}{L}\right)^2 + \left(\tfrac{N_\uparrow(t)}{L}\right)^2 + \left(\tfrac{N_\downarrow(t)}{L}\right)^2 + \left(\tfrac{N_d(t)}{L}\right)^2
        \right]^m.
\end{equation}

The same doublon-dependent probabilities can be coarse-grained to obtain the charge and spin variants.
For the masked charge alphabet $\{H,S,D\}$, the counts are
\begin{equation}
    N_H(t)=N_0(t), \quad N_S(t)=L-N_H(t)-N_D(t)=(N_{\uparrow}+N_{\downarrow})-2N_d(t), \quad N_D(t)=N_d(t)\,,
\end{equation}
leading to
\begin{equation}\label{eq:IPR_charges}
    \mathrm{IPR}_{m,\rm est}^{\mathrm{charge}}(t)
        := \left[ \left(\tfrac{N_H(t)}{L}\right)^2
        + \left(\tfrac{N_S(t)}{L}\right)^2
        + \left(\tfrac{N_D(t)}{L}\right)^2\right]^m\,.
\end{equation}

For the masked spin alphabet $\{\varnothing,U,D\}$, the empty and doublon sites are combined into a single spinless category with
\begin{equation}
    N_{\varnothing}(t) = N_0(t) + N_d(t)
        = L - (N_{\uparrow} + N_{\downarrow}) + 2N_d(t)\,,
\end{equation}
while the singly occupied states remain spin-resolved as
$N_U(t)=N_{\uparrow}-N_d(t)$ and $N_D(t)=N_{\downarrow}-N_d(t)$.
This yields
\begin{equation}
    \label{eq:IPR_spins}
    \mathrm{IPR}_{m, \rm{est}}^{\mathrm{spin}}(t)
        := \left[ \left(\tfrac{N_{\varnothing}(t)}{L}\right)^2
        + \left(\tfrac{N_U(t)}{L}\right)^2
        + \left(\tfrac{N_D(t)}{L}\right)^2 \right]^m\,.
\end{equation}

IPRs for a $4$-site marginal, consisting of four qubits forming a square including the centre of the $5\times5$ lattice, as well as their estimates obtained from the IID approximation, are shown in Fig.~\ref{fig:mipr}.
This very simple IID approximation provides a good estimate for the charge IPR but not for the full or spin IPR.
Similar to other observables, marginal IPR approaches the asymptotic value more slowly for larger values of $U$.

The IPR on either the full system or on a marginal is also the probability that a pair of samples, each drawn independently from the same system, match as bit strings on all sites.  We can express this as the probability of an event  $\textrm{IPR}_m = \textrm{Pr}(\textrm{spins and charges match})$, with the understanding that it refers to a single region $m$ of sites and the ``match'' is an event on two samples from the system.  With this we can apply the definition of conditional probability $P(A|B) = P(A,B)/P(B)$ to interpret IPR ratios,
\begin{alignat}{9}
    \label{eq:ipr_ratio_full_charge}
    \operatorname{Pr}(\text{spins match}|\text{charges match}) &= \dfrac{\mathrm{IPR_m}}{\mathrm{IPR}_m^{\mathrm{charge}}}\,,
    \\
    \label{eq:ipr_ratio_full_spin}
    \operatorname{Pr}(\text{charges match}|\text{spins match}) &= \dfrac{\mathrm{IPR}_m}{\mathrm{IPR}_m^{\mathrm{spin}}}\,.
\end{alignat}

The conditional probabilities allow for examination of thermalisation in the various sectors of spin and charge separately.
For example, the decay of $\textrm{Pr}(\textrm{spins match} | \textrm{charges match})$ quantifies the mixing of spins that is not due to mixing between modes of different charges, i.e. it isolates the thermalisation of spin states of single occupancy sites.
Similarly, the decay of $\textrm{Pr}(\textrm{charges match} | \textrm{spins match})$ quantifies transport in the charge sector --- a smaller value of $\textrm{Pr}(\textrm{charges match} | \textrm{spins match})$  means that a site with a charge carrier is equally likely to be a doublon or holon, indicating thermalisation in the positions of charge carriers. These interpretations are also supported by considering the additive contributions these IPR ratios make to the classical R\'enyi-2 entropy $S_2 = -\ln(\textrm{IPR})$.

\begin{figure}[!htbp]
  \centering

  \vspace{-0.5em}

  \includegraphics[width=0.98\linewidth]{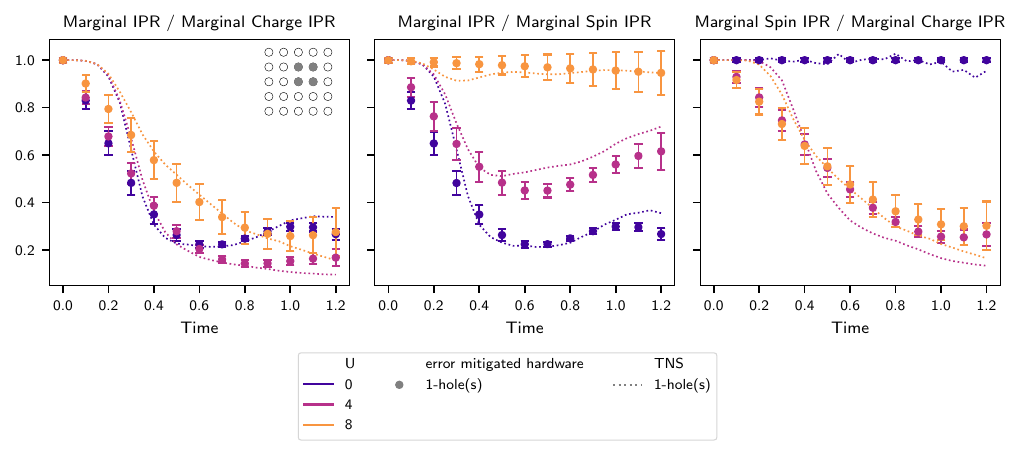}

  \caption{\justifying
    Ratios of marginal inverse participation ratios ($\mathrm{IPR}_m^{(\mathrm{full})}(t)/\mathrm{IPR}_m^{(\mathrm{charge})}(t)$, $\mathrm{IPR}_m^{(\mathrm{full})}(t)/\mathrm{IPR}_m^{(\mathrm{spin})}(t)$, and $\mathrm{IPR}_m^{(\mathrm{spin})}(t)/\mathrm{IPR}_m^{(\mathrm{charge})}(t)$) defined in \cref{eq:IPR_m}
    and shown in Fig.~\ref{fig:mipr} for the $5\times5$ system, with all
    interaction strengths $U$ overlaid in each panel.
    At $U=0$ the $\mathrm{IPR}_m^{(\mathrm{spin})}(t)/\mathrm{IPR}_m^{(\mathrm{charge})}(t)$ ratio remains close to unity, indicating that spin and charge delocalise at similar rates in the non-interacting case.
    For $U>0$ the ratio $\mathrm{IPR}_m^{(\mathrm{spin})}(t)/\mathrm{IPR}_m^{(\mathrm{charge})}(t)$ decays significantly below one, showing that the spin IPR approaches its baseline value more rapidly than the charge IPR, and hence that the spin sector locally equilibrates faster than the charge sector.
  }
  \label{fig:ipr_ratios_allU}
\end{figure}

Results for the IPR ratios and the associated predictions are shown in Fig.~\ref{fig:ipr_ratios_allU}.
The first plot demonstrates spin dynamics and thermalisation in the directions of spin on single occupancy sites. 
We see a relatively weak dependence on $U$, consistent with the prediction that spin dynamics proceeds even in the presence of strong interactions.
The second plot demonstrates the transport of charge --- given that a site is a doublon or a holon, is it equally likely to be either of those, or does it show a high probability to be one or the other? 
We see at large $U$ the charge carrier ``remembers'' its identity, i.e. if a holon is seen somewhere, then a doublon is unlikely to be seen on that site in another sample at the same time.

The ratio $\mathrm{IPR}_m^{(\mathrm{spin})}(t)/\mathrm{IPR}_m^{(\mathrm{charge})}(t)$, shown on the third plot of~\cref{fig:ipr_ratios_allU}, can be interpreted using the IID approximation above.
Since $\mathrm{IPR}_m^{(\mathrm{spin})}/\mathrm{IPR}_m^{(\mathrm{charge})}\approx (N_HN_D) / (N_\uparrow N_\downarrow)$, this ratio is larger than one when the number of holons and doublons is larger than the number of singly occupied sites and smaller than one in the opposite case.
This ratio stays close to 1 in the non-interacting case and is less than one for positive values of $U$, agreeing with the fact that interactions reduce the formation of holon-doublon pairs.

Finally we explore the different IPR signals as a function of system's size. We find that the qualitative trends (weak dependence on doping and a slow down of thermalisation with increasing interaction) remain, but the quality of the signal deteriorates. This is shown in \cref{fig:ipr_system_size}.

\subsection{Holon Pair Dynamics} \label{sec:holon_pair}

\begin{figure}[!htbp]
    \centering
    \includegraphics[width=1\linewidth]{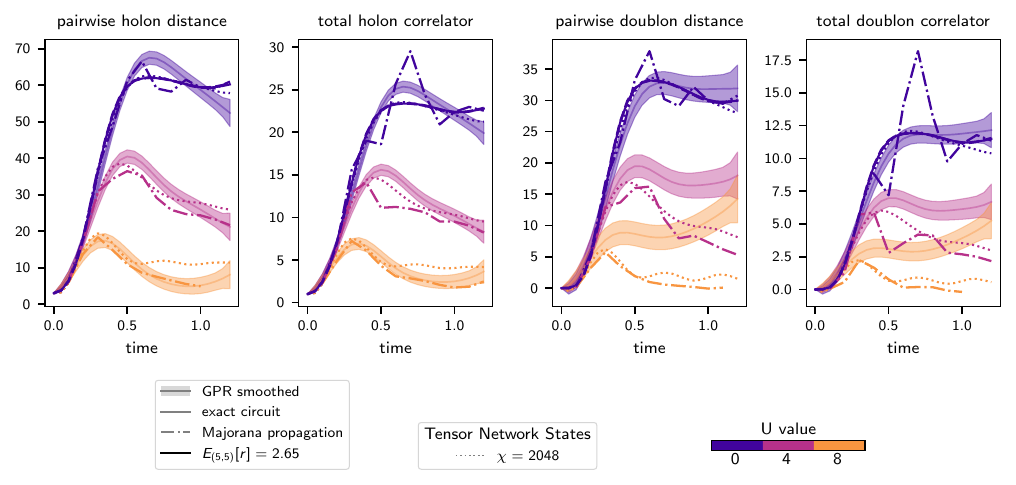}
    \caption{\justifying Dynamics of the pairwise holon distance $\sum_{i <j} r_{ij} \langle h_i h_j\rangle$, pairwise doublon distance $\sum_{i <j} r_{ij} \langle d_i d_j\rangle$, total holon correlator $\sum_{i <j}  \langle h_i h_j\rangle$ and total doublon correlator $\sum_{i <j}  \langle d_i d_j\rangle$ for the $5\times5$ \neel ordered state with a pair of holons placed $r=3$ apart. The hardware data is shown after GPR smoothing for the sake of clarity.}\label{fig:pairwise_holon_distance}
\end{figure}
\begin{figure}[!htbp]
    \centering
    \includegraphics[width=1\linewidth]{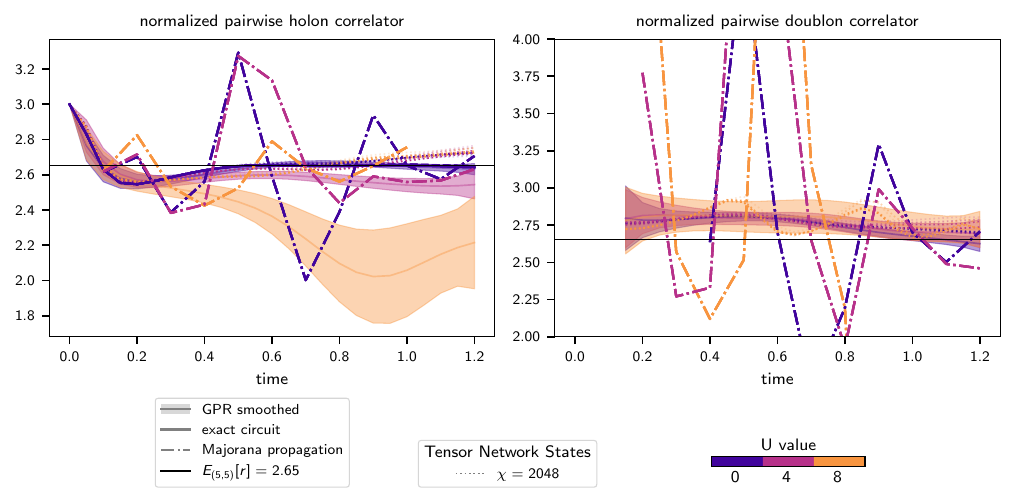}
    \caption{\justifying Dynamics of the average pairwise holon distance $\langle r_h\rangle $ (left) and average pairwise doublon distance $\langle r_d\rangle $ (right) of a $5\times5$ \neel ordered state with a pair of holons placed $r=3$ apart. The hardware data is shown after GPR smoothing for the sake of clarity.} \label{fig:average_pairwise_distance}
\end{figure}

Of particular interest in Fermi-Hubbard type models is the pairwise behaviour of holons due to their joint interaction with the background spin-order \cite{weng1999mean, wang1989effective, schrieffer1989dynamic, nielsen2022nonequilibrium, ji2021coupling}. The anti-ferromagnetic order of the system is predicted to be disturbed by the motion of a holon, creating a region of spin disturbance around each holon, which is predicted to introduce interactions between holons mediated by this disturbance. Because of the proliferation of holon-doublon pairs in these simulations, it is difficult to observe the relative motion of individual pairs of holons. However, we can investigate the dynamics of the collective pairwise behaviour of holons.

We consider input states which contain a pair of separated holons. We observe the dynamics of the average pairwise distance between holons:
\begin{equation}
\langle r_h \rangle  :=\frac{\sum_{i<j} r_{ij} \langle h_i h_j\rangle}{\sum_{i<j} \langle h_i h_j\rangle} ,
\end{equation}
where $h_i = (1-n_{i\uparrow})(1-n_{i\downarrow})$ and $r_{i,j}$ is the euclidean distance between site $i$ and $j$,
and similarly, the average pairwise distance between doublons:
\begin{equation}
\langle r_d \rangle := \frac{\sum_{i<j} r_{ij} \langle d_i d_j\rangle}{\sum_{i<j} \langle d_i d_j\rangle},
\label{eqn:rd}
\end{equation} where $d_i = n_{i\uparrow} n_{i\downarrow}$. For an $n\times  m$ square grid, the average Euclidean distance $r$ between a pair of randomly chosen non-overlapping points is given by the average distance between distinct cells:
\begin{equation}
\mathbb{E}_{n,m}[r]:=
\frac{
2\sum_{dx=1}^{n-1}\sum_{dy=1}^{m-1}(n-dx)(m-dy)\sqrt{dx^{2}+dy^{2}}
\;+\;m\sum_{dx=1}^{n-1} dx\,(n-dx)
\;+\;n\sum_{dy=1}^{m-1} dy\,(m-dy)
}{\binom{nm}{2}},
\end{equation}
where here the first term counts instances of pairwise distances not aligned along the $x$ or $y$ axis, and the second and third terms count all instances of pairwise distances aligned along an $x$ or $y$ axis.

Fig.~\ref{fig:average_pairwise_distance} shows the average pairwise distances between holons and between doublons when a pair of holons is introduced to a \neel ordered background on a $5\times5$ system.
We see that for $\langle r_d \rangle$ there is a strong tendency, at every $U$, for the dynamics to approach the average distance between two randomly located points in a $5\times5$ lattice, $E_{5,5}[r]=2.65$ -- illustrated in the right hand inset in Fig.~\ref{fig:average_pairwise_distance}. This suggests that holon-doublon pairs are appearing in a spatially homogeneous way.

The $\langle r_h \rangle$, on the other hand, behaves quite differently, with the hardware signal starting at $3$, corresponding to the distance between the initial holon pair, and then trending away from $\mathbb{E}_{5,5}[r]$ by an increasing degree with the strength of $U$. The fact that this behaviour does not appear for the average pairwise doublon distance is suggestive that this trend is not a consequence of an inhomogeneity in the background holon proliferation, but rather due to the evolution of the average relative position of the pair of holons consistently introduced at $t=0$. This implies that the hardware signal shows that, as $U$ increases, the pair of holons has a stronger tendency to move towards one another. We note that the hardware signal and TDVP signal disagree quite dramatically, even at early times, and predict qualitatively different physical behaviour -- the TDVP data suggests that $U$ has little to no impact on the average holon distance, and that the average holon distance trends towards random placement of holons for all $U$. We note that for $U=0$, the circuit dynamics match both the exact circuit and the exact underlying FLO dynamics at late times, whereas TDVP fails to do so for this signal. We note that the dramatic variation in size of the error bars on these signals is due to normalisation of $\langle r_h\rangle$ and $\langle r_d \rangle$, which can diverge when there is a small number of doublons or holons in the system. This sensitivity may contribute to the large divergence from TDVP -- and we expect that, at mid to late times, Trotter error will play a significant role in the $U=8$ curve.

In Figure~\ref{fig:pairwise_holon_distance} we show the numerator and denominator of $\langle r_h \rangle$ ($\langle r_d \rangle$) -- the pairwise holon (doublon) distance and total holon (doublon) correlator respectively. Strikingly we see a dramatic divergence between hardware data and TDVP for doublons, yet general agreement in their ratio. This suggests a divergence in their predicted proliferation rate, but an agreement in the prediction of homogeneity of proliferation. In contrast, the holon variants more closely conform to both TDVP and MP than their doublon counterparts, but diverge in their ratio. Despite the strong general agreement between MP and TDVP in pairwise holon (doublon) distance and total holon (doublon), MP shows no consistent trend corroborating either TDVP or hardware in $\langle r_h \rangle$. All of this suggests that the target signal is too sensitive to resolve accurately and it is not definitive which method (hardware or TDVP) gives a more accurate qualitative account of the physics.


\clearpage

\section{Quantum Computing Hardware}\label{sec:hardware}

All experimental results were produced using superconducting quantum computers provided by Google. All devices have degree four connectivity arranged in a diamond lattice. The native two-qubit gates employed are CZ gates using tunable couplers.

\begin{figure*}[!htbp]
    \centering
        \includegraphics[width=0.85\linewidth]{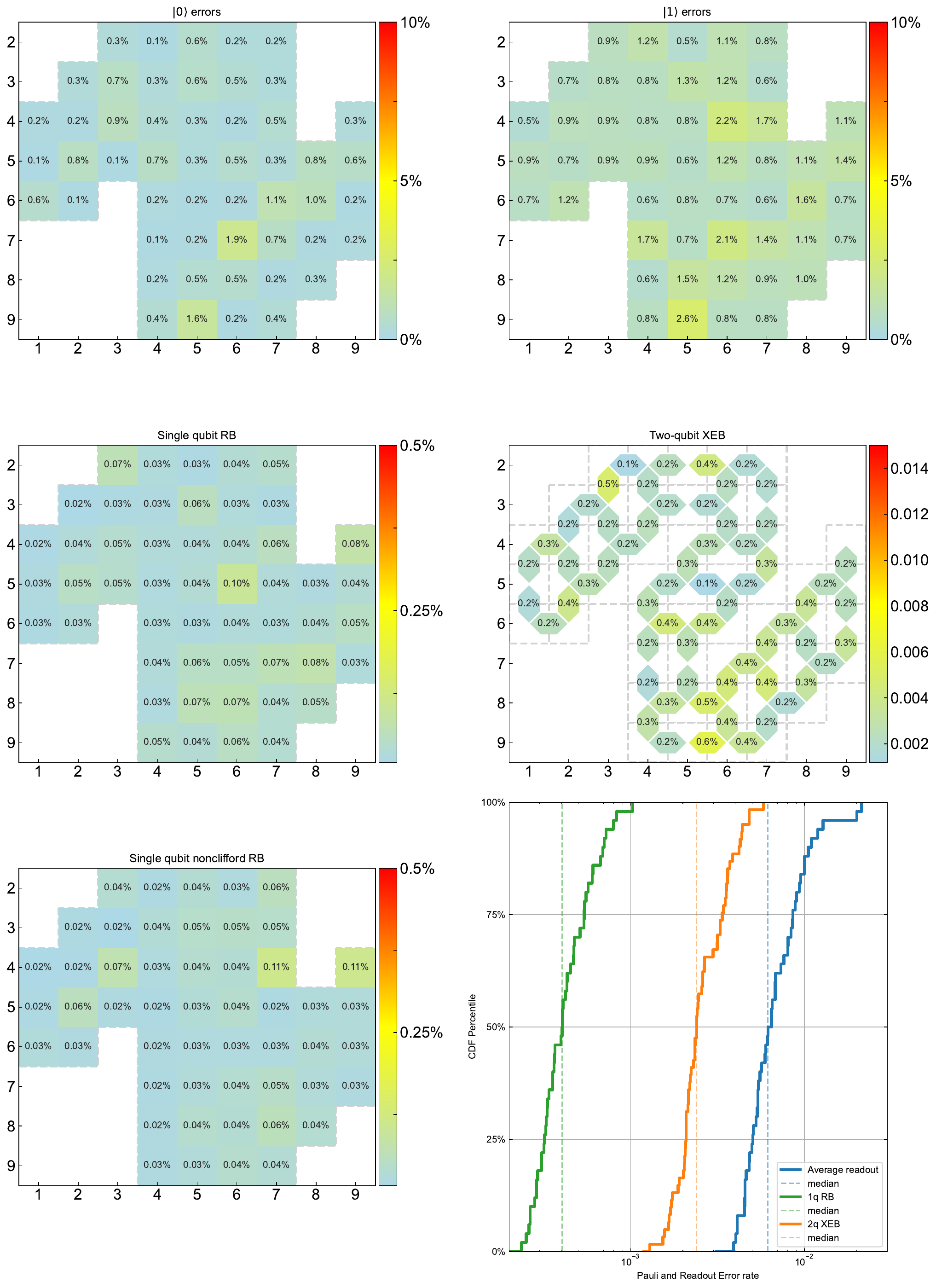}
    \hfill
   \caption{\justifying Benchmark of the qubits and gates used on the 72 qubit superconducting quantum chip using randomised benchmarking (RB) and cross-entropy benchmarking (XEB).} \label{fig:72chip}
\end{figure*}

\begin{figure*}[!htbp]
    \centering
        \includegraphics[width=0.85\linewidth]{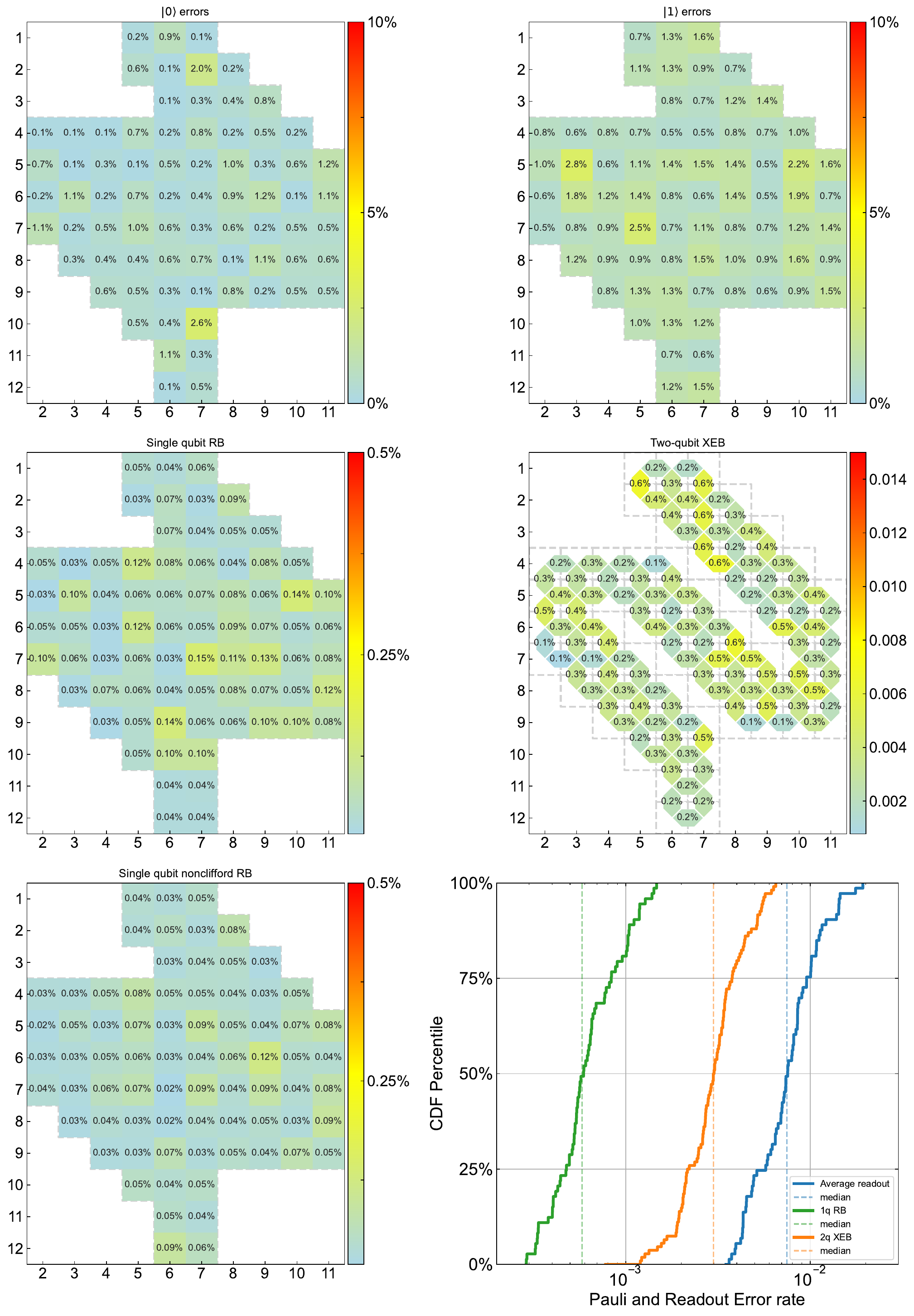}
    \caption{\justifying Benchmark of the qubits and gates used on the 105 qubit superconducting quantum chip using randomised benchmarking (RB) and cross-entropy benchmarking (XEB).}
    \label{fig:105chip}
    
\end{figure*}

The experiments were run on $72$-qubit and $105$-qubit Willow quantum processors. The qubit layouts, the two-qubit gates calibrated, and typical gate fidelities for a given calibration are shown in Figs.~\ref{fig:105chip} and~\ref{fig:72chip}. Gate errors are reported as Pauli errors as described in \cite{arute2019quantum}. Both devices were accessed using Google's Quantum Engine. Device performance was comparable to that reported in \cite{google2025quantum}.

\clearpage

\section{Circuit Descriptions} \label{sec:circuits}

\subsection{Fermion to Qubit Mapping}

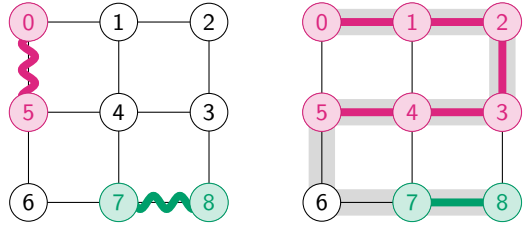
\begin{figure}[!htbp]
    \centering
\begin{tikzpicture}[every node/.style={font=\sffamily},scale=0.6]

    \begin{scope}[shift={(-6.5,0)}]

    \foreach \x in {0,1,2} {
        \draw (2*\x,0)--(2*\x,-4);
        \draw (0,-2*\x)--(4,-2*\x);
    }

    \draw[snake it, ibmMagenta, line width=3] (0,0)--(0,-2);
    \draw[snake it, armygreen, line width=3] (2,-4)--(4,-4);

    \foreach \x in {0,1,2} {
        \foreach \y[evaluate={\xi=int((-1)^\y*\x + (1-(-1)^\y))},evaluate={\l=int(3*\y+\x)}] in {0,1,2} {
        \node at (2*\xi,-2*\y) [circle,draw,fill=white,minimum size=5mm] {};
        \node at (2*\xi,-2*\y) {\l};
        }
    }

    \node at (0,0) [circle,draw=ibmMagenta,minimum size=5.1mm,fill=ibmMagenta!20!white] {};
    \node at (0,0) [ibmMagenta] {0};
    \node at (0,-2) [circle,draw=ibmMagenta,minimum size=5.1mm,fill=ibmMagenta!20!white] {};
    \node at (0,-2) [ibmMagenta] {5};

    \node at (2,-4) [circle,draw=armygreen,minimum size=5.1mm,fill=armygreen!20!white] {};
    \node at (2,-4) [armygreen] {7};
    \node at (4,-4) [circle,draw=armygreen,minimum size=5.1mm,fill=armygreen!20!white] {};
    \node at (4,-4) [armygreen] {8};
    \end{scope}

    \draw[rounded corners, line width=10, opacity=0.15] (0,0)--(4,0)--(4,-2)--(0,-2)--(0,-4)--(4,-4);

    \foreach \x in {0,1,2} {
        \draw (2*\x,0)--(2*\x,-4);
        \draw (0,-2*\x)--(4,-2*\x);
    }

    \draw[ibmMagenta,line width=3] (0,0)--(4,0)--(4,-2)--(0,-2);
    \draw[armygreen,line width=3] (2,-4)--(4,-4);

    \foreach \x in {0,1,2} {
        \foreach \y[evaluate={\xi=int((-1)^\y*\x + (1-(-1)^\y))},evaluate={\l=int(3*\y+\x)}] in {0,1,2} {
        \node at (2*\xi,-2*\y) [circle,draw,fill=white,minimum size=5mm] {};
        \node at (2*\xi,-2*\y) {\l};
        }
    }

    \node at (0,0) [circle,draw=ibmMagenta,minimum size=5.1mm,fill=ibmMagenta!20!white] {};
    \node at (0,0) [ibmMagenta] {0};
    \node at (0,-2) [circle,draw=ibmMagenta,minimum size=5.1mm,fill=ibmMagenta!20!white] {};
    \node at (0,-2) [ibmMagenta] {5};
    \node at (2,0) [circle,draw=ibmMagenta,minimum size=5.1mm,fill=ibmMagenta!20!white] {};
    \node at (2,0) [ibmMagenta] {1};
    \node at (2,-2) [circle,draw=ibmMagenta,minimum size=5.1mm,fill=ibmMagenta!20!white] {};
    \node at (2,-2) [ibmMagenta] {4};
    \node at (4,0) [circle,draw=ibmMagenta,minimum size=5.1mm,fill=ibmMagenta!20!white] {};
    \node at (4,0) [ibmMagenta] {2};
    \node at (4,-2) [circle,draw=ibmMagenta,minimum size=5.1mm,fill=ibmMagenta!20!white] {};
    \node at (4,-2) [ibmMagenta] {3};

    \node at (2,-4) [circle,draw=armygreen,minimum size=5.1mm,fill=armygreen!20!white] {};
    \node at (2,-4) [armygreen] {7};
    \node at (4,-4) [circle,draw=armygreen,minimum size=5.1mm,fill=armygreen!20!white] {};
    \node at (4,-4) [armygreen] {8};
\end{tikzpicture}
    \caption{\justifying How the support of hopping terms maps from fermionic modes (left) to qubits (right) under Jordan-Wigner with the snake ordering (highlighted on the qubit side). In the qubit picture, a hopping term is supported on the qubits corresponding to the modes and all qubits between them in the ordering. Horizontal hopping terms (green) are supported on 2 qubits, but vertical hops (magenta) can be supported on longer chains of qubits.
    }
    \label{fig:snake}
\end{figure}

We use the Jordan-Wigner (JW) mapping to encode the fermionic system.
Each mode is assigned to a qubit and Fock states are represented by computational basis states with $\ket{1}$ ($\ket{0}$) denoting that a qubit's mode is occupied (unoccupied). The modes are assigned an ordering, the same applying to the qubits, and the creation/annihilation operators map to qubit operators
\begin{equation}
c_j^\dagger = \sigma^+_j\prod_{i<j}Z_i,\quad c_j = \sigma^-_j\prod_{i<j}Z_i,\quad n_j=\frac{1}{2}(1-Z_j),
\end{equation}
with $\sigma^\pm_j=(X\mp iY)/2$.
This structure ensures that the encoded fermionic operators satisfy the canonical anticommutation relations.
We use the ``snake'' ordering for modes on a square lattice where modes are ordered along rows from left-to-right and right-to-left in an alternating manner -- see Fig.~\ref{fig:snake}.
As we do not include any simulations of hopping between spin sectors, one can equivalently think of the encoding of the spin-up and spin-down sectors one after the other on a single JW chain, or otherwise think of them as encoded on two separate JW representations -- this does not practically change any of the circuits.

\subsubsection{Encoded Interactions}
Hopping interactions in the presence of a magnetic flux are represented as
\begin{equation}
e^{i\phi_{ij}}c_i^\dagger c_j+e^{-i\phi_{ij}}c^\dagger_jc_i=\frac{1}{2}(\cos\phi_{ij}(X_iX_j+Y_iY_j)+\sin\phi_{ij}(Y_iX_j-X_iY_j))\prod_{i<k<j}Z_{k}.
\end{equation}
where the Peierls phases $\phi_{ij}$ for each edge define the magnetic flux as detailed in \cref{sec:model_details}. In practice we only set $\phi_{ij}\in\{0,\pi\}$, meaning
\begin{equation}
e^{i\phi_{ij}}c_i^\dagger c_j+e^{-i\phi_{ij}}c^\dagger_jc_i=\begin{cases}
    \frac{1}{2}(X_iX_j+Y_iY_j)\prod_{i<k<j}Z_{k}&:\;\phi_{ij}=0,\\
    -\frac{1}{2}(X_iX_j+Y_iY_j)\prod_{i<k<j}Z_{k}&:\;\phi_{ij}=\pi.\\
\end{cases}
\end{equation}
In each case the hopping term is implemented with the same circuit (see \cref{sec:gate decomp}) except for a sign flip on the time parameter $t$ in the $\phi_{ij}=\pi$ case.

The onsite Coulomb terms of the Fermi-Hubbard Hamiltonian are represented on qubits as
\begin{equation}
n_{i,\uparrow}n_{i,\downarrow}=\frac{1}{4}(1-Z_{i,\uparrow}-Z_{i,\downarrow}+Z_{i,\uparrow}Z_{i,\downarrow}).
\end{equation}
When summed over every spin pair, $\sum_i(1-Z_{i,\uparrow}-Z_{i,\downarrow})$ is a constant when restricted to a fixed particle number subspace, so this part may be ignored and it suffices to encode the Coulomb term as
\begin{equation}
n_{i,\uparrow}n_{i,\downarrow}=\frac{1}{4}Z_{i,\uparrow}Z_{i,\downarrow}.
\end{equation}


 \subsubsection{Layout On Device}
\begin{figure}[!htbp]
    \centering
    \includegraphics[width=0.6\linewidth]{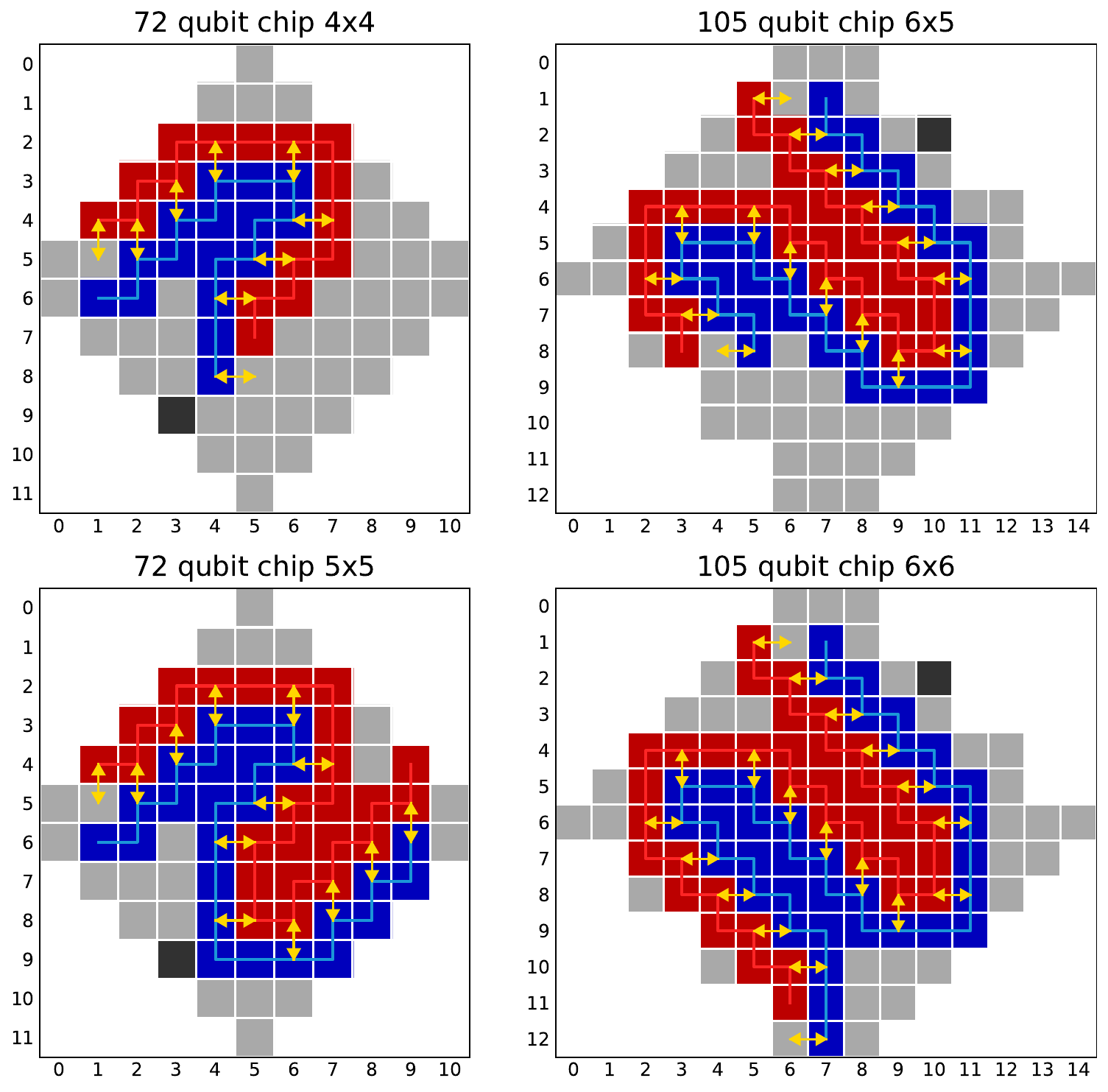}
    \caption{\justifying Layout of qubits on the chips used. Qubits highlighted in red (blue) correspond to encoded spin-up (spin-down) fermionic modes. Dead qubits are highlighted in black. The Jordan-Wigner ordering is indicated by the lines running through each spin sector. The yellow arrows show where FSWAPs are applied to bring together qubits corresponding to the spin-up and spin-down parts of a site in preparation for evolving under the Coulomb interactions. FSWAPs suffice instead of SWAPs, as the encoded Coulomb interaction is a $ZZ$ Pauli on the two qubits.}
    \label{fig:bigzags}
\end{figure}
Despite the grid structure of the simulated system, the operations for hopping between neighbouring sites (discussed in \cref{sec:trotter step}) only act between qubits adjacent in the JW ordering. This means that the qubits for each spin sector need only be connected to their neighbours in the ordering, with the additional restriction that qubit pairs corresponding to spin pairs be nearby to minimise the cost of bringing them together for the 2-qubit Coulomb interaction. Fig.~\ref{fig:bigzags} shows the on-chip qubit layouts used in this work.

\subsection{Second-Order Trotter Circuit}\label{sec:trotter step}

Our second-order Trotter step of time $t$ has the basic structure
\begin{equation}
    S_2(t) = \prod_{h\in \mathcal H}^\rightarrow e^{-ih\frac{t}{2}}\prod_{o\in\mathcal O}e^{-iot}\prod_{h\in\mathcal H}^\leftarrow e^{-ih\frac{t}{2}},
\end{equation}
where $\mathcal H$ is the set of hopping interactions in the Hamiltonian and $\mathcal O$ is the set of Coulomb interactions.
The first round of hopping interactions is applied using an FSWAP network, which will be explained below and the second round of hopping interactions is applied in reverse order as indicated by the arrows.
The Coulomb terms are implemented in parallel, see \cref{sec:gate decomp} for the circuit. Due to the layout, not all spin pairs are adjacent with respect to the device connectivity, so swaps must be applied to bring them together (see Fig.~\ref{fig:bigzags}). It suffices to use FSWAPs to do this because all that is required is the transformation of a $ZZ$ on two non-adjacent qubits to one on two adjacent qubits. This reduces the gate cost as the FSWAP can be applied with fewer gates than a qubit SWAP. After this, the Coulomb terms are applied and the FSWAPs are undone.

\subsubsection{Swap Network}
As noted in Fig.~\ref{fig:snake}, horizontal hopping terms are supported on only two qubits and can be implemented efficiently (see \cref{sec:gate decomp}) while vertical hopping terms may involve long strings of $Z$ operators, which can be costly. One way around this is fermionic swap networks \cite{kivlichan18}, in which FSWAP operations
\begin{equation}
\textup{FSWAP}=\begin{pmatrix}1&0&0&0\\0&0&1&0\\0&1&0&0\\0&0&0&-1\end{pmatrix}=\textup{SWAP}\cdot\textup{CZ},
\end{equation}
are used to rearrange fermionic modes to shorten these $Z$ strings.

FSWAP transforms Paulis under conjugation like
\begin{equation}
XI\leftrightarrow ZX,\quad YI\leftrightarrow ZY,\quad ZI\leftrightarrow IZ,
\end{equation}
and in the fermionic picture, when acting on  qubits adjacent in the ordering, it transforms encoded fermionic operators like
\begin{equation}
c_j\leftrightarrow c_{j+1},\quad c^\dagger_j\leftrightarrow c^\dagger_{j+1},\quad n_j\leftrightarrow n_{j+1},
\end{equation}
swapping the positions of the encoded modes.
From this, one sees that if two modes are made adjacent via FSWAPs, then the encoded hopping term between them is transformed into a two-qubit term with no $Z$ string.

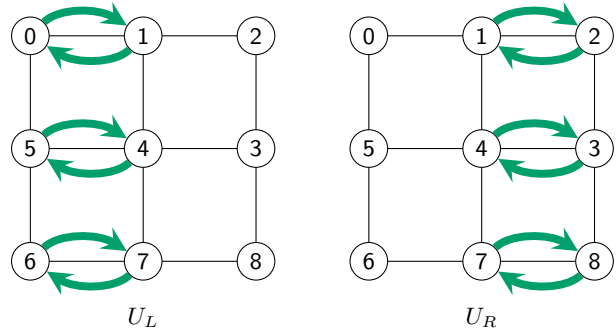
\begin{figure}[!htbp]
    \centering
\begin{tikzpicture}[every node/.style={font=\sffamily},scale=0.75]
    \foreach \x in {0,1,2} {
        \draw (0,-2*\x)--(4,-2*\x);
        \draw (2*\x,0)--(2*\x,-4);
    }

    \foreach \x in {0,1,2} {
        \foreach \y[evaluate={\xi=int((-1)^\y*\x + (1-(-1)^\y))},evaluate={\l=int(3*\y+\x)}] in {0,1,2} {
        \node (\l) at (2*\xi,-2*\y) [circle,draw,fill=white,minimum size=5mm] {};
        \node at (2*\xi,-2*\y) {\l};
        }
    }

    \begin{scope}[->,armygreen,>=stealth, line width=3]
        \draw (0) .. controls  (.5,.5) and (1.2,.5) .. (1);
        \draw (5) .. controls  (.5,.5-2) and (1.2,.5-2) .. (4);
        \draw (6) .. controls  (.5,.5-4) and (1.2,.5-4) .. (7);
        \draw (1) .. controls  (1.5,-.5) and (.8,-.5) .. (0);
        \draw (4) .. controls  (1.5,-.5-2) and (.8,-.5-2) .. (5);
        \draw (7) .. controls  (1.5,-.5-4) and (.8,-.5-4) .. (6);
    \end{scope}

    \node at (2,-5) {$U_L$};

\begin{scope}[shift={(6,0)}]
 \foreach \x in {0,1,2} {
        \draw (0,-2*\x)--(4,-2*\x);
        \draw (2*\x,0)--(2*\x,-4);
    }

    \foreach \x in {0,1,2} {
        \foreach \y[evaluate={\xi=int((-1)^\y*\x + (1-(-1)^\y))},evaluate={\l=int(3*\y+\x)}] in {0,1,2} {
        \node (\l) at (2*\xi,-2*\y) [circle,draw,fill=white,minimum size=5mm] {};
        \node at (2*\xi,-2*\y) {\l};
        }
    }

    \begin{scope}[->,armygreen,>=stealth, line width=3]
        \draw (1) .. controls  (2+.5,.5) and (2+1.2,.5) .. (2);
        \draw (4) .. controls  (2+.5,.5-2) and (2+1.2,.5-2) .. (3);
        \draw (7) .. controls  (2+.5,.5-4) and (2+1.2,.5-4) .. (8);
        \draw (2) .. controls  (2+1.5,-.5) and (2+.8,-.5) .. (1);
        \draw (3) .. controls  (2+1.5,-.5-2) and (2+.8,-.5-2) .. (4);
        \draw (8) .. controls  (2+1.5,-.5-4) and (2+.8,-.5-4) .. (7);
    \end{scope}

    \node at (2,-5) {$U_R$};
\end{scope}
\end{tikzpicture}
    \caption{\justifying Action of $U_L$ and $U_R$ on a $3\times3$ grid. The green arrows indicate FSWAPs.
    }
    \label{fig:ULUR}
\end{figure}

Given a lattice of dimensions $L_x$ and $L_y$, we define as $U_L$ ($U_R$) (see Fig.~\ref{fig:ULUR}) the circuit that FSWAPs even (odd) columns with those on the right, starting with column $0$ ($1$). As discussed in \cref{sec:gate decomp}, where a hopping term needs to be applied to the pairs of modes being swapped, this can be done by simultaneously applying both hop and FSWAP without increasing the circuit cost. For any given Trotter step size $t$, the hopping terms are implemented — for time $t/2$ — according to the following:

\textbf{Even $L_x$:}
\begin{enumerate}
\item Implement all vertical hopping terms between adjacent modes.
\item Apply $U_L$
\item Implement all vertical hopping terms between adjacent modes.
\item Apply $U_R$
\item Repeat from 2. terminating when all hopping terms have been implemented.
\end{enumerate}

\textbf{Odd $L_x$:}
\begin{enumerate}
\item Implement all vertical hopping terms between adjacent modes.
\item Apply $U_L$.
\item Apply $U_R$.
\item Repeat from 1. terminating when all hopping terms have been implemented.
\end{enumerate}

In both cases, the FSWAPs in the $U_L$ and $U_R$ circuits should be replaced with a merged FSWAP and hop if the hopping term between the mode pair has not yet been implemented.

\begin{figure}[!htbp]
    \centering
\begin{tikzpicture}[scale=0.85,>=stealth]
\begin{scope}[scale=0.48,every node/.style={scale=0.6,font=\fontsize{14}{16.8},font=\sffamily}]

\draw[dashed] (0,1.1)--(4*6.2+4,1.1);
\node[fill=white,font=\large] at (2*6.2+2,1.1) {Step 1 $(t_1)$};

\begin{scope}[black!20!white]
    \foreach \x in {0,1,2} {
        \draw[line width=2] (0,-2*\x)--(4,-2*\x);
        \draw (2*\x,0)--(2*\x,-4);
    }

    \draw[line width=2] (4,0)--(4,-2);
    \draw[line width=2] (0,-2)--(0,-4);
\end{scope}
    \foreach \x in {0,1,2} {
        \foreach \y[evaluate={\xi=int((-1)^\y*\x + (1-(-1)^\y))},evaluate={\l=int(3*\y+\x)}] in {0,1,2} {
        \node (\l) at (2*\xi,-2*\y) [circle,draw,fill=white,minimum size=5mm] {};
        \node at (2*\xi,-2*\y) [] {\l};
        }
    }

    \begin{scope}[ibmMagenta,line width=1.8,opacity=0.75]
        \draw[snake it] (2)--(3);
        \draw[snake it] (5)--(6);
    \end{scope}

    \draw[line width=2,->,>=stealth] (4.6,-2)--(5.6,-2);

\begin{scope}[shift={(6.2,0)}]
\begin{scope}[black!20!white]
    \foreach \x in {0,1,2} {
        \draw[line width=2] (0,-2*\x)--(4,-2*\x);
        \draw (2*\x,0)--(2*\x,-4);
    }

    \draw[line width=2] (4,0)--(4,-2);
    \draw[line width=2] (0,-2)--(0,-4);
\end{scope}
    \foreach \x in {0,1,2} {
        \foreach \y[evaluate={\xi=int((-1)^\y*\x + (1-(-1)^\y))},evaluate={\l=int(3*\y+\x)}] in {0,1,2} {
        \node (\l) at (2*\xi,-2*\y) [circle,draw,fill=white,minimum size=5mm] {};
        \node at (2*\xi,-2*\y) [] {\l};
        }
    }

    \begin{scope}[ibmMagenta,line width=1.8]
        \draw[snake it] (0)--(1);
        \draw[snake it] (5)--(4);
        \draw[snake it] (6)--(7);
    \end{scope}

    \begin{scope}[->,armygreen,>=stealth, line width=1.8]
        \draw (0) .. controls  (.5,.5) and (1.2,.5) .. (1);
        \draw (5) .. controls  (.5,.5-2) and (1.2,.5-2) .. (4);
        \draw (6) .. controls  (.5,.5-4) and (1.2,.5-4) .. (7);
        \draw (1) .. controls  (1.5,-.5) and (.8,-.5) .. (0);
        \draw (4) .. controls  (1.5,-.5-2) and (.8,-.5-2) .. (5);
        \draw (7) .. controls  (1.5,-.5-4) and (.8,-.5-4) .. (6);
    \end{scope}

    \draw[line width=2,->,>=stealth] (4.6,-2)--(5.6,-2);

\end{scope}

\begin{scope}[shift={(2*6.2,0)}]
\begin{scope}[black!20!white]
    \foreach \x in {0,1,2} {
        \draw[line width=2] (0,-2*\x)--(4,-2*\x);
        \draw (2*\x,0)--(2*\x,-4);
    }

    \draw[line width=2] (4,0)--(4,-2);
    \draw[line width=2] (0,-2)--(0,-4);
\end{scope}
    \foreach \x in {0,1,2} {
        \foreach \y[evaluate={\l=int(3*\y+\x)}] in {0,1,2} {
        \node (\x\y) at (2*\x,-2*\y) [circle,draw,fill=white,minimum size=5mm] {};
        }
    }
    \node (1) at (0,0) {1};
    \node (0) at (2,0) {0};
    \node (2) at (4,0) {2};
    \node (4) at (0,-2) {4};
    \node (5) at (2,-2) {5};
    \node (3) at (4,-2) {3};
    \node (7) at (0,-4) {7};
    \node (6) at (2,-4) {6};
    \node (8) at (4,-4) {8};

    \begin{scope}[ibmMagenta,line width=1.8]
        \draw[snake it] (10)--(20);
        \draw[snake it] (11)--(21);
        \draw[snake it] (12)--(22);
    \end{scope}

    \begin{scope}[->,armygreen,>=stealth, line width=1.8]
        \draw (10) .. controls  (2.5,.5) and (3.2,.5) .. (20);
        \draw (11) .. controls  (2.5,.5-2) and (3.2,.5-2) .. (21);
        \draw (12) .. controls  (2.5,.5-4) and (3.2,.5-4) .. (22);
        \draw (20) .. controls  (3.5,-.5) and (2.8,-.5) .. (10);
        \draw (21) .. controls  (3.5,-.5-2) and (2.8,-.5-2) .. (11);
        \draw (22) .. controls  (3.5,-.5-4) and (2.8,-.5-4) .. (12);
    \end{scope}

    \draw[line width=2,->,>=stealth] (4.6,-2)--(5.6,-2);
\end{scope}

\begin{scope}[shift={(3*6.2,0)}]
\begin{scope}[black!20!white]
    \foreach \x in {0,1,2} {
        \draw[line width=2] (0,-2*\x)--(4,-2*\x);
        \draw (2*\x,0)--(2*\x,-4);
    }

    \draw[line width=2] (4,0)--(4,-2);
    \draw[line width=2] (0,-2)--(0,-4);
\end{scope}
    \foreach \x in {0,1,2} {
        \foreach \y[evaluate={\l=int(3*\y+\x)}] in {0,1,2} {
        \node (\x\y) at (2*\x,-2*\y) [circle,draw,fill=white,minimum size=5mm] {};
        }
    }
    \node at (0,0) {1};
    \node at (2,0) {2};
    \node at (4,0) {0};
    \node at (0,-2) {4};
    \node at (2,-2) {3};
    \node at (4,-2) {5};
    \node at (0,-4) {7};
    \node at (2,-4) {8};
    \node at (4,-4) {6};

    \draw[ibmMagenta, snake it,line width=1.8] (01)--(02);
    \draw[ibmMagenta, snake it,line width=1.8] (20)--(21);
    \draw[line width=2,>=stealth,dotted] (4.6,-2)--(5.6,-2);
\end{scope}

\begin{scope}[shift={(4*6.2,0)}]
\begin{scope}[black!20!white]
    \foreach \x in {0,1,2} {
        \draw[line width=2] (0,-2*\x)--(4,-2*\x);
        \draw (2*\x,0)--(2*\x,-4);
    }

    \draw[line width=2] (4,0)--(4,-2);
    \draw[line width=2] (0,-2)--(0,-4);
\end{scope}
    \foreach \x in {0,1,2} {
        \foreach \y[evaluate={\l=int(3*\y+\x)}] in {0,1,2} {
        \node (\x\y) at (2*\x,-2*\y) [circle,draw,fill=white,minimum size=5mm] {};
        }
    }
    \node at (0,0) {1};
    \node at (2,0) {2};
    \node at (4,0) {0};
    \node at (0,-2) {4};
    \node at (2,-2) {3};
    \node at (4,-2) {5};
    \node at (0,-4) {7};
    \node at (2,-4) {8};
    \node at (4,-4) {6};

    \begin{scope}[ibmMagenta,line width=1.8]
        \draw[snake it] (00)--(10);
        \draw[snake it] (01)--(11);
        \draw[snake it] (02)--(12);
    \end{scope}

    \begin{scope}[->,armygreen,>=stealth, line width=1.8]
        \draw (00) .. controls  (.5,.5) and (1.2,.5) .. (10);
        \draw (01) .. controls  (.5,.5-2) and (1.2,.5-2) .. (11);
        \draw (02) .. controls  (.5,.5-4) and (1.2,.5-4) .. (12);
        \draw (10) .. controls  (1.5,-.5) and (.8,-.5) .. (00);
        \draw (11) .. controls  (1.5,-.5-2) and (.8,-.5-2) .. (01);
        \draw (12) .. controls  (1.5,-.5-4) and (.8,-.5-4) .. (02);
    \end{scope}

    \draw[line width=2,dotted] (4.6,-2)--(5.6,-2);
    \draw[line width=2,->,>=stealth] (4.6-6.2,-2)--(5.6-6.2,-2);

\end{scope}

\begin{scope}[shift={(0,-6.2)}]
\begin{scope}[black!20!white]
    \foreach \x in {0,1,2} {
        \draw[line width=2] (0,-2*\x)--(4,-2*\x);
        \draw (2*\x,0)--(2*\x,-4);
    }

    \draw[line width=2] (4,0)--(4,-2);
    \draw[line width=2] (0,-2)--(0,-4);
\end{scope}
    \foreach \x in {0,1,2} {
        \foreach \y[evaluate={\l=int(3*\y+\x)}] in {0,1,2} {
        \node (\x\y) at (2*\x,-2*\y) [circle,draw,fill=white,minimum size=5mm] {};
        }
    }
    \node at (0,0) {2};
    \node at (2,0) {1};
    \node at (4,0) {0};
    \node at (0,-2) {3};
    \node at (2,-2) {4};
    \node at (4,-2) {5};
    \node at (0,-4) {8};
    \node at (2,-4) {7};
    \node at (4,-4) {6};

    \begin{scope}[->,armygreen,>=stealth, line width=1.8]
        \draw (10) .. controls  (2.5,.5) and (3.2,.5) .. (20);
        \draw (11) .. controls  (2.5,.5-2) and (3.2,.5-2) .. (21);
        \draw (12) .. controls  (2.5,.5-4) and (3.2,.5-4) .. (22);
        \draw (20) .. controls  (3.5,-.5) and (2.8,-.5) .. (10);
        \draw (21) .. controls  (3.5,-.5-2) and (2.8,-.5-2) .. (11);
        \draw (22) .. controls  (3.5,-.5-4) and (2.8,-.5-4) .. (12);
    \end{scope}
    \draw[line width=2,->,>=stealth] (4.6,-2)--(5.6,-2);
    \draw[line width=2,->,>=stealth,dotted] (4.6-6.2,-2)--(5.6-6.2,-2);
\end{scope}

\begin{scope}[shift={(6.2,-6.2)}]
\begin{scope}[black!20!white]
    \foreach \x in {0,1,2} {
        \draw[line width=2] (0,-2*\x)--(4,-2*\x);
        \draw (2*\x,0)--(2*\x,-4);
    }

    \draw[line width=2] (4,0)--(4,-2);
    \draw[line width=2] (0,-2)--(0,-4);
\end{scope}
    \foreach \x in {0,1,2} {
        \foreach \y[evaluate={\l=int(3*\y+\x)}] in {0,1,2} {
        \node (\x\y) at (2*\x,-2*\y) [circle,draw,fill=white,minimum size=5mm] {};
        }
    }
    \node at (0,0) {2};
    \node at (2,0) {0};
    \node at (4,0) {1};
    \node at (0,-2) {3};
    \node at (2,-2) {5};
    \node at (4,-2) {4};
    \node at (0,-4) {8};
    \node at (2,-4) {6};
    \node at (4,-4) {7};

    \draw[ibmMagenta, snake it,line width=2] (01)--(02);
    \draw[ibmMagenta, snake it,line width=2] (20)--(21);
    \draw[line width=2,->,>=stealth] (4.6,-2)--(5.4,-2);
\end{scope}

\begin{scope}[shift={(2*6.2,-6.2)}]
\begin{scope}[black!20!white]
    \foreach \x in {0,1,2} {
        \draw[line width=2] (0,-2*\x)--(4,-2*\x);
        \draw (2*\x,0)--(2*\x,-4);
    }

    \draw[line width=2] (4,0)--(4,-2);
    \draw[line width=2] (0,-2)--(0,-4);
\end{scope}
    \foreach \x in {0,1,2} {
        \foreach \y[evaluate={\l=int(3*\y+\x)}] in {0,1,2} {
        \node (\x\y) at (2*\x,-2*\y) [circle,draw,fill=white,minimum size=5mm] {};
        \node (\x\y) at (2*\x,-2*\y) [circle,draw=ibmViolet,minimum size=8mm,line width=2] {};
        }
    }
    \node at (0,0) {2};
    \node at (2,0) {0};
    \node at (4,0) {1};
    \node at (0,-2) {5};
    \node at (2,-2) {3};
    \node at (4,-2) {4};
    \node at (0,-4) {8};
    \node at (2,-4) {6};
    \node at (4,-4) {7};

    \draw[line width=2,->] (4.8,-2)--(5.6,-2);
\end{scope}

\begin{scope}[shift={(3*6.2,-6.2)}]
\begin{scope}[black!20!white]
    \foreach \x in {0,1,2} {
        \draw[line width=2] (0,-2*\x)--(4,-2*\x);
        \draw (2*\x,0)--(2*\x,-4);
    }

    \draw[line width=2] (4,0)--(4,-2);
    \draw[line width=2] (0,-2)--(0,-4);
\end{scope}
    \foreach \x in {0,1,2} {
        \foreach \y[evaluate={\l=int(3*\y+\x)}] in {0,1,2} {
        \node (\x\y) at (2*\x,-2*\y) [circle,draw,fill=white,minimum size=5mm] {};
        }
    }
    \node at (0,0) {2};
    \node at (2,0) {0};
    \node at (4,0) {1};
    \node at (0,-2) {3};
    \node at (2,-2) {5};
    \node at (4,-2) {4};
    \node at (0,-4) {8};
    \node at (2,-4) {6};
    \node at (4,-4) {7};

    \draw[ibmMagenta, snake it,line width=2] (01)--(02);
    \draw[ibmMagenta, snake it,line width=2] (20)--(21);
    \draw[line width=2,->,>=stealth] (4.6,-2)--(5.6,-2);
\end{scope}

\begin{scope}[shift={(4*6.2,-6.2)}]
\begin{scope}[black!20!white]
    \foreach \x in {0,1,2} {
        \draw[line width=2] (0,-2*\x)--(4,-2*\x);
        \draw (2*\x,0)--(2*\x,-4);
    }

    \draw[line width=2] (4,0)--(4,-2);
    \draw[line width=2] (0,-2)--(0,-4);
\end{scope}
    \foreach \x in {0,1,2} {
        \foreach \y[evaluate={\l=int(3*\y+\x)}] in {0,1,2} {
        \node (\x\y) at (2*\x,-2*\y) [circle,draw,fill=white,minimum size=5mm] {};
        }
    }
    \node at (0,0) {2};
    \node at (2,0) {0};
    \node at (4,0) {1};
    \node at (0,-2) {3};
    \node at (2,-2) {5};
    \node at (4,-2) {4};
    \node at (0,-4) {8};
    \node at (2,-4) {6};
    \node at (4,-4) {7};

    \begin{scope}[->,armygreen,>=stealth, line width=1.8]
        \draw (10) .. controls  (2.5,.5) and (3.2,.5) .. (20);
        \draw (11) .. controls  (2.5,.5-2) and (3.2,.5-2) .. (21);
        \draw (12) .. controls  (2.5,.5-4) and (3.2,.5-4) .. (22);
        \draw (20) .. controls  (3.5,-.5) and (2.8,-.5) .. (10);
        \draw (21) .. controls  (3.5,-.5-2) and (2.8,-.5-2) .. (11);
        \draw (22) .. controls  (3.5,-.5-4) and (2.8,-.5-4) .. (12);
    \end{scope}
    \draw[line width=2,dotted] (4.6,-2)--(5.6,-2);
\end{scope}

\begin{scope}[shift={(0,-2*6.2)}]
\begin{scope}[black!20!white]
    \foreach \x in {0,1,2} {
        \draw[line width=2] (0,-2*\x)--(4,-2*\x);
        \draw (2*\x,0)--(2*\x,-4);
    }

    \draw[line width=2] (4,0)--(4,-2);
    \draw[line width=2] (0,-2)--(0,-4);
\end{scope}
    \foreach \x in {0,1,2} {
        \foreach \y[evaluate={\l=int(3*\y+\x)}] in {0,1,2} {
        \node (\x\y) at (2*\x,-2*\y) [circle,draw,fill=white,minimum size=5mm] {};
        }
    }
    \node at (0,0) {2};
    \node at (2,0) {1};
    \node at (4,0) {0};
    \node at (0,-2) {3};
    \node at (2,-2) {4};
    \node at (4,-2) {5};
    \node at (0,-4) {8};
    \node at (2,-4) {7};
    \node at (4,-4) {6};

    \begin{scope}[ibmMagenta,line width=1.8]
        \draw[snake it] (00)--(10);
        \draw[snake it] (01)--(11);
        \draw[snake it] (02)--(12);
    \end{scope}

    \begin{scope}[->,armygreen,>=stealth, line width=1.8]
        \draw (00) .. controls  (.5,.5) and (1.2,.5) .. (10);
        \draw (01) .. controls  (.5,.5-2) and (1.2,.5-2) .. (11);
        \draw (02) .. controls  (.5,.5-4) and (1.2,.5-4) .. (12);
        \draw (10) .. controls  (1.5,-.5) and (.8,-.5) .. (00);
        \draw (11) .. controls  (1.5,-.5-2) and (.8,-.5-2) .. (01);
        \draw (12) .. controls  (1.5,-.5-4) and (.8,-.5-4) .. (02);
    \end{scope}

    \draw[line width=2,->,>=stealth] (4.6,-2)--(5.6,-2);
    \draw[line width=2,->,>=stealth,dotted] (4.6-6.2,-2)--(5.6-6.2,-2);

\end{scope}

\begin{scope}[shift={(6.2,-2*6.2)}]
\begin{scope}[black!20!white]
    \foreach \x in {0,1,2} {
        \draw[line width=2] (0,-2*\x)--(4,-2*\x);
        \draw (2*\x,0)--(2*\x,-4);
    }

    \draw[line width=2] (4,0)--(4,-2);
    \draw[line width=2] (0,-2)--(0,-4);
\end{scope}
    \foreach \x in {0,1,2} {
        \foreach \y[evaluate={\l=int(3*\y+\x)}] in {0,1,2} {
        \node (\x\y) at (2*\x,-2*\y) [circle,draw,fill=white,minimum size=5mm] {};
        }
    }
    \node at (0,0) {1};
    \node at (2,0) {2};
    \node at (4,0) {0};
    \node at (0,-2) {4};
    \node at (2,-2) {3};
    \node at (4,-2) {5};
    \node at (0,-4) {7};
    \node at (2,-4) {8};
    \node at (4,-4) {6};

    \draw[ibmMagenta, snake it,line width=2] (01)--(02);
    \draw[ibmMagenta, snake it,line width=2] (20)--(21);
    \draw[line width=2,->] (4.6,-2)--(5.6,-2);
\end{scope}

\begin{scope}[shift={(2*6.2,-2*6.2)}]
\begin{scope}[black!20!white]
    \foreach \x in {0,1,2} {
        \draw[line width=2] (0,-2*\x)--(4,-2*\x);
        \draw (2*\x,0)--(2*\x,-4);
    }

    \draw[line width=2] (4,0)--(4,-2);
    \draw[line width=2] (0,-2)--(0,-4);
\end{scope}
    \foreach \x in {0,1,2} {
        \foreach \y[evaluate={\l=int(3*\y+\x)}] in {0,1,2} {
        \node (\x\y) at (2*\x,-2*\y) [circle,draw,fill=white,minimum size=5mm] {};
        }
    }
    \node (1) at (0,0) {1};
    \node (0) at (2,0) {2};
    \node (2) at (4,0) {0};
    \node (4) at (0,-2) {4};
    \node (5) at (2,-2) {3};
    \node (3) at (4,-2) {5};
    \node (7) at (0,-4) {7};
    \node (6) at (2,-4) {8};
    \node (8) at (4,-4) {6};

    \begin{scope}[ibmMagenta,line width=1.8]
        \draw[snake it] (10)--(20);
        \draw[snake it] (11)--(21);
        \draw[snake it] (12)--(22);
    \end{scope}

    \begin{scope}[->,armygreen,>=stealth, line width=1.8]
        \draw (10) .. controls  (2.5,.5) and (3.2,.5) .. (20);
        \draw (11) .. controls  (2.5,.5-2) and (3.2,.5-2) .. (21);
        \draw (12) .. controls  (2.5,.5-4) and (3.2,.5-4) .. (22);
        \draw (20) .. controls  (3.5,-.5) and (2.8,-.5) .. (10);
        \draw (21) .. controls  (3.5,-.5-2) and (2.8,-.5-2) .. (11);
        \draw (22) .. controls  (3.5,-.5-4) and (2.8,-.5-4) .. (12);
    \end{scope}

    \draw[line width=2,->,>=stealth] (4.6,-2)--(5.6,-2);
\end{scope}

\begin{scope}[shift={(3*6.2,-2*6.2)}]
\begin{scope}[black!20!white]
    \foreach \x in {0,1,2} {
        \draw[line width=2] (0,-2*\x)--(4,-2*\x);
        \draw (2*\x,0)--(2*\x,-4);
    }

    \draw[line width=2] (4,0)--(4,-2);
    \draw[line width=2] (0,-2)--(0,-4);
\end{scope}
    \foreach \x in {0,1,2} {
        \foreach \y[evaluate={\l=int(3*\y+\x)}] in {0,1,2} {
        \node (\x\y) at (2*\x,-2*\y) [circle,draw,fill=white,minimum size=5mm] {};
        }
    }
    \node at (0,0) {1};
    \node at (2,0) {0};
    \node at (4,0) {2};
    \node at (0,-2) {4};
    \node at (2,-2) {5};
    \node at (4,-2) {3};
    \node at (0,-4) {7};
    \node at (2,-4) {6};
    \node at (4,-4) {8};

    \begin{scope}[ibmMagenta,line width=1.8]
        \draw[snake it] (00)--(10);
        \draw[snake it] (01)--(11);
        \draw[snake it] (02)--(12);
    \end{scope}

    \begin{scope}[->,armygreen,>=stealth, line width=1.8]
        \draw (00) .. controls  (.5,.5) and (1.2,.5) .. (10);
        \draw (01) .. controls  (.5,.5-2) and (1.2,.5-2) .. (11);
        \draw (02) .. controls  (.5,.5-4) and (1.2,.5-4) .. (12);
        \draw (10) .. controls  (1.5,-.5) and (.8,-.5) .. (00);
        \draw (11) .. controls  (1.5,-.5-2) and (.8,-.5-2) .. (01);
        \draw (12) .. controls  (1.5,-.5-4) and (.8,-.5-4) .. (02);
    \end{scope}

    \draw[line width=2,dotted] (4.6,-2)--(5.6,-2);

\end{scope}

\begin{scope}[shift={(0*6.2,-3*6.2)}]
\draw[dashed] (0,1.1)--(4*6.2+4,1.1);
\node[fill=white,font=\large] at (2*6.2+2,1.1) {Step 2 $(t_2)$};
\begin{scope}[black!20!white]
    \foreach \x in {0,1,2} {
        \draw[line width=2] (0,-2*\x)--(4,-2*\x);
        \draw (2*\x,0)--(2*\x,-4);
    }

    \draw[line width=2] (4,0)--(4,-2);
    \draw[line width=2] (0,-2)--(0,-4);
\end{scope}
    \foreach \x in {0,1,2} {
        \foreach \y[evaluate={\l=int(3*\y+\x)}] in {0,1,2} {
        \node (\x\y) at (2*\x,-2*\y) [circle,draw,fill=white,minimum size=5mm] {};
        }
    }
    \node at (0,0) {0};
    \node at (2,0) {1};
    \node at (4,0) {2};
    \node at (0,-2) {5};
    \node at (2,-2) {4};
    \node at (4,-2) {3};
    \node at (0,-4) {6};
    \node at (2,-4) {7};
    \node at (4,-4) {8};

        \draw[ibmMagenta, snake it,line width=2] (20)--(21);
        \draw[ibmMagenta, snake it,line width=2] (01)--(02);

    \node at (1,.64) [ibmMagenta,font=\fontsize{12}{14}] {\textbf{merged hops}};

    \draw[line width=2,->,>=stealth] (4.6,-2)--(5.6,-2);
    \draw[line width=2,->,>=stealth,dotted] (4.6-6.2,-2)--(5.6-6.2,-2);
\end{scope}

\begin{scope}[shift={(6.2,-3*6.2)}]
\begin{scope}[black!20!white]
    \foreach \x in {0,1,2} {
        \draw[line width=2] (0,-2*\x)--(4,-2*\x);
        \draw (2*\x,0)--(2*\x,-4);
    }

    \draw[line width=2] (4,0)--(4,-2);
    \draw[line width=2] (0,-2)--(0,-4);
\end{scope}
    \foreach \x in {0,1,2} {
        \foreach \y[evaluate={\xi=int((-1)^\y*\x + (1-(-1)^\y))},evaluate={\l=int(3*\y+\x)}] in {0,1,2} {
        \node (\l) at (2*\xi,-2*\y) [circle,draw,fill=white,minimum size=5mm] {};
        \node at (2*\xi,-2*\y) [] {\l};
        }
    }

    \begin{scope}[ibmMagenta,line width=1.8]
        \draw[snake it] (0)--(1);
        \draw[snake it] (5)--(4);
        \draw[snake it] (6)--(7);
    \end{scope}

    \begin{scope}[->,armygreen,>=stealth, line width=1.8]
        \draw (0) .. controls  (.5,.5) and (1.2,.5) .. (1);
        \draw (5) .. controls  (.5,.5-2) and (1.2,.5-2) .. (4);
        \draw (6) .. controls  (.5,.5-4) and (1.2,.5-4) .. (7);
        \draw (1) .. controls  (1.5,-.5) and (.8,-.5) .. (0);
        \draw (4) .. controls  (1.5,-.5-2) and (.8,-.5-2) .. (5);
        \draw (7) .. controls  (1.5,-.5-4) and (.8,-.5-4) .. (6);
    \end{scope}

    \draw[line width=2,->,>=stealth] (4.6,-2)--(5.6,-2);

\end{scope}

\begin{scope}[shift={(2*6.2,-3*6.2)}]
\begin{scope}[black!20!white]
    \foreach \x in {0,1,2} {
        \draw[line width=2] (0,-2*\x)--(4,-2*\x);
        \draw (2*\x,0)--(2*\x,-4);
    }

    \draw[line width=2] (4,0)--(4,-2);
    \draw[line width=2] (0,-2)--(0,-4);
\end{scope}
    \foreach \x in {0,1,2} {
        \foreach \y[evaluate={\l=int(3*\y+\x)}] in {0,1,2} {
        \node (\x\y) at (2*\x,-2*\y) [circle,draw,fill=white,minimum size=5mm] {};
        }
    }
    \node (1) at (0,0) {1};
    \node (0) at (2,0) {0};
    \node (2) at (4,0) {2};
    \node (4) at (0,-2) {4};
    \node (5) at (2,-2) {5};
    \node (3) at (4,-2) {3};
    \node (7) at (0,-4) {7};
    \node (6) at (2,-4) {6};
    \node (8) at (4,-4) {8};

    \begin{scope}[ibmMagenta,line width=1.8]
        \draw[snake it] (10)--(20);
        \draw[snake it] (11)--(21);
        \draw[snake it] (12)--(22);
    \end{scope}

    \begin{scope}[->,armygreen,>=stealth, line width=1.8]
        \draw (10) .. controls  (2.5,.5) and (3.2,.5) .. (20);
        \draw (11) .. controls  (2.5,.5-2) and (3.2,.5-2) .. (21);
        \draw (12) .. controls  (2.5,.5-4) and (3.2,.5-4) .. (22);
        \draw (20) .. controls  (3.5,-.5) and (2.8,-.5) .. (10);
        \draw (21) .. controls  (3.5,-.5-2) and (2.8,-.5-2) .. (11);
        \draw (22) .. controls  (3.5,-.5-4) and (2.8,-.5-4) .. (12);
    \end{scope}

    \draw[line width=2, dotted] (4.6,-2)--(5.6,-2);
\end{scope}

\end{scope}
\end{tikzpicture}
    \caption{\justifying Illustration of the first Trotter step and the start of the second on a $3\times 3$ lattice. Identical operations are applied to both spin sectors, so only spin-up is shown. Green arrows indicate FSWAPs, magenta wavy lines indicate hopping interactions.
    All hopping terms are for time $t_1/2$ ($t_2/2$) in step 1 (2), except when merging is noted.
    Hops and FSWAPs on the same lattice indicate simultaneous implementation and the violet circles indicate Coulomb interactions for time $t_1$ ($t_2$) with the corresponding spin-down modes. Labels indicate the location of the encoded fermionic modes at each stage. The final stage of step 1 and the first stage of step 2 are merged, so the hopping interactions are applied for time $(t_1+t_2)/2$.
   }
    \label{fig:swapnet}
\end{figure}
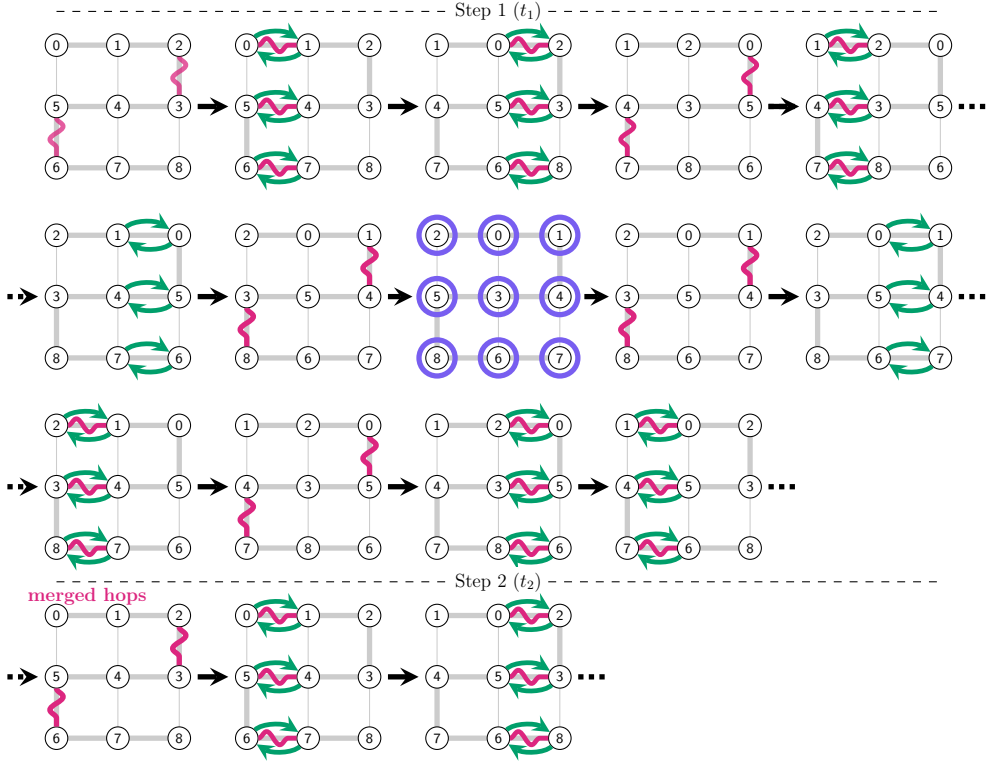

After the Coulomb interactions, the second round of hopping terms is implemented by applying the same gates as the first round, only in reverse order.
See Fig.~\ref{fig:swapnet} for an example illustration of a full step.

This schedule results in fewer swaps than a typical swap network \cite{kivlichan18}, which restores the modes to their original positions at the end.
Here, the modes are left in a jumbled state and the second round of hops is applied by reversing the schedule of the first round, restoring the initial layout before the following step.
This jumbling does not affect the Coulomb interactions, as every spin-pair moves together.

\subsubsection{Trotter Step Merging}\label{sec:step merging}
The symmetric structure of a second-order Trotter step offers an opportunity to reduce gate cost by merging repeated actions across steps.
In particular, a Trotter step ends with the application of all vertical hopping terms adjacent in the ordering, which is then immediately followed by the same action at the start of the following step. In a circuit with $N$ steps with step $j\in(1,\dots, N)$ running for time $t_j$ we omit these from the end of every step $j<N$ and at the beginning of each step $k>1$ we have these terms evolve for time $(t_k+t_{k-1})/2$ rather than $t_k/2$, reducing the number of gates required.

This is illustrated in Fig.~\ref{fig:swapnet}.

\subsection{Gate Decompositions}\label{sec:gate decomp}

All operations required for the Trotter circuit can be implemented with only single-qubit Pauli rotations
and CZ gates, which are native to the hardware. We use the following notation for single-qubit rotations:
\begin{equation}
    R_\sigma(\theta)=e^{-i\sigma\frac{\theta}{2}}.
\end{equation}

\subsubsection{Interactions}
We use the ``fsim'' gate~\cite{fsim} as a primitive in our circuits:
$$
\text{fsim}(\theta)=\begin{pmatrix}
    1 & 0 & 0 & 0 \\
    0 & \cos(\theta) & -i\sin(\theta) & 0 \\
    0 & -i\sin(\theta) & \cos(\theta) & 0 \\
    0 & 0 & 0 & 1 \\
\end{pmatrix}
$$
which decomposes into the native gate set, acting on qubits $i$ and $j$, as
$$
\Qcircuit @C=1em @R=1.0em {
\lstick{i}  & \gate{R_x(-\frac\pi 2)} & \gate{R_y(-\frac\pi 2)} & \ctrl{1} & \gate{R_x(-\theta)} & \ctrl{1} & \gate{R_y(\frac\pi 2)} & \gate{R_x(\frac\pi 2)} & \qw \\
\lstick{j}  & \gate{R_x(-\frac\pi 2)} & \qw & \ctrl{0} & \gate{R_x(\theta)} & \ctrl{0} & \qw & \gate{R_x(\frac\pi 2)} & \qw
}
$$
The operation is symmetric under the exchange of qubits.

FSWAP is implemented with the circuit
$$
\Qcircuit @C=1em @R=1.0em {
\lstick{i} & \multigate{1}{\text{fsim}(\frac\pi 2)} & \gate{R_z(\frac\pi 2)} & \qw\\
\lstick{j} & \ghost{\text{fsim}(\frac\pi 2)} & \gate{R_z(\frac\pi 2)} & \qw
}
$$
evolution under a hopping term for time $t$ between two modes adjacent in the JW ordering is implemented with
$$
\Qcircuit @C=1em @R=1.0em {
\lstick{i} & \multigate{1}{\text{fsim}(-t)} & \qw\\
\lstick{j} & \ghost{\text{fsim}(-t)} & \qw
}
$$
and a hopping term merged with an FSWAP
$$
\Qcircuit @C=1em @R=1.0em {
\lstick{i}  & \gate{R_z(\frac{3\pi} 2)} & \multigate{1}{\text{fsim}(-t-\frac\pi 2)} & \qw\\
\lstick{j}  & \gate{R_z(\frac{3\pi} 2)} & \ghost{\text{fsim}(-t-\frac\pi 2)}& \qw
}
$$
in cases where a magnetic field flips the sign of a hopping term, we use the same circuit, only with the sign of $t$ flipped.
Evolution under the Coulomb term for time $t$ with strength $U$ is implemented with
$$
\Qcircuit @C=1em @R=1.0em {
\lstick{i}  & \gate{R_y(\frac\pi 2)} & \ctrl{1} & \gate{R_x(\frac {U t}{2})} & \ctrl{1} & \gate{R_y(-\frac\pi 2)} & \qw \\
\lstick{j}   & \qw & \ctrl{0} & \qw & \ctrl{0} & \qw &  \qw
}
$$

\subsubsection{Singlet Initial States}\label{sec:singlets}

Singlets between lattice sites $A$ and $B$ are constructed with the following circuit
$$
\Qcircuit @C=1em @R=1.0em {
\lstick{A_\uparrow}   & \gate{H} & \qw      & \qw      & \ctrl{1} & \gate{H} & \gate{X} & \qw \\ 
\lstick{A_\downarrow} & \gate{H} & \ctrl{1} & \qw      & \ctrl{0} & \qw      & \qw    &  \qw \\ 
\lstick{B_\uparrow}   & \gate{H} & \ctrl{0} & \gate{H} & \ctrl{1} & \qw      & \qw    &  \qw \\ 
\lstick{B_\downarrow} & \gate{H} & \qw      & \qw      & \ctrl{0} & \gate{H} & \gate{X} & \qw 
}
$$

Where $A_\uparrow$ is the qubit representing spin-up for the site $A$ and so on. Due to the arrangement of the up and down qubits on the device, the qubits required for some singlets are not necessarily adjacent on the quantum processor. As a result, SWAP or FSWAP gates must be inserted. FSWAPS can only be used when any adjacent singlets have not already been entangled and all qubits involved in the swap are in the $\ket 0$ state.


\subsection{Total Gate Costs}

Shortening the swap network by exploiting the symmetry of the second-order Trotter step leads to significant savings in gate costs versus a standard swap network, which returns modes to their original positions as in \cite{kivlichan18}. \Cref{tab:gate costs} compares 2-qubit gate count and total gate depth for these two methods. For fairness, any gate savings from merging between Trotter steps as described in \cref{sec:step merging} are included when costing the standard swap network as well as our circuit.

\begin{table}[t]
    \centering
    \begin{tabular}{c|c c|c c|c c c }
      \multicolumn{1}{c}{} & \multicolumn{2}{c}{This Work} & \multicolumn{2}{c}{Standard Swap Network} & \multicolumn{3}{c}{3 Trotter steps} \\[0.3em]
         \multicolumn{1}{c|}{System Size}  & 2q Gate Count & Depth & 2q Gate Count & Depth & 2q Gate count & Depth & \# qubits  \\
         \hline
         \multicolumn{1}{c|}{$4\times4$} & $400N+20$ & $64N+6$ & $516N+32$ & $72N+6$ & $1220$ & $198$ & $34$\\
         \multicolumn{1}{c|}{$5\times5$} & $862N+40$ & $88N+6$ & $1022N+40$ & $96N+6$ & $2626$ & $270$ & $51$  \\
         \multicolumn{1}{c|}{$6\times5$} & $1040N+52$ & $88N+6$ & $1236N+48$ & $96N+6$ & $3172$ & $270$ & $62$\\
         \multicolumn{1}{c|}{$6\times6$} & $1440N+52$ & $96N+6$ & $1756N+72$ & $104N+6$ & $4372$ & $294$ & $74$\\
\end{tabular}
    \caption{\justifying Total number of 2-qubit quantum gates and gate depth for a circuit with $N$ Trotter layers and concrete numbers for our deepest circuits with 3 Trotter steps. Costs are given after an automated optimisation pass, which may remove additional gates from our manually optimised circuits. This optimisation compiles the circuit into alternating layers of 1 and 2-qubit gates, so 2-qubit gate depth can be taken as half the depth. Measurement at the end of the circuit requires a final layer of single-qubit gates, which is included in these costs. Note that these numbers are the same for all initial states except for the singlet initial state, which requires two more layers of 2-qubit gates.}
    \label{tab:gate costs}
\end{table}



\subsection{Comparison with Compact Encoding and native fsim}
We include here a comparison between the circuit costs of our experiment against potential alternatives for simulating 3 second-order Trotter layers of a square Fermi-Hubbard model. Specifically, we consider the comparative cost of employing a different fermionic encoding -- the compact encoding~\cite{derby2021compact} -- and also the potential cost of using variable-angle FSim gates.

\begin{figure}[!htbp]
    \centering
    \includegraphics[width=\columnwidth]{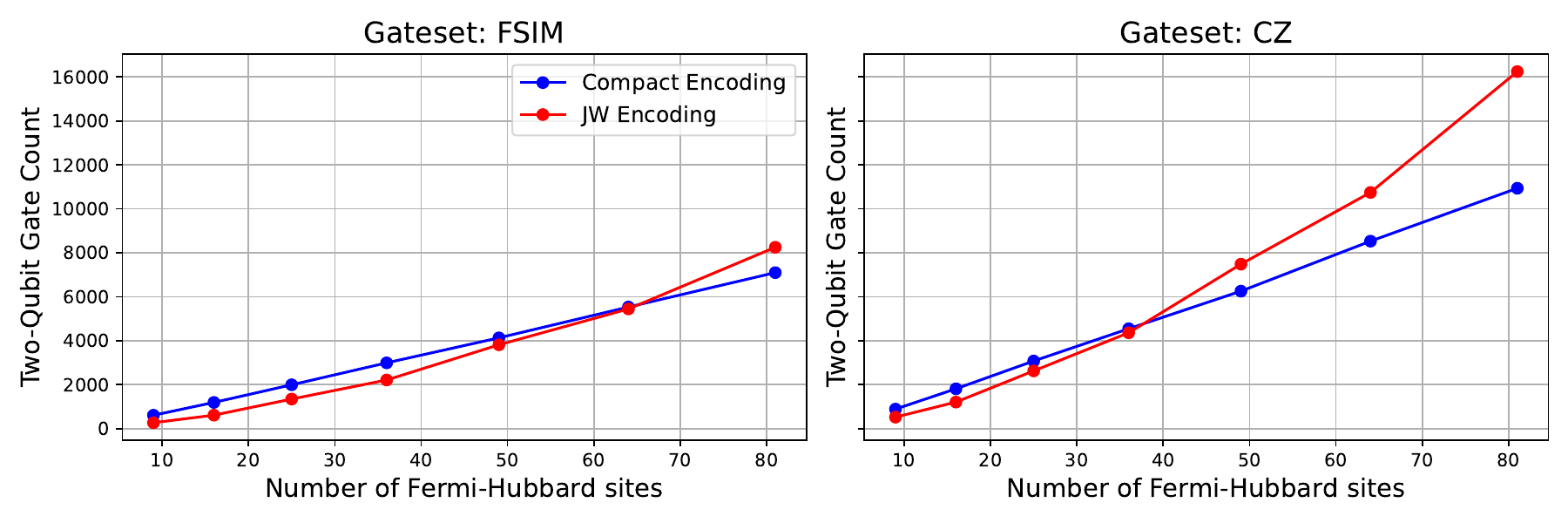}
    \caption{\justifying Two-Qubit Gate Counts for square lattices with 3 second-order Trotter steps, comparing compact and JW encodings.
    }
    \label{fig:two_qubit_counts_comparison}
\end{figure}

\begin{figure}[!htbp]
    \centering
    \includegraphics[width=\columnwidth]{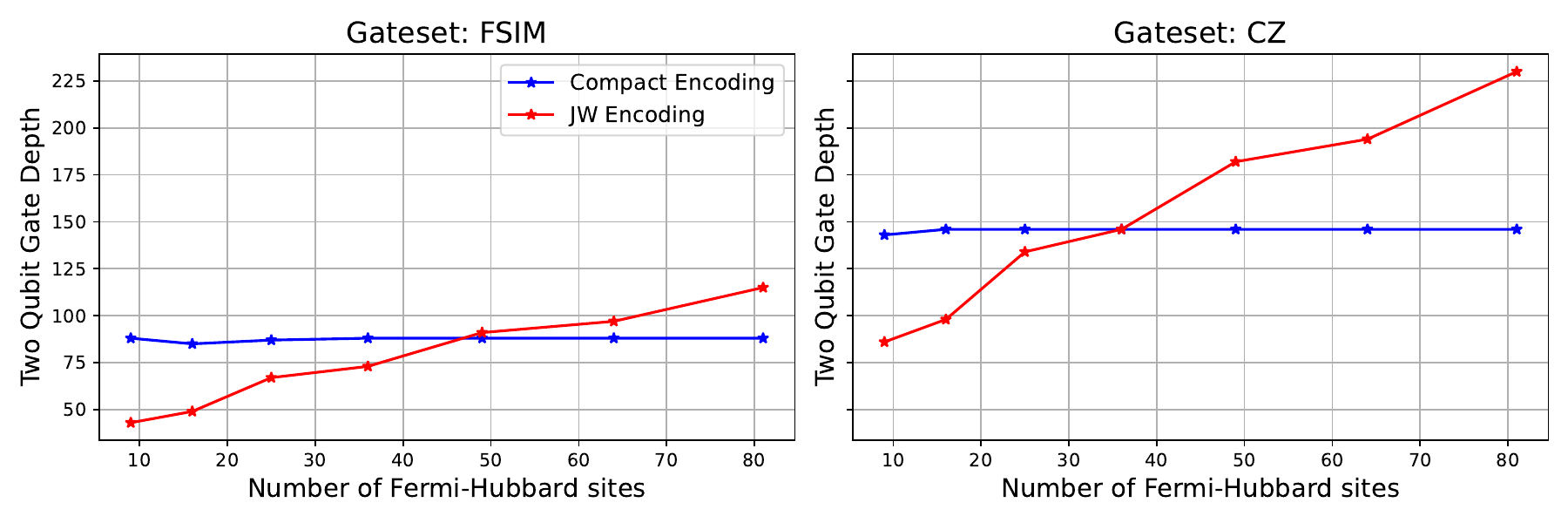}
    \caption{\justifying Two-qubit gate depth for square lattices with 3 Trotter steps, comparing compact and JW encodings.}
    \label{fig:two_qubit_depth_comparison}
\end{figure}

\begin{figure}[!htbp]
    \centering
    \includegraphics[width=0.5\columnwidth]{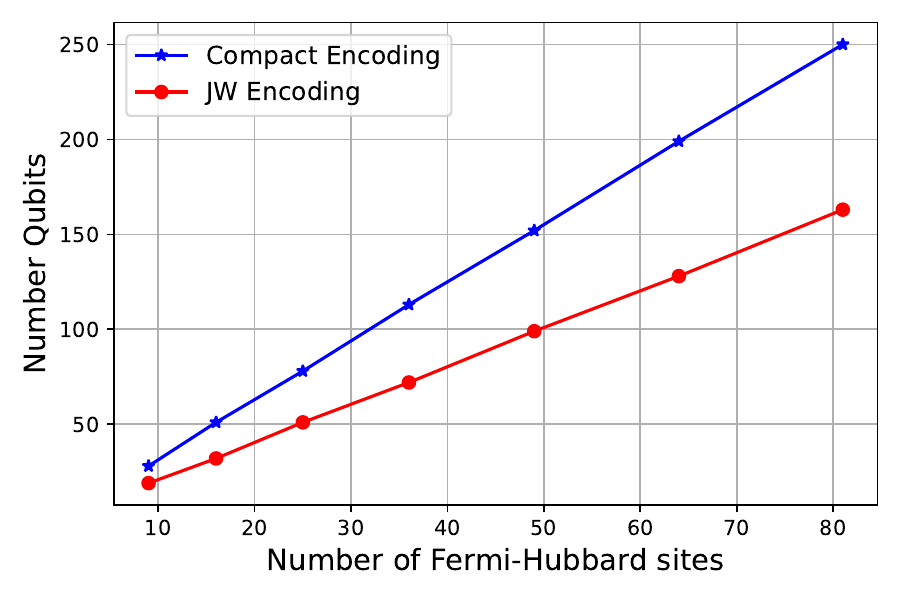}
    \caption{\justifying Number of qubits required for square lattices with 3 Trotter steps, comparing compact and JW encodings.}
    \label{fig:num_qubits_comparison}
\end{figure}

One of the major circuit overheads of our experiment is deploying the fermionic swap networks on the JW encoding which causes circuit depth and gate count for our circuits to respectively scale linearly and quadratically with system size.
There exist local fermionic encodings whose operator mappings -- an example being the compact encoding. Here we demonstrate that the JW encoding remains the more cost-effective choice at the scales of these experiments. However, we note that we are right at the cusp of a cross-over point for $6\times6$ in terms of both two-qubit gate cost and circuit depth when using CZ gates, as illustrated in the right-hand insets of  Figs.~\ref{fig:two_qubit_counts_comparison} and~\ref{fig:two_qubit_depth_comparison}. We note, however, as illustrated in Fig.~\ref{fig:num_qubits_comparison}, that were we to use the compact encoding for $6\times6$, we would require more qubits than were available to us on the device at the time of this work.

The native gateset employed for this experiment was the CZ gate. While the Willow quantum processor can, in principle, have a variable-angle fsim gate tuned up we chose to keep the CZ gate due to its high fidelity. However, looking towards future experiments, we include here an illustration of the potential improvements an arbitrary angle fsim gate can produce in terms of gate depth and count, as seen in the left-hand insets of  Figs.~\ref{fig:two_qubit_counts_comparison} and~\ref{fig:two_qubit_depth_comparison}. In this case, the crossover point for gate count is an $8 \times 8$ FH model, and for gate depth a $7\times7$ FH model -- requiring $200$ and $150$ qubits respectively.  We note that these figures do not change dramatically even if the variable angles are kept to a small number of fixed values, such as, for example, a native FSWAP and CZ. This motivates work targeting a high-quality tune-up of a small number of different angles of the fsim gate on the device.

\clearpage

\section{Error Mitigation Techniques}\label{sec:error_mitigation}

A number of known error mitigation techniques were tested and compared to optimise the performance of the experiment. These included Pauli and readout twirling, post-selection, training on fermionic linear optics (TFLO), Gaussian process regression (GPR), averaging over symmetries, and probabilistic error amplification / zero-noise extrapolation (PEA+ZNE). We also introduce a novel error-mitigation method based on reweighting samples, called Maximum Entropy Shot-Reweighting. In this section, we outline these methods and present a comparative analysis of their performance. In the final experiments, we used all of these methods except for PEA+ZNE, which we found performed less well than TFLO and used significantly more resources. For most experiments, unless otherwise stated, we used Pauli and readout twirling, post-selection, TFLO, GPR, and averaging over symmetries. We used sample reweighting when TFLO and GPR were not available (e.g.\ when not computing an observable).

\begin{figure}[!htbp]
  \begin{center}
    \includegraphics[width=0.95\textwidth]{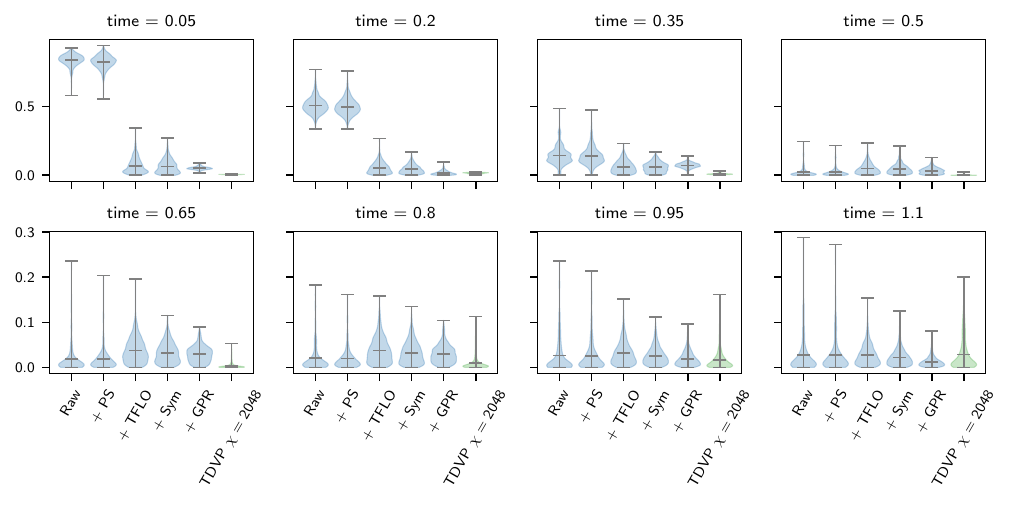}
  \end{center}
  \caption{\justifying Error profile across all single and two qubit $Z$ expectation values across different
  stages of the error mitigation pipeline and for TDVP at $U=0$ for a $6 \times 5$ system. PS: postselection; Sym: symmetry averaging. 
  }
  \label{fig:error_mitigation_violins}
\end{figure}

The overall effect of the various methods that we implemented is shown in Fig.~\ref{fig:error_mitigation_violins}. We find that error mitigation is less effective at higher times, which we attribute to the observables being less polarised, so the noisy observables are closer to the true values than at lower times. Nevertheless, in all cases, the combination of TFLO and GPR is able to significantly reduce the worst-case errors.

\subsection{Noise averaging and suppression} \label{sec:pauli_twirling}
To simplify the noise model for further error mitigation and also suppress it as much as possible on the hardware level, we implemented different noise averaging and suppression techniques; namely, Pauli twirling, dynamical decoupling and read-out twirling.

Pauli twirling \cite{bennett96} is a simple and effective technique to suppress and simplify error models in quantum circuits.
The main assumption going into it is that the main source of errors are two-qubit gates and that these are all Clifford gates (such as the CNOT or CZ gate).
Each such Clifford $C$ gate gets pre-multiplied with a randomly chosen 2-qubit Pauli gate $P$ and post-multiplied by an appropriate 2-qubit Pauli gate $Q$ such that the ideal gate $C$ remains unchanged, i.e. such that $P C Q = C$.
However, in the presence of noise, this sandwiching has two advantageous effects: It simplifies the noise model by making the transfer matrix of the noise gate diagonal in the Pauli basis, and it turns coherent errors into incoherent ones such that their effect accumulates with the square root of the number of noisy gates instead of linearly.

Besides Pauli twirling, we also implemented naive dynamical decoupling using spin echoes \cite{Mi_2021}.
This has the effect of flipping the qubit state and thus causing phase errors accumulated during one half of the idle period to cancel those from the other half.

For the read-out stage, Pauli twirling is modified to read-out twirling with random $X$ gates \cite{vdBerg2022modelfree}. This is particularly advantageous because readout on superconducting qubit hardware is affected by biased errors.
Instead of pre-multiplying the readout operation with randomly chosen Pauli gates, we use randomly chosen $X$ gates and save the bitstring corresponding to the bitflip pattern effected by the $X$ gates.
The effect of the random $X$ gates can then be undone in post-processing by xor-ing the measurement results with the saved bitstring.

The Pauli and readout twirling leave the two- and single-qubit structure of our circuits invariant---they only insert new single-qubit gates adjacent to pre-existing single-qubit gates.
This enables simple compression by merging all runs of adjacent single qubits into one general, single-qubit \texttt{cirq.PhasedXZGate} such that the final circuits are alternating layers of CZ gates and \texttt{cirq.PhasedXZGate}s.
This fixed-layer structure further enables the use of parametric circuits.
Instead of submitting one circuit per value of $U$, $t$ and twirling instance to the Google Quantum Cloud API we can submit only one parametric circuit for each number of Trotter layers and system size and all information about the twirling protocol, evolution time $t$ and interaction strength $U$ enters only into the parameters of the \texttt{PhasedXZGate}s and can be submitted via the API for parametric circuits.
We found that this massively sped up our experimental runs and brought down run-times from 10+ hours for a full run down to less than an hour.

\subsection{Post-Selection}\label{sec:post_selection}
\begin{figure}[!htbp]
  \centering
    \includegraphics{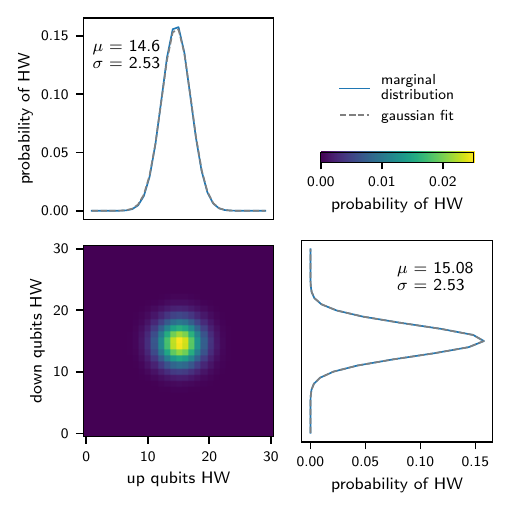}
    \caption{\justifying The distribution of Hamming weights on the up and down qubits is well described by a Gaussian centred at $(14.6, 15)$ (the correct Hamming weight in the absence of hardware errors). 2.4\% of shots have exactly Hamming weight $(14, 15)$.
        The data presented here is from the same $6 \times 5$ run also presented
        in Fig.~\ref{fig:central_hole_6x5_global_observables}. For $6 \times 6$ runs
        the probability of a sample being in the initial Hamming weight sector
        further drops to $1.9\%$.
    }
    \label{fig:hammingweights}
\end{figure}

The symmetries present in the Hamiltonian imply that particle number is conserved in each spin sector of the model. For all input states in the simulations, the particle number in each spin sector is a good quantum number, allowing post-selection on these conserved quantities. We note that deviation from the true sector follows a Gaussian profile, as shown in Fig.~\ref{fig:hammingweights}. This hints that errors have a local effect on the desired output state, as expected, and hence that local observables will be less affected by noise than if the error were well modelled by a completely depolarising channel.

\begin{figure*}[!htbp]
    \centering
    \includegraphics[width=0.9\linewidth]{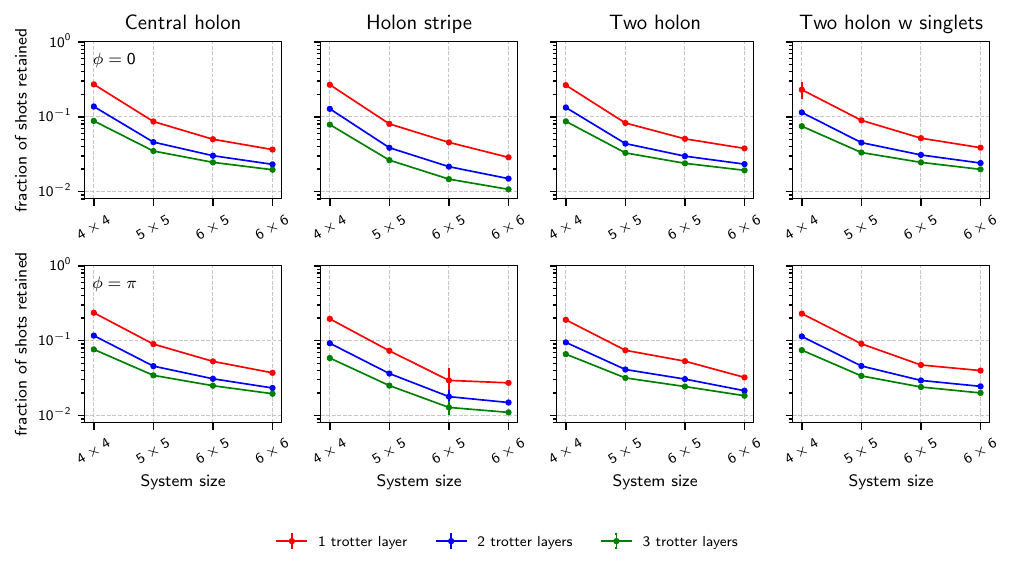}
    \caption{\justifying The median fraction of accepted shots after post-selecting on the correct Hamming weight per spin sector. The median is taken across all times and all values of $U$ and demonstrates how the fraction of retained shots changes as we increase the number of Trotter layers and the system size.}
    \label{fig:postselection_rate}
\end{figure*}

For different system sizes and initial states, we observe differing performance when using the post-selection to improve the signal in calculated expectation values. The differences become particularly notable when using these expectation values in the rest of our error mitigation pipeline. Generally, we see that full post-selection on the correct particle number per spin sector is useful for the smallest system size $4\times4$, whereas for the larger system sizes, the signal obtained improves when keeping shots with up to $8$ Hamming weight errors, or more.
Nevertheless, it is insightful to see how the acceptance fraction of shots retained during post-selection changes as we increase the number of Trotter layers and system size. This is explored in Fig.~\ref{fig:postselection_rate}.

The best post-selection hyper-parameter, the maximum Hamming weight error to accept per shot, is estimated by benchmarking the performance of the TFLO error-mitigated expectation values on the $U=0$ data for all initial states and system sizes. In Fig.~\ref{fig:train_and_test_error}, we show how the mean, median and maximum absolute error change as we vary the post-selection hyper-parameter. We see that the training error ($U\approx0$) shows a similar trend to the test error ($U = 4, 8$), which motivates using the training error in other systems as a proxy for expected performance overall when the test error is not accessible.

\subsection{Training on Fermionic Linear Optics (TFLO)}\label{sec:tflo}
\begin{figure*}[!htbp]
    \centering
    \includegraphics[width=0.9\linewidth]{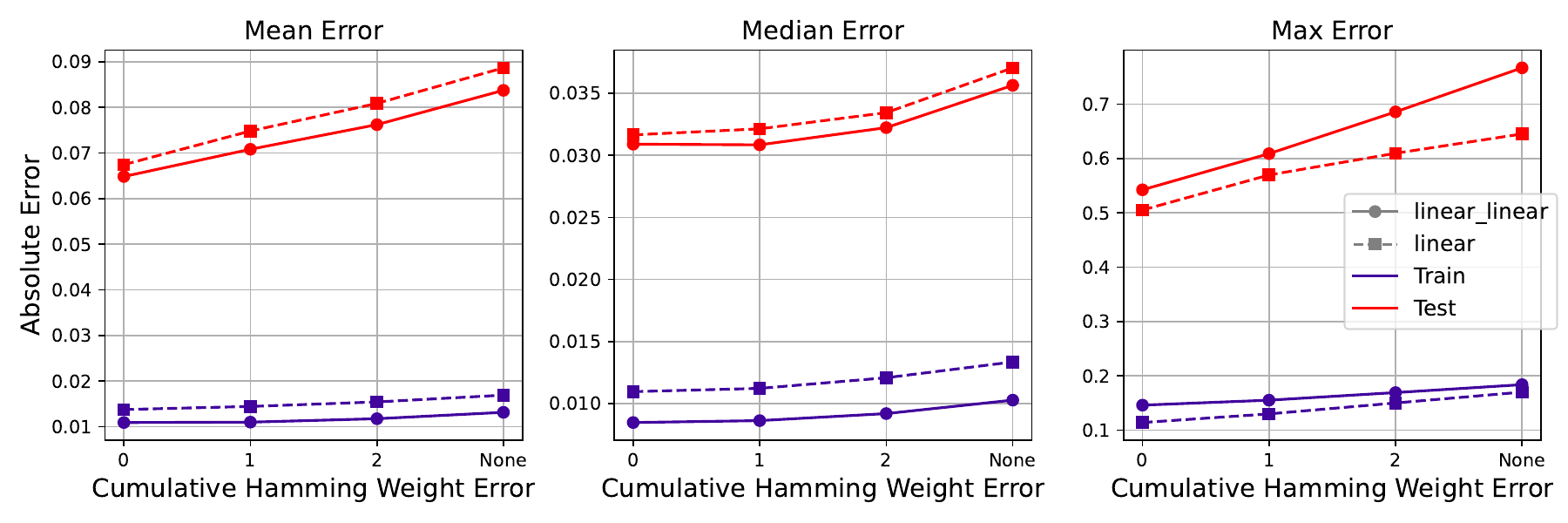}
    \caption{\justifying The mean, median and maximum training error ($U\approx0$) and test error ($U = 4, 8$) over all single body and two body Pauli-$Z$ expectation values mitigated with different levels of post-selection, and additional TFLO, GPR and symmetry averaging. This data is produced by the $4\times4$ experiment with the central hole initial state. A comparison of the linear-linear and linear ansatz deployed in TFLO is shown, while varying the cumulative hamming weight errors accepted in the shots used to give the expectation values.}
    \label{fig:train_and_test_error}
\end{figure*}

Learning based error mitigation, using data from classically simulable quantum circuits, has emerged as a useful strategy to improve the signal obtained from noisy quantum computers \cite{Czarnik2021, montanaro2021, Bultrini2023, Lowe2021, Sopena2021}. The dynamics of the non-interacting ($U=0$) Fermi-Hubbard model can be efficiently simulated by fermionic linear optics (FLO) \cite{terhal2002}. The TFLO error mitigation method \cite{montanaro2021} leverages this fact by training a noise model on the output of noisy quantum simulations of the non-interacting Fermi-Hubbard model against the noiseless output of an efficient FLO simulation.
The arguably simplest model to use is an affine linear fit of the form
\begin{equation}
  O_{\textnormal{exact}} = m O_{\textnormal{noisy}} + b
  \label{eq:ordinary_tflo}
\end{equation}
which is the ansatz suggested in \cite{montanaro2021, Czarnik2021} and found to be effective in \cite{stanisic2022observing} in the context of mitigating VQE expectation values. In simulable small-scale experiments, we additionally found a significant dependence of the noisy value on the evolution time $\evolutiontime$ (see Fig.~\ref{fig:tflo} below). We hence modified the fitting ansatz to use the form
\begin{equation}
  O_{\textnormal{exact}} = m O_{\textnormal{noisy}} + c \evolutiontime + b
  \label{eq:time_dep_tflo}
\end{equation}
which we found to work slightly better in the context of time dynamics simulations (see Figs.~\ref{fig:error_mitigation_comparison} and~\ref{fig:train_and_test_error}). We call the former method ``linear'' and the latter method ``linear-linear''.

\begin{figure}[!htbp]
    \includegraphics{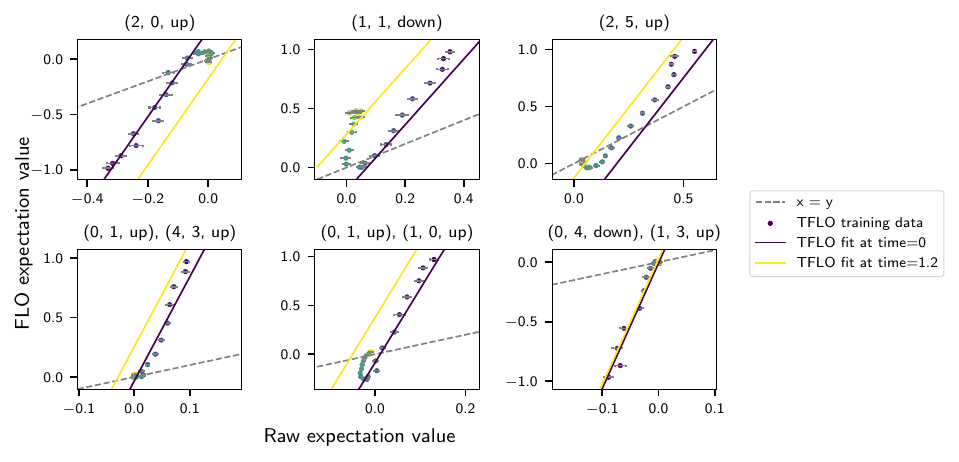}
    \caption{\justifying A few examples of the linear-linear fits performed for TFLO error mitigation for $6 \times 5$.
    The top row shows a selection of single-qubit $Z$ expectation values, while the bottom row shows some two-qubit $Z$ expectation values.
    We see that the fit slope for the two-qubit fits is larger than for the single-qubit fits, and that the product of the single-qubit slopes often matches the two-qubit slope reasonably well.}
    \label{fig:tflo}
\end{figure}

To do error mitigation with this model, we performed a linear least squares fit with \cref{eq:time_dep_tflo} using the device data at $U=0$ as $O_{\textnormal{noisy}}$ and the FLO simulation data as $O_{\textnormal{exact}}$.
For each observable, the optimal fit parameters $c^*, m^*$ and $b^*$ are then used together with the $U \neq 0$ data to predict mitigated values according to \cref{eq:time_dep_tflo}.
The TFLO training set is generated by using the exact same time dynamics simulation parameters as in the non-interacting cases, while setting $U\approx0$ in the circuits, such that the compiler does not optimise away the gate.

As discussed above, TFLO relies on the assumption that we can fit the exact observables to the output of a noiseless FLO simulation. Whilst this can often be done using a controlled least-squares fit, the case we are considering is unsuitable for this approach because of the relative stability of the observables over time, which leads to a signal-to-noise ratio that is too small. We therefore chose to adopt a more sophisticated, three-stage fitting procedure:
\begin{enumerate}
\item We determine the best TFLO rescaling parameters via a standard least-square fit procedure on those observables with a sufficiently large SNR.
\item After grouping the observables by (i) the number of qubits they act on, and (ii) whether they act in multiple Fermi-Hubbard sites, we compute a prior distribution over the optimal fit parameters.
\item For each of the groups defined in 2., we use the prior distribution computed above to perform a linear least-square fit with regularisation.
\end{enumerate}
The process above allows us to obtain good fit parameters even in the presence of low SNR in the time series.

\subsection{Non-Gaussian circuits - classical simulation}
\label{sec:nearflo}
One type of experiment highlighted in the main text considers an initial singlet covering state. This is a non-Gaussian or magic state~\cite{hebenstreit2019all, liu2022many,howard2017application,cudby2023gaussian,reardon2024improved,dias2024classical,hakkaku2022quantifying} because it cannot be generated a FLO unitary~\cite{terhal2002}, and in general the simulation cost grows exponentially with the number of singlets present.

Since fermionic Gaussian states are those of non-interacting fermions corresponding to the $U=0$ regime, this is implicitly assumed for the rest of this section.

The singlet creation operator on vacuum is given by $\frac{c^{\dagger}_{a,\uparrow} c^{\dagger}_{b,\downarrow} - c^{\dagger}_{a,\downarrow} c^{\dagger}_{b,\uparrow}}{\sqrt{2}}$ for physical sites $a$ and $b$. However, we note that the initial state can be written as a product state across $N_{\text{singlets}}$ singlets and $N_{\text{lonely}}$ lonely sites (which can be empty, fully ($\uparrow \downarrow$) or partially filled with $\uparrow$ or $\downarrow$) according to \cref{eqn:singlet_init_state}
\begin{equation}
    \centering
    \ket{\psi}_{\text{init}} = \bigotimes_{k=1}^{N_{\text{singlets}}} \ket{\text{singlet}}_{s_k} \bigotimes_{l=1}^{N_{\text{lonely}}} \ket{\text{lonely}}_{s_l} 
    \label{eqn:singlet_init_state}
\end{equation}
where $s_k$ is the pair of sites from the $k^{th}$ singlet and $s_l$ a lonely site. The observables needed for TFLO error mitigation are linear combinations of $\langle Z_i \rangle$ and $\langle Z_i Z_j \rangle$ correlators $\forall i,j$ modes, which are enough to perform TFLO corrections for all physical observables considered in the main text except for the Wilson lines. These require higher weight Z strings, and this computation is omitted since the exponential cost with respect to $N_{\text{singlets}}$ becomes cumbersome. \\

\textbf{Densities} \\
\label{app:density_near_flo}
The observables $\langle \hat{n}_i \rangle, \forall i$ fermionic modes is
\begin{align}
    \centering
    \bra{\psi_{\text{init}}(t)} c^{\dagger}_i c_i \ket{\psi_{\text{init}}(t)} &= \bra{\psi_{\text{init}}}e^{iHt} c^{\dagger}_i c_i e^{-iHt} \ket{\psi_{\text{init}}} \\
    &= \sum_{\alpha, \beta} R_{\alpha,i} R^*_{\beta,i} \bra{\psi_{\text{init}}} c^{\dagger}_{\alpha} c_{\beta} \ket{\psi_{\text{init}}} \\
    &= \sum_{\alpha} |R_{\alpha,i}|^2
\end{align}
where $H$ is the Hamiltonian defined in \cref{eq:FH_Ham_intro} and $R = e^{-iht}$ where $h$ is the lattice adjacency matrix. Because of the initial state structure, for a non-vanishing overlap in the sum, the coefficients $\alpha$ and $\beta$ must be equal, since flipping two qubits within or across physical sites results in an orthogonal state. \\

\textbf{Density-Density correlators} \\
Here, we are concerned with 
\begin{align}
    \centering
    \bra{\psi_{\text{init}}(t)} c^{\dagger}_i c_i c^{\dagger}_j c_j \ket{\psi_{\text{init}}(t)} &= \bra{\psi_{\text{init}}}e^{iHt} c^{\dagger}_i c_i c^{\dagger}_j c_j e^{-iHt} \ket{\psi_{\text{init}}} \\
    &= \sum_{\alpha, \beta, \gamma, \eta} R_{\alpha,i} R^*_{\beta,i} R_{\gamma,j} R^*_{\eta, j} \bra{\psi_{\text{init}}} c^{\dagger}_{\alpha} c_{\beta} c^{\dagger}_{\gamma} c_{\eta} \ket{\psi_{\text{init}}}
\end{align}
Since $\psi_{\text{init}}$ is not a Gaussian state, we need to consider the $(2L)^P$ mode combinations in the sum with operator weight $P$. For this special case of $P=4$, the computational complexity can be reduced to scale as $O(L^2)$ due to the structure of the initial state.

Note that $c^{\dagger}_{\alpha} c_{\beta} c^{\dagger}_{\gamma} c_{\eta}$ flips four qubits and the structure of a singlet state across sites $a$ and $b$ is 
\begin{equation}
    \centering
    \frac{1}{\sqrt{2}} (\ket{1_{i,\uparrow} 0_{i,\downarrow} 0_{j,\uparrow} 1_{j,\downarrow}}  + \ket{0_{i,\uparrow} 1_{i,\downarrow} 1_{j,\uparrow} 0_{j,\downarrow}}).
\end{equation}
The non-vanishing overlap of the singlet state with itself is to flip all four qubits. Similarly, any lonely site has non-vanishing self-overlap when nothing is flipped. Therefore, we only need to consider the cases where $\alpha, \beta, \gamma$ and $\eta$ land on a single or a pair of physical sites that are part of the same singlet. 

\textbf{Four distinct indices} \\
Using this argument, one concludes that all non-trivial combinations of distinct $\alpha, \beta, \gamma$ and $\eta$ must be within a single singlet. For this case, the correlator becomes
\begin{align}
    \centering
    \langle n_i n_j \rangle_{\text{4 distinct}} (t) &= \sum_{(i, j)}^{N_{\text{singlets}}} \Bigg( \sum_{\alpha,\beta,\gamma,
    \eta \in \text{perm}[i\uparrow, i\downarrow, j\uparrow, j\downarrow]} R_{\alpha,i} R^*_{\beta,i} R_{\gamma,j} R^*_{\eta, j} \nonumber \\
    &\times \frac{1}{2} \bra{\Omega} (c^{\dagger}_{i,\uparrow} c^{\dagger}_{j, \downarrow} - c^{\dagger}_{i,\downarrow}c^{\dagger}_{j,\uparrow})^{\dagger} c^{\dagger}_{\alpha} c_{\beta} c^{\dagger}_{\gamma} c_{\eta} (c^{\dagger}_{i,\uparrow} c^{\dagger}_{j, \downarrow} - c^{\dagger}_{i,\downarrow}c^{\dagger}_{j,\uparrow}) \ket{\Omega}\Bigg)
\end{align}
for the vacuum state $\ket{\Omega}$. `Perm' specifies permutations of a list and $(i,j)$ are site pairs of the singlet covering.  \\
    
\textbf{Two distinct indices}  \\
There are two cases to consider here, $\alpha=\beta, \gamma=\eta$ and $\alpha=\eta, \beta=\gamma$. For both, the distinct indices are either
\begin{align}
   & 1. \text{ within the same singlet or lonely site.} \nonumber \\
   & 2. \text{ across two singlets or two lonely sites.}\nonumber \\
   & 3. \text{ between a singlet and a lonely site.} \nonumber \\
\label{eqn:conditions_two_distinct}
\end{align}

The two-body correlator is therefore
\begin{align}
\centering
    \langle n_i n_j \rangle_{\text{2 distinct}} (t) &= \sum_{(i,j)} \Bigg( \sum_{\alpha, \gamma \in \text{perm}[i\uparrow, i\downarrow, j\uparrow, j\downarrow]} |R_{\alpha,i}|^2 |R_{\gamma,j}|^2 \bra{\Omega} \mathcal{C}^{\dagger}_{i,j} n_{\alpha} n_{\gamma} \mathcal{C}_{i,j}\ket{\Omega}  \nonumber \\
    &+\sum_{\alpha, \beta \in \text{perm}[i\uparrow, i\downarrow, j\uparrow, j\downarrow]} R_{\alpha,i} R_{\beta,i}^* R_{\beta,j} R^*_{\alpha,j} \bra{\Omega} \mathcal{C}^{\dagger}_{i,j} n_{\alpha} (1-n_{\beta}) \mathcal{C}_{i,j}\ket{\Omega}  \Bigg)
\end{align}
where $(i,j)$ is a site index pair across all the conditions outlined in \cref{eqn:conditions_two_distinct} and $\mathcal{C}_{ij}$ is the creation operator of the type of initial state inhabiting sites $i$ and $j$.

\textbf{Single distinct index}
This is similar to the case of the number operator discussed in \cref{app:density_near_flo}, i.e. 
\begin{equation}
    \centering
    \langle n_i n_j \rangle_{\text{1 distinct}}(t) = \sum_{\alpha} \sum_{i}^{2L} |R_{\alpha,i}|^2 |R_{\alpha,j}|^2\bra{\psi_{\text{init}}} n_i \ket{\psi_{\text{init}}}
\end{equation}
where $i$ is the mode index.

Finally, the case where $c^{\dagger}_{\alpha} c_{\beta} c^{\dagger}_{\gamma} c_{\eta}$ has three unique indices is trivially zero since there is either a double creation/annihilation on the same mode. Therefore,
\begin{equation}
    \centering
     \langle n_i n_j \rangle (t) =  \langle n_i n_j \rangle_{\text{1 distinct}}(t) +  \langle n_i n_j \rangle_{\text{2 distinct}}(t) +  \langle n_i n_j \rangle_{\text{4 distinct}}(t)
\end{equation}
which scales as $O(L^2)$.

\subsection{Sampled bitstring probabilities at zero interaction strength}
\label{sec:bitstring_probabilities}
In this section, we discuss the classical computations of overlap probabilities of the form $|\langle z |U | 0^{\otimes n}\rangle|^2$ for some bitstring $z$ (sampled in the computational basis, from the device, TN MPS state or exact FLO-sampler), which are required for cross-entropy benchmarking analyses discussed in \cref{sec:cross-entropy}. Here, $U$ denotes both the creation of the initial state and subsequent time evolution. 
Consider any initial state considered in the main text in the most general form,
\begin{equation}
    \ket{\psi_{\text{init}}} = \sum_m \alpha_m \left(\prod_{i=1}^{W} c^{\dagger}_{m_i}\right) \ket{\Omega}
\end{equation}
where index $m_i$ runs over the modes making up the Fock state (indexed $m$) with complex coefficient $\alpha_m$ and $\ket{\Omega}$ is the vacuum state. $W$ is the hamming weight of all the Fock states making up $\ket{\psi_{\text{init}}}$. For any FLO initial state considered (e.g. Néel state with central hole) there is a single term in the summation and the overlap inherits the efficient classical simulatability of conventional fermionic linear optics. The initial state with singlet covering has exponentially many terms in the summation with respect to the number of singlets present. 

We can write a sampled bitstring $z$ as a single Slater determinant
\begin{equation}
\centering
\ket{z} = \prod_{i=1}^W c^{\dagger}_{m'_i} \ket{\Omega} 
\end{equation}
such that the overlap probability with the time-evolved state is

\begin{align}
     \Big| \langle z | \psi(t)\rangle \Big|^2 &= \Big|\sum_{m} \alpha_m \bra{\Omega} \Big(\prod_{i=1}^{W} c^{\dagger}_{m'_i} \Big)^{\dagger}  e^{-iHt} \Big(\prod_{i=1}^{W} c^{\dagger}_{m_i} \Big) \ket{\Omega}\Big|^2 \\
     &= \Big|\sum_{m}\alpha_m  \Bigg( \sum_{q_1, \dots, q_W} ( \prod_{i=W}^1 {R}^{*}_{q_i, m'_i}) \Bigg) \bra{\Omega} c_{q_W} \dots c_{q_1} c^{\dagger}_{m_1} \dots c^{\dagger}_{m_W} \ket{\Omega}\Big|^2 \\
     &= \Big|\sum_{m}  \alpha_m  \Bigg( \sum_{q_1, \dots, q_W} ( \prod_{i=W}^1 {R}^{*}_{q_i, m'_i}) \Bigg) \text{sgn} \Big(\sigma_{q1, \dots, q_W}^{m_1, \dots, m_W} \Big)\Big|^2 \\
     &= \Bigg|\sum_{m}  \alpha_m \text{det} 
\begin{pmatrix} 
    \textcolor{gray}{(m_1)} & {R}^*_{m_1,m'_1} & {R}^*_{m_1,m'_2} & \dots & {R}^*_{m_1, m'_W} \\
   \textcolor{gray}{\vdots} & \dots & \dots & \vdots \\
   \textcolor{gray}{(m_W)} & {R}^*_{m_W,m'_1} & {R}^*_{m_W,m'_2} & \dots & {R}^*_{m_W, m'_W}
\end{pmatrix}\Bigg|^2
\label{eqn:sampled_bitstring_probability}
\end{align}
where ${R} = e^{-iht}$. The values $m_1$ and $m_W$ in grey denote the row indices; to help visualise their positions relative to each other, and do not take part in the overall determinant.  $\text{sgn}(\sigma_{A}^B)$ is equal to $(-1)^{P_{AB}}$, where $P_{AB}$ is the number of swaps needed to permute the set of indices $A$ into that of $B$. 

For the singlet covering case, $|\alpha_m| = (\frac{1}{\sqrt{2}})^{N_{\text{singlets}}}$ $\forall m$ so the main bottleneck for computing this probability is the exponentially growing number of Fock states with respect to $N_{\text{singlets}}$. 

\subsection{Sampling from Trotterised and exact FLO simulations}
\label{sec:flo_sampling}
For several experiments, such as that of the holon stripes in~\cref{sec:stripe}, holon/doublon pairwise distance dynamics in~\cref{sec:holon_pair} and crucially the cross-entropy benchmarking analyses in~\cref{sec:cross-entropy}, FLO samples were required to provide the ground truth for the $U=0$ experiments. These samples were generated via the Bravyi-K{\"o}nig sampling algorithm~\cite{bravyi2011classical} developed in the FLOYao package~\cite{jan_lukas_bosse_2022_7303997}.

\subsection{Gaussian Process Regression (GPR)}\label{sec:GPR}
In the ideal case, all time series signals considered in this work are smooth functions
of $\evolutiontime$, but hardware, twirling and shot noise lead to discontinuous
time series---even after all other error mitigation is done. To further use the
prior information about the smoothness of our time series, we use GPR \cite{Rasmussen_2004, prml} and
fit a Gaussian process model to the noisy time series. Concretely, we used
a squared exponential kernel with length scale $l = 0.4$ and noise strength $\sigma = 1.5$.
These values of $l$ and $\sigma$ were found by maximizing the log-marginal-likelihood
of these hyper-parameters on a wide selection of observable time series and noticing that
the result of this maximisation clustered around these values of the hyper-parameters. In all the figures in this work that include GPR, we show the 68$\%$ one-sigma region as a shaded region.

\subsection{Symmetry averaging} \label{sec:symmetry_averaging}
The symmetries of the system provide a natural avenue to mitigate errors in the calculation. First of all, we expect the measured expectation values to be invariant under some symmetries, depending on the initial state, lattice structure and observable of interest. Moreover, the Hamiltonian itself will be invariant under both the spatial symmetries of the lattice and the spin-reflection symmetry $R c_{i,\uparrow} R^\dagger = c_{i, \downarrow}$. Up to Trotter error, this invariance will carry through to the circuits used to simulate the Hamiltonian.
As a consequence of this, we can increase the effective shot-count for the expectation values of multi-qubit $Z$ strings (and any expectation value derived from them) by averaging each multi-qubit $Z$ expectation value with its symmetry-related partners.
Moreover, we can exploit the fact that some observables need to be identical due to lattice symmetries to reduce Trotter error. Indeed, since said observables might be affected differently by Trotter error, by averaging across them, we can expect to reduce the Trotter error itself.

\subsection{Error propagation}
The time series Figs.~\ref{fig:staggered_magnetisation_scales},~\ref{fig:central_hole_5x5_global_observables}, \ref{fig:central_hole_6x5_global_observables}, \ref{fig:central_hole_6x6_global_observables}, etc. show error bars computed using Gaussian error propagation.
To that end, the bars on the initial raw expectation values obtained from the shots are calculated using

\begin{equation}  \sigma_{\text{raw}}^2 =
 \frac{1}{\ntwirling(\ntwirling - 1)} \sum_{i=1}^{\ntwirling} (\hat \mu_i - \mu)^2+ \frac{1}{\ntwirling^2 \nshots} \sum_{i=1}^{\ntwirling} \sigma_i^2
\label{eq:twirling-error-bar}
\end{equation}

where $\ntwirling$ is the number of twirling instances, $\nshots$ is the number of shots taken per twirling instance, $\mu_i$ and $\sigma_i^2$ are the sample mean and sample variance per twirling instance. Finally, $\mu = \frac{1}{\ntwirling} \sum_{i=1}^{\ntwirling} \mu_i$ is the total mean.

\subsubsection{Pauli twirling and error bars}
\label{sec:twirling-and-error-bars} Determining error bars in conjunction with Pauli twirling requires some care, in order to correctly account for the iterated expectations. In particular, one can show that the naïve approach of collecting all $\ntwirling \times \nshots$ into a single set and then computing the expectation of an observable $O$ as \begin{equation}
\hat\mu_O = \frac{1}{\ntwirling \times \nshots} \sum_{i, j=1}^{\ntwirling, \nshots} O(x_{i,j})
\end{equation}
and the variance of $\hat \mu_O$ in the usual way as

 \begin{equation}
 \hat \sigma^2_{\hat \mu_O} = \frac{1}{\ntwirling \times \nshots (\ntwirling \times \nshots - 1)} \sum_{i, j=1}^{\ntwirling, \nshots} (O(x_{i,j}) - \hat \mu_O)^2
 \label{eq:naive_twirling_error}
 \end{equation}

is inaccurate, since in non-trivial cases (such as when $\nshots = 1$ or all twirling instances are the same) it will underestimate the overall error. This can be easily demonstrated by considering the case in which we take a large number of shots ($\nshots$) taken from a small set of $\ntwirling$ twirling instances that produce very different distributions. In this case, the error computed via \cref{eq:naive_twirling_error} will be very small, despite the fact that the data has only been collected from very few twirling instances, which all give very different estimates for $\mu_O$. There are two key contributions that need to be correctly accounted for to get an accurate estimate $\hat \sigma^2_{\hat \mu_O}$ for the variance of $\hat \mu_O$. First, the variance coming from sampling different twirling instances (given by the first term in \cref{eq:twirling-error-bar}), and second, the contribution coming from the shot noise in each circuit (the second term in \cref{eq:twirling-error-bar}). By employing the law of total variance and going through the derivations, we then obtain the formula \cref{eq:twirling-error-bar} used in this work.

\subsubsection{Verification via bootstrapping}
We chose to use the \texttt{uncertainties} Python package \cite{uncertainties} to compute error propagation. Whilst the package implements automatic Gaussian error propagation for many basic mathematical functions, the validity of its propagation calculations depends on the assumption that \begin{inparaenum} \item the input distributions are all Gaussian \item all data transformation functions are (or at least well-approximated by) affine linear functions. \end{inparaenum} Moreover, it is important to note that the \texttt{uncertainties} package only tracks and propagates the individual uncertainty of each variable, but does not take correlations into account. In order to verify the accuracy of Gaussian error propagation, we have compared the results to those obtained via the alternative method of bootstrapping \cite[Chap. 8]{Wasserman2007}. Bootstrapping is a technique used to estimate the distribution (or properties thereof) of an estimator by replacing the full distribution of the in-going (unknown) random variable with the empirical distribution obtained from the measured samples. In our work, the estimator is the mitigated and processed observable values.
\begin{figure}[!htbp]
  \begin{center}
  \includegraphics{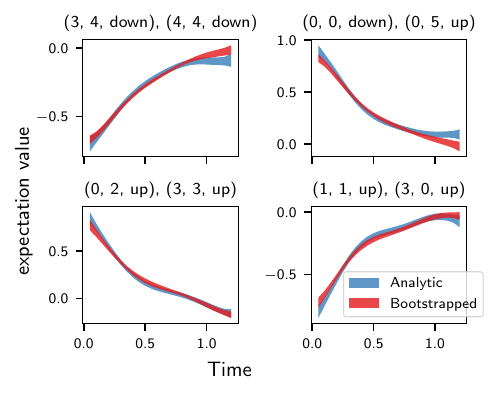}
  \end{center}
  \caption{\justifying Comparison of error bars computed via bootstrapping vs Gaussian
    error propagation for a random choice of 4 observables. Again, we use the
    same $6 \times 5$ data set also presented in
    Fig.~\ref{fig:central_hole_6x5_global_observables} for this example.
  }
  \label{fig:bootstrapping-vs-error-propagation}
\end{figure}
When implementing the bootstrap method in conjunction with twirling, a similar argument to that discussed in \cref{sec:twirling-and-error-bars} holds, meaning that one first needs to sample the $\ntwirling$ circuits with replacement from the pool of twirling instances and then sample $\nshots$ shots with replacement from each of the sampled circuits. Only in this way can one obtain a meaningful dataset that can be processed through the whole error mitigation and data analysis pipeline. This procedure needs to be repeated multiple times to allow us to estimate the standard deviations of the final mitigated observables. As shown in Fig.~\ref{fig:bootstrapping-vs-error-propagation}, 
while bootstrapping largely agrees with the error bars derived from Gaussian error propagation, there are some meaningful discrepancies. We attribute these to the fitting approach used in TFLO (fitting noisy values to exact values, rather than vice versa). Accordingly, the reported error bars should be interpreted with caution. 


\subsection{Maximum entropy shot-reweighting}
\label{sec:shot_reweighting}

In this section, we apply the principle of maximum entropy~\cite{jaynes1957information} to smoothly reweight samples according to the minimal adjustment in information gained by knowing some corrected/improved local expectation values. We consider a list of sampled bit strings $\{x_1,...,x_{n_{shots}}\}$, where each $x_i \in \{0,1\}^n$ , which we use to estimate some observables $\{O_j\}_{j=1}^{n_{observables}}$,
$$ \tilde{O}_j = \frac{1}{n_{shots}} \sum_{i=1}^{n_{shots}} O_j(x_i) $$
and suppose for this class of observables we have a correction procedure for these estimates: $\tilde{O}_j \rightarrow \hat{O}_j$.   For example, the observables $\{\hat{O}_j\}$ could be 1-local and 2-local Paulis obtained from TFLO.

A natural question is: can we \emph{reconcile} our list of samples with the corrected expectation values?  Here, we propose to smoothly reweight the samples according to the principle of maximum entropy.   Introduce a reweighting function $p$, which is a normalised $\sum_i p(x_i) = 1$ probability mass function on the set $\{x_1,...,x_{n_{shots}}\}$ and is intended to satisfy:
\begin{equation}\sum_{i = 1}^m p(x_i) O_j(x_i) = \hat{O}_j \label{eq:shotmitigationconstraints}\end{equation}
for each corrected observable in the specified set $\{O_j\}_{j=1}^{n_{observables}}$. Defining $\vec{O} = (O_1,...,O_{n_{observables}})$, this equation \cref{eq:shotmitigationconstraints} has a solution if and only if $\hat{\vec{O}}$ is a normalised convex combination of the vectors $\{\vec{O}(x_i)\}_{i=1}^{n_{shots}}$. To specify a unique solution, we consider the principle of maximum entropy, and seek a reweighting distribution $p$ that maximises entropy $H(\{p\}) = -\sum_{i=1}^m p(x_i) \ln p(x_i)$ subject to the linear constraints from the corrected observables.

When exact solutions for $p$ in \cref{eq:shotmitigationconstraints} exist (for a particular set of samples and corrected observables) the maximum entropy problem is feasible, and if all $p(x)$ are bounded away from the extreme points $0 < p(x) < 1$, the optimisation problem has a solution in terms of Lagrange multipliers $\{\lambda_j\}_{j=1}^{n_{observables}}$, one $\lambda_j$ for each observable $O_j$
$$p(x) = \frac{1}{Z(\vec{\lambda})} e^{ \vec{\lambda}\cdot \vec{O}(x)} \quad , \quad Z(\lambda) = \sum_{i =1}^{n_{shots}} e^{ \vec{\lambda}\cdot \vec{O}(x_i)}$$
The condition $0 < p(x) < 1$ for all $x$ is necessary for the $\vec{\lambda}$ to be finite.
This is typical for the case $n_{shots} > n_{observables}$, and then this solution can greatly reduce the number of parameters involved in the distribution, down from the number of samples to the number of observables that are being corrected.

An important distinction is that by only reweighting samples from the device, we are not replacing the quantum measurement distribution with a maximum entropy approximation based on local observables, rather we are ``tilting'' the empirical distribution with importance weights that reflect the corrected local observables,
\begin{equation}p_\textrm{mitigated} \propto e^{ \vec{\lambda}\cdot \vec{O}(x)} p_\textrm{device}(x).\end{equation}
Therefore, the reweighted samples can be regarded as samples from the distribution above, up to the fact that the solutions for $\vec{\lambda}$ are also estimated from a finite set of samples.  For fixed $\vec{\lambda}$ (assumed to converge on the sample set), we can compute the standard error in the usual way for samples with importance weights,
$$\langle F \rangle = \sum_{i=1}^{n_{shots}} p(x_i) F(x_i) \quad , \quad \textrm{SE}^2 = \frac{\sum_{i=1}^{n_{shots}} p(x_i)(F(x_i) - \langle F\rangle)^2}{n_\textrm{eff} -1} \quad , \quad n_\textrm{eff} = \frac{1}{\sum_{i=1}^{n_{shots}} p(x_i)^2 }$$

The effective number of mitigated shots $n_\textrm{eff}$ accounts for the fact that concentrating most of the weight on a small number of samples increases the uncertainty in computing further expectation values. When the corrected observable values are far from the predictions of the unmitigated data, the reweighting procedure concentrates the weight on relatively few of the samples.  This is determined by the magnitude of the additive correction, so in practice the concentration of weights (large $||\vec{\lambda}||$) is most apparent when the expectation values of (traceless) local observables are large (e.g. at early times in the quantum circuit).    

Lastly, the optimal $\{\lambda_j\}$ satisfying the constraints  \cref{eq:shotmitigationconstraints} can be found by gradient descent iterations, with
$$\nabla \vec{\lambda} =  \hat{\vec{O}} - \langle \vec{O}\rangle_{\vec{\lambda}}$$
with iterations $\lambda_j \rightarrow \lambda_j + \eta  \; \nabla \lambda_j$ where $\eta$ is a positive parameter.  Note that the expectation in the current model $\langle \vec{O}\rangle_{\vec{\lambda}}$ can be evaluated efficiently since it only involves a sum over $n_{shots}$ samples. To justify this, consider the quantity which is analogous to the free energy in statistical mechanics,
\[
    F = \log Z(\lambda),
\]
which has the  property that its derivatives with respect to $\{\lambda_j\}$ are
exactly the expectation values of the \(O_j\),
\[
    \frac{\partial F(\vec{\lambda})}{\partial  \lambda_j} = \frac{1}{Z(\vec{\lambda}) }\sum_{i=1}^{n_{shots}}  O_j(x_i) e^{\vec{\lambda} \cdot \vec{O}(x_i)}
    = \langle O_j\rangle_{\vec{\lambda}}
\]
so $\nabla F(\vec{\lambda}) = \langle \vec{O} \rangle_{\vec{\lambda}}$.  Consider the entropy
$$H(\vec{\lambda}) =  -\sum_{i=1}^{n_{shots}} p(x_i) \ln p(x_i) = -\sum_{i=1}^{n_{shots}}\frac{e^{\vec{\lambda} \cdot \vec{O}(x_i)} }{Z(\vec{\lambda})}\left(\vec{\lambda}\cdot \vec{O}(x_i) - \ln Z(\vec{\lambda}) \right) = F(\vec{\lambda}) - \vec{\lambda}\cdot \vec{O}(x_i)$$
For any $\vec{\lambda}$ define
\begin{equation}\hat{H}(\vec{\lambda}) = F(\vec{\lambda}) - \vec{\lambda}\cdot \hat{\vec{O}}\label{eq:shot_mitigation_entropy}\end{equation}
which satisfies
\begin{equation}\nabla \hat{H}(\vec{\lambda}) = \langle \vec{O}\rangle_{\vec{\lambda}}-\hat{\vec{O}}   \label{eq:shot_mitigation_entropy_gradient}\end{equation}
 so that $\nabla \hat{H}(\vec{\lambda}) = 0$ implies the expectation value constraints are satisfied.  If $\vec{\lambda}^*$ is the optimal solution, then $\hat{H}(\vec{\lambda}) - \hat{H}(\vec{\lambda}^*)$ is the KL divergence $D_{\textrm{KL}}(p_{\vec{\lambda^*}}||p_{\vec{\lambda}})$ since
$$D_{\textrm{KL}}(p_{\vec{\lambda}^*}||p_{\vec{\lambda}}) = \sum_{i=1}^{n_{shots}} p_{\vec{\lambda}^*} \ln \left(\frac{p_{\vec{\lambda}^*}}{p_{\vec{\lambda}}} \right) = (\vec{\lambda}^* - \vec{\lambda})\cdot \hat{\vec{O}} + F(\vec{\lambda}) - F(\vec{\lambda}^*) = \hat{H}(\vec{\lambda}) - \hat{H}(\vec{\lambda}^*) $$
therefore $\hat{H}(\vec{\lambda})$ is also the cross entropy to the optimal distribution.  We can explicitly verify that $\hat{H}$ is convex by relating the Hessian of $\hat{H}$ to the covariance matrix,
$$ \frac{\partial ^2 \hat H(\vec{\lambda})}{\partial  \lambda_k \partial  \lambda_j} = \langle O_k O_j\rangle_{\vec{\lambda}} - \langle O_k\rangle_{\vec{\lambda}} \langle O_j\rangle_{\vec{\lambda}}$$
therefore $\nabla^2 \hat{H} = \textrm{Cov}_{p_{\vec{\lambda}}}(\vec{O}) \succeq 0$ and so $\hat{H}$ is convex.  This ensures that the gradient descent will reduce the cross entropy $\hat{H}(\vec{\lambda})$ to the ideal solution at each step, $\hat{H}\left(\vec{\lambda} - \eta \nabla \hat{H}\right) < \hat{H}(\vec{\lambda})$, and so the gradient iterations in \eqref{eq:shot_mitigation_entropy_gradient} converge efficiently.

In our implementation, we used all single-qubit $Z$ expectation values, together with certain two-qubit $ZZ$ expectation values (those across adjacent sites in the lattice in the same spin sectors, and between spin sectors on each site), as the known exact expectation values $\{\hat O_j\}$. These are either computed exactly, when $U=0$, or via TFLO when $U\neq 0$. The optimal Lagrange multipliers $\lambda_j$ are found by performing gradient descent on the entropy $\hat H(\lambda)$ given by \cref{eq:shot_mitigation_entropy} with the gradient of \cref{eq:shot_mitigation_entropy_gradient}.
As mentioned, when the expectation values are strongly polarised (e.g.\ at early times), the norm $\Vert \vec \lambda \Vert$ can become large, and so we introduce an ad-hoc regularisation by adding $+ c \Vert \vec \lambda \Vert^2$ to the cost function and modifying the gradient accordingly. Another challenge arises because the TFLO-corrected local observables are not exact, so they are not guaranteed to be globally consistent. In that case, no feasible solution to the optimisation problem exists, and the regularisation leads to an approximate solution.

\subsection{Probabilistic Error Amplification (PEA) and Zero Noise Extrapolation (ZNE)} \label{sec:PEA}

We implemented a variant of PEA introduced in \cite{kim2023evidence} in which a local Pauli noise model (under Pauli twirling) is learned by fitting the decay rates of tomographic signals \cite{vandenBerg2023PEC}. An amplified variant of this noise model is then simulated in the circuit by the stochastic inclusion of Pauli operators. Zero-noise extrapolation of the amplified signal is then applied via either a linear or an exponential fit.
Only the CZ gate noise is modelled. We assume local noise confined to the support of the target gates. We identify all instances of distinct layers of parallel CZ gates in the target circuit and learn the local noise model for each such distinct layer.

The protocol for learning the noise is to assume that a \emph{Pauli-twirled} noise channel $\Lambda$ is associated with a layer of gates.
The Pauli twirling is done with the same protocol described in \cref{sec:pauli_twirling}.
The noise model is the formal exponentiation of a sparse Lindbladian $\Lambda = e^{\mathcal{L}}$ and
\begin{align}
	\mathcal{L}(\rho) &= \sum_{k \in \mathcal{K}} \lambda_k(P_k \rho P_k - \rho) \\
	\Lambda(\rho) &= \prod_{k \in \mathcal{K}} (\omega_k \cdot+ (1-\omega_k)P_k \cdot P^{\dagger}_k) \rho, \label{eq:paulichannel}
\end{align}
where $\mathcal{K}$ indexes the set of Pauli operators generating the Lindbladian  and $\omega_k = (1+e^{-2 \lambda_k})/2$. In line with the approach of \cite{vandenBerg2023PEC,kim2023evidence}, we assume that the generators of the model are local with respect to the device topology, so $\mathcal{K}$ is chosen to contain all single-qubit Pauli operators and all two-qubit Pauli operators for each pair of qubits sharing an edge in the hardware graph.

The model parameters $\lambda_k$ are estimated by measuring Pauli fidelities for a chosen measurement basis $\{B_j\}_{j \in \mathcal{B}}$ and solving a linear system of equations. The fidelities
\begin{align}
	f_j =\frac{1}{2^n} \text{Tr}[B^{\dagger}_j \Lambda(B_j)],
\end{align}
are estimated through standard benchmarking techniques \cite{vandenBerg2023PEC}, which extract them from an exponential decay.

The measurement bases $\{B_j\}_{j \in \mathcal{B}}$ are used to define a matrix
$M = \mathcal{M}(\mathcal{B},\mathcal{K})$ with entries
	\begin{align}
		M_{i,j} &= \begin{cases}
			0, & [B_i, P_j] = 0\\
			1, & [B_i, P_j] \neq 0.
		\end{cases}
	\end{align}
We then estimate the $\lambda_k$ by solving
\begin{align}
		\lambda (f) := \underset{\lambda \geq 0}{\text{argmin}} \frac{1}{2} \| M \lambda +\log (f)/2 \|^{2}_2 ,
\end{align}
where the logarithm here is entrywise and $\lambda$ and $f$ are vectors of the noise parameters and learned fidelities. To ensure that the matrix $M$ is non-singular, it suffices to choose $\mathcal{B}=\mathcal{K}$, and, as shown in \cite{vandenBerg2023PEC}, fidelities for this set of local observables can be measured using just 9 distinct measurement settings.

To amplify the effect of noise and produce the ZNE fitting data, we simulate increasing the noise rates on the device as $\lambda_k \to G \lambda_k$ with gain factors $G\in[1.0, 1.1, 1.34, 1.58]$, by randomly inserting Pauli gates according to equation~\cref{eq:paulichannel} with appropriately modified probabilities $\omega_k$. To produce the ZNE fitting data, we sample $500$ random instances per logical circuit, where the randomisation incorporates both twirling and error amplification, and take $1024$ samples per circuit. The number of instances was chosen because we found it sufficient to get well-converged expectation values, while the number of shots was chosen to match the overhead incurred by changing the circuit parameters that we run at.
To ensure that all circuits are run within a reasonable time frame from when the noise profiling takes place and to reduce the compilation overhead, we employ parameterised circuits as already described in \cref{sec:pauli_twirling} as well as the same Pauli twirling protocol.

\subsection{Performance Comparisons}
In order to assess the best performing error mitigation strategy, we benchmark variants of TFLO and ZNE on the $4\times4$ system. For these circuits, we have access to exact expectation values, enabling us to directly measure the performance of each error mitigation strategy. As our benchmark experiment, we chose the setup consisting of a central holon and evolution under the Hamiltonian excluding a magnetic flux.

\begin{figure*}[!htbp]
    \centering
    \includegraphics[width=\linewidth]{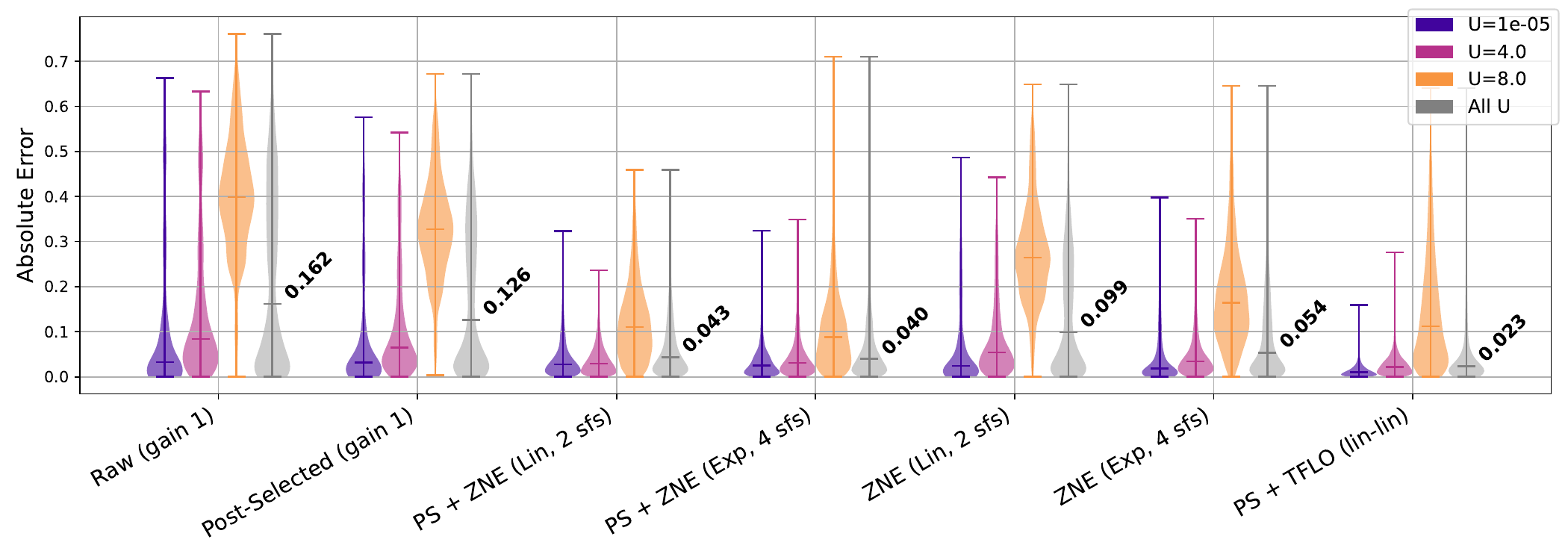}
    \caption{\justifying Comparison of error distributions across different ZNE implementations benchmarked against TFLO. The distribution of absolute errors is shown for the exponential and linear ansatz deployed in ZNE. The number of scale factors (sfs) was chosen to give the best ZNE performance. Coloured violins correspond to individual interaction strengths $U$, while the grey violins show the pooled error over all $U$ values. Medians over the pooled $U$ distributions are annotated. The different ZNE fitting procedures are described in the surrounding text.
    }
    \label{fig:zne_mitigation_comparison}
\end{figure*}

In Fig.~\ref{fig:zne_mitigation_comparison}, we compare different ZNE fitting procedures
against each other, and also against TFLO and no error mitigation or only post-selection.
``PS + ZNE'' always refers to ZNE applied to post-selected shots, ``Lin.'' and ``Exp.''
refer to extrapolation fits with a linear and exponential function, respectively, while
``$<n>$ sfs'' refers to the number of gain factors included in the extrapolation fit. It is clear that TFLO outperforms all variants of ZNE that we tried in terms of median error (by almost a factor of 2), and reduces the raw error by almost a factor of 8.

\begin{figure*}[!htbp]
    \centering
    \includegraphics[width=\linewidth]{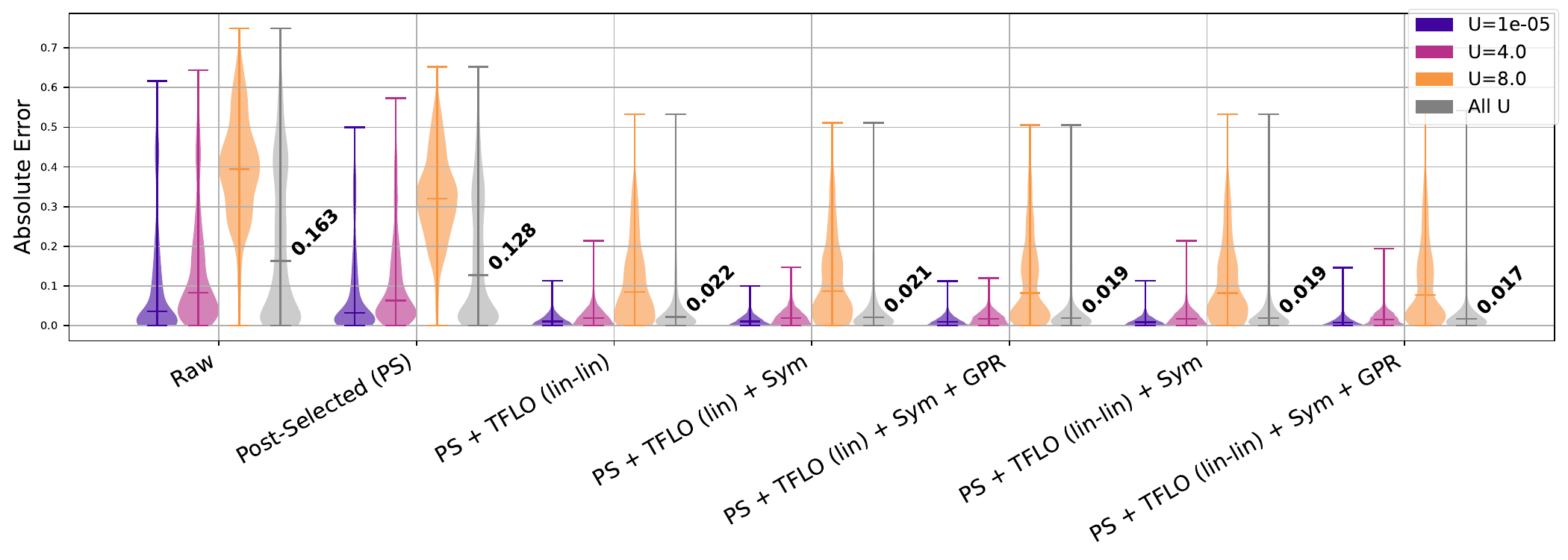}
    \caption{\justifying Comparison of error distributions across different TFLO implementations when applied to mitigating all single and two-body Pauli $Z$ expectation values for the dynamics of the $4\times4$ model with a central holon on a \neel background. The distribution of absolute errors is shown for each method, highlighting the effect of post-selection, symmetrisation, and TFLO mitigation. Coloured violins correspond to individual interaction strengths $U$, while the grey violins show the pooled error over all $U$ values. Medians over the pooled $U$ distributions are annotated.}
    \label{fig:error_mitigation_comparison}
\end{figure*}

In Fig.~\ref{fig:error_mitigation_comparison}, we show the distribution of absolute errors over all single-body $Z$ and all two-body $ZZ$ expectation values, for a variety of error mitigation methods, and their combinations.
As before, ``+'' refers to an error mitigation method being applied on top of the preceding ones, ``lin-lin.'' to the TFLO fit being done according to \cref{eq:time_dep_tflo} and ``lin.'' according to \cref{eq:ordinary_tflo}, respectively.
Small differences between apparently identical conditions in Fig.~\ref{fig:zne_mitigation_comparison} and~\ref{fig:error_mitigation_comparison} are due to results corresponding to different experimental runs. We see that the use of TFLO achieves a significant reduction in error (by up to more than 9x compared with raw data), including for $U=4,8$.
In addition, GPR provides a further small improvement. The lin-lin variant of TFLO was the most effective.

It is also worth noting that PEA+ZNE comes with a significant circuit overhead: To get reasonably well-converged twirling and error amplification at the higher scaling factors, we found that $\sim 500$ twirling instances are necessary.
Multiplying this by the four scaling factors used to get stable extrapolation fits, we end up with a $100\times$ circuit overhead over the $20$ Pauli twirling instances that we used at gain level 1 for TFLO.
Given the minor performance differences between the best-performing TFLO and the best-performing PEA+ZNE implementation, we hence opted to use TFLO as the much cheaper error mitigation method for all our experiments.


\clearpage

\section{Tensor Network Simulations}
\label{sec:tensor_networks}

We implemented state-of-the-art tensor network simulations using TeNPy \cite{tenpy2024}, employing the \texttt{TwoSiteTDVPEngine} for the time-dependent variational principle (TDVP) algorithm, which can capture the precise time evolution with sufficient resources.
\par

We studied alternative tensor network routines before concluding that TDVP was the most suitable for the system under study.
These studies will be presented in future work~\cite{alam2025fermionic}, but in brief, they contained: direct contraction of the quantum circuit \cite{Gray_2021_hyperoptimized}; simulation of the time-dependent density matrix renormalisation group within the MPS/MPO framework~\cite{White_2004_real, Ronca_2017_timestep}; and an explicit fourth-order Runge–Kutta integrator applied to the fermionic MPS representation with compression at each step \cite{provazza_fast_2024}.
\par

TDVP can simulate the real- or imaginary-time evolution of quantum many-body systems represented by MPS.
MPSs are characterised by their bond dimension, $\chi$, which limits the size of indices connecting the tensors within, which effectively constrains the correlations that can be expressed, i.e., the entanglement the network can represent is limited, and may be less than that of the target system~\cite{Schollwock_2011_dmrg}.
TDVP progresses by iteratively evolving the Schr\"{o}dinger equation for time step $\Delta t$, and at each step projecting the state onto the tangent space of the MPS manifold, i.e., the subset of states which may be represented within the given bond dimension.
Smaller $\Delta t$ yield more precise evolution, at the expense of higher runtime~\cite{haegeman2011time, Haegeman_2016_unifying}. For a more extensive review of time dynamics methods for MPS, see the review by Paeckel \emph{et al.}~\cite{Paeckel_2019_review}. In addition to this cornerstone algorithm, there is an extensive, ever-evolving tensor network literature, often tailored to specific studies. For example, tensor networks combined with Monte Carlo methods~\cite{chen2025tensor} have been applied to highly disordered systems. Tensor networks combined with machine learning have been used to study molecular open quantum systems~\cite{schroder2019tensor}. Finally, PEPS~\cite{verstraete2004renormalization} and MERAs~\cite{vidal2008class} hold promise for capturing more entanglement in higher-dimensional systems, but also pose many numerical challenges. For instance, contracting a PEPS is computationally $\#$P-complete~\cite{schuch2007computational}. In this context, approximate PEPS contraction algorithms based on Belief Propagation/Simple Update have proven particularly effective in simulating large 2D systems with low coordination number~\cite{Begusic_2024_fast, patra2024efficient, Tindall_2024_efficient}. More recently, a variety of tensor network methods for simulating (noisy) quantum circuits have emerged and have shown the great promise~\cite{thompson2025, Ayral_2023_dmrg, Sander_2025_quantum, Guo_2019_general, Noh_2020_efficient}.

\par

The two most common variants of TDVP are the one- and two-site evolution, which differ in how much of the MPS is updated at each time step.
One-site TDVP evolves each tensor individually, while two-site TDVP evolves pairs of neighbouring sites by forming local clusters of enlarged bond dimension, before truncating the bond to the given $\chi$ through the singular value decomposition (SVD) algorithm. Two-site TDVP is known to be more accurate at the expense of higher runtime and memory requirements.
\par

\begin{figure}[!htbp]
    \centering
    \includegraphics[width=0.6\linewidth]{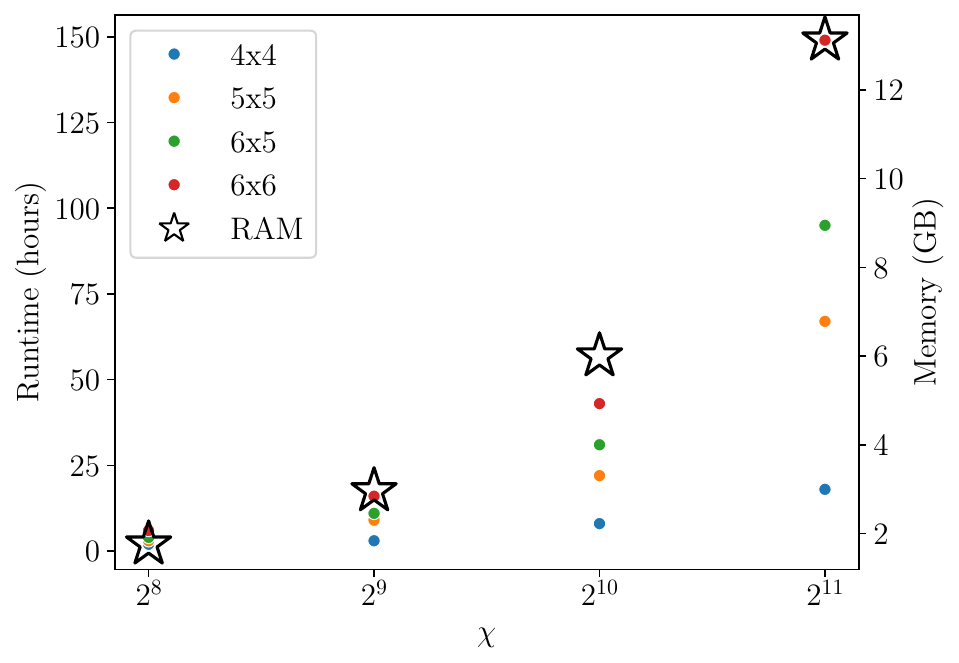}
    \caption{\justifying
        Time and memory requirements for
        TDVP simulations corresponding to experimental data.
        The left-hand y-axis shows the wallclock runtime corresponding to coloured points in the chart, while the right-hand y-axis indicates the RAM requirements of the largest system size ($6\times6$) at each bond dimension, shown by the black stars.
    }
    \label{fig:tn_resources}
\end{figure}

The computational cost of TDVP is cubic in the bond dimension, and also depends on the physical dimension $d$ of each of the $N$ sites in the system, i.e., the runtime of each time step scales like $\mathcal{O}(N d^3 \chi^3)$.
To ensure the tensor network simulations are as accurate as possible, we use two-site TDVP throughout this work and choose a cautious value of $\Delta t = 0.01$ to minimize the TDVP projection error when starting the dynamics from a low bond dimension initial state. However, we note that the tensor network simulations could be made more accurate with a smaller $\Delta t$ and a larger $\chi$ than was feasible to investigate here. However, investigations on small-sized systems with accessible ground truth found that using $\Delta t=0.001$ did not substantially affect accuracy.
In Fig.~\ref{fig:tn_resources} we report the maximum requirements of computing the TDVP instances reported throughout, noting the considerable increase in runtime with respect to bond dimension, as expected.
\par

To achieve exact evolution of a system with $N$ sites, an exact MPS simulation requires  $\chi=d^{\frac{N}{2}}$, where $d$ is the dimension of each site \cite{perez2006matrix}.
In our case for fermions, each site can be unoccupied, doubly occupied, or occupied by either an up or down spin, so $d=4$.
For the largest systems studied in this work, we have a lattice of size $6 \times 6$, i.e. $N=36$, which demands $\chi=4^{18} = 2^{36}$, which is intractable in both memory and runtime requirements. 
Even taking into account the U(1)$\times$ U(1) symmetry conservation of $N_{\uparrow}$ and $N_{\downarrow}$ electrons (which was preserved in our simulations), the maximum bond dimension for an exact MPS calculation at half-filling $N_{\uparrow}=N_{\downarrow}=N/2$ at $N=36$ is still $\chi=\frac{1}{4}\sqrt{\binom{N}{N_{\uparrow}}\binom{N}{N_{\downarrow}}}\approx 2^{31}$.
However, we can still retrieve meaningful approximate simulations with a much smaller bond dimension.
\par

For all hardware data reported in the main text, we implement corresponding instances of TDVP with $\chi = 2^n $ for $n \in [8, 11]$.
The largest system we study is of size $6 \times6$ with $\chi=2^{11}$, which took 6.5 days to run and required 13 GB of RAM.
In all simulations, we used a time step $\Delta t=0.01$ to evolve the dynamics incrementally, balancing the algorithm's accuracy with practical considerations of compute runtime.
We also fix the truncation threshold within the singular value decomposition subroutine to $10^{-10}$.
We performed our simulations on the Google Cloud Platform with virtual machines of type \texttt{E2-highmem-4}, i.e., on 2.25 GHz Intel processors with four vCPUs and 32 GB RAM per instance~\cite{gcp_machine_types}.
\par


\clearpage

\section{Majorana Propagation Simulations} \label{sec:majorana_propagation}

In order to further test the ability of classical methods to simulate the system we are considering accurately, we implemented the Majorana propagation (MP) algorithm \cite{miller2025}.
This method relies on implementing Heisenberg evolution through the circuit on a set of polynomials of Majorana operations that represent the operators of interest: the expectation values are then given by the overlap between the Heisenberg-evolved observable and the initial state.

\begin{figure*}[!htbp]
    \centering
    \includegraphics[width=\linewidth]{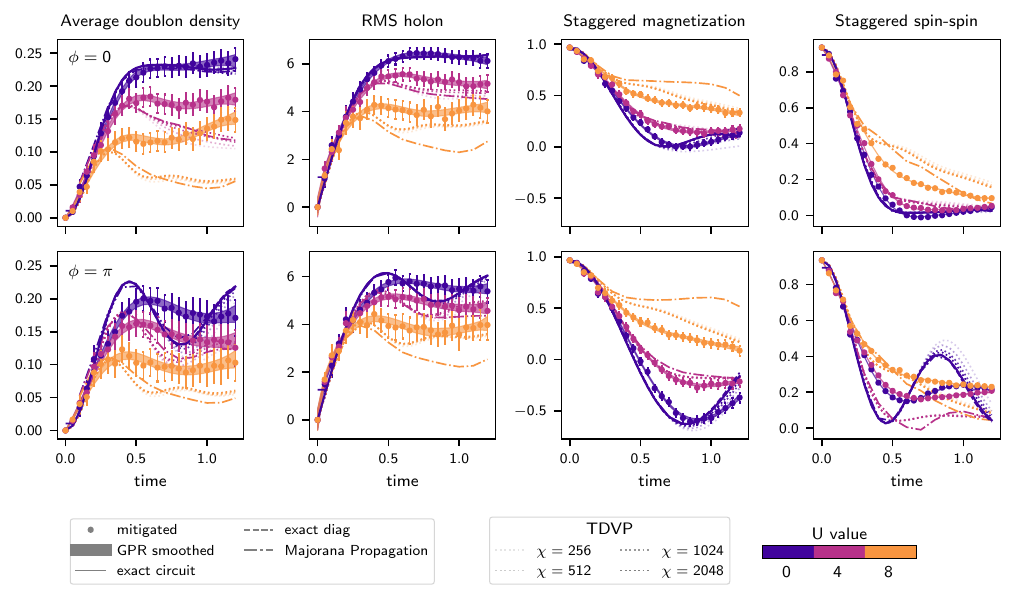}
    \caption{\justifying Evolution of average doublon density, RMS holon, staggered magnetisation and staggered spin-spin correlation for the $6\times5$ lattice with a central hole for $\phi=0$ (top row) and $\phi=\pi$ (bottom row), obtained from TFLO mitigated experimental results using $62$ qubits and $3172$ two-qubit gates. Here we also plot the results obtained using tensor network simulations at different bond dimensions, and exact circuit and TDS evolution when possible $(U=0, 4, 8)$.}
    \label{fig:central_hole_6x5_global_observables}
\end{figure*}

\begin{figure*}[!htbp]
    \centering    \includegraphics[width=\linewidth]{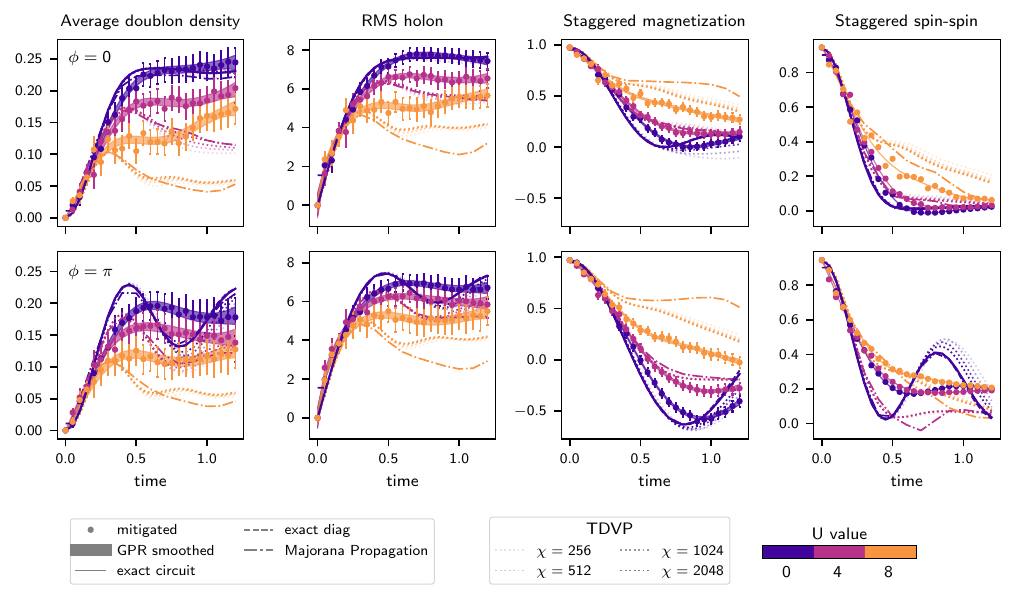}
    \caption{\justifying Evolution of average doublon density, RMS holon, staggered magnetisation and staggered spin-spin correlation for the $6\times6$ lattice with a central hole for $\phi=0$ (top row) and $\phi=\pi$ (bottom row), obtained from TFLO mitigated experimental results using $74$ qubits and $4372$ two-qubit gates. Here we also plot the results obtained using tensor network simulations at different bond dimensions, and exact circuit and TDS evolution when possible $(U=0, 4, 8)$.}
    \label{fig:central_hole_6x6_global_observables}
\end{figure*}

It is important to note that (i)~the majority of gates in our circuits (FSWAP and hopping) leave the Majorana weight unchanged, and (ii)~that many observables of physical interest (including spins, charges and related quantities) have low Majorana weight.
This suggests that the Majorana propagation algorithm should be expected to perform well in the simulation of the time series shown in Fig.~\ref{fig:staggered_magnetisation_scales}, \ref{fig:central_hole_5x5_global_observables}, \ref{fig:central_hole_6x5_global_observables} and~\ref{fig:central_hole_6x6_global_observables}.

The MP algorithm increases the number of terms to track with each gate application.
To keep the number of terms feasible, it is necessary to regularly truncate the Majorana polynomial during the computation.
In order to control this, we truncate the Majorana polynomial by discarding those terms for which either the Majorana weight -- the number of single Majorana operators in the monomial -- exceeds a certain weight threshold, or the coefficient is below a given coefficient threshold.

We implement this algorithm following the implementation of the related Pauli propagation algorithm~\cite{PauliPropagation1,PauliPropagation2} in~\cite{pauli_propagation_jl} and building upon the  \texttt{Yao.jl} quantum simulation framework~\cite{YaoFramework2019}.

\begin{figure}[!htbp]
  \begin{center}
    \includegraphics{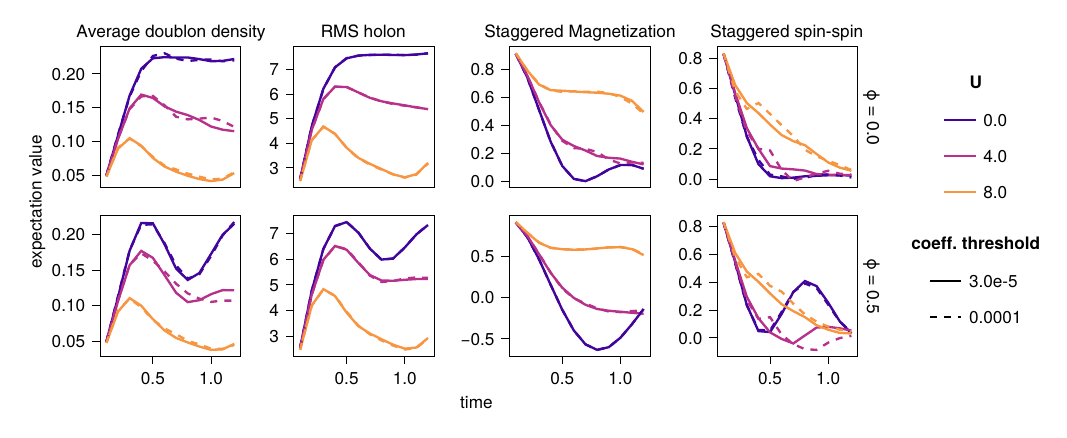}
  \end{center}
  \caption{\justifying
      \mpsims time series for coefficient truncation thresholds for the $6 \times 6$
      system with the \neel state with a central hole.
      The data with the coefficient threshold $3 \times 10^{-5}$ is also shown in Fig.~\ref{fig:central_hole_6x6_global_observables} and we obtained very similar results for the $5\times 5$ and $6\times 5$ lattices
}
  \label{fig:mp_sim_convergence}
\end{figure}

Naively, the inclusion of singlets in the initial state would increase the simulation cost exponentially with the number of singlets, because each circuit that prepares the singlet term doubles the number of Majorana monomials we need to track.
But by exploiting the fact that -- at least in the Heisenberg picture -- the state preparation circuits are the last ones to act on the qubits involved, we can discard all terms orthogonal to the zero-state immediately after applying a singlet preparation circuit.
This keeps the number of Majorana monomials we need to track constant and implies that including singlets in the initial state does not increase the hardness of the MP simulation.
We hence expect that simulations with the other initial states considered in this work can be done to the same accuracy as those shown in Fig.~\ref{fig:mp_sim_convergence}.

\begin{figure}[!htbp]
  \begin{center}
    \includegraphics{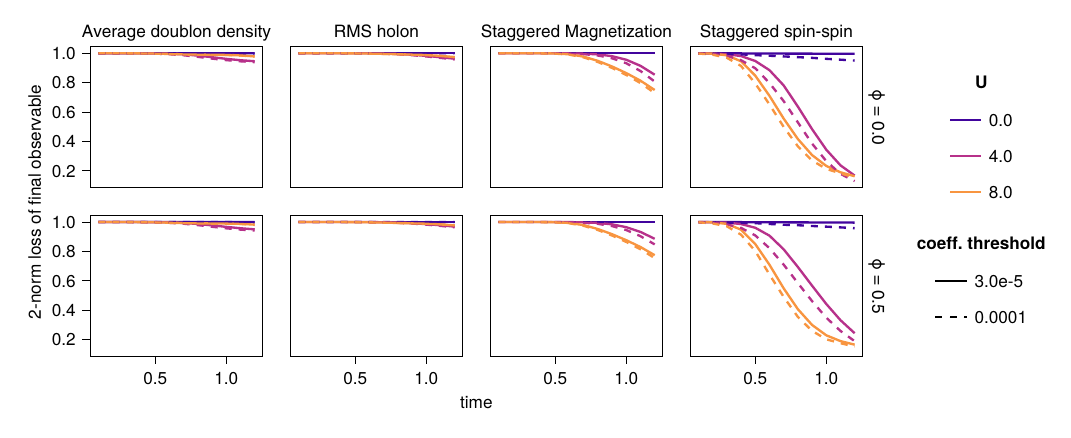}
  \end{center}
  \caption{\justifying
    Ratio of the $\ell^2$-norm of the observable before and after Heisenberg evolving it through the circuit for the same time series shown in Fig.~\ref{fig:mp_sim_convergence}.
    In the absence of truncation, the unitary evolution would preserve the $\ell^2$-norm of the observable perfectly, making the $\ell^2$-norm loss shown here a loose proxy for the accuracy of the simulation.
  }
  \label{fig:mp_sim_norm_loss_google}
\end{figure}


Fig.~\ref{fig:mp_sim_convergence} shows the performance of our \mpsims for different coefficient and weight thresholds, up to a weight threshold of 10 and a coefficient threshold of $3 \times 10^{-5}$.
It is apparent that the simulations for the average number of doublons, RMS holon and staggered magnetisation are not significantly affected by the change in coefficient threshold, suggesting that the time series presented in the main text are converged -- a necessary condition for quantitative accuracy.
This is further exemplified by fig.~\ref{fig:mp_sim_norm_loss_google} where we show the loss of $\ell^2$-norm incurred by the truncation throughout the circuit for the same time series shown in fig.~\ref{fig:mp_sim_convergence}.
We see that, except for the staggered spin-spin correlator, we lose only relatively little $\ell^2$-norm due to the truncation, hinting that the \mpsim results for all observables except the staggered spin-spin correlator are fairly accurate.

However, we found that that the $\ell^2$-norm incurred by truncation ceased to be a good proxy for the accuracy of the \mpsim for observables with higher Majorana weight.
For the \mpsims of the different holon and doublon correlators shown in fig. \ref{fig:pairwise_holon_distance} we observed less than 5 \% $\ell^2$-norm loss even at the latest times and for $U=8$, but the accuracy of the simulations very low as can be seen from the $U=0$ data in fig. \ref{fig:pairwise_holon_distance}.
Overall, we take this to mean that the convergence with the truncation thresholds is a better proxy for the accuracy of the \mpsims than the $\ell^2$-norm loss is. 
Finding better accuracy proxies that don't require running multiple simulations at different truncation accuracies is an open research question.


\begin{figure}[!htbp]
  \centering
  \begin{minipage}[t]{0.49\linewidth}
      \centering
      \includegraphics[width=\linewidth]{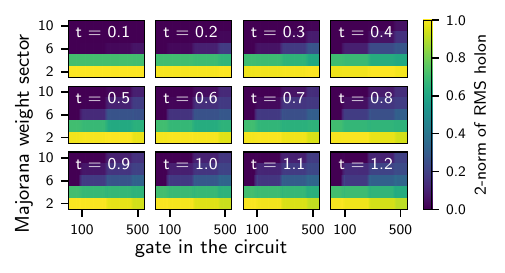}
      \\[-0.5em]
      {\small (a)}
  \end{minipage}%
  \hfill
  \begin{minipage}[t]{0.49\linewidth}
      \centering
      \includegraphics[width=\linewidth]{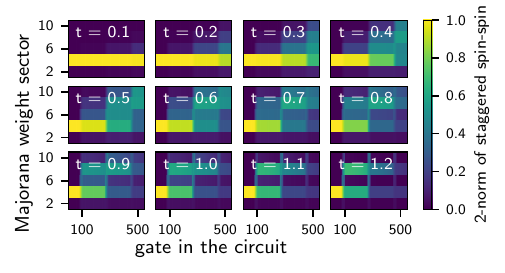}
      \\[-0.5em]
      {\small (b)}
  \end{minipage}
  \caption{\justifying
    Spreading of the observables over the Majorana weight sectors as they Heisenberg evolve through the circuit for RMS holon (left) and the spin-spin correlator (right) at all simulated times for $U=8$ and $\phi =0$. Note that for the MP simulations we wrote the circuit entirely in the fermionic language, i.e.\ as a sequence of hopping and onsite evolutions instead of a hardware native sequence of single qubit and CZ gates. This means the gate count for MP simulations is much lower than for the hardware circuits, even though they describe equivalent unitaries.
  }
  \label{fig:mp_sim_evolution}
\end{figure}

Fig.~\ref{fig:mp_sim_evolution} shows the impact of the Heisenberg evolution through the circuit of the spread of Majorana polynomials over the different Majorana weights.
It is important to note that the weight 2 and 4 sectors can be simulated exactly without truncation: this implies that, for all observables of interest, in the $U=0$ case, the simulations are essentially intact.
For higher values of $U$, however, this changes depending on the circuit of interest.
In particular, while the number of RMS holons remains mostly supported in the weight-2 sector throughout the evolution, for the staggered spin-spin correlation function, this changes dramatically, with support rapidly spreading from the weight-4 sector to higher-weight sectors.

\begin{figure}[!htbp]
  \begin{center}
    \includegraphics{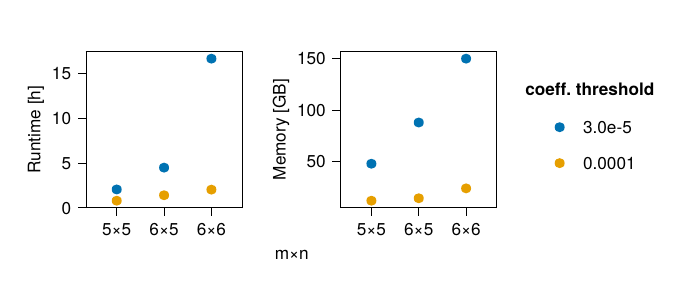}
  \end{center}
  \caption{\justifying
    Maximum memory and runtime requirements across all observables, and time
    steps for the \mpsims.
}
  \label{fig:mp_sim_resources}
\end{figure}


\clearpage


\section{Linear Cross-Entropy Benchmarking and Model Testing}
\label{sec:cross-entropy}

Cross-entropy benchmarking is by now a well-established method for validation of NISQ devices~\cite{arute2019quantum,morvan2024phase,zlokapa2023boundaries,cheng2025generalized,decross2025computational}, using either exponential classical resources to compute ideal amplitudes or efficient classical simulation of special ensembles of circuits (such as Clifford circuits~\cite{chen2023linear} or number-conserving circuits with $\mathcal{O}(1)$~\cite{kaneda2025large}).  Existing applications of this method rely on approximate Haar random states as output, e.g. produced through ergodic circuit ensembles~\cite{cheng2025generalized} or time averaging in an analogue simulator~\cite{mark2023benchmarking}. Here, we adapt the linear cross-entropy toolkit to highly entangled (but not at all random) output states of the Fermi-Hubbard simulation results described in the main text.  We extend linear XEB to the setting of free fermion evolutions by using FLO determinants to efficiently compute output amplitudes, along with related classical simulation methods to determine the time-dependent collision probability of the ideal circuit. This allows us to perform XEB verification without using exponential-time classical resources.

The goal of our benchmarking protocol is to determine whether the device samples are reflected with a higher-than-average probability in the ideal \emph{untrotterised} output state of the circuit, leading to a fidelity estimate that combines the effects of device noise and Trotterisation to benchmark the entire digital implementation of the time dynamics.

\notocline\subsection{Linear XEB for FLO circuits} The standard linear XEB quantity for random circuit sampling is
$$
F = 2^n \langle p_{\textrm{ideal}}\rangle_{\textrm{device}} - 1
$$
in which $\langle p_{\textrm{ideal}}\rangle_{\textrm{device}}$ is estimated by summing the ideal circuit probabilities $p_{\textrm{ideal}}(z) = |\langle z | U | 0^n\rangle|^2$ of sufficiently many bit string sample $z$ observed from the device. If the device were noiseless, then this expectation value would be the collision probability of the circuit output in the $Z$ basis
$$
\langle p_{\textrm{ideal}}\rangle_{\textrm{ideal}} = \sum_{z \sim p_{\textrm{ideal}}} p_{\textrm{ideal}}(z) = \sum_{z} p_{\textrm{ideal}}(z)^2 \equiv C_\textrm{ideal}
$$
This computational basis collision probability $C_\textrm{ideal}$ for approximately Haar random circuits in Hilbert space dimension $D$ is tightly concentrated on $C_\textrm{ideal} = 2/D$, and so a noiseless implementation of random quantum circuits will find $F = 1$.   In contrast, the collision probability for the uniform distribution is $C_{\textrm{uniform}} = 1/D$, which has $F = 0$.  The main reason that $F$ is scaled and shifted this way is so that it corresponds directly to an overall circuit fidelity under the global depolarising noise model,
$$
\rho_\textrm{device} = F |\psi_\textrm{ideal}\rangle\langle \psi_{\textrm{ideal}}| + (1-F) (I/D)
$$
The ansatz above implies $p_\textrm{device}(z) = F p_\textrm{ideal}(z) + (1-F) (1/D)$ and so the estimator becomes
$$
\langle p_{\textrm{ideal}}\rangle_{\textrm{device}} = \sum_{z \sim p_{\textrm{ideal}}} p_\textrm{ideal} = \sum_{z} p_\textrm{ideal}(z)(F p_\textrm{ideal}(z) + (1-F)(1/D)) = F(2/D) + (1-F)(1/D) = (1 + F)/D
$$
which explains how the definition of $F$ corresponds to a circuit fidelity.   Summarising, the method requires (1) classical simulation of the ideal output probabilities to compute the estimator of $\langle p_{\textrm{ideal}}\rangle_{\textrm{device}}$, and (2) knowledge of the collision probability for the ideal output state, in order to compute the correct rescaling and shifting to relate $\langle p_{\textrm{ideal}}\rangle_{\textrm{device}}$ to a circuit fidelity $F$ under the global depolarizing noise ansatz.

For $U = 0$ circuits with appropriate input states, we can efficiently compute amplitudes for the ideal physical time evolution (for more details see \cref{sec:bitstring_probabilities}), so we can compute the estimator $\langle p_{\textrm{ideal}}\rangle_{\textrm{device}}$ using samples from the device.   To understand how to scale it into a fidelity under the global depolarising model, we need to estimate the time-dependent collision probability
$$C(t) = \sum_z p_\textrm{ideal}(z,t)^2$$
which we can approximate at each point in time using a Monte Carlo estimator for $C = \langle p_{\textrm{ideal}}\rangle_{\textrm{ideal}}$ that combines efficient strong and weak simulation of FLO circuits.    Now, if we assume the global depolarising noise ansatz with fidelity $F$,
$$
\langle p_{\textrm{ideal}}\rangle_{\textrm{device}} = \sum_z p_\textrm{ideal}(z)(F p_\textrm{ideal}(z) + (1-F)/D) = F C + (1-F)/D
$$
This motivates defining the general analogue of the linear XEB quantity to be
\begin{equation}F = \frac{D \langle p_{\textrm{ideal}}\rangle_{\textrm{device}} - 1}{C D - 1}.\label{eq:XEBF}\end{equation}
Following the role of $D$ in the derivation above, when symmetries like particle number and spin conservation in the Fermi-Hubbard model constrain ideal $Z$ basis measurements to a subset of  bit strings, we take $D$ to be the size of this subset of allowed measurement outcomes.  A caution is that just as with linear XEB for random circuit sampling, $F$ can be interpreted as a fidelity under the assumption of global depolarising noise, but more generally it can be seen as a normalised linear overlap between distributions since $\langle p_\textrm{ideal}\rangle_\textrm{device} = \langle p_\textrm{ideal},p_\textrm{device}\rangle$, and values $F > 1$ are possible in principle, which which we address in the next section.  

\begin{figure}[!htbp]
  \begin{center}
    \includegraphics[width=0.95\textwidth]{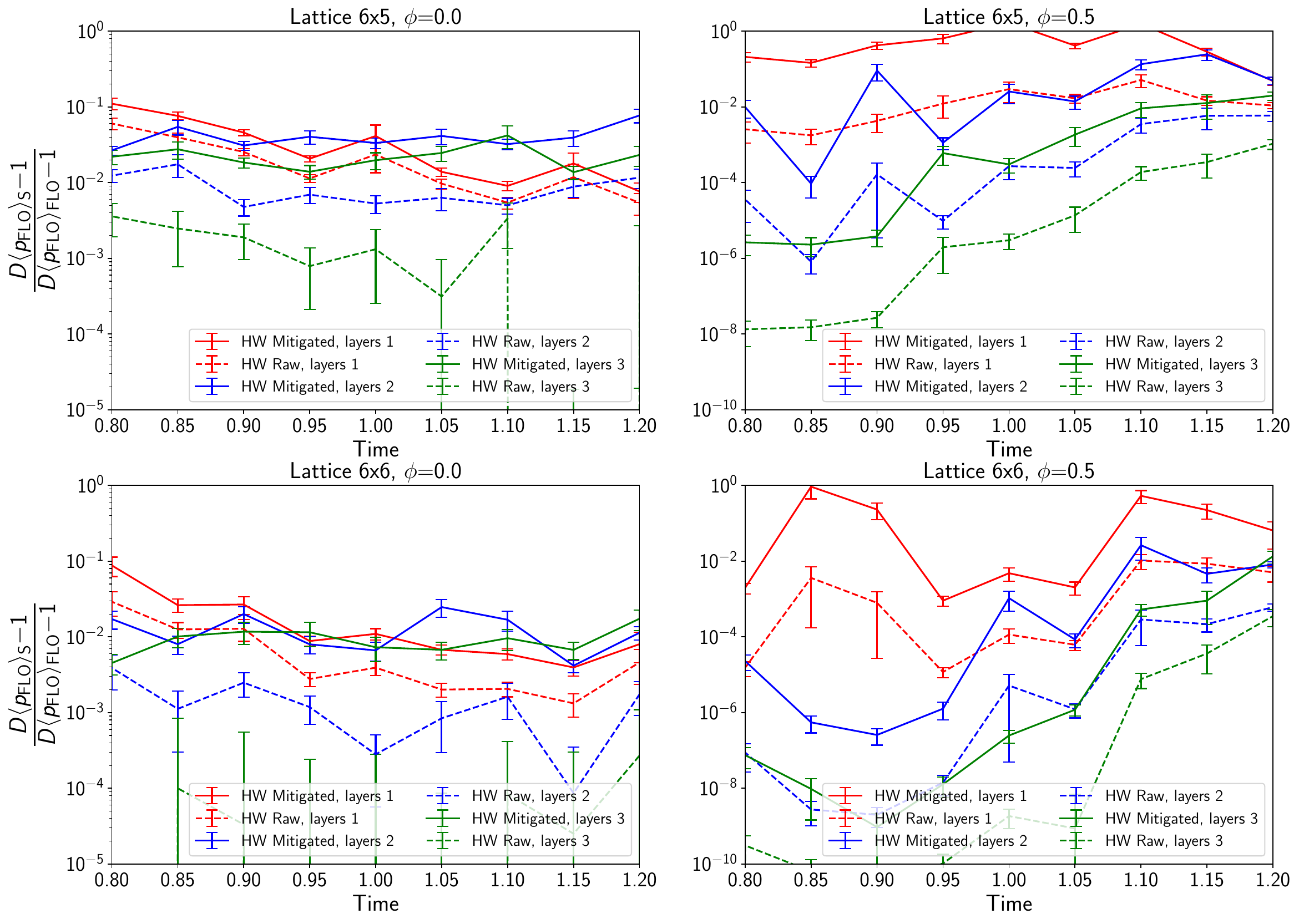}
  \end{center}
  \caption{\justifying Estimates of the normalised linear overlap in \eqref{eq:XEBF}, which corresponds to a circuit fidelity under the global depolarising noise model using linear XEB applied to FLO circuits. In all cases, post-selected samples from the device are compared to ideal amplitudes from untrotterised FLO, and so this quantity represents the total overlap of the digital implementation with the ideal state, including device noise and Trotter error. \label{fig:noisetestingXEB} }
\end{figure}

Linear XEB results for unmitigated samples, mitigated samples (produced by the maximum entropy shot reweighting of \cref{sec:shot_reweighting}) and different numbers of Trotter layers are shown in Fig.~\ref{fig:noisetestingXEB}.   Since the method is designed around the benchmarking of highly entangled states, we exclude short times when $C \gg D^{-1}$ in \eqref{eq:XEBF}, and present results for times $0.8,...,1.2$, when the collision entropy of the ideal distribution is nearly saturated.   The mitigated results show fidelities around the 1\% level at late times, and also enable a comparison of simulations with different numbers of Trotter layers.  In the $\phi = 0$ results, we can discern a negative slope for the $n_{\textrm{layers}} =1$ results (indicating an exponential decay in fidelity with evolution time) while the $n_{\textrm{layers}} = 2,3$ do not show this decay trend and achieve a higher fidelity at time $t = 1.2$.

\notocline\subsection{Linear Cross-Entropy Model Testing}  
\begin{figure}[!htbp]
  \begin{center}
    \includegraphics[width=0.95\textwidth]{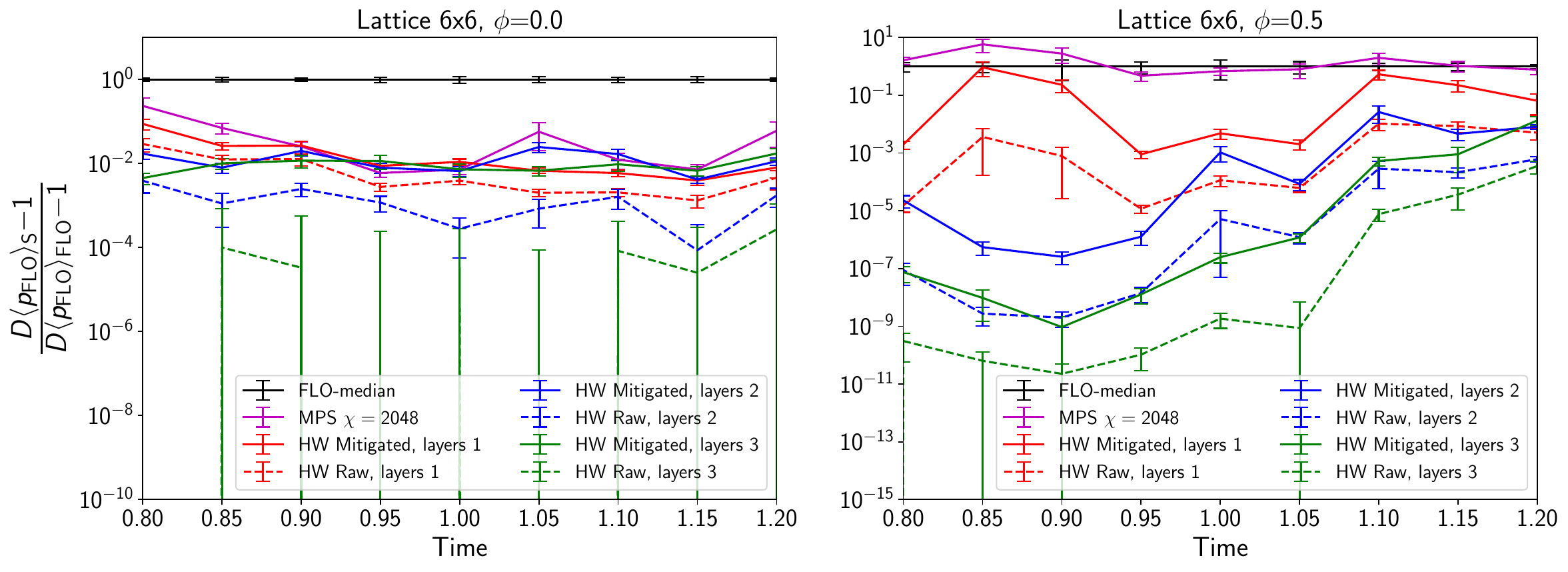}
  \end{center}
  \caption{\justifying Estimates of the  normalised linear overlap or linear XEB fidelity including a comparison of both the experimental data and the MPS as a sample generator, both compared to ideal FLO.  The MPS generally achieves a larger value of $F$ in \eqref{eq:XEBF}, but as described in the text this is an inconclusive result due to being unable to normalise the test across different sample generators.  \label{fig:noisetestingXEBMPS} }
\end{figure}

In the context of random quantum circuit sampling (RCS),  linear XEB has been used to score classical simulations such as large scale tensor networks against random circuits~\cite{liu2024verifying}, as well as classical algorithms for ``spoofing'' XEB~\cite{barak2020spoofing,gao2024limitations}.   This approach seems natural (replace the quantum computer with a competing classical sample generator), but a drawback is that the linear estimator $\langle p_\textrm{model}\rangle_\textrm{generator}$ is not a proper scoring rule when one fixes the model (in this case the ideal circuit) and changes the generator.  In other words, in both RCS and our setting, there are sample generators which achieve higher scores than samples from the ideal circuit.  An extreme case is a sample generator which always outputs one fixed string $x$ which has a larger than average likelihood $p(x)$ - such a generator can achieve a linear XEB score larger than $1$, in both RCS and our setting.  
To understand how this estimator  $\langle p_\textrm{model}\rangle_\textrm{generator}$ can be normalised, we can view it as an inner product, 
$$\langle p_\textrm{model}\rangle_\textrm{generator} = \sum_x p_\textrm{model}(x) p_\textrm{generator}(x) = \langle p_\textrm{model},p_\textrm{generator}\rangle$$
where here $\langle \cdot , \cdot \rangle$ is the inner product between the distributions as vectors in high dimensional Euclidean space.   Therefore the linear estimator depends on the angle between these vectors, but also on their 2-norms, $\langle p_\textrm{model}\rangle_\textrm{generator}= \langle p_\textrm{model},p_\textrm{generator}\rangle = \|p_\textrm{model}\|_2 \|p_\textrm{generator}\|_2 \cos \theta$.  The central limitation is that $\|p_\textrm{device}\|_2$ cannot be determined accurately when bit string collisions (seeing the same string twice) are rare or non-existent in the set of samples.   Suppose one compares the linear estimator with a fixed ideal model for both the quantum device and the MPS as sample generators, 
$$\frac{\langle p_\textrm{FLO}\rangle_\textrm{device}}{\langle p_\textrm{FLO}\rangle_\textrm{MPS}} = \frac{\|p_\textrm{device}\|_2 \cos \theta_{\textrm{FLO},\textrm{device}}}{\|p_\textrm{MPS}\|_2 \cos \theta_{\textrm{FLO},\textrm{MPS}}}$$
and we generically expect $\|p_\textrm{MPS}\|_2 \geq \|p_\textrm{device}\|_2$ , since the MPS accesses a limited Hilbert space and the device is subject to noise (recall that $\|(1,0,...,0)\|_2 = 1$ and $\|(D^{-1},...,D^{-1})\|_2 = D^{-1/2}$, so the 2-norm decreases as the distribution becomes more delocalized or uniform).    This explains why an approximate simulator with a more concentrated distribution can ``spoof'' the results.   Note the situation is asymmetric: as long as we believe $\|p_\textrm{MPS}\|_2 \geq \|p_\textrm{device}\|_2$, then $\langle p_\textrm{FLO}\rangle_\textrm{device} > \langle p_\textrm{FLO}\rangle_\textrm{MPS}$ is meaningful, but $\langle p_\textrm{FLO}\rangle_\textrm{device} < \langle p_\textrm{FLO}\rangle_\textrm{MPS}$ is inconclusive.  In Fig.~\ref{fig:noisetestingXEBMPS} we report the same normalised linear overlaps shown in Fig.~\ref{fig:noisetestingXEB}, but now with the inclusion of the MPS as sample generator.   We see the MPS samples generally find a larger inner product with the FLO distribution than the experiment, so this result is inconclusive for the reasons just discussed.

\begin{figure}
  \begin{center}
    \includegraphics[width=0.45\textwidth]{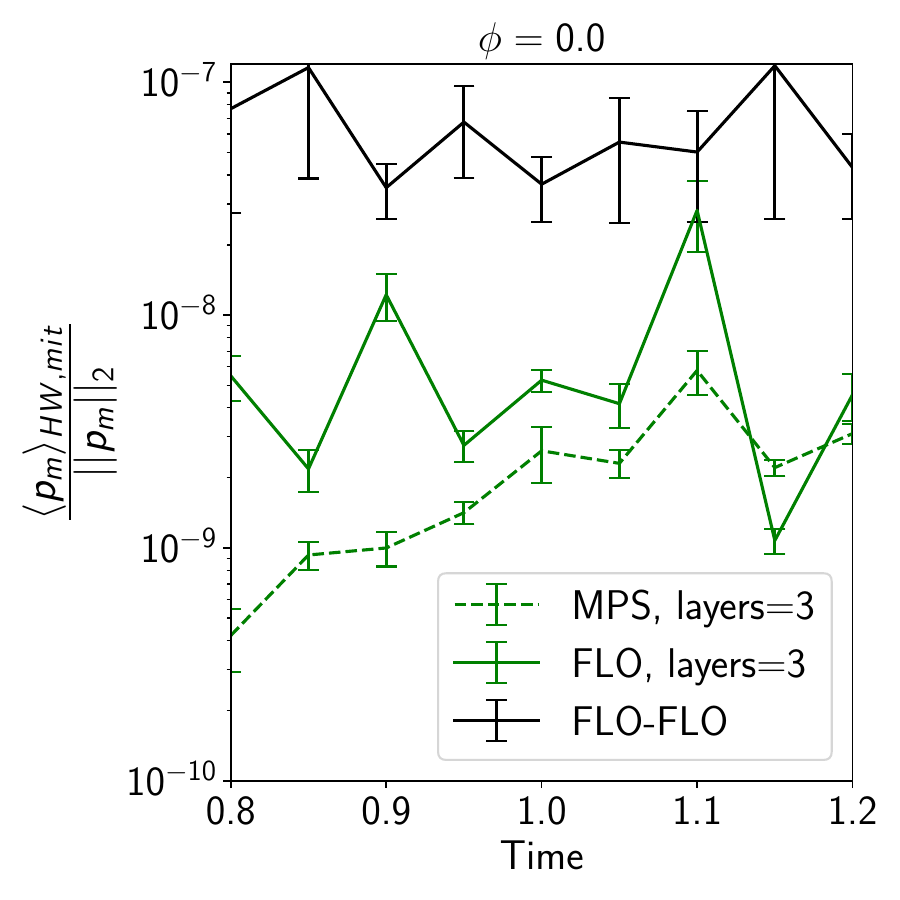}
    \includegraphics[width=0.45\textwidth]{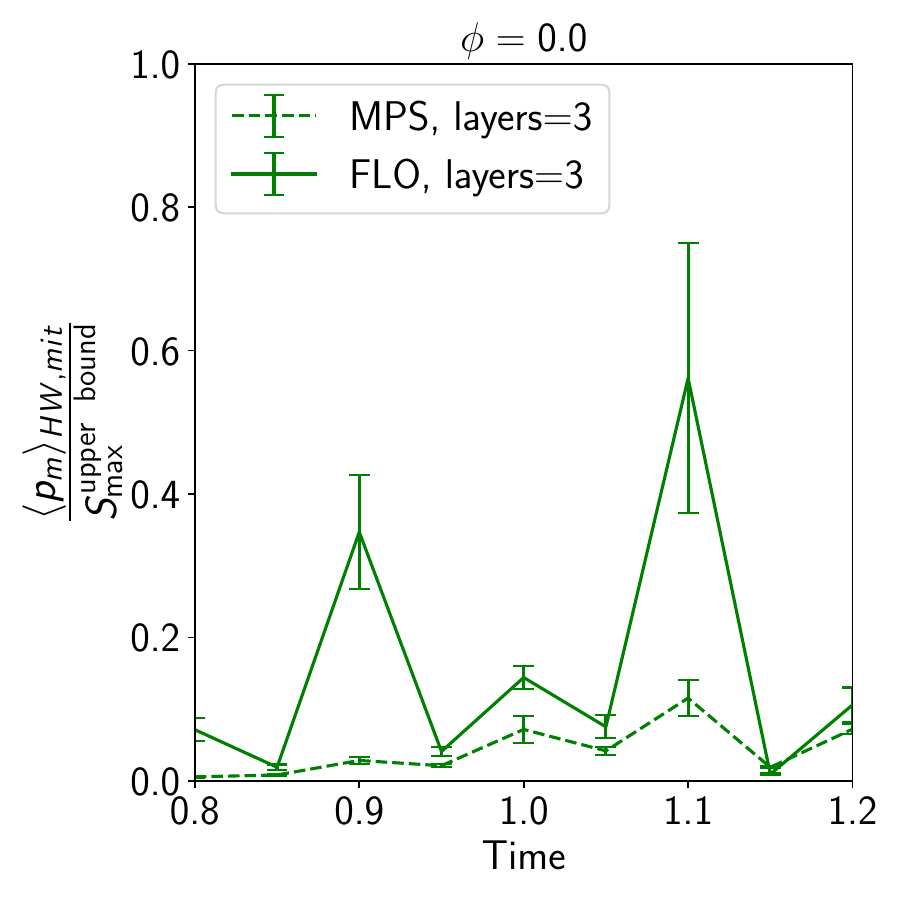}
  \end{center}
  \caption{\justifying Spherical scores for FLO and the largest MPS with $\chi = 2048$ for an experiment with $6\times 5$ sites, parameters $U = 0, \phi = 0.0 $, with a singlets plus center holons initial state.  Both FLO and MPS are compared to the same set of MESR mitigated experimental samples.  The black line on the left is $\|p_\textrm{FLO}\|_2$, which is an upper bound on the maximum possible score.  The plot on the right shows the spherical scores of FLO and the MPS, normalised by this best known upper bound. \label{fig:sphericalScoreXEB}
  }
  \end{figure}

Another conceptually distinct use of the linear estimator $\langle p_\textrm{model}\rangle_\textrm{generator}$ is to fix the generator (in our case it will be hardware samples), and compare likelihoods for those samples under different models (e.g. ideal FLO and approximate MPS).  For a fixed sample generator and set of models, each model can be assigned a spherical score,
$$S(p_\textrm{model}) = \frac{\langle p_\textrm{model} \rangle_{\textrm{generator}}}{\|p_\textrm{model}\|_2}$$
The spherical score is a proper scoring rule over models: it is uniquely maximized when the model matches the generator, and so it can't be spoofed.  A larger spherical score for a given model means that is more aligned with the generator distribution, as a vector in high dimensional Euclidean space.
The spherical scores for FLO and MPS likelihoods modelling experimental samples are shown in Fig. ~\ref{fig:sphericalScoreXEB}.  The ratio of the spherical score of two models is the ratio of $\cos(\theta)$ between each model and the generator, so a ratio of $\frac{S(m_2)}{S(m_1)} = r$ implies that model $m_2$ is $r$ times closer to the generator in cosine-angle than model $m_1$.  The maximum possible score is $\|p_\textrm{HW}\|_2$, which is intractable, so to give further context to these scores we compute $\|p_\textrm{FLO}\|_2 = S_{\max}^{\textrm{upper bound}}$ as an upper bound on the best possible score, since  $\|p_\textrm{FLO}\|_2 > \|p_\textrm{HW}\|_2$ under the assumption that the device samples are more delocalized than the ideal state.  

\clearpage

\section{Trotter Error}\label{sec:trotter_error}

\begin{figure*}[!htbp]
  \centering

\includegraphics[width=\linewidth]{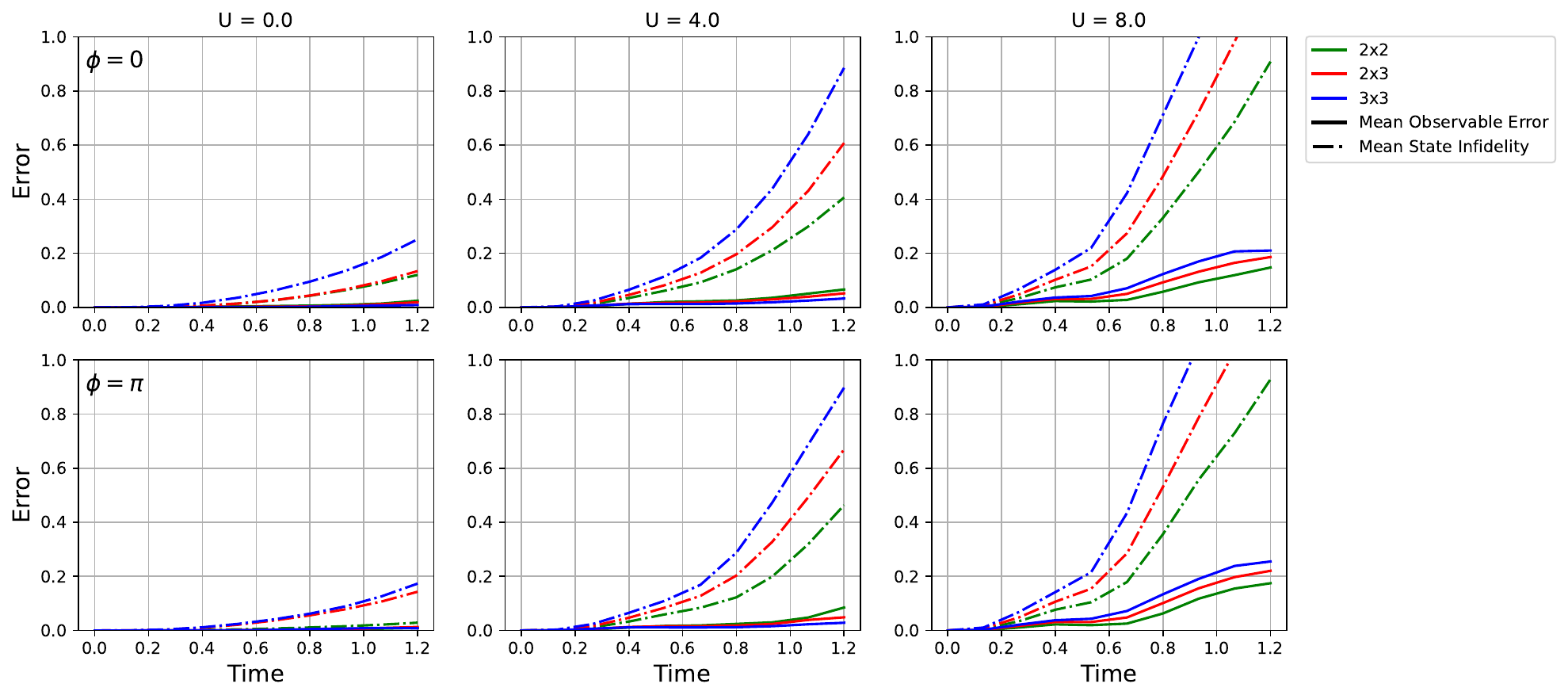}
  \caption{\justifying Scaling of average error in state infidelity (dashed) and observable expectation value (solid) with time for coupling strength $U=0, 4, 8$ and magnetic field $\phi=0, \pi$ for lattice sizes $L_x \times L_y = 2\times2, 2\times 3$ and $3\times3$.}
  \label{fig:trotter-mean}
\end{figure*}

\begin{figure*}[!htbp]
  \centering

\includegraphics[width=\linewidth]{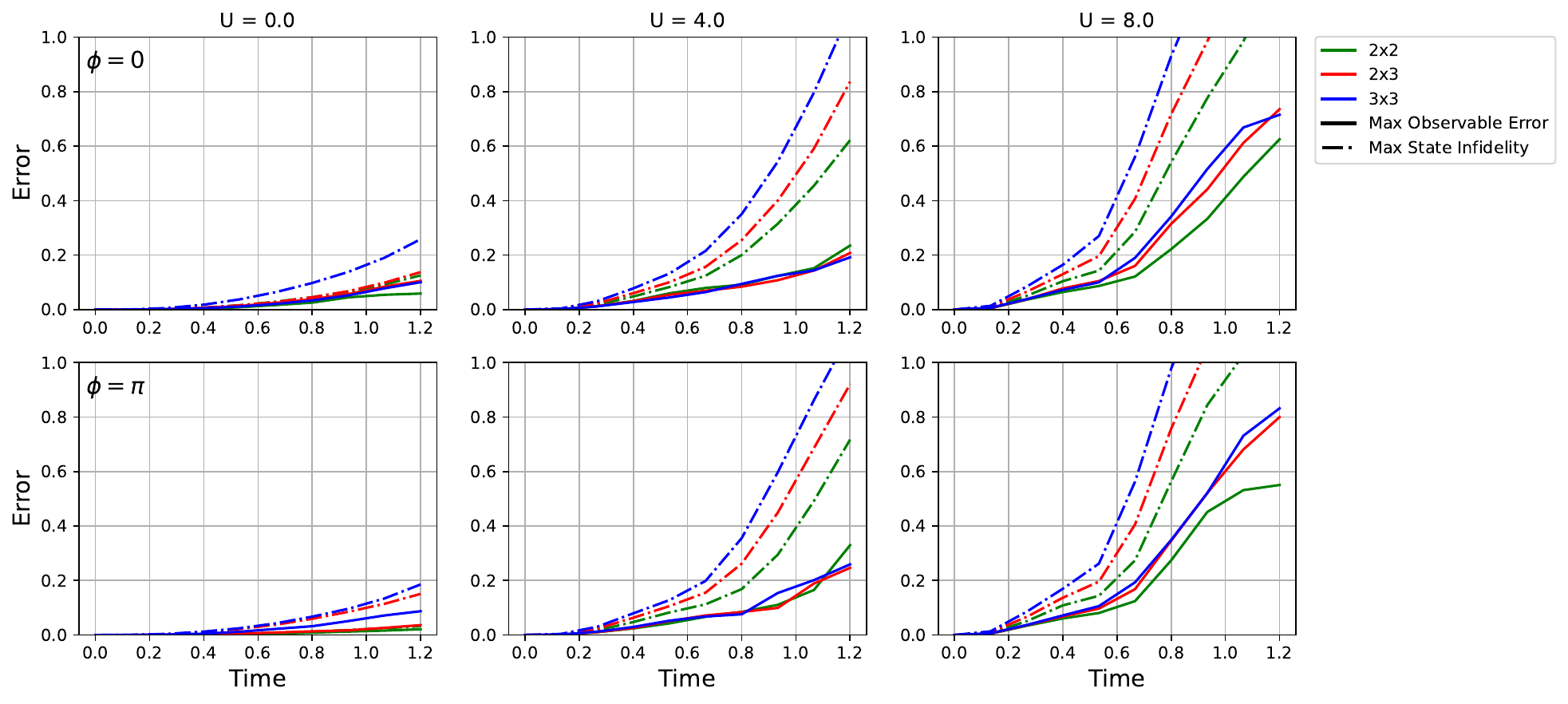}
  \caption{\justifying Scaling of maximum error in state infidelity (dashed) and observable expectation value (solid) with time for coupling strength $U=0, 4, 8$ and magnetic field $\phi=0, \pi$ for lattice sizes $L_x \times L_y = 2\times2, 2\times 3$ and $3\times3$.}
  \label{fig:trotter-max}
\end{figure*}

\begin{figure*}[!htbp]
  \centering

\includegraphics[width=\linewidth]{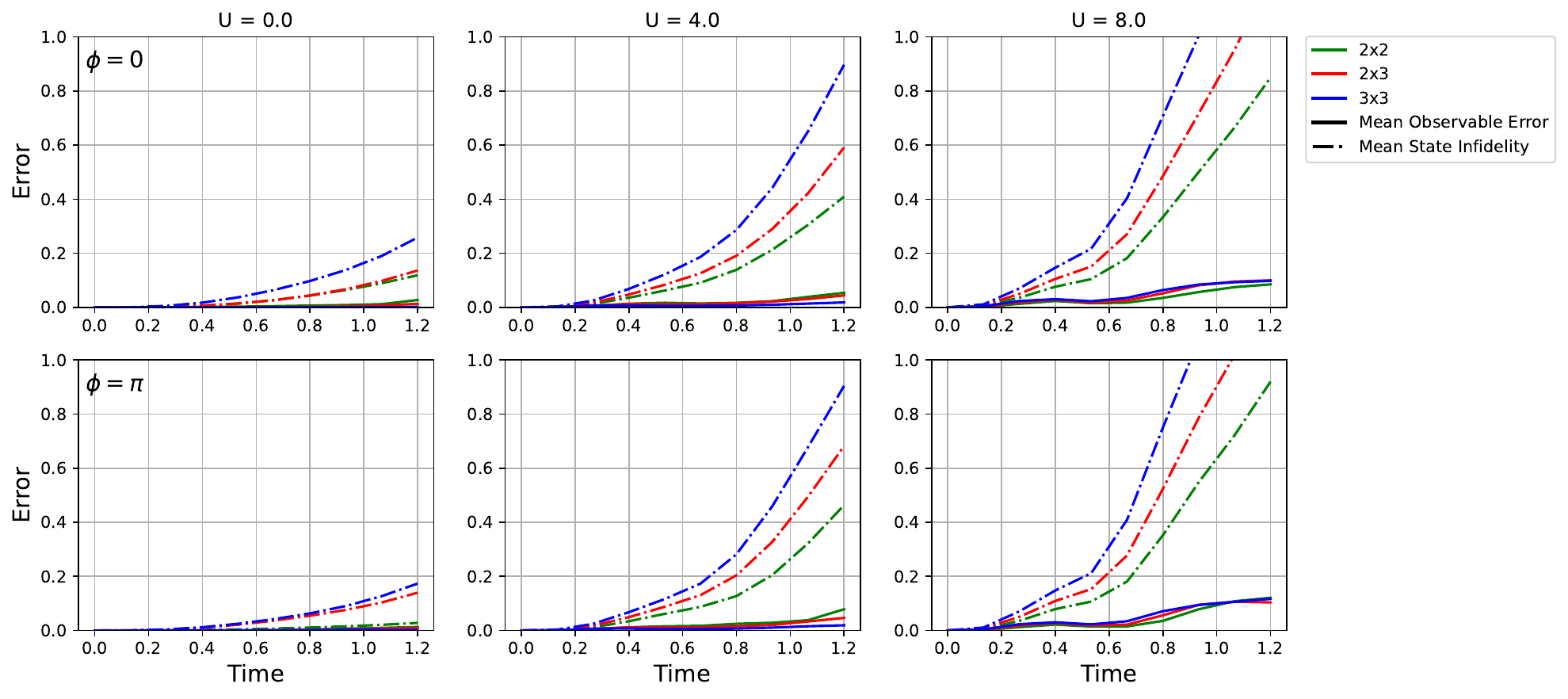}
  \caption{\justifying Scaling of state infidelity (dashed) and of maximum error in observable expectation value (solid) with time for coupling strength $U=0, 4, 8$ and magnetic field $\phi=0, \pi$ for lattice sizes and observable weights: $L_x \times L_y = 2\times2$ and weight-$4$, $2\times 3$ and weight-$6$, $3\times3$ and weight-$8$. }
  \label{fig:trotter-mean-scaling}
\end{figure*}

\begin{figure*}[!htbp]
  \centering

\includegraphics[width=\linewidth]{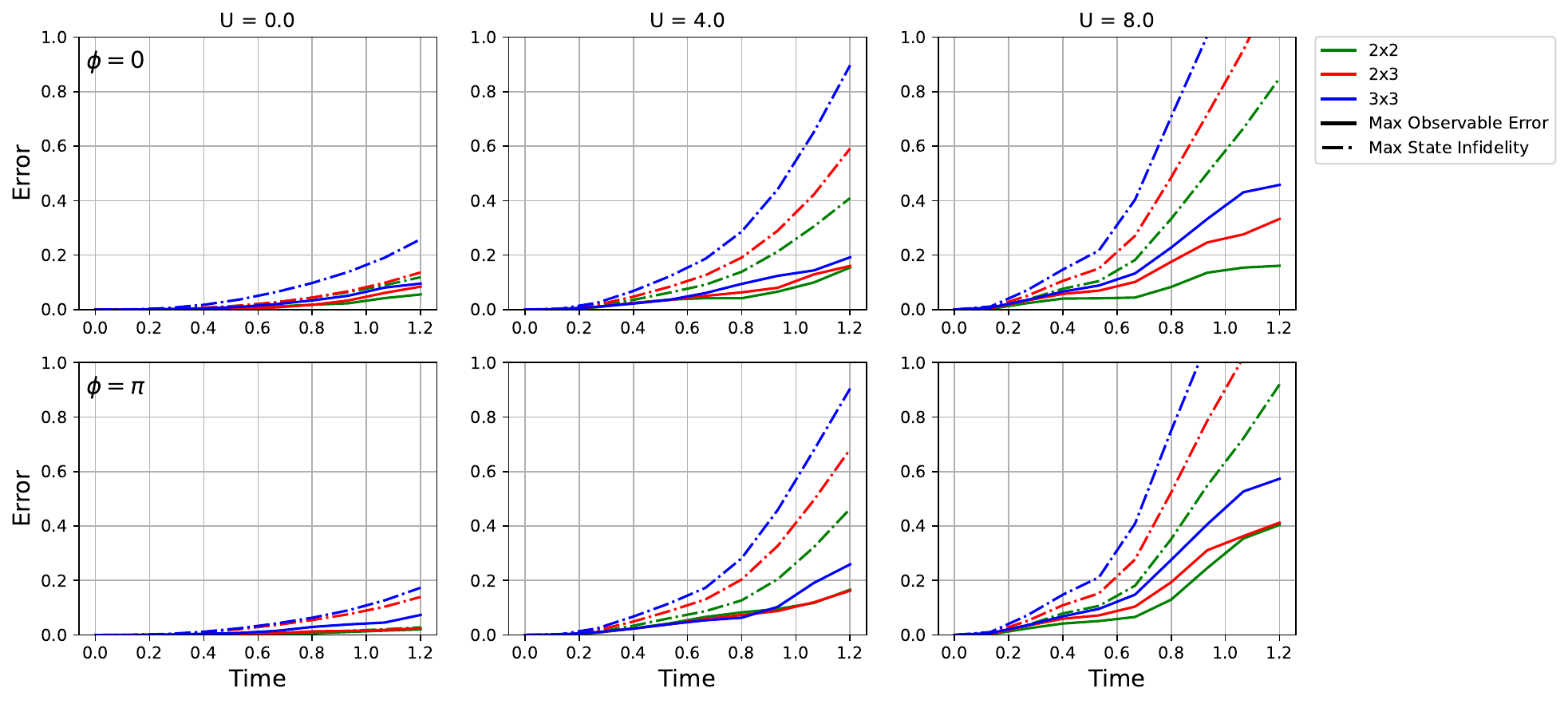}
  \caption{\justifying Scaling of state infidelity (dashed) and of maximum error in observable expectation value (solid) with time for coupling strength $U=0, 4, 8$ and magnetic field $\phi=0, \pi$ for lattice sizes and observable weights $L_x \times L_y = 2\times2$ and weight-$4$, $2\times 3$ and weight-$6$, $3\times3$ and weight-$8$. }
  \label{fig:trotter-max-scaling}
\end{figure*}

We evaluate the impact of Trotterisation on state infidelity and the expectation values of observables for small lattice sizes. We use three second-order Trotter steps and look at times between $t=0$ and $t=1.2$ and coupling strengths $U=0,4,8$. We compute the exact time-evolved state $\ket{\phi(t)} = e^{-i t H} \ket{\phi(0)}$  and the Trotterised time-evolved state, $\ket{\phi'(t)} = \prod_{j=1}^3 S_2(t/3)\ket{\phi(0)}$. We then calculate
\begin{equation}\label{eq:state-error-trotter}
	2 \sqrt{1 - |\langle\phi(t) | \phi' (t)\rangle|^{2}}
\end{equation}
for a variety of initial states $\ket{\phi(0)}$. We look at a N\'eel ordered state, hole-doping in the centre and upper left corner of the lattice, an upper and lower side hole and a vertical holon stripe.

For all Pauli-$Z$ observables between weight-$1$ and weight-$4$, we compute the error in expectation value
\begin{equation}\label{eq:trotter-obs-error}
	|\bra{\phi(t)}P\ket{\phi(t)} - \bra{\phi'(t)}P\ket{\phi'(t)}|.
\end{equation}
We then compute the average of both metrics over all states and observables at each time, the result of which is shown in Fig.~\ref{fig:trotter-mean} and also the maximum of both metrics over all states and observables, as shown in Fig.~\ref{fig:trotter-mean}.

We find that for $U=0$ and $U=4$ with and without magnetic flux, the average Trotter error on up to weight $4$ observables remains below $10\%$ for all times, with no clear increasing trend as a function of system size. However, for $U=8$, the Trotter error begins to become more dramatic, increasing to up to $\%25$.

There are good reasons to suspect that Trotter error for this model should not dramatically increase with system size for a given fixed time. This is because Trotter error is upper bounded by local commutator relations  \cite{Childs_2019,Childs_2021}, which do not increase with system size for intensive observables. It has also been shown that this system size independence of Trotter error can hold for relatively large Trotter steps, based on the separate study of Trotterised dynamics as Floquet dynamics \cite{Heyl_2019}.

Additionally, we repeat this analysis for observables, which scale with system size. We compute the mean and maximum Trotter error at each time for all weight-$4$ observables for size $2 \times2$, all weight-$6$ observables for size $2 \times3$ and all weight-$8$ observables for size $3 \times 3$. We do this for the initial state, which corresponds to hole doping at the centre of the lattice. The results of this analysis can be found in Fig.~\ref{fig:trotter-mean-scaling} and Fig.~\ref{fig:trotter-max-scaling}.

For $U=0$ and $U=4$, with and without magnetic flux, the average error Trotter error in observables remains below $10\%$, and there is still no obvious trend in system size.

\begin{figure*}[!htbp]
    \centering
    \includegraphics[width=\linewidth]{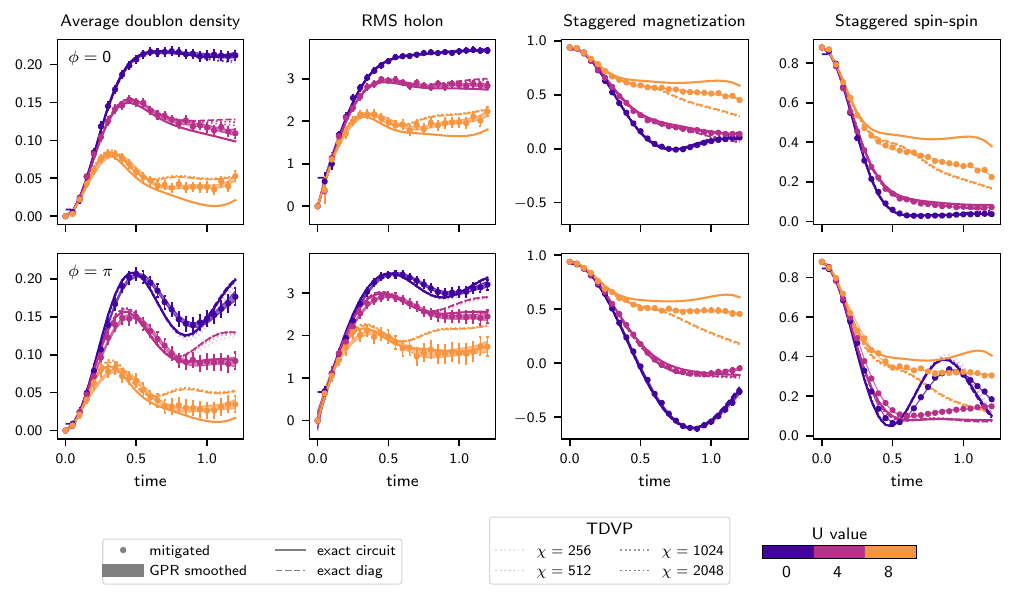}
    \caption{\justifying Evolution of average doublon density, RMS holon, staggered magnetisation and staggered spin-spin correlation for the $4\times4$ lattice with a central hole for $\phi=0$ (top row) and $\phi=\pi$ (bottom row), obtained from TFLO mitigated experimental results using $34$ qubits and $1220$ two-qubit gates. Here we also plot the results obtained using tensor network simulations at different bond dimensions, and exact circuit and TDS evolution when possible $(U=0, 4, 8)$.}
    \label{fig:central_hole_4x4_global_observables}
\end{figure*}

Although the above analyses are not reasonably possible for a $4\times4$ system, we can simulate both the exact circuit and target continuous dynamics of the Fermi-Hubbard model for a $4\times4$ system with some effort. We have also run our experiments on a $4\times4$ system. So we can compare the performance of the Trotter circuit to the target dynamics, and compare these to the hardware signal, TDVP and MP. This is shown for a few of the global target observables considered in this work in Fig.~\ref{fig:central_hole_4x4_global_observables}.

We see that for $U=0$, the Trotter error is negligible at all times, and all signals approximately agree. For $U=4$, Trotter error begins to play an important role at approximately $t=0.9$, and for $U=8$ Trotter error begins to play an important role at about $t=0.5$

\FloatBarrier


\clearpage

\section{Additional holon stripe discussion}
\label{app:additional_stripe_info} 
\begin{figure}[ht] 
\centering
\includegraphics[width=\textwidth]{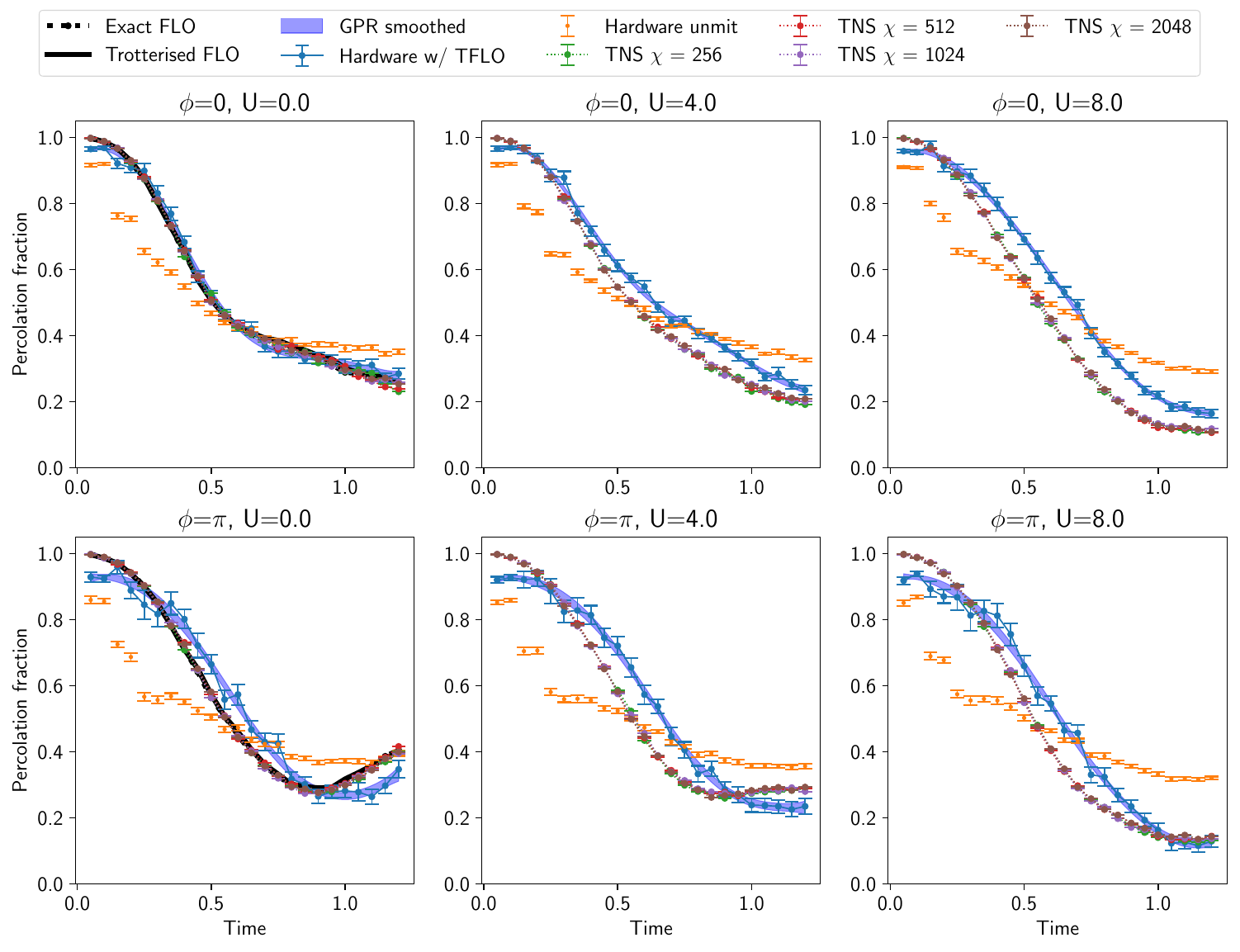}
\caption{\justifying $4\times4$, The unmitigated hardware data is postselected by antiferromagnetic order measured by AFM $\in [-1,0]$ (see \cref{eq:afm}, number of doublons $\mathcal{D}=[0,1,2]$ present and particle number conservation. TFLO is then applied to this set of shots to produce the Hardware w/ TFLO line, and finally GPR is applied to give the 68$\%$ one sigma region (for more details, see \cref{sec:post_selection}, \cref{sec:GPR} and \cref{sec:tflo}).}
\label{fig:4x4_percolation_postselection}
\end{figure}

\begin{figure}[!htbp]
\centering
\includegraphics[width=\textwidth]{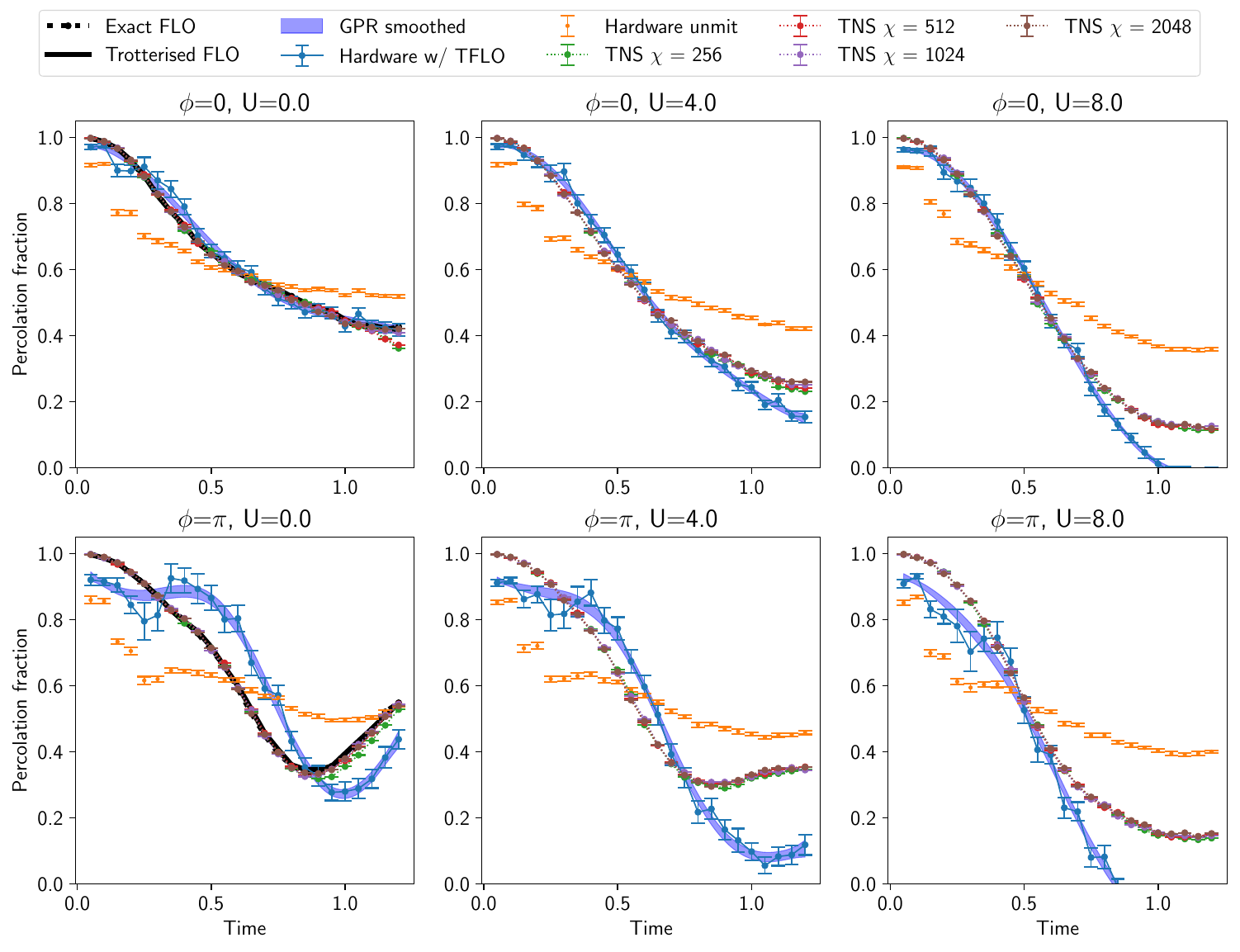}
\caption{\justifying $4\times4$, The unmitigated hardware data is postselected by only particle number conservation. TFLO is then applied to this set of shots to produce the Hardware w/ TFLO line, and finally GPR is applied to give the 68$\%$ one sigma region (for more details, see \cref{sec:post_selection}, \cref{sec:GPR} and \cref{sec:tflo}).}
\label{fig:4x4_percolation_allshots}
\end{figure}

In \cref{sec:stripe}, we presented the time evolution of the $5\times 5$ lattice with an initial state of a holon stripe in a background which is exactly N\'eel in the squeezed space (i.e. when the holon stripe is removed) with post-selection on doublon number and antiferromagnetic-ness. We include a discussion here on the sensitivity of the percolation observable to post-selection of shots and noise. By comparing the results with and without post-selection, we see that the proliferation of background holons confounds the percolation signal as a consequence of this sensitivity. This is demonstrated by comparing Fig.~\ref{fig:4x4_percolation_postselection} vs Fig.~\ref{fig:4x4_percolation_allshots}.

\begin{figure}[!htbp]
\centering
\includegraphics[width=\textwidth]{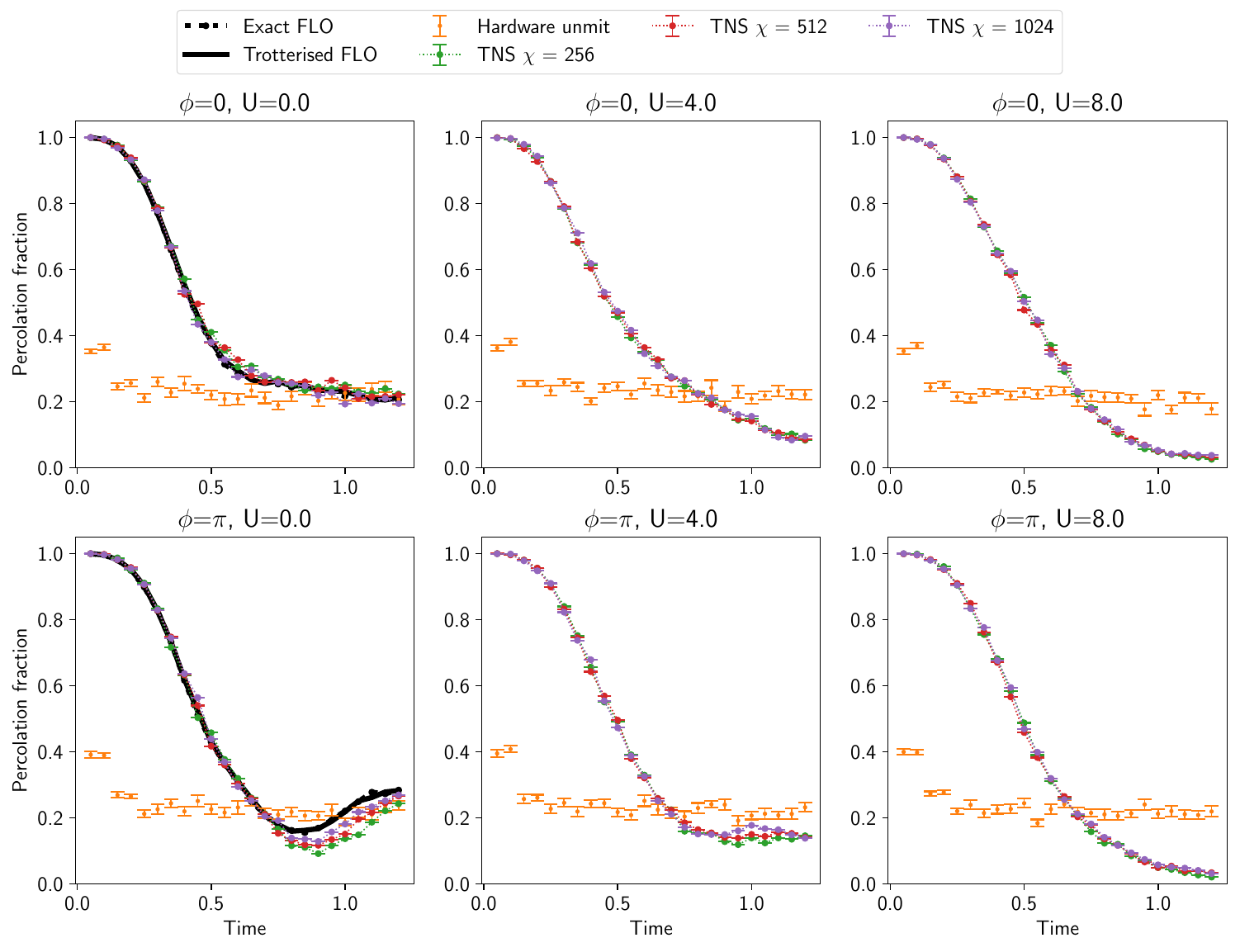}
\caption{\justifying $6\times6$ percolation fraction. The unmitigated hardware data is postselected by antiferromagnetic order measured by AFM $\in [-1,0]$ (see \cref{eq:afm}), number of doublons $\mathcal{D}=[0,1,2,3,4,5,6]$ present and particle number conservation. At this system size we can clearly see a deviation between the $U=0$ free fermion FLO simulation vs TNS, most prominent for $\phi=\pi$ flux experiment, where the TNS converges slower towards exact FLO with increasing $\chi$ compared to smaller system sizes.}
\label{fig:6x6_percolation_postselection}
\end{figure}

For the sake of completeness we include a handful of plots illustrating the same experiment for $4 \times 4$ (Fig. \ref{fig:4x4_percolation_postselection}, Fig. \ref{fig:4x4_percolation_allshots}) and $6 \times 6$ (Fig. \ref{fig:6x6_percolation_postselection}) 
lattices with and without post-selection respectively. Details of how these figures are produced coincide with the details given in \cref{sec:stripe}. We see that as our system size scales, the signal performance degrades rapidly, with TFLO failing for $6\times6$.
In \cref{sec:SPAM_percolation}, we include a discussion of the fragility of the percolation observable to noise by employing a simple measurement error model on the post-selected TDVP samples.






\notocline\subsection{Effects of a simple bitflip noise model on percolation}
\label{sec:SPAM_percolation}
\begin{figure}[!htbp]
\centering
\includegraphics[width=\textwidth]{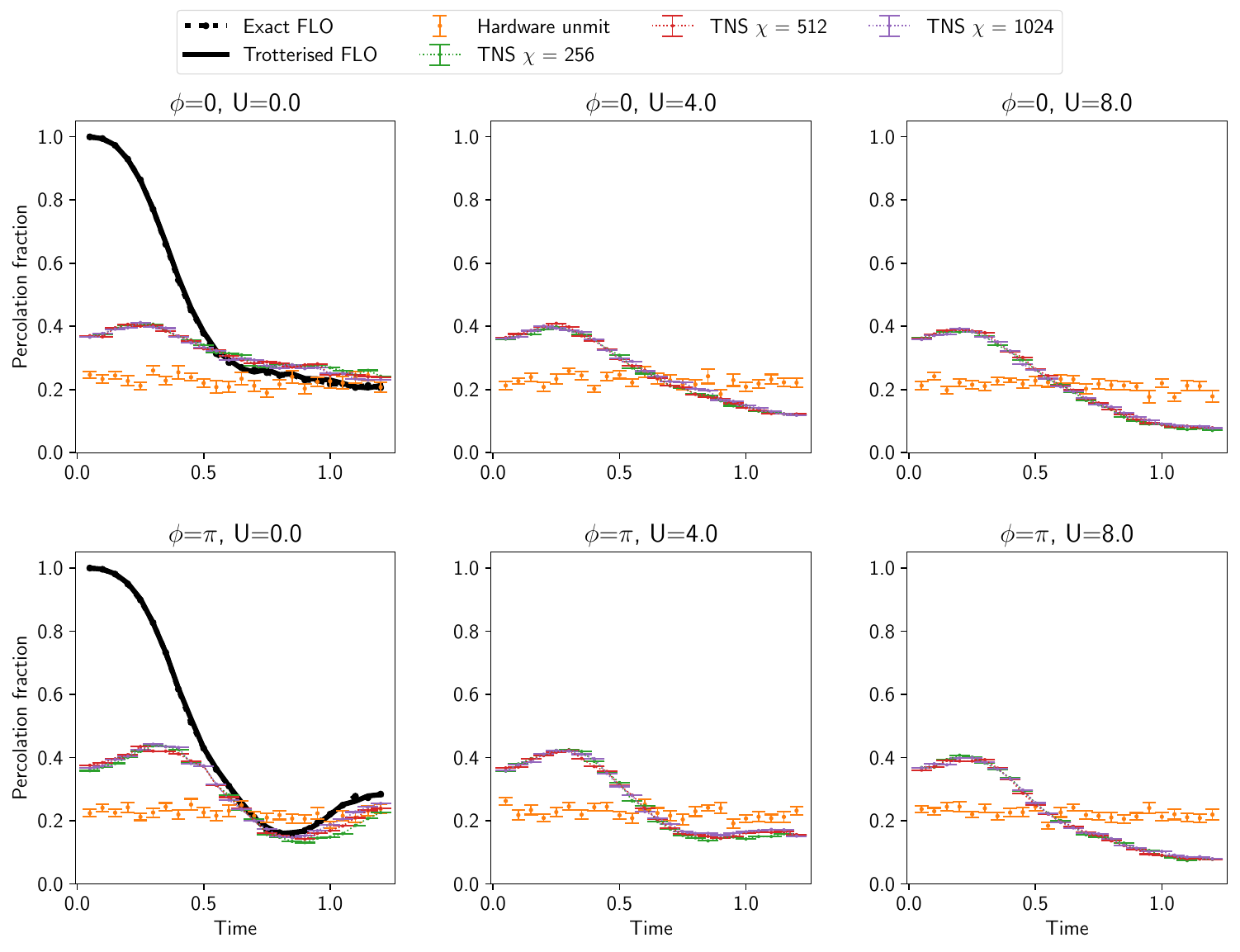}
\caption{\justifying $6 \times 6$ lattice percolation fraction. The TNS results are changed with the application of a bitflip error  model with probability $p=0.1$ to TNS samples. Shots are post-selected to AFM range [-1,0] and doublon numbers $\mathcal{D}=[0,1,2,3,4,5,6]$. }
\label{fig:6x6_TN_bitfliperror_percolation_pt1}
\end{figure}

\begin{figure}[!htbp]
\centering
\includegraphics[width=\textwidth]{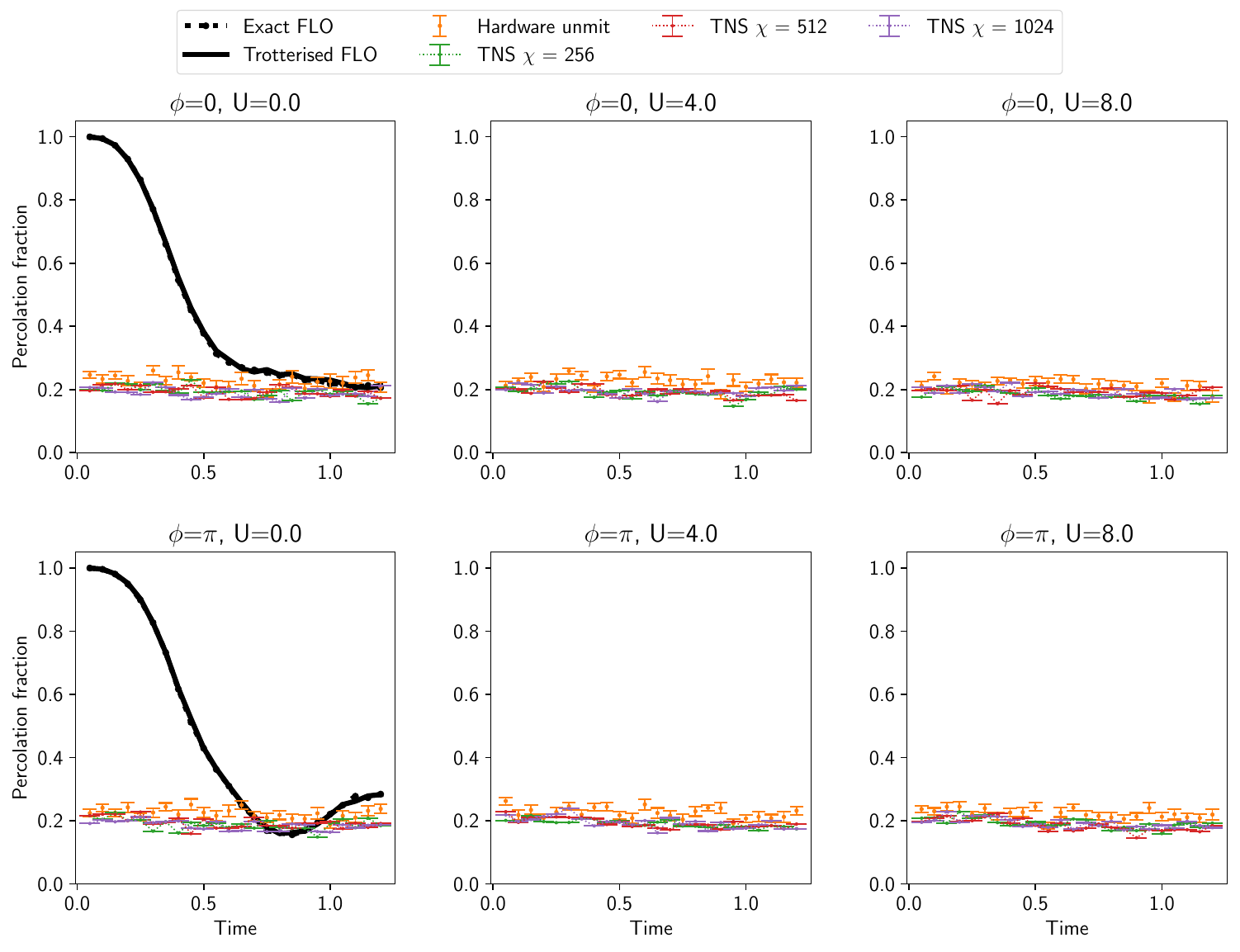}
\caption{\justifying $6 \times 6$ lattice percolation fraction. The TNS results are changed with the application of a bitflip error  model with probability $p=0.5$ to TNS samples. Shots are post-selected to AFM range [-1,0] and doublon numbers $\mathcal{D}=[0,1,2,3,4,5,6]$. }
\label{fig:6x6_TN_bitfliperror_percolation_pt5}
\end{figure}

In the previous sections discussing percolations, the hardware results are visually much flatter than those of TNS and FLO. Here, we demonstrate that a very simple bitflip error model can introduce dramatic flattening of the percolation curve, thereby showing that the non-local percolation observable is very fragile to basic SPAM noise. To implement this, for the TNS computational basis samples, we simply applied a bitflip with probability $p$ and recomputed the percolation fraction. The results are shown in Fig.~\ref{fig:6x6_TN_bitfliperror_percolation_pt1}  and Fig.~\ref{fig:6x6_TN_bitfliperror_percolation_pt5} with $p=0.1$, $p=0.5$ respectively. The results uncover a dramatic change in the TNS results, as the lines flatten to become comparable to those of the unmitigated hardware results.

\clearpage

\section{Illustrative Plots of Performance and Scaling}\label{sec:further_hw_plots}
For completeness, here we collate further data retrieved from the quantum hardware. These plots are meant to illustrate the general behaviour of the simulations at various sizes.
The plots herein all relate to initial states with a central hole and the dynamics with and without a magnetic flux. In all cases, the data refers to experimental data, which has then been mitigated with TFLO and GPR. The plots show a collection of global observables: average doublon density (\cref{eq:doublon_density}); RMS holon (\cref{eq:holon_RMS}); staggered magnetisation (\cref{eq:stag_mag}); and staggered spin-spin (\cref{eq:stag_spin_spin}). We also show the dynamics of spin densities, the correlation between nearest neighbours, and the doublon density deviation from the system mean.

\notocline\subsection{\texorpdfstring{$4\times4$}{4x4}}
\begin{figure*}[!htbp]
    \centering
    \includegraphics[width=\linewidth]{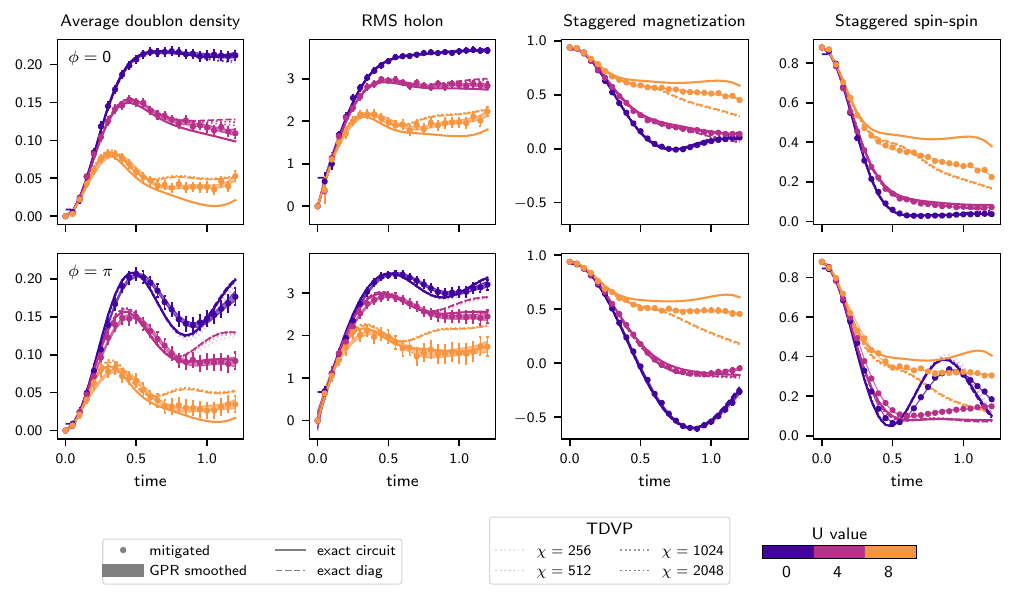}
    \caption{\justifying Evolution of average doublon density, RMS holon, staggered magnetisation and staggered spin-spin correlation for the $4\times4$ lattice with a central hole for $\phi=0$ (top row) and $\phi=\pi$ (bottom row), obtained from TFLO mitigated experimental results using $34$ qubits and $1220$ two-qubit gates. Here we also plot the results obtained using tensor network simulations at different bond dimensions, and exact circuit and TDS evolution when possible $(U=0, 4, 8)$.}
    \label{fig:central_hole_4x4_global_observables_alt}
\end{figure*}

\begin{figure*}[!htbp]
    \centering
    \includegraphics[width=\linewidth]{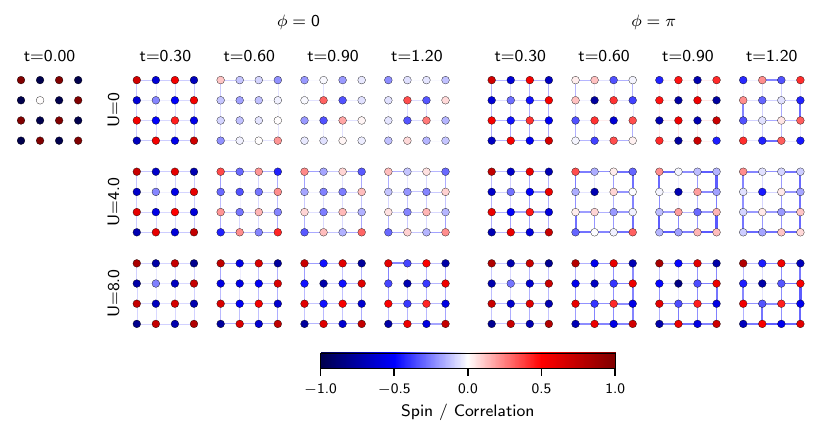}
    \caption{\justifying Spin densities at each site (circles) and correlations between nearest neighbours (links) at selected times for the $4\times 4$ lattice with a central hole for $\phi=0$ (left) and $\phi=\pi$ (right). Observe the enhancement of spin correlations in the Dirac semimetal regime ($\phi=\pi$). }
    \label{fig:central_hole_4x4_spins}
\end{figure*}

\begin{figure*}[!htbp]
    \centering
    \includegraphics[width=\linewidth]{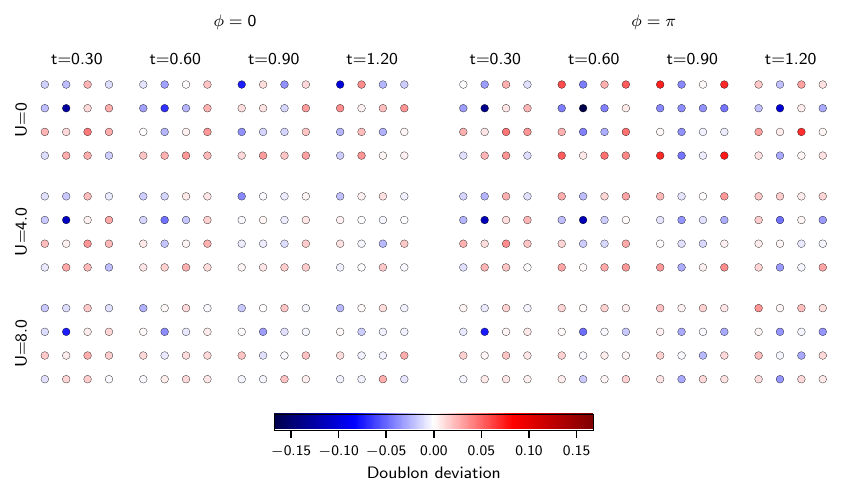}
    \caption{\justifying Doublon density deviation from the mean at selected times for the $4\times4$ lattice with a central hole for $\phi=0$ (left) and $\phi=\pi$ (right).}
    \label{fig:central_hole_4x4_doublon}
\end{figure*}
\newpage
\FloatBarrier

\notocline\subsection{\texorpdfstring{$5\times5$}{5x5}}
\begin{figure*}[!htbp]
    \centering
    \includegraphics[width=\linewidth]{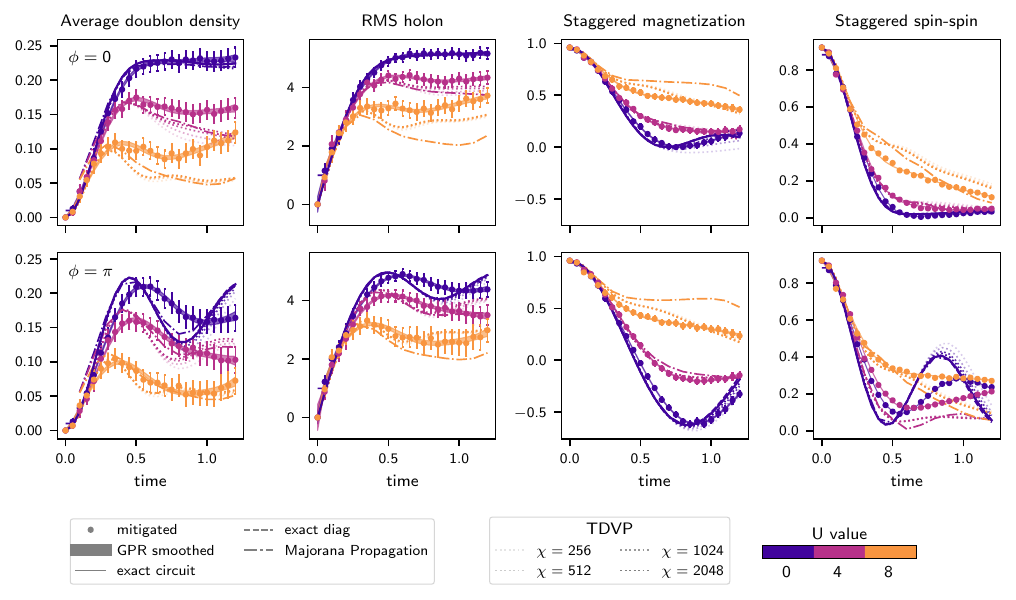}
    \caption{\justifying Evolution of average doublon density, RMS holon, staggered magnetisation and staggered spin-spin correlation for the $5\times5$ lattice with a central hole for $\phi=0$ (top row) and $\phi=\pi$ (bottom row), obtained from TFLO mitigated experimental results using $51$ qubits and $2626$ two-qubit gates. Here we also plot the results obtained using tensor network simulations at different bond dimensions, and exact circuit and TDS evolution when possible $(U=0, 4, 8)$.}
    \label{fig:central_hole_5x5_global_observables_alt}
\end{figure*}

\begin{figure*}[!htbp]
    \centering
    \includegraphics[width=\linewidth]{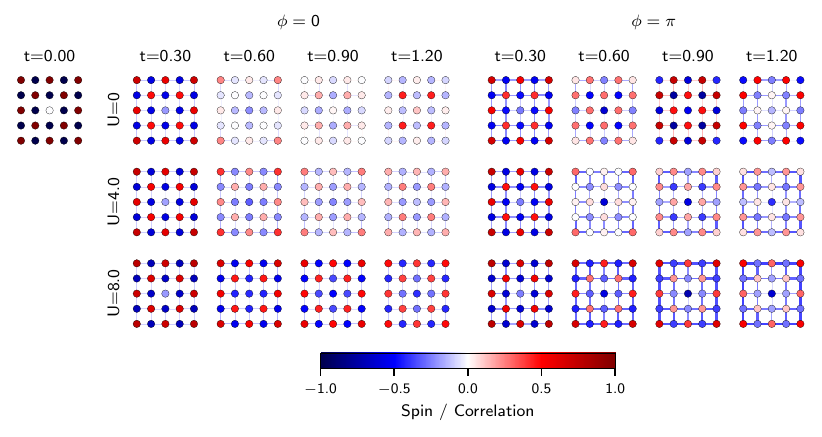}
    \caption{\justifying Spin densities at each site (circles) and correlations between nearest neighbours (links) at selected times for the $5\times 5$ lattice with a central hole for $\phi=0$ (left) and $\phi=\pi$ (right). Observe the enhancement of spin correlations in the Dirac semimetal regime ($\phi=\pi$). }
    \label{fig:central_hole_5x5_spins_temp}
\end{figure*}

\begin{figure*}[!htbp]
    \centering
    \includegraphics[width=\linewidth]{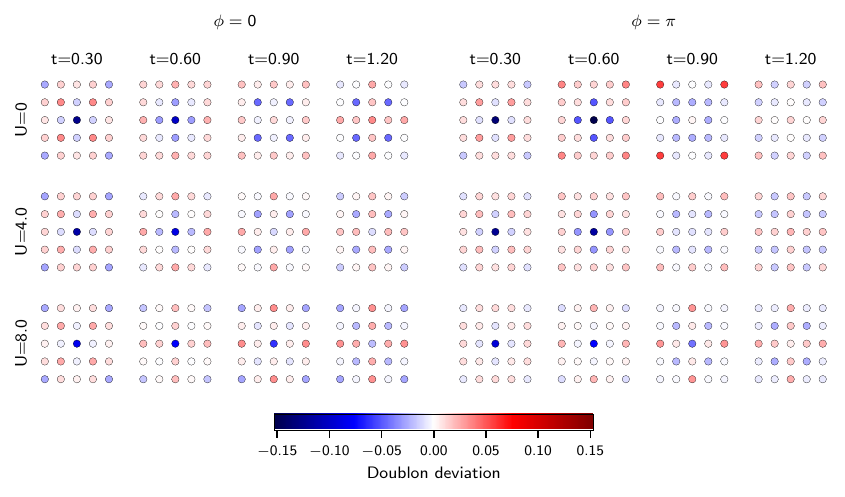}
    \caption{\justifying Doublon density deviation from the mean at selected times for the $5\times5$ lattice with a central hole for $\phi=0$ (left) and $\phi=\pi$ (right).}
    \label{fig:central_hole_5x5_doublon}
\end{figure*}
\newpage
\FloatBarrier

\notocline\subsection{\texorpdfstring{$6\times5$}{6x5}}

\begin{figure*}[!htbp]
    \centering
    \includegraphics[width=\linewidth]{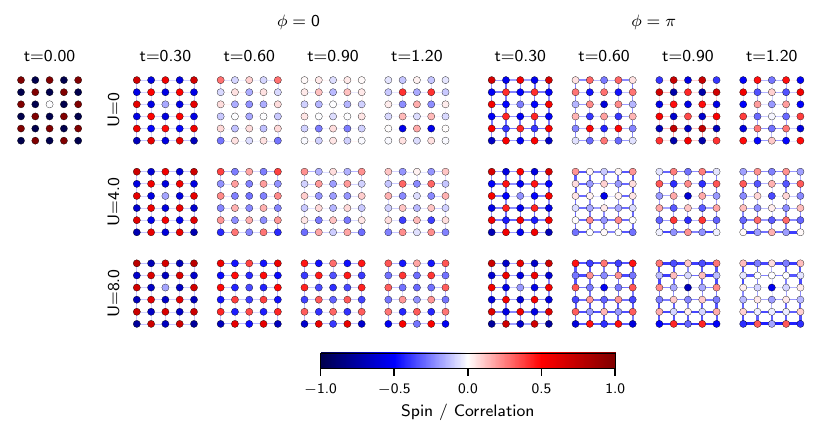}
    \caption{\justifying Spin densities at each site (circles) and correlations between nearest neighbours (links) at selected times for the $6\times 5$ lattice with a central hole for $\phi=0$ (left) and $\phi=\pi$ (right). Observe the enhancement of spin correlations in the Dirac semimetal regime ($\phi=\pi$). }
    \label{fig:central_hole_6x5_spins}
\end{figure*}

\begin{figure*}[!htbp]
    \centering
    \includegraphics[width=\linewidth]{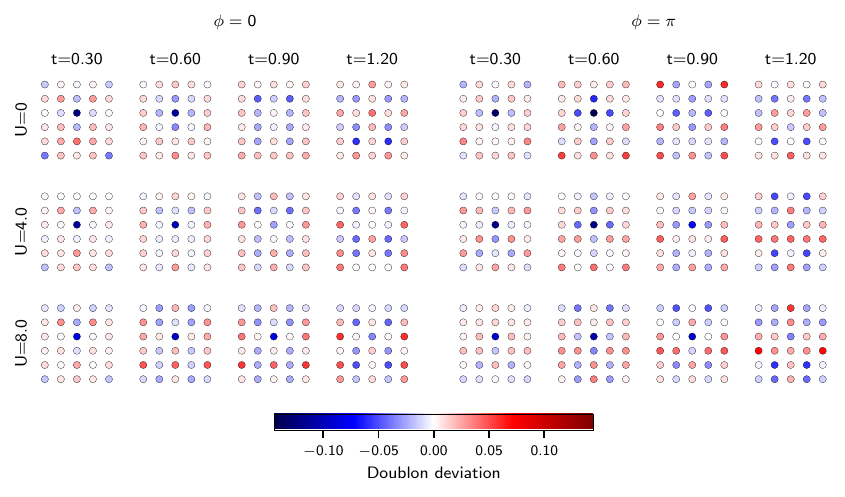}
    \caption{\justifying Doublon density deviation from the mean at selected times for the $6\times5$ lattice with a central hole for $\phi=0$ (left) and $\phi=\pi$ (right).}
    \label{fig:central_hole_6x5_doublon}
\end{figure*}
\newpage
\FloatBarrier

\notocline\subsection{\texorpdfstring{$6\times6$}{6x6}}

\begin{figure*}[!htbp]
    \centering
    \includegraphics[width=\linewidth]{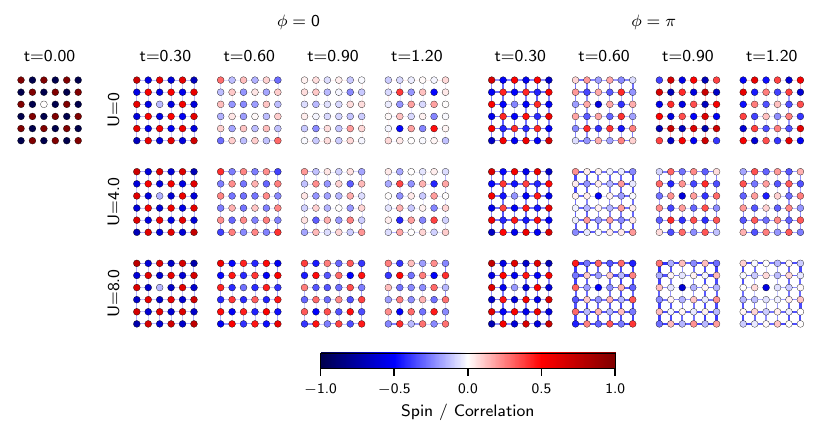}
    \caption{\justifying Spin densities at each site (circles) and correlations between nearest neighbours (links) at selected times for the $6\times 6$ lattice with a central hole for $\phi=0$ (left) and $\phi=\pi$ (right). Observe the enhancement of spin correlations in the Dirac semimetal regime ($\phi=\pi$). }
    \label{fig:central_hole_6x6_spins}
\end{figure*}

\begin{figure*}[!htbp]
    \centering
    \includegraphics[width=\linewidth]{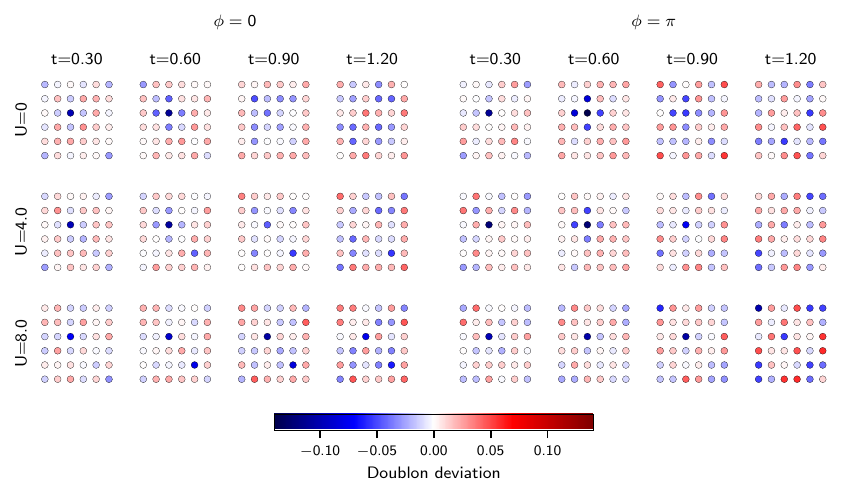}
    \caption{\justifying Doublon density deviation from the mean at selected times for the $6\times6$ lattice with a central hole for $\phi=0$ (left) and $\phi=\pi$ (right).}
    \label{fig:central_hole_6x6_doublon}
\end{figure*}
\FloatBarrier


\clearpage

\bibliographystyle{apsrev4-2}
\bibliography{references}

\end{document}